\documentclass[12pt]{Latex/PhDthesisPSnPDF}

\title{On Aspects of Infinite Derivatives Field Theories \& Infinite Derivative Gravity }

\author{Ali (Ilia) Teimouri \\ MSc by Research in Quantum Fields \& String Theory (Swansea University) }

\collegeordept{Physics \\ Department of Physics}
\university{Lancaster University}

\crest{\includegraphics[width=4cm]{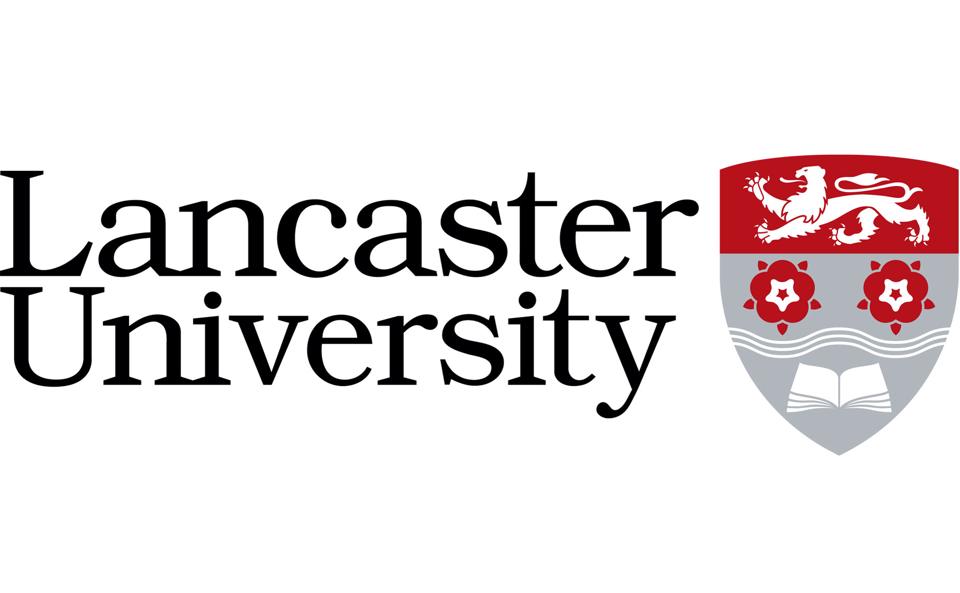}}

\renewcommand{\submittedtext}{A thesis submitted to Lancaster University for the degree of }
\degree{Doctor of Philosophy in the Faculty of Science and Technology}
\degreedate{January 2018}
\newcommand{\bbox}{\bar{\Box}}
\newcommand{\bn}{\bold{n}}

\newcommand{\ba}{\begin{eqnarray}}
\newcommand{\ea}{\end{eqnarray}}
\newcommand{\be}{\begin{equation}}
\newcommand{\ee}{\end{equation}}
\newcommand{\bi}{\begin{itemize}}
\newcommand{\ei}{\end{itemize}}
\newcommand{\al}{\alpha}
\newcommand{\bt}{\beta}

\newcommand{\la}{\lambda}
\newcommand{\ka}{\kappa}

\newcommand{\sa}{\sigma}

\newcommand{\Ga}{\Gamma}

\newcommand{\cF}{{\cal F}}
\newcommand{\cG}{{\cal G}}
\newcommand{\cH}{{\cal H}}

\newcommand{\cL}{{\cal L}}

\newcommand{\cP}{{\cal P}}
\newcommand{\cR}{{\cal R}}

\newcommand{\+}{^{\dagger}}

\newcommand{\p}{\partial}

\newcommand{\n}{\nabla}

\newcommand{\Ra}{\Rightarrow}

\newcommand{\LF}{\left(}
\newcommand{\RF}{\right)}
\newcommand{\LT}{\left[}
\newcommand{\RT}{\right]}

\newcommand{\kb}{\bar{k}}

\newcommand{\non}{\nonumber\\}

\newcommand{\lie}{\pounds}

\usepackage{graphicx}
\usepackage[normalem]{ulem}

\begin{document}


\renewcommand\baselinestretch{1.5}
\baselineskip=18pt plus1pt


\maketitle  








\begin{abstracts}        

Infinite derivative theory of gravity is a modification to the general theory
of relativity. Such modification maintains the massless graviton as the only
true physical degree of freedom and avoids ghosts. Moreover, this class of
modified gravity can address classical singularities.

In this thesis  some essential aspects of an infinite derivative theory of gravity are studied. Namely, we  considered the Hamiltonian formalism, where the true physical degrees of freedom for infinite derivative scalar models and infinite derivative gravity are obtained. Furthermore, the Gibbons-Hawking-York boundary term for the infinite derivative theory of gravity was obtained. Finally, we considered the thermodynamical aspects of the infinite derivative theory of gravity over different backgrounds. Throughout the thesis, our methodology is applied to general relativity, Gauss-Bonnet and $f(R)$ theories of gravity as a check and validation. 
 
\end{abstracts}



\frontmatter

\begin{dedication} 

\textbf{\textit{To my parents: Sousan and Siavash.}}

\end{dedication}



\begin{acknowledgements}      
I am grateful and indebted to my parents; without their
help I could never reach this stage of my life. They are the sole people
whom supported me unconditionally all the way from the beginning. I\ hope
I\ could make them proud.

I\ am thankful to all the people who have guided me through the course of my PhD, my supervisor: Dr Jonathan Gratus
and my departmental collaborators: Prof Roger Jones, Dr Jaroslaw Nowak, Dr John McDonald and Dr David Burton.

I am grateful to my colleagues: Aindri\'u Conroy, James Edholm, Saleh Qutub and Spyridon Talaganis. I am  specially thankful to Aindri\'u Conroy and Spyridon Talaganis, from whom I have learnt a lot. I\ shall thank Mikhail Goykhman for all the fruitful discussions and his valuable feedbacks.        
 
I shall also thank the long list of my friends and specially Sofia who made Lancaster a joyful place for me, despite of the everlasting cold and rainy weather. 

Last but not least, I shall also thank Dr Anupam Mazumdar for all the bitter experience. Nevertheless,
I\ have learnt a lot about ethics and sincerity.

\end{acknowledgements}




\begin{declaration}        

This thesis is my own work and no portion of the work referred to in this thesis has been submitted in support of an application for another degree or qualification at this or any other institute of learning.

\end{declaration}


\begin{quote}

\textit{`I would rather have a short life with width rather than a narrow one with length."}\\
Avicenna

\end{quote}


\setcounter{secnumdepth}{3} 
\setcounter{tocdepth}{3}    
\tableofcontents            


\listoffigures  
\chapter{Relevant Papers by the Author}
\begin{center}\uline{Chapter 3}\end{center}
\begin{itemize}
\item S.~Talaganis and \textbf{A.~Teimouri},
  \textit{``Hamiltonian Analysis for Infinite Derivative Field Theories and Gravity,'}'
  arXiv:1701.01009 [hep-th].
\end{itemize}
\begin{center}\uline{Chapter 4}\end{center}
\begin{itemize}
\item  \textbf{A.~Teimouri}, S.~Talaganis, J.~Edholm and A.~Mazumdar,
  \textit{``Generalised Boundary Terms for Higher Derivative Theories of Gravity,'}'
  JHEP {\bf 1608}, 144 (2016)  
  [arXiv:1606.01911 [gr-qc]].
\end{itemize}
\begin{center}\uline{Chapter 5}\end{center}
\begin{itemize}
\item A.~Conroy, A.~Mazumdar and \textbf{A.~Teimouri},
  \textit{``Wald Entropy for Ghost-Free, Infinite Derivative Theories of Gravity,'}'
  Phys.\ Rev.\ Lett.\  {\bf 114}, no. 20, 201101 (2015), Erratum: [Phys.\ Rev.\ Lett.\  {\bf 120}, no. 3, 039901 (2018)]   
  [arXiv:1503.05568 [hep-th]].
 
 \item A.~Conroy, A.~Mazumdar, S.~Talaganis and \textbf{A.~Teimouri},
  \textit{``Nonlocal gravity in D dimensions: Propagators, entropy, and a bouncing cosmology,''}
  Phys.\ Rev.\ D {\bf 92}, no. 12, 124051 (2015)  
  [arXiv:1509.01247 [hep-th]]. 
  
  \item S.~Talaganis and \textbf{A.~Teimouri},
  \textit{``Rotating black hole and entropy for modified theories of gravity,''}
  arXiv:1705.02965 [gr-qc].
  
  \item \textbf{A.~Teimouri},
  \textit{``Entropy of non-local gravity,''}
  arXiv:1705.11164 [gr-qc].
  
  \item A.~Conroy, T.~Koivisto, A.~Mazumdar and \textbf{A.~Teimouri},
  \textit{``Generalized quadratic curvature, non-local infrared modifications of gravity and Newtonian potentials,'}'
  Class.\ Quant.\ Grav.\  {\bf 32}, no. 1, 015024 (2015)  
  [arXiv:1406.4998 [hep-th]].

\end{itemize}



\mainmatter




\chapter{Introduction}\label{intro}
General theory of relativity (GR), \cite{Einstein:1916vd},  can be regarded as a revolutionary step
towards understanding one of the most controversial topics of theoretical
physics: gravity. The impact of GR is outstanding. Not only does it relate
the geometry of space-time to the existence of the matter in a very startling
way, but it also passed, to this day, all experimental and observational
tests it has undergone. However, like many other theories, GR\ is not perfect \cite{Hawking:1976ra}.
At classical level, it is suffering from black hole and cosmological singularities;
and at quantum level, the theory is not renormalisable and  also not complete
in the ultraviolet (UV) regime. In other words, at short distances (high
energies) the theory blows up. 

It should be noted that non-renormalisability \textit{is not} necessarily an indication that a theory is not UV complete. In fact, non-renormalisability can indicate a breakdown of perturbation theory at the energies of the order of the mass scale of the non-renormalisable operators, with a UV-complete but non-renormalisable theory at higher energies (\textit{e.g.} loop quantum gravity). Renormalisability is  however desirable as it allows the theory to be consistent and calculable at all energies, thus renormalisability may help to formulate of a UV complete theory, but its absence dos not necessarily mean the theory is UV incomplete.

As of today, obtaining a successful theory of quantum gravity \cite{Veltman:1975vx,DeWitt:2007mi,DeWitt:1967yk,DeWitt:1967ub,DeWitt:1967uc} remains an
open problem. At microscopic level, the current standard model (SM) of particle
physics describes the weak, strong and electromagnetic interactions. The
interactions in SM are explained upon quantisation of gauge field theories \cite{mdSchwartz}.
On the other side of the spectrum, at macroscopic level, GR describes the
gravitational interaction based on a classical gauge field theory. Yet, generalisation of 
the gauge field theory to describe gravity at the quantum level is an open problem.  Essentially,
quantising GR\ leads to a non-renormalisable theory \cite{DeWitt:1975ys}. On the other hand also,
the generalisation of the SM, with the current understanding of the gauge
groups, provides no description of gravity. 

Renormalisation plays a crucial role in formulating a consistent
theory of quantum gravity \cite{Wilson:1974mb}. So far, efforts on this direction were not
so successful. Indeed, as per now, quantum gravity is not renormalisable
by power counting. This is to say that, quantum gravity is UV\ divergent. The superficial degree of divergence for a given Feynman diagram can be written as \cite{peskin}, 
\begin{equation}
D=d+\Big[n\bigg(\frac{d-2}{2}\bigg)-d\Big]V-\bigg(\frac{d-2}{2}\bigg)N
\end{equation}
where $d$ is the dimension of space-time, $V$ is the number of vertices, $N$ is the number of external lines in a diagram, and there are $n$ lines meeting at each vertex. The quantity that multiplies $V$ in above expression is just the dimension of the coupling constant (for example for a theory like $\lambda\phi^n$, where $\lambda $ is the coupling constant). There are three rules governing the renormalisability \cite{peskin}: 

\begin{enumerate}
\item When the coupling constant has positive mass dimension, the theory is super-renormalisable.
\item When the coupling constant is dimensionless the theory is renormalisable. \item When the coupling constant has negative mass dimension the theory is non-renormalisable.  
\end{enumerate}
The gravitational coupling, which we know as
the Newton's constant, $G_{N}=M^{-2}_{P}$, is dimensionful (where $M_{P}$
is the Planck mass) with negative mass dimension,  whereas,  the coupling constants of gauge theories,
such as $\alpha$ of quantum electrodynamics (QED) \cite{Lepage:1980fj}, are dimensionless. 

Moreover, in
perturbation theory and in comparison with gauge theory, after each loop
order, the superficial UV\ divergences in quantum gravity becomes worse \cite{tHooft:1974toh,Goroff:1985th,Goroff:1985sz}. Indeed, in each graviton
loop there are two more powers of loop momentum (that is to say that there are two more powers in energy expansion, \textit{i.e.} 1-loop has order $(\partial g)^{4}$, 2-loop has order $(\partial g)^{6}$ and etc.), this is to atone dimensionally
for the two powers of $M_P$ in the denominator of the gravitational coupling.
Instead of the logarithmic divergences of gauge theory, that are renormalisable
via a finite set of counterterms, quantum gravity contains an infinite set
of counterterms. This makes gravity, as given by the Einstein-Hilbert (EH) action,
an effective field theory, useful at scales only much less the the Planck
mass.

Non-local theories may provide a promising path towards quantisation of gravity.
Locality in short means that a particle is only affected by its neighbouring
companion \cite{mdSchwartz,Conroy:2017uds}. Thus, non-locality simply means that a particle's behaviour is
no longer constrained to its close neighbourhood but it also can be affected
by interaction far away. Non-locality can be immediately seen in many approaches
to quantise gravity, among those, string theory (ST) \cite{polch1,polch2,Eliezer:1989cr} and loop quantum gravity (LQG)
\cite{Ashtekar:2007tv,Rovelli:1997yv} are well known. Furthermore, in string field theory (SFT) \cite{Witten:1985cc,Siegel:1988yz}, non-locality presents
itself, for instance in p-adic strings \cite{Freund:1987kt} and zeta strings \cite{Dragovich:2007wb}. Thus, it is reasonable
to ask wether non-locality is essential to describe gravity.     

ST~\footnote{It shall be mention that ST on its own has no problem in quantising gravity, as it is fundamentally a 2-dimensional CFT, which is completely a healthy theory. However, ST\ is known to work well only for small string coupling constant. Thus, ST successfully describes weakly interacting gravitons, but it is less well developed to describe strong gravitational field.} is known to treat the divergences and attempts to provide a  finite
theory of quantum gravity \cite{polch1,polch2}. This is done by introducing a length scale, corresponding
to the string tension, at which particles are no longer point like. ST\,
takes strings as a replacement of particles and count them as the most fundamental
objects in nature. Particles after all are the excitations of the strings.
There have been considerable amount of progress in unifying the fundamental
forces in ST. This was done most successfully for weak, strong and electromagnetic
forces. As for gravity, ST relies on supergravity (SUGRA) \cite{sugrabook}, to treat the
divergences. This is due to the fact  that supersymmetry soothes some of the UV divergences of quantum field theory, via cancellations between bosonic and fermionic loops, hence the UV divergences of quantum gravity become milder in SUGRA. For instance, ST
cures  the two-loop UV divergences; comparing this with the UV\ divergences
of GR at two-loop order shows that ST\ is astonishingly useful. However,
SUGRA and supersymmetric theories in general  have their own shortcomings.
For one thing, SUGRA theories  are not testable experimentally, at very least for the next
few decades. 

An effective theory of gravity, which one derives from ST (or otherwise) permits for higher-derivative terms. Before discussing higher derivative terms in the context of gravity one can start by considering a simpler problem of an effective field theory for a scalar field. 
In the context of ST, one may find an action of the following form, 
\begin{equation}
S=\int d^Dx\Big[\frac{1}{2}\phi K(\Box)\phi-V(\phi)\Big],
\end{equation}
where $K(\Box)$ denotes the kinetic operator and it contains infinite series
of higher derivative terms. The d'Alembertian operator is given by $\Box=g^{\mu\nu}\nabla_{\mu}\nabla_{\nu}$.
Finally, $V(\phi)$ is the interaction term. The choice of $K(\Box)$ depends
on the model one studies, for instance in the p-adic \cite{Freund:1987kt,Freund:1987ck,Brekke:1988dg,Frampton:1988kr} or random lattice \cite{Douglas:1989ve,Brezin:1990rb,Ghoshal:2006te,Gross:1989ni,Biswas:2004qu}, the
form of the kinetic operator is taken to be $K(\Box)=e^{-\Box/M^2}$, where
$M^2$ is the appropriate mass scale proportional to the string tension. The
choice of $K(\Box)$ is indeed very important. For instance for $K(\Box)=e^{-\Box/M^2}$,
which is an entire function \cite{comvar}, one obtains a ghost free propagator. That is
to say that there is no field with negative kinetic energy. In other words,
the choice of an appropriate $K(\Box)$ can prevent introducing extra un-physical
states in the propagator. 

Furthermore, ST \cite{Witten:1985cc,Tong:2009np} serves two types of perturbative corrections to a given background,
namely the string loop corrections and the string world-sheet corrections.
The latter is also known as alpha-prime ($\alpha'$) corrections. In
terminology, $\alpha'$ is inversely proportional to the string tension and
is equal to the sting length squared ($\alpha'=l^{2}_{s}$) and thus we shall
know that it is working as a scale. Schematically the $\alpha'$ correction to a  Lagrangian
is given by, 
\begin{equation}
L=L^{(0)}+\alpha'L^{(1)}+\alpha'^{2}L^{(2)}+\cdots,
\end{equation}
 where $L^{(0)}$ is the leading order Lagrangian and the rest are the sub-leading
corrections. This nature of the ST permits to have corrections to GR. In
other words, it had been suggested that a successful action of quantum gravity
shall contain, in addition to the EH term, corrections that
are functions of the metric tensor with more than two derivatives. The assumption
is that these corrections are needed if one wants to cure non-renormalisability
of the EH action \cite{Tong:2009np}. An example of such corrections  can be schematically
written as, 
\begin{equation}
l^{2}_{s}(a_1 R^{2}+a_2 R_{\mu\nu}R^{\mu\nu}+a_2 R_{\mu\nu\lambda\sigma}R^{\mu\nu\lambda\sigma})+\cdots
\end{equation}  
where $a_i$ are appropriate coefficients.
After all, higher derivative terms in the action above, would have a
minimal influence on the low energy regime and so the classical experiments
remain unaffected. However, in the high energy domain they would dictate
the behaviour of the theory. For instance, such corrections lead to  stabilisation of the
divergence structure and finally the power counting renormalisability. Moreover, higher derivative gravity focuses specifically on studying the problems of consistent higher derivative expansion series of gravitational terms and can be regarded as a possible approach to figure out the full theory of gravity. 

In this thesis, we shall consider infinite derivatives theories \cite{Biswas:2014yia,Biswas:2009nx,Biswas:2010yx,Moffat:1990jj,Biswas:2012ka,Bluhm:1991tq,Lidsey:2007wa,Barnaby:2006hi,Barnaby:2007yb,Joukovskaya:2007nq,Arefeva:2007wvo,Arefeva:2007xdy,Arefeva:2008zru,Calcagni:2007ru,Mulryne:2008iq,Calcagni:2008nm,Vernov:2010ui,Galli:2010qx}. These theories
are a sub-class of non-local theories. In the context of gravity, infinite
derivative theories are constructed by infinite series of higher-derivative
terms. Those terms contain more than two derivatives of the metric tensor.
Infinite derivative theories of gravity (IDG) gained an increasing amount
of attention on recent years as they address the Big Bang singularity problem
\cite{Biswas:2005qr,Biswas:2010zk,Biswas:2012bp,Craps:2014wga,Calcagni:2013vra,Koshelev:2012qn} and they also have other interesting cosmological \cite{Khoury:2006fg,Biswas:2006bs,Dimitrijevic:2013ofa,Briscese:2012ys,Briscese:2013lna,Deser:2007jk,Nojiri:2007uq,Krasnov:2007ky} implementations. Particularly,
as studied in \cite{Biswas:2005qr}, IDG can provide a cosmological non-singular bouncing solution
where the Big Bang is replaced with Big Crunch. Subsequently, further progress
made in \cite{Biswas:2010zk,Biswas:2012bp} to discover inflationary scenarios linked to IDG. Moreover, such
IDG can modify the Raychaudari equations \cite{Conroy:2014dja}, such that one obtains a non-singular
bouncing cosmology without violating the null energy conditions. Additionally,
at microscopic level, one may consider small black holes with mass much smaller
than the Planck mass and observe that IDG prevents singularities in the Newtonian
limit where the gravitational potential is very weak \cite{Biswas:2011ar}. After all, many infinite
derivative theories were proposed in different contexts. \cite{Tomboulis:1980bs,Tomboulis:1983sw,Tomboulis:1997gg,Tomboulis:2015gfa,Modesto:2011kw,Modesto:2014lga,Barvinsky:2012ts,Barvinsky:2014lja,Anselmi:2008ry,Anselmi:2013chz,Moffat:2010bh,Modesto:2010uh,Addazi:2015dxa,Addazi:2015ppa} 

It has been shown by Stelle \cite{Stelle:1976gc,Stelle:1977ry} that, gravitational actions which include terms
quadratic in the curvature tensor are renormalisable. Such action was written
as, 
\begin{equation}
S=\int d^{4}x\sqrt{-g}[\alpha R+\beta R^{2}+\gamma R_{\mu\nu}R^{\mu\nu}],
\end{equation} 
the appropriate choice of coupling constants, $\alpha,\beta$ and $\gamma$
leads to a renormalisable theory. Even though such theory is
renormalisable, yet, it suffers from ghost. It shall be noted that one does
not need to add the Riemann squared term to the action above, as in four dimensions it would be the Gauss-Bonnet theory, 
\begin{equation}
S_{GB}=\int d^{4}x\sqrt{-g}[R^{2}-4R_{\mu\nu}R^{\mu\nu}+R_{\mu\nu\lambda\sigma}R^{\mu\nu\lambda\sigma}],
\end{equation} 
which is an Euler topological invariant and it should be noted that such modification does not add any local dynamics to the graviton (because it is topological). For Stelle's action \cite{Stelle:1976gc,Stelle:1977ry} the GR propagator
is modified schematically as, 
\begin{equation}
\Pi=\Pi_{GR}-\frac{\mathcal{P}^{2}}{k^{2}+m^{2}},
\end{equation}
where it can be seen that there is an extra pole with a negative residue
in the spin-2 sector of the propagator (where $\mathcal{P}$ denotes spin projector operator). This concludes that the theory admits
a massive spin-2 ghost. In literature this is known as the Weyl ghost. We
shall note that the propagator in 4-dimensional GR is given by, 
 \begin{equation}
\Pi_{GR}=\frac{\mathcal{P}^{2}}{k^{2}}-\frac{\mathcal{P}^{0}_{s}}{2k^{2}},
\end{equation}
this shows that even GR\ has a negative residue at the $k^2=0$ pole, and
thus a ghost. Yet this pole merely corresponds to the physical graviton
and so is not harmful. The existence of ghosts is something that one shall
take into account. At classical level they indicate that there is vacuum
instability and at quantum level they indicate that unitarity is violated.

In contrast, there are other theories of modified gravity that may be ghost
free, yet they are not renormalisable. From which, $f(R)$ theories \cite{Sotiriou:2008rp} are the
most well known. The action for $f(R)$ theory is given by, 
\begin{equation}
S=\frac{1}{2}\int d^{4}x\sqrt{-g}f(R),
\end{equation}
where $f(R)$ is the function of Ricci scalar. The most famous sub-class of
$f(R)$ theory is known as Starobinsky model \cite{Starobinsky:1980te} which has implications in primordial
inflation. Starobinsky action is given by, 
\begin{equation}
S=\frac{1}{2}\int d^{4}x\sqrt{-g}(M^{2}_{P}R+c_0R^{2}),
 \end{equation}
where $c_0$ is constant. The corresponding propagator is given by, 
\begin{equation}
\Pi=\Pi_{GR}+\frac{1}{2}\frac{\mathcal{P}^{0}_{s}}{k^{2}+m^{2}},
\end{equation}
where there is an additional propagating degree of freedom in the scalar
sector of the propagator, yet this spin-0 particle is not a ghost and is
non-tachyonic for $m^{2}\geq 0$.

\cite{VanNieuwenhuizen:1973fi} proposed a ghost-free tensor Lagrangian and its application in gravity. After that, progress were made by \cite{Siegel:2003vt,Tseytlin:1995uq,Biswas:2005qr} to construct a ghost-free action
in IDG\ framework. Such attempts were made to mainly address the cosmological
and black hole singularities. Further development were made by \cite{Barvinsky:2012ts,Barvinsky:2014lja,Moffat:2010bh,Modesto:2010uh,Biswas:2011ar,Biswas:2013kla} to obtain
a ghost-free IDG\ . This is to emphasis that ghost-freedom and renormalisability
are important attributes when it comes to constructing a successful theory of
quantum gravity. 
\\\\
\textit{Infinite derivative theory of gravity}
\\\\

Among various modifications of GR, infinite derivative theory of gravity
is    the promising theory in the sense that it is ghost-free,
tachyonic-free and renormalisable, it also addresses the singularity problems.
A covariant, quadratic in curvature, asymptotically free theory of gravity
which is ghost-free and tachyon-free around constant curvature backgrounds
was proposed by \cite{Biswas:2011ar}. 
  
We shall mention that asymptotic freedom means that the coupling constant
decreases as the energy scale increases and vanishes at short distances.
This is for the case that the coupling constant of the theory is small enough
and so the theory can be dealt with perturbatively. As an example, QCD is
an asymptotically free theory \cite{mdSchwartz}. 

Finally, the IDG modification of gravity can be written as \cite{Biswas:2013kla},
\begin{eqnarray}\label{idgxxx}
S&=&S_{EH}+S_{UV}\nonumber\\
&=& \frac{1}{2}\int d^{4}x\sqrt{-g}\Big[M^{2}_{P}R+RF_1(\bar\Box)R+R_{\mu\nu}F_2(\bar\Box)R^{\mu\nu}+R_{\mu\nu\lambda\sigma}F_3(\bar\Box)R^{\mu\nu\lambda\sigma}\Big],\nonumber\\
\end{eqnarray}
where $S_{EH}$ denotes the Einstein-Hilbert action and $S_{UV}$ is the IDG
modification of GR. In above notation, $M_P$ is the Planck mass, $\bar\Box\equiv
\Box/M^2$ and $M$ is the mass scale at which the non-local modifications
become important. The $F_i$'s are functions of the d'Alembertian operator
and given by, 
\begin{equation}
F_i(\bar\Box)=\sum^{\infty}_{n=0}f_{i_{n}}\bar\Box^{n}.
\end{equation}
As we shall see later in section \ref{liniliaz} one can perturb the action around Minkowski background. To do so, one uses the definition of the Riemannian curvatures in linearised regime and thus obtains,
\begin{eqnarray}\label{lin-act-000}
S_{(2)}&=& \frac{1}{32\pi G^{(D)}_{N}} \int d^D x \bigg[\frac{1}{2}h_{\mu
\nu} \Box
a(\bbox) h^{\mu
\nu}+h_{\mu}^{\sigma} b(\bbox) \partial_{\sigma} \partial_{\nu} h^{\mu \nu}
\big.
\big.\nonumber \\&+&h c(\bbox)\partial_{\mu} \partial_{\nu}h^{\mu \nu} +
\frac{1}{2}h
\Box
d(\bbox)h \big.
\big.+ h^{\lambda \sigma} \frac{f(\bbox)}{2\Box}\partial_{\sigma}\partial_{\lambda}\partial_{\mu}\partial_{\nu}h^{\mu
\nu}\bigg], 
\end{eqnarray}  
where \cite{Biswas:2014tua}, 
\begin{equation}
a(\bbox)=1+M^{-2}_{P}({\cal F}_2(\bbox)\Box+4{\cal F}_3(\bbox)\Box),
\end{equation}
\begin{equation}
b(\bbox)=-1-M^{-2}_{P}({\cal F}_2(\bbox)\Box+4{\cal F}_3(\bbox)\Box),
\end{equation}
\begin{equation}
c(\bbox)=1-M^{-2}_{P}(4{\cal F}_1(\bbox)\Box+{\cal F}_2(\bbox)\Box),
\end{equation}
\begin{equation}
d(\bbox)=-1+M^{-2}_{P}(4{\cal F}_1(\bbox)\Box+{\cal F}_2(\bbox)\Box),
\end{equation}
\begin{equation}
f(\bbox)=2M^{-2}_{P}(2{\cal F}_1(\bbox)\Box+{\cal F}_2(\bbox)\Box+2{\cal
F}_3(\bbox)\Box).\end{equation}
The field equations can be  expressed in terms of the inverse propagator as,
\begin{equation}
\Pi^{-1\rho\sigma}_{\mu\nu}h_{\rho\sigma}=\kappa T_{\mu\nu},
\end{equation}
by writing down the spin projector operators in D-dimensional Minkowski space and then moving to momentum space the graviton propagator for IDG\ gravity can be obtained in the form of, 
 \begin{equation}
\Pi^{(D)}(-k^{2})=\frac{\mathcal P^{2}}{k^{2}a(-k^{2})}+\frac{\mathcal P^{0}_{s}}{k^{2}[a(-k^{2})-(D-1)c(-k^{2})]}.
\end{equation}
We note that, $\mathcal P^{2}$ and $\mathcal P^{0}_{s}$ are tensor and scalar
spin projector operators respectively. Since we do not wish to introduce
any extra propagating degrees of freedom apart from the massless graviton,
we are going to take $f(\bbox)=0$ that implies $a(\bbox)=c(\bbox)$. Thus, 
\begin{equation}
\Pi^{(D)}(-k^{2})=\frac{1}{k^2a(-k^2)} \left( \mathcal{P}^{2}-\frac{1}{D-2}\mathcal{P}_{s}^{0}
\right).
\end{equation}
To this end, the form of $a(-k^2)$ should be such that it does not introduce
any new propagating
degree of freedom, 
it was argued in Ref.~\cite{Biswas:2011ar,Biswas:2005qr} that the form of
$a(\Box)$ should be an {\it entire function}, so as not to introduce
any pole in the complex plane, which would result in additional degrees of
freedom in the momentum space. 
In fact, the IDG\ action is ghost-free under the constraint \cite{Biswas:2013kla}, 
\begin{equation}
2F_1(\bar\Box)+F_2(\bar\Box)+2F_3(\bar\Box)=0
\end{equation}
around Minkowski background. In other words, the above constraint ensures that
the massless graviton remains the only propagating degree of freedom and
no extra degrees of freedom are being introduced. More specifically, if one
chooses the graviton propagator to be constructed by an exponential of an
entire function \cite{Biswas:2005qr}~\footnote{The appearance of $a(-k^2)$ in the propagator
definition is the consequence of having infinite derivative modification
in the gravitational action. \cite{Biswas:2005qr} }, 
\begin{equation}\label{holahob}
a(-k^2)=e^{k^2/M^{2}},
\end{equation}
  the propagators becomes (for $D=4$), 
\begin{equation}
\Pi(-k^2)=\frac{1}{k^2e^{k^2/M^{2}}}\Big(\mathcal{P}^{2}-\frac{1}{2}\mathcal{P}^{0}_{s}\Big)=\frac{1}{e^{k^2/M^{2}}}\Pi_{GR}.
\end{equation}
Indeed, the choice of the exponential of an entire function prevents the
production of new poles. For an exponential entire function, the propagator
becomes exponentially suppressed in the UV\ regime while in the infrared
(IR) regime one recovers the physical graviton propagator of GR \cite{Talaganis:2014ida,Conroy:2015nva}. Recovery of GR at IR\ regime takes place as $k^2\rightarrow 0$ that leads to $a(0)=1$ and thus, 
\begin{equation}
\lim_{k^2\rightarrow 0}\Pi(-k^2)=\frac{1}{k^2}\Big(\mathcal{P}^{2}-\frac{1}{2}\mathcal{P}^{0}_{s}\Big),
\end{equation}
which is the GR\ propagator. 
Furthermore,
the IDG action given in (\ref{idgxxx}) can resolve the singularities presented
in GR, at classical level \cite{Biswas:2005qr}, upon choosing the exponential given in (\ref{holahob}) (which represents the infinite derivatives);  (consult appendix \ref{NewtonianPotential}). Such theory is known to also treat the UV\ behaviour,
leading to the convergent of Feynman diagrams \cite{Talaganis:2014ida}. 
\\\\
The IDG\ theory given by the action (\ref{idgxxx}) is motivated and established
fairly recently. Thus, there are many features in the context of IDG that
must be studied. In this thesis we shall consider three important aspects
of infinite derivative theories: The Hamiltonian analysis, the generalised
boundary term  and thermodynamical implications.
\\\\
\\\\
\textit{Hamiltonian formalism}
\\\\

Hamiltonian analysis is a powerful tool when it comes to studying the stabilities
and instabilities of a given theory. It furthermore can be used to calculate
the number of the degrees of freedom for the theory of interest. Stabilities
of a theory can be investigated using the Ostrogradsky's theorem \cite{Woodard:2015zca}.

Let us consider the following Lagrangian density, 
\begin{equation}
\mathcal{L}=\mathcal{L}(q,\dot{q},\ddot{q}),
\end{equation} 
where ``dot" denotes time derivative and so such Lagrangian density is a
function of position, $q$, and its first and second derivatives, in this sense   $\dot{q}$
is  velocity and $\ddot{q}$ is  acceleration. In order to study the classical
motion of the system, the action must be stationary under arbitrary variation
of $\delta q$. Hence,  the condition that must be satisfied are given by
the Euler-Lagrange equations: 
\begin{equation}
\frac{\partial \mathcal{L}}{\partial q}-\frac{d}{dt}\Big(\frac{\partial \mathcal{L}}{\partial
\dot q}\Big)+\frac{d^{2}}{dt^{2}}\Big(\frac{\partial \mathcal{L}}{\partial
\ddot q}\Big)=0.
\end{equation}
The acceleration can be uniquely solved by position and velocity if and only
if $\frac{\partial^2 \mathcal{L}}{\partial
\ddot q^{2}}$ is invertible. In other words, when $\frac{\partial^2 \mathcal{L}}{\partial
\ddot q^{2}}\neq 0$, the theory is called non-degenerate. If $\frac{\partial^2
\mathcal{L}}{\partial
\ddot q^{2}}= 0$, then the acceleration can not be uniquely determined. Indeed,
non-degeneracy of the Lagrangian permits to use the initial data, $q_0,\dot{q}_0,\ddot{q}_0$
and $\dddot{q}_0$, and determine the solutions. Now let us define the following,
\begin{equation}
Q_1=q, \qquad\ P_1=\frac{\partial \mathcal{L}}{\partial
\dot q}-\frac{d^{}}{dt^{}}\Big(\frac{\partial \mathcal{L}}{\partial
\ddot q}\Big),
\end{equation}
\begin{equation}
Q_2=\dot q, \qquad\ P_2=\frac{\partial \mathcal{L}}{\partial
\ddot q},
\end{equation}
where $P_i$'s are the canonical momenta. In this representation the acceleration
can be written in terms of $Q_1$, $Q_2$ and $P_2$ as $\ddot{q}=f(Q_1,Q_2,P_2)$.
The corresponding Hamiltonian density would then take the following form,
\begin{equation}
\mathcal{H}=P_1 Q_1+P_2f(Q_1,Q_2,P_2)-\mathcal{L}(Q_1,Q_2,f),
\end{equation}
for such theory the vacuum decays into both positive and negative energy,
and thus the theory is instable, this is because the Hamiltonian density, $\mathcal{H}$,  is linear in the canonical momentum $ P_1$, \cite{Woodard:2015zca}. Such instabilities are called Ostrogradsky
instability. 

Higher derivative theories are known to suffer from such instability \cite{Eliezer:1989cr}.
From the propagator analysis, the instabilities are due to the presence of ghost in theories
that contain two or more derivatives. In gravity, the
four-derivative gravitational action proposed by Stelle \cite{Stelle:1976gc} is an example where
one encounters Ostrogradsky
instability. 

Ostrogradsky
instability is built upon the fact that for the highest momentum operator,
which is associated with the highest derivative of the theory, the energy
is given linearly, as opposed to quadratic. Yet in the case of IDG, one employs a specific expansion of infinite derivatives, which traces back to the Taylor expansion of $F(\bbox)$, this leads to a graviton propagator that admits no ghost degree of freedom and thus one can overcome the instabilities. We mentioned that, IDG\ theory is ghost-free and
there is no extra degrees of freedom. In this regard, we shall perform the
Hamiltonian analysis for the IDG gravity \cite{Mazumdar:2017kxr} to make sure that the theory is
not suffering from Ostrogradsky
instability. Such analysis is performed in Chapter \ref{dou}.
\clearpage
\textit{Boundary term}  
\\\\

Given an action and a well posed variational principle, it is possible to
associate a boundary term to the corresponding theory \cite{York:1972sj}. In GR, when varying
the EH action, the surface contribution shall vanish if the
action is to be stationary \cite{Gibbons:1976ue}. The surface contribution that comes out of the
variation of the action is constructed by variation of the metric tensor
(\textit{i.e.} $\delta g_{\mu\nu}$) and variation of its derivatives (\textit{i.e.}
$\delta(\partial_{\sigma}g_{\mu\nu})$). However, imposing $\delta g_{\mu\nu}=0$ and
fixing the variation of the derivatives of the metric tensor are not enough
to eliminate the surface contribution. 

To this end, Gibbons, Hawking and York (GHY) \cite{York:1972sj,Gibbons:1976ue} proposed a modification to the EH
action, such that the variation of the modification cancels the term containing
$\delta (\partial_{\sigma}g_{\mu\nu})$ and so imposing $\delta g_{\mu\nu}=0$
would be sufficient to remove the surface contribution. Such modification
is given by, 
\begin{equation}
S=S_{EH}+S_{GHY}\sim\int d^{4}x\sqrt{-g}R+2\oint d^3x\sqrt{h}K,
\end{equation}
where $S_{GHY}$ is the GHY boundary term, $K$ is the trace
of the extrinsic curvature on the boundary and $h$ is the determinant of
the induced metric defined on the boundary. Indeed, $S_{GHY}$ is essential to
make the GR's action as given by the $S_{EH}$ stationary.
It shall
be noted that boundary terms are needed for those space-times that have well defined boundary. As an example, in the case of black holes, GHY term is defined on the horizon of the black hole (where the geometrical boundary of the black hole is located).   

Additionally, $S_{GHY}$ possesses other important features. For instance, in
Hamiltonian formalism, GHY action plays an important role when it comes to
calculating the Arnowitt-Deser-Misner (ADM) energy \cite{Arnowitt:1959ah}. Moreover, in Euclidean
semiclassical approach, the black hole entropy is given entirely by the GHY
term \cite{Hawking:1995fd}. It can be concluded that, given a theory, obtaining a correct boundary
is vital in understanding the physical features of the theory. 

To understand the physics of IDG better, we shall indeed find the boundary
term associated with the theory. Thus, in chapter \ref{chahar} we generalise the GHY
boundary term for the IDG action \cite{Teimouri:2016ulk} given by (\ref{idgxxx}). The exitance of
infinite series of covariant derivative in the IDG theory requires us to
take a more sophisticated approach. To this end, ADM\ formalism and in
particular coframe slicing was utilised. As we shall see later,
our method recovers GR's boundary term when $\Box \rightarrow 0$.
\\\\\\\\
\textit{Thermodynamics}
\\\\

Some of the most physically interesting solutions of GR are black holes. The laws that are governing
the black holes' thermodynamics are known to be analogous to those that are
obtained by the ordinary laws of thermodynamics. So far, only limited family
of black holes are known, they are stationary  asymptotically flat solutions
to Einstein equations. These solutions are given by \cite{gravitationwein},
 
\begin{center}
\begin{tabular}{ c|c|c } 

  & Non-rotating ($J=0$) & Rotating ($J\neq0$)\\ \hline
 Uncharged ($Q=0$) & Schwarzschild & Kerr\\ \hline
 Charged ($Q\neq0$) & Reissner-Nordstr\"{o}m & Kerr-Newman 
 
\end{tabular}
\end{center}

where $J$ denotes the angular momentum and $Q$ is the electric charge. The
reader shall note that a static background is a stationary one, and as a result
a rotating solution is also stationary yet not static. Moreover, electrically
charged black holes are solutions of  Einstein-Maxwell equations and we will
not consider them in this thesis. 

Let us summarise the thermodynamical laws that are governing the black hole
mechanics. The four laws of black hole thermodynamics are put forward by Bardeen, Carter, and Hawking \cite{Bardeen:1973gs}.
They are: 
\begin{enumerate}
\item \textit{Zeroth law}: states that the surface gravity of a stationary
black hole is uniform over the entire event horizon ($H$). \textit{i.e.} 
\begin{equation}
\kappa =const \quad\text{on}\quad H.
\end{equation}
\item \textit{First law}: states that the change in mass ($M$), charge ($Q$),
angular momentum ($J$) and surface area ($A$) are related by: 
\begin{equation}
\frac{\kappa}{8\pi}\delta A=\delta M+\Phi\delta Q-\Omega\delta J
\end{equation}
where we note that $A=A(M,Q,J)$, $\Phi$ is the electrostatic potential
and  $\Omega$ is the angular velocity. 
\item \textit{Second law}: states that the surface area of a black hole can
never decrease, \textit{i.e.}
\begin{equation}
\delta A\geq 0
\end{equation}
given the null energy condition is satisfied. 
\item \textit{Third law}: states that the surface gravity of a black hole
can not be reduced to zero within a finite advanced time, conditioning that
the stress-tensor energy is bounded and satisfies the weak energy condition.
\end{enumerate}
Hawking discovered that the quantum processes lead to a thermal flux of
particles from black holes, concluding that they do indeed behave as thermodynamical
systems \cite{Hawking:1974sw}. To this end, it was found that black holes possesses a well defined
temperature given by, 
\begin{equation}
T=\frac{\hbar\kappa}{2\pi}, 
\end{equation}     
this is known as the Hawking's temperature. Given this and the first law
imply that the entropy of a black hole is proportional to the area of its
horizon and thus the well known formula of \cite{Bekenstein:1973ur}, 
\begin{equation}
S\equiv\frac{A}{4\hbar G_N},
\end{equation}
from the second law we must also conclude that the entropy of an isolated
system can never decrease. It is important to note that Hawking radiation
implies that the black hole area decreases which is the violation of the
second law, yet one must consider the process of black hole evaporation as
a whole. In other words, the total entropy, which is the sum of the radiation
of the black hole entropies, does not decrease. 

So far we reviewed the entropy which corresponds to GR\ as it is described
by the EH action. Deviation from GR\ and moving to higher order
gravity means getting corrections to the entropy. Schematically we can write
(for $f(R)$ and Lovelock entropies \cite{Jacobson:1993xs}),
\begin{equation}
S\sim \frac{A}{4G_N}+\text{higher curvature corrections},
\end{equation}
as such the first law holds true for the modified theories of gravity including
the IDG\ theories. Yet in some cases the second law can be violated by means
of having a decrease in entropy (for instance Lovelock gravity \cite{Jacobson:1993xs}). Indeed, to this day the nature
of these violations are poorly understood. In other words it is not yet established
wether $\delta(S_{BH}+S_{outside})\geq0$ holds true. The higher corrections
of a given theory are needed to understand the second law better. To this
end, we shall obtain the entropy for number of backgrounds and regimes \cite{Conroy:2015wfa,Conroy:2015nva,Talaganis:2017evj,Teimouri:2017xqn} in
chapter \ref{panj} for IDG\ theories. 
\section{Summary of results in literature}

In this part, we shall present series of studies made in the IDG framework,
yet they are not directly the main focus of this thesis.
\\\\
\textit{UV quantum behaviour} 
\\\\

The perturbation around Minkowski background led to obtaining the linearised
action and subsequently the linearised field equations for action (\ref{idgxxx}),
the relevant Bianchi identity was obtained and the corresponding propagator
for the IDG\ action was derived. Inspired by this developments, an infinite
derivative scalar toy model was proposed by \cite{Talaganis:2014ida}. Such action is given by (note that this action can be generalised to include quadratic terms as well, yet the purpose of the study in \cite{Talaganis:2014ida} was to consider a toy model which can be handled technically),
\begin{equation}\label{spyrosaction}
S=\int d^{4}x\bigg[\frac{1}{2}\phi\Box a(\bar\Box)\phi+\frac{1}{4M_P}(\phi\partial_\mu\phi\partial^\mu\phi+\phi\Box
\phi a(\bar\Box)\phi-\phi\partial_\mu
\phi a(\bar\Box)\partial_\mu
\phi)\bigg], 
\end{equation}
for above action, 1-loop and 2-loop computations were performed and it was
found that counter terms can remove the momentum cut-off divergences. Thus,
it was concluded that the corresponding Feynman integrals are convergent.
It has been also shown by \cite{Talaganis:2014ida} that, at 2-loops the theory is UV\ finite. Furthermore,
a method was suggested for rendering arbitrary n-loops to be finite. Also consult \cite{Biswas:2014tua,Talaganis:2017tnr}. It should be noted that (\ref{spyrosaction}) is a toy model with cubic interactions and considered in \cite{Talaganis:2014ida} due to its simplicity, however it is possible to consider quadratic interactions too and thus generalise the action.  
\\\\
\textit{Scattering amplitudes}
\\\\

One of the most interesting aspect of each theory in the view of high energy
particle physics is  studying the behaviour of the cross sections corresponding
to the scattering processes \cite{Elvang:2013cua}. A theory can not be physical if the cross section
despairs at high energies. This is normally the case for theories with more than
two derivatives. However, it has been shown by \cite{Talaganis:2016ovm} that, infinite derivative
scalar field theories can avoid this problem. This has been done by dressing
propagators and vertices where the external divergences were eliminated when
calculating the scattering matrix element. This is to say that, the cross
sections within the infinite derivative framework remain finite.
\\\\
\textit{Field equations}
\\\\

In \cite{Biswas:2013cha}, the IDG\ action given in (\ref{idgxxx}) was considered. The full non-linear
field equations were obtained using the variation principle. The corresponding
Bianchi identities were verified and finally the linearised field equations
were calculated around Minkowski background. In similar fashion \cite{Conroy:2017uds} obtained
the linearised field equations around the de-Sitter (dS) background.
\\\\
\textit{Newtonian potential}
\\\\

Authors of \cite{Biswas:2011ar} studied the Newtonian potential corresponding to the IDG action, given
in (\ref{idgxxx}), in weak field regime. In linearised field equations taking
$a(\bar\Box)=e^{-\bar\Box}$ leads to the following Newtonian potential (See Appendix \ref{NewtonianPotential} for derivation), 
\begin{equation}
\Phi(r)=-\frac{\kappa m_g\text{Erf}(\frac{Mr}{2})}{8\pi r},
\end{equation}
where $m_g$ is the mass of the object which generates the gravitational potential
and $\kappa=8\pi G_{N}$. In the limit where $r\rightarrow\infty$ one recovers
the Minkowski space-time. In contrast, when $r\rightarrow0$, the Newtonian
potential becomes constant. This is where IDG\ deviates from GR\ for good,
in other words, at short distances the singularity of the $1/r$ potential
is replaced with a finite constant.  

Similar progress was made by \cite{Edholm:2016hbt}. In the context of IDG\ the Newtonian
potential was studied for a more generalised choice of entire function, \textit{i.e.}
$a(\bar\Box)=e^{\gamma(\bar\Box)}$, where $\gamma$ is an entire function.
It was shown that at large distances the Newtonian potential goes as $1/r$
and thus in agreement with GR, while at short distances the potential is non-singular.  

Later on, \cite{Conroy:2017nkc} studied the Newtonian potential for a wider class of IDG. Such
potentials were found to be oscillating and non-singular, a seemingly feature
of IDG. \cite{Conroy:2017nkc} showed that for an IDG\  theory constrained to allow defocusing
of null rays and thus the geodesics completeness, the Newtonian potential
can be made non-singular and be in agreement with GR\ at large distances.

\cite{Conroy:2017nkc} concluded that, in the context of higher derivative theory of gravity,
null congruences can be made complete, or can be made defocused upon satisfaction
of two criteria at microscopic level: first, the graviton propagator shall
have a scalar mode, comes with one additional root, besides the massless
spin-2 and secondly, the IDG gravity must be, at least, ghost-free or tachyon-free.
\clearpage
\textit{Singularities}
\\\\

GR allows space-time singularity, in other words, null geodesic congruences
focus in the presence of matter. \cite{Conroy:2016sac} discussed the  singularity freedom in
the context of IDG\ theory. To this end, the Raychaudari equation corresponding
to the IDG\ was obtained and the bouncing cosmology scenarios were studied. The latest progress in this
direction outlined the requirements for defocusing condition for null congruences
around dS and Minkowski backgrounds. 
\\\\\\\\\\
\textit{Infrared modifications} 
\\\\

\cite{Conroy:2014eja} considered an IDG\ action where the non-local modifications are accounted
in the IR\ regime. The infinite derivative action considered in \cite{Conroy:2014eja} contains an infinite
power series of inverse d'Alembertian operators. As such they are given by,
\begin{equation}
G_i(\bar\Box)=\sum^{\infty}_{n=1}c_{i_{n}}\bar\Box^{-n}.
\end{equation}
The full non-linear field equations for this action was obtained and the
corresponding Bianchi identities were presented. The form of the Newtonian
potential in this type of gravity was calculated. Some of the cosmological
of implications, such as dark energy,  of this theory were also studied \cite{Maggiore:2014sia}. 
\section{Organisation of thesis}
The content of this thesis is organised as follows: 

\begin{description}
\item[Chapter 2:] In this chapter the infinite derivative theory of gravity (IDG) is introduced and derived. This serves as a brief review on the derivation of the theory which would be the focus point of this thesis. 
\item[Chapter 3:] Hamiltonian analysis for an infinite derivative gravitational
action, which is constructed by Ricci scalar and covariant derivatives, is performed. First, the relevant Hamiltonian constraints (\textit{i.e.} primary/secondary and first class/ second class) are defined and a formula for calculating the number of degrees of freedom is proposed. Then, we applied the analysis to number of theories. For instance, a scalar field model and the well known $f(R)$ theory. In the case of gravity we employed ADM\ formalism and applied the regular Hamiltonian analysis to identify the constraints and finally to calculate the number of degrees of freedom.
\item[Chapter 4:] In this chapter the generalised GHY boundary term for the infinite derivative theory of gravity is obtained. First, the ADM\ formalism is reviewed and the coframe slicing is introduced. Next, the infinite derivative action is written in terms of auxiliary fields. After that, a generalised formula for obtaining the GHY boundary term is introduced. Finally, we employ the generalised GHY formulation to the infinite derivative theory of gravity and obtain the boundary term.
\item[Chapter 5:] In this chapter thermodynamical aspects of the infinite derivative theory of gravity are studied. We shall begin by reviewing the Wald's prescription on entropy calculation. Then, Wald's approach is used to obtain the entropy for  IDG theory over a generic spherically symmetric background. Such entropy is then analysed in the weak field regime. Furthermore, the entropy of IDG action obtained over the $(A)dS$ background. As a check we used an approximation to recover the entropy of the well known Gauss-Bonnet theory from the $(A)dS$ background. We then study the entropy over a rotating background. This had been done by generalising  the Komar integrals, for theories containing Ricci scalar, Ricci tensor and their derivatives. Finally, we shall obtain the entropy of a higher derivative gravitational theory where the action contains inverse d'Alembertian operators (\textit{i.e.} non-locality).

\item[Conclusion:] In the final part of this thesis, we summarise the results of our study and discuss the findings. Furthermore, the future work is discussed in this section.  

\item[Appendices:] We start by giving the notations, conventions and useful formulas relevant to this thesis. Furthermore, the detailed computations, relevant to each chapter, were presented so the reader can easily follow them.
  
\end{description}

\chapter{Overview of infinite derivative gravity}\label{yek}
In this chapter we shall summarise the derivation of the infinite derivative
gravitational (IDG) action around flat background. In following chapters we study different aspects
of this gravitational action. 
\section{Derivation of the IDG action}\label{ch2}
The most general, quadratic in curvature, and generally covariant gravitational
action in four dimensions \cite{Biswas:2013kla} can be written as, 
\begin{eqnarray}
S&=&S_{EH}+S_{UV},\\
S_{EH}&=&\frac{1}{2}\int d^{4}x\sqrt{-g} M^{2}_{P}R,\\
S_{UV}&=& \frac{1}{2}\int d^{4}x\sqrt{-g}\Big(R_{\mu_1\nu_1\lambda_1\sigma_1}\mathcal{O}^{\mu_1\nu_1\lambda_1\sigma_1}_{\mu_2\nu_2\lambda_2\sigma_2}R^{\mu_2\nu_2\lambda_2\sigma_2}\Big),\label{uvaction}
\end{eqnarray}
where $S_{EH}$ is the Einstein-Hilbert action and $S_{UV}$ denotes the higher
derivative modification of the GR in ultraviolet sector. The operator $\mathcal{O}^{\mu_1\nu_1\lambda_1\sigma_1}_{\mu_2\nu_2\lambda_2\sigma_2}$
retains general covariance. 

Expanding (\ref{uvaction}), the total action becomes, 
\begin{eqnarray}\label{actionexpansion}
S&=&\frac{1}{2}\int dx^{4}\sqrt{-g}\Big[M^{2}_{P}R+RF_1(\Box)R+RF_2(\Box)\nabla_\nu\nabla_\mu
R^{\mu\nu}+R_{\mu\nu} F_3(\Box)R^{\mu\nu}\nonumber\\
&+&R^{\nu}_{\ \mu}F_4(\Box)\nabla_\nu\nabla_\lambda R^{\mu\lambda}+R^{\lambda\sigma}F_5(\Box)\nabla_\mu\nabla_\sigma
\nabla_\nu\nabla_\lambda R^{\mu\nu}+RF_6(\Box)\nabla_\mu\nabla_\nu\nabla_\lambda
\nabla_\sigma
R^{\mu\nu\lambda\sigma
}\nonumber\\
&+&R_{\mu\lambda}F_7(\Box)\nabla_\nu\nabla_\sigma R^{\mu\nu\lambda\sigma}+R^{\rho}_{\
\lambda}F_8(\Box)\nabla_\mu\nabla_\sigma\nabla_\nu\nabla_\rho R^{\mu\nu\lambda\sigma}\nonumber\\
&+&R^{\mu_1\nu_1}F_9(\Box)\nabla_{\mu_1}\nabla_{\nu_1}\nabla_\mu\nabla_\nu\nabla_\lambda\nabla_\sigma
R^{\mu\nu\lambda\sigma}+R_{\mu\nu\lambda\sigma}F_{10}(\Box)R^{\mu\nu\lambda\sigma}\nonumber\\
&+&R^{\rho}_{\ \mu\nu\lambda}F_{11}(\Box)\nabla_\rho\nabla_\sigma R^{\mu\nu\lambda\sigma}+R_{\mu\rho_1\nu\sigma_1}F_{12}(\Box)\nabla^{\rho_1}\nabla^{\sigma_1}\nabla_\rho\nabla_\sigma
R^{\mu\rho\nu\sigma}\nonumber\\
&+&R^{\nu_1\rho_1\sigma_1}_{\mu}F_{13}(\Box)\nabla_{\rho_1} \nabla_{\sigma_1}\nabla_{\nu_1}\nabla_\nu\nabla_\lambda\nabla_\sigma
R^{\mu\nu\lambda\sigma}\nonumber\\
&+&R^{\mu_1\nu_1\rho_1\sigma_1}F_{14}(\Box)\nabla_{\rho_1} \nabla_{\sigma_1}\nabla_{\nu_1}\nabla_{\mu_1}\nabla_{\mu}\nabla_\nu\nabla_\lambda\nabla_\sigma
R^{\mu\nu\lambda\sigma}\Big],
\end{eqnarray} 
it shall be noted that we performed integration by parts where it was appropriate. Also, $F_i$'s are analytical functions of d'Alembertian operator ($\Box=g^{\mu\nu}\nabla_{\mu}\nabla_\nu$).
Around Minkowski background the operator would be simplified to: $\Box=\eta^{\mu\nu}\partial_{\mu}\partial_\nu$.
The functions $F_i$'s are given explicitly by, 
\begin{equation}\label{analyticalfunction}
F_i(\bar\Box)=\sum^{\infty}_{n=0}f_{i_n}\bar\Box^n,
\end{equation}
where $\bar\Box\equiv\Box/M^{2}$. In this definition, $M$ is the mass-scale
at which the non-local modifications become important at UV scale. Additionally, $f_{i_n}$
are the appropriate coefficients of the sum in (\ref{analyticalfunction}).

Making use of the antisymmetric properties of the Riemann tensor, 
\begin{equation}
R_{(\mu\nu)\rho\sigma}=R_{\mu\nu(\rho\sigma)}=0,
\end{equation}
and the Bianchi identity, 
\begin{equation}\label{antisymriemann}
\nabla_\alpha R^{\mu}_{\ \nu\beta\gamma}+\nabla_\beta R^{\mu}_{\ \nu\gamma\alpha}+\nabla_\gamma
R^{\mu}_{\ \nu\alpha\beta}=0,
\end{equation}
the action given in (\ref{actionexpansion}), reduces to,
\begin{eqnarray}\label{simaction}
S&=&\frac{1}{2}\int dx^{4}\sqrt{-g}\Big[M^{2}_{P}R+RF_1(\Box)R+R_{\mu\nu}
F_3(\Box)R^{\mu\nu}
+RF_6(\Box)\nabla_\mu\nabla_\nu\nabla_\lambda
\nabla_\sigma
R^{\mu\nu\lambda\sigma
}\nonumber\\
&+&R_{\mu\nu\lambda\sigma}F_{10}(\Box)R^{\mu\nu\lambda\sigma}
+R^{\nu_1\rho_1\sigma_1}_{\mu}F_{13}(\Box)\nabla_{\rho_1} \nabla_{\sigma_1}\nabla_{\nu_1}\nabla_\nu\nabla_\lambda\nabla_\sigma
R^{\mu\nu\lambda\sigma}\nonumber\\
&+&R^{\mu_1\nu_1\rho_1\sigma_1}F_{14}(\Box)\nabla_{\rho_1} \nabla_{\sigma_1}\nabla_{\nu_1}\nabla_{\mu_1}\nabla_{\mu}\nabla_\nu\nabla_\lambda\nabla_\sigma
R^{\mu\nu\lambda\sigma}\Big].
\end{eqnarray} 
Due to the perturbation around Minkowski background, the covariant derivatives
become partial derivatives and can commute around freely. As an example,
(see Appendix \ref{simplificationexampleidg})\begin{eqnarray}
RF_6(\Box)\nabla_\mu\nabla_\nu\nabla_\lambda
\nabla_\sigma
R^{\mu\nu\lambda\sigma
}&=&\frac{1}{2}RF_6(\Box)\nabla_\mu\nabla_\nu\nabla_\lambda
\nabla_\sigma
R^{\mu\nu\lambda\sigma
}\nonumber\\&+&\frac{1}{2}RF_6(\Box)\nabla_\mu\nabla_\nu\nabla_\lambda
\nabla_\sigma
R^{\mu\nu\lambda\sigma
}.
\end{eqnarray}
By commuting the covariant derivatives we get, 
\begin{eqnarray}
RF_6(\Box)\nabla_\mu\nabla_\nu\nabla_\lambda
\nabla_\sigma
R^{\mu\nu\lambda\sigma
}&=&\frac{1}{2}RF_6(\Box)\nabla_\nu\nabla_\mu\nabla_\lambda
\nabla_\sigma
R^{\mu\nu\lambda\sigma
}\nonumber\\&+&\frac{1}{2}RF_6(\Box)\nabla_\mu\nabla_\nu\nabla_\lambda
\nabla_\sigma
R^{\mu\nu\lambda\sigma
}.
\end{eqnarray}
Finally, it is possible to relabel the indices and obtain, 
\begin{equation}
RF_6(\Box)\nabla_\mu\nabla_\nu\nabla_\lambda
\nabla_\sigma
R^{\mu\nu\lambda\sigma
}=RF_6(\Box)\nabla_\nu\nabla_\mu\nabla_\lambda
\nabla_\sigma
R^{(\mu\nu)\lambda\sigma
}=0,
\end{equation}
which vanishes due to antisymmetric properties of the Riemann tensor as mentioned
in (\ref{antisymriemann}).

After all the relevant simplifications, we can write the IDG\ action as,
\begin{eqnarray}\label{mainaction}
S= \frac{1}{2}\int d^{4}x\sqrt{-g}\Big(M^{2}_{P}R+RF_1(\bbox)R+R_{\mu\nu}
F_2(\bbox)R^{\mu\nu}+R_{\mu\nu\lambda\sigma}F_{3}(\bbox)R^{\mu\nu\lambda\sigma}
\Big).\nonumber\\
\end{eqnarray}
 This is an infinite derivative modification to the GR. 
\chapter{Hamiltonian analysis}\label{dou} 
 In this chapter, we shall perform a Hamiltonian analysis on the IDG action
given in (\ref{mainaction}). Due to the technical complexity, the analysis
are being performed on a simpler version of this action by dropping the $R_{\mu\nu}
F_2(\bbox)R^{\mu\nu}$ and $R_{\mu\nu\lambda\sigma}F_{3}(\bbox)R^{\mu\nu\lambda\sigma}$
terms. In our analysis, we obtain the true dynamical degrees of freedom.
We shall note that not including $R_{\mu\nu}
F_2(\bbox)R^{\mu\nu}$ and $R_{\mu\nu\lambda\sigma}F_{3}(\bbox)R^{\mu\nu\lambda\sigma}$
to the IDG action does not change the dynamics of the theory we are considering and thus the degrees of freedom would not be changed. Consult \cite{Talaganis:2014ida} for propagator analysis and the degrees of freedom. In fact, $R_{\mu\nu}
F_2(\bbox)R^{\mu\nu}$ and $R_{\mu\nu\lambda\sigma}F_{3}(\bbox)R^{\mu\nu\lambda\sigma}$ terms exist as a matter of generality.  We will proceed, by first shortly reviewing the Hamiltonian analysis, provide
the definitions for {\it primary, secondary, first-class} and {\it second-class
constraints} \cite{Anderson:1951ta,Dirac:1958sq,Dirac1,Dirac:1958sc,Wipf:1993xg}  and write down the formula for counting the number of degrees
of freedom. We then provide some scalar toy models as examples and show how
to obtain the degrees of freedom in those models.  After setting up the preliminaries
and working out the toy examples, we turn our attention to the IDG action
and perform the analysis, finding the constraints and finally the number
of degrees of freedom. 

Hamiltonian analysis can be used as a powerful tool to investigate the stability
and boundedness of a given theory. It is well known that, higher derivative
theories, those that contain more than two derivatives, suffer from Ostrogradsky's
instability \cite{Woodard:2015zca}. Having infinite number of covariant derivatives in the IDG\
action however makes the Ostrogradsky's analysis redundant, as one can employ a specific expansion which traces back to the Taylor expansion of $F(\bbox)$ that leads to a ghost free graviton propagator. 

In the late
1950s, the $3+1$ decomposition became appealing; Richard Arnowitt, Stanley
Deser and Charles W.
Misner (ADM) \cite{Arnowitt:1962hi,Gourgoulhon:2007ue} have shown that it is possible
to decompose four-dimensional space-time
such that one foliates the arbitrary region $\mathcal{M}$ of the space-time
manifold with a family of spacelike hypersurfaces 
$\Sigma_{t}$, one for each instant in time.
In this chapter, we shall show how by using the ADM decomposition, and finding
the relevant constraints, one can obtain the number of degrees of freedom.
It will be also shown, that how the IDG action can admit finite/infinite
number of the degrees of freedom.
\section{Preliminaries}
Suppose
we have an action that depends on time evolution. We can write down the equations
of motion by imposing the stationary conditions on the action and then use
variational method.
Consider the following action, 
\begin{equation}
I=\int \mathcal{L} (q,\dot{q})dt\,,
\end{equation}
the above action is expressed as a time integral and $\mathcal{L}$ is the
Lagrangian density depending on the position $q$ and the velocity $\dot{q}$.
The variation of the action leads to  the equations of motion known as Euler-Lagrange
equation,
\begin{equation}
\frac{d}{dt}\bigg(\frac{\partial\mathcal{L}}{\partial\dot{q}}\bigg)-\frac{\partial\mathcal{L}}{\partial
q}=0\,,
\end{equation}
we can expand the above expression, and write, 
\begin{equation}
\ddot{q}\frac{\partial^{2}\mathcal{L}}{\partial \dot{q}\partial \dot{q}}=\frac{\partial\mathcal{L}}{\partial
q}-\dot{q}\frac{\partial^{2}\mathcal{L}}{\partial q\partial \dot{q}}\,,
\end{equation}
the above equation yields an acceleration, $\ddot{q}$, which can be uniquely
calculated by position and velocity at a given time,
if and only if $\frac{\partial^{2}\mathcal{L}}{\partial \dot{q}\partial \dot{q}}$
is invertible. In other words, if the determinant  of the matrix $\frac{\partial^{2}\mathcal{L}}{\partial\dot{q}\partial
\dot{q}}\neq 0$, \textit{i.e.} non vanishing, then the theory is called {\it non-degenerate}.
If the determinant is zero, then the acceleration can \textit{not}
be uniquely determined by position and the velocity. The latter system is
called {\it singular} and leads to {\it constraints} in the phase space \cite{Wipf:1993xg,Henneaux}.
\subsection{Constraints for a singular system}  

In order to formulate the Hamiltonian we need to first define the canonical
momenta, 
\begin{equation}
p=\frac{\partial\mathcal{L}}{\partial\dot{q}}\,.
\end{equation}
The non-invertible matrix $\frac{\partial^{2}\mathcal{L}}{\partial
\dot{q}\partial \dot{q}}$ indicates that not all the velocities can be written
in terms of the canonical momenta, in other words, not all the momenta are
independent, and there are some relation between the canonical coordinates \cite{Anderson:1951ta,Dirac:1958sq,Dirac1,Dirac:1958sc,Wipf:1993xg},
 such
as,
\begin{equation}
\varphi(q,~p)=0\, \Longleftrightarrow {\it primary~constraints}\,,
\end{equation}
known as \textit{primary constraints}. Take $\varphi(q,~p)$
for instance, if we have {\it vanishing canonical momenta}, then we have
{\it primary constraints}. The \textit{primary constraints} hold without
using the equations of motion. The {\it primary constraints} define a submanifold
smoothly embedded in a phase space, which is also known as the \textit{primary
constraint surface}, 
$\Ga_p$. We can now define the Hamiltonian density as, 
\begin{equation}\label{genham}
\mathcal{H}=p\dot{q}-\mathcal{L}\,.
\end{equation}
If the theory admits {\it primary constraints}, we will have to redefine
the Hamiltonian density,  and write the \textit{total} Hamiltonian density
as, 
\begin{equation}\label{totalH}
\mathcal{H}_{tot}=\mathcal{H}+\lambda^{a}(q,p)\varphi_{a} (q,p)\,,
\end{equation}
where now $\lambda^{a}(q,p)$ is called the {\it Lagrange multiplier}, and
$\varphi_{a} (q,p)$ are linear combinations of the primary constraints~\footnote{We
should point out that the total Hamiltonian density is the sum of the canonical
Hamiltonian density and terms which are products of Lagrange multipliers
and the primary constraints. The time evolution of the primary constraints,
should it be equal to zero, gives  the {\it secondary constraints} and those
{\it secondary constraints} are evaluated by computing the Poisson bracket
of the {\it primary constraints} and the total Hamiltonian density. In the
literature, one may also come across the \textit{extended} Hamiltonian density,
which is the sum of the canonical Hamiltonian density and terms which are
products of Lagrange multipliers and the first-class constraints, see~\cite{Henneaux}.}.
The Hamiltonian equations of motion are the time evolutions, in which the
Hamiltonian density remains invariant under arbitrary variations of $\delta
p$, $\delta q$ and $\delta \lambda$ ; 
\begin{eqnarray}
\dot{p}=-\frac{\delta \mathcal{H}_{tot}}{\delta q}=\{q,\cH_{tot}\}\,,\\
\dot{q}=-\frac{\delta \mathcal{H}_{tot}}{\delta p}=\{p,\cH_{tot}\}\,.
\end{eqnarray}
As a result, the Hamiltonian equations of motion can be expressed in terms
of the Poisson bracket.
In general, for canonical coordinates, $(q^i,~p_i)$,
on the phase space, given two functions $f(q,~p)$ and $g(q,~p)$, the Poisson
bracket can be defined as 
\begin{equation}\label{poisdef}
\{f,g\}=\sum^{n}_{i=1}\Big(\frac{\partial f}{\partial q^{i}}\frac{\partial
g}{\partial p_i}-\frac{\partial f}{\partial p_i}\frac{\partial
g}{\partial q^{i}}\Big)\,,
\end{equation} 
where $q_i$ are the generalised coordinates, and $p_i$ are the generalised
conjugate momentum, and $f$ and $g$ are any function of phase space coordinates.
Moreover, $i$ indicates the number of the phase space variables.

Now, any quantity is \textit{weakly} vanishing when it is numerically restricted
to be zero on a submanifold $\Ga$ of the phase space, but does not vanish
throughout the phase space. In other words, a function $F(p,q)$ defined in
the neighbourhood of $\Ga$ is called \textit{weakly zero}, if
\begin{equation} 
F(p,q) |_{\Ga}=0\Longleftrightarrow F(p,q) \approx 0 \,,
\end{equation} 
where $\Ga$ is the \textit{constraint surface} defined on a submanifold of
the phase space. Note that the notation ``$\approx$'' indicates that the
quantity is \textit{weakly} vanishing; this is a standard Dirac's terminology,
where $F(p,q)$ shall vanish on the constraint surface, $\Ga$, but not necessarily
throughout the phase space.

When a theory admits \textit{primary constraints}, we must ensure that the
theory is consistent by essentially 
checking whether the primary constraints are preserved under time evolution
or not. In other words, we demand that, on the constraint surface $\Ga_p$,
\begin{equation}\label{bvbv}
\dot\varphi |_{\Ga_p}=\{\varphi,\mathcal{H}_{tot}\}|_{\Ga_p}=0\,~~~\Longleftrightarrow
\dot\varphi =\{\varphi,\mathcal{H}_{tot}\}\approx 0 \,.
\end{equation}
That is,
\begin{equation} 
\dot\varphi =\{\varphi,\mathcal{H}_{tot}\}\approx 0\,~~~\Longrightarrow {\it
secondary~constraint} \,.
\end{equation} 
By \textit{demanding} that Eq.~\eqref{bvbv} (not identically) be zero on
the constraint surface $\Ga_p$ yields a \textit{secondary constraint}~\cite{Anderson:1951ta,Matschull:1996up},
and the theory is consistent. In case, whenever Eq.~\eqref{bvbv} fixes a
Lagrange multiplier, then there will be no \textit{secondary constraints}.
The \textit{secondary constraints} hold when the equations of motion are
satisfied, but need not hold if they are not satisfied. However, if Eq.~\eqref{bvbv}
is identically zero, then there will be no {\it secondary constraints}. All
constraints ({\it primary and secondary}) define a smooth submanifold of
the phase space called the 
\textit{constraint surface}: $\Ga_1 \subseteq \Ga_p$.  
A theory can also admit \textit{tertiary constraints}, and so on and so forth
\cite{Henneaux}. We can verify whether the theory is consistent by checking
if the {\it secondary constraints} are preserved under time evolution or
not. 

Note that $\mathcal{H}_{tot}$ is the total Hamiltonian density defined by
Eq.~(\ref{totalH}). 
To summarize, if a canonical momentum is vanishing, we have a \textit{primary
constraint}, while \textit{enforcing} that the time evolution of the {\it
primary constraint}  vanishes on the constraint surface, $\Ga_1$ give rise
to a \textit{secondary constraint}.
\subsection{First and second-class constraints}

Any theory that can be formulated in Hamiltonian formalism gives rise to
Hamiltonian constraints. Constraints in the context of Hamiltonian formulation
can be thought of as reparameterization; while the invariance is preserved
\footnote{For example, in the case of gravity, constraints are obtained by
using the ADM formalism that is reparameterizing the theory under spatial
and time coordinates. Hamiltonian constraints generate time diffeomorphism,
see~\cite{sudarshan}.}. The most important step in Hamiltonian analysis is
the classification of the constrains. By definition, we call a function $f(p,q)$
to be
{\it  first-class} if its Poisson brackets with all other constraints vanish
{\it weakly}. A function which is not {\it first-class} is called
{\it second-class}~\footnote{One should mention that the {\it primary/secondary}
and {\it first-class/second-class} classifications overlap. A {\it primary
constraint} can be {\it first-class or second-class} and a {\it secondary
constraint} can also be {\it first-class} or {\it second-class}.}. On the
constraint surface $\Ga_1$, this is mathematically expressed as
\begin{eqnarray}
\left. \{f(p,q),\varphi\}\right|_{\Ga_1}\approx0 ~~\Longrightarrow~~ \textit{first-class}\,,\\
\left. \{f(p,q),\varphi\}\right|_{\Ga_1}\not\approx0 ~~\Longrightarrow~~
\textit{second-class}\,.
\end{eqnarray}
We should point out that we use the ``$\approx$'' sign as we are interested
in whether the Poisson brackets of $f(p,q)$ with all other constraints vanish
on the constraint surface $\Ga_1$ or not. Determining whether they vanish
globally, \textit{i.e.}, throughout the phase space, is not necessary for
our purposes.


\subsection{Counting the degrees of freedom}

Once we have the physical canonical variables, and we have fixed the number
of {\it first-class and/or second-class} constraints, we can use the following
formula to count the number of the physical degrees of freedom~\footnote{Note
that the phase space is composed of all positions and velocities together,
while the configuration space consists of the position only. }, see~\cite{Henneaux},

\begin{equation}\label{dofcount}
\mathcal{N}=\frac{1}{2}(2\mathcal{A}-\mathcal{B}-2\mathcal{C})=\frac{1}{2}\mathcal{X}
\end{equation}
 where
 \begin{itemize}
 \item{$\mathcal{N}=$ number of physical degrees of freedom}
 \item{ $\mathcal{A}=$  number of \textit{configuration space variables}}

 \item{$\mathcal{B}=$  number of \textit{{second-class constraints}}} 
 \item{$\mathcal{C}=$  number
of \textit{{first-class constraints}}}
\item{$\mathcal{X}=$  number of \textit{independent canonical variables}}
\end{itemize}
\section{Toy models}
In this section we shall use Dirac's prescription and provide the relevant
constraints for some toy models and then obtain the number of degrees of
freedom. Our aim will be to study some very simplistic time dependent models
before extending our argument to a covariant action. 
\subsection{Simple homogeneous case}
Let us consider a very simple time dependent  action, 
\begin{equation}
I=\int\dot \phi^{2}dt\,,
\end{equation}
where $\phi$ is some time dependent variable, and $\dot{\phi}\equiv\partial_0\phi$.
For the above action the canonical 
momenta is~\footnote{We are working around Minkowski background with mostly
plus, \textit{i.e.}, $(-,+,+,+)$.}
\begin{equation}\label{jgjfh}
p=\frac{\partial \mathcal{L}}{\partial\dot\phi}=2\dot\phi\,.
\end{equation}
If the canonical momenta is not vanishing, \textit{i.e.} $p\neq 0$,  then there is
no constraints, and hence no classification, \textit{i.e.} ${\cal B}=0$ in Eq.~(\ref{dofcount}),
and so will be, ${\cal C}=0$. The number of degrees of freedom is then given
by the total number of the independent canonical variables: 
\begin{equation}
\mathcal{N}=\frac{1}{2}\mathcal{X}=\frac{1}{2}(p_,~\phi)=\frac{1}{2}(1+1)=1\,.
\end{equation}
Therefore, this theory contains only one physical degree of freedom. A simple
generalization of a time-dependent variable to infinite derivatives
can be given by:
\begin{eqnarray}\label{timedepact}
I&=&\int dt \phi\mathcal{F}\left(-\frac{\partial^{2}}{\partial t^{2}}\right)\phi\nonumber\\
&=&\int dt\bigg(c_0\phi^{2}+c_1\phi\left(-\frac{\partial^{2}}{\partial t^{2}}\right)\phi+c_2\phi\left(-\frac{\partial^{2}}{\partial
t^{2}}\right)^{2}\phi+c_3\phi\left(-\frac{\partial^{2}}{\partial
t^{2}}\right)^{3}\phi+\cdots\bigg)\nonumber\\
&=&\int dt\bigg(c_0\phi^{2}-c_1\phi\phi^{(2)}+c_2\phi\phi^{(4)}-c_3\phi\phi^{(6)}+\cdots\bigg),
\end{eqnarray}
where $\phi=\phi(t)$, and ${\cal F}$ could take a form, like:
\begin{equation}
\mathcal{F}\left(-\frac{\partial^{2}}{\partial t^{2}}\right)=\sum^{\infty}_{n=0}c_n\left(-\frac{\partial^{2}}{\partial
t^{2}}\right)^{n}\,.
\end{equation}
The next step is to find the conjugate momenta, so that we can use the generalised
formula \cite{Woodard:2015zca},
\begin{eqnarray}\label{generalisedconjugatemomenta}
&&p_1=\frac{\partial\mathcal{L}}{\partial\dot{\phi}}-\frac{d}{dt}\bigg(\frac{\partial\mathcal{L}}{\partial\ddot{\phi}}\bigg)+\bigg(\frac{d}{dt}\bigg)^{2}\bigg(\frac{\partial\mathcal{L}}{\partial\dddot{\phi}}\bigg)-\cdots,
\quad \ \nonumber\\
&&p_2=\frac{\partial\mathcal{L}}{\partial\ddot{\phi}}-\frac{d}{dt}\bigg(\frac{\partial\mathcal{L}}{\partial\dddot{\phi}}\bigg)+\bigg(\frac{d}{dt}\bigg)^{2}\bigg(\frac{\partial\mathcal{L}}{\partial\ddddot{\phi}}\bigg)-\cdots,\quad
\nonumber\\
&&\vdots
\end{eqnarray} 
Now the conjugate momenta for action Eq.~(\ref{timedepact})
as, \begin{eqnarray}
&&p_1=c_1\dot\phi-c_{2} \phi^{(3)}+c_{3}\phi^{(5)}-c_{4}\phi^{(7)}+\cdots\nonumber\\
&&p_2=-c_1\phi+c_2\phi^{(2)}-c_3\phi^{(4)}+c_4\phi^{(6)}-\cdots\nonumber\\
&&p_3=-c_2\phi^{(1)}+c_3\phi^{(3)}-c_4\phi^{(5)}+\cdots\nonumber\\
&&p_4=c_2\phi^{}-c_3\phi^{(2)}+c_4\phi^{(4)}-\cdots\nonumber\\
&&\vdots
\end{eqnarray}
and, so on and so forth. For  Eq. (\ref{timedepact}), we can count the
number of the  degrees of freedom essentially
by identifying the independent number of  canonical variables, that is,
\begin{equation}
\mathcal{N}=\frac{1}{2}\mathcal{X}=\frac{1}{2}(\phi,~p_{1},~p_{2},~\cdots)=\frac{1}{2}(1+1+1+\cdots)=\infty\,.
\end{equation}
An infinite number of canonical variables corresponding
to an infinite number of time derivatives acting on a time-dependent variable
leads to
a theory that contains infinite number of degrees of freedom. 
\subsection{Scalar Lagrangian with covariant derivatives}

As a warm up exercise, let us consider the following action, 
\begin{equation}
I=\int d^{4}x\Big(c_0\phi^{2}+c_1\phi\bbox\phi\Big)\,,
\end{equation}
where $\phi$ is a generic scalar field of mass dimension 2; and $\bbox\equiv\Box/M^{2}$,
where $M$ is the scale of new physics
beyond the Standard Model, $\Box$ is d'Alembertian operator of the form $\Box=\eta^{\mu\nu}\nabla_\mu\nabla_\nu$,
where $\eta_{\mu\nu}$ is the Minkowski metric, and $c_0,~c_1$ are constants.
We can always 
perform integration by parts on the second term, and rewrite
\begin{equation}
c_1\phi\bbox\phi=\frac{c_1}{M^{2}}\partial_\mu\phi\partial^{\mu}\phi=-\frac{c_1}{M^{2}}\partial_0\phi\partial^0\phi+\frac{c_1}{M^{2}}\partial_i\phi\partial^{i}\phi\,,
\end{equation}
therefore the canonical momenta can be expressed, as 
\begin{equation}
\pi=\frac{\partial \mathcal{L}}{\partial\dot\phi}=2\frac{c_1}{M^{2}}\dot\phi\,,
\end{equation}
where we have used the notation $\dot\phi\equiv\partial_0\phi$, also note
that $\partial_0\phi\partial^0\phi=-\partial_0\phi\partial_0\phi$.  The next
step is to write down the Hamiltonian density, as: 
\begin{eqnarray}
&&\mathcal{H}=\pi\dot\phi-\mathcal{L}=2\frac{c_1}{M^{2}}\dot\phi^{2}-c_0\phi^{2}-c_1\phi\bbox\phi\nonumber\\&&=2\frac{c_1}{M^{2}}\dot\phi^{2}-c_0\phi^{2}+\frac{c_1}{M^{2}}(-\dot\phi^{2}+\partial_i\phi\partial^{i}\phi)
\\ \nonumber 
&&=-c_0\phi^{2}+\frac{c_1}{M^{2}}\dot\phi^{2}+\frac{c_1}{M^{2}}\partial_i\phi\partial^{i}\phi\,.
\end{eqnarray}  
Again, if $\pi \neq 0$, or for instance, $c_1\neq 0$, then  there are no
constraints. 
The number of degrees of freedom for the action will be given by,
\begin{equation}
\frac{1}{2}\mathcal{X}=\frac{1}{2}\{
(p,\phi) \}=1.
\end{equation}
It can be seen from the examples provided that going to higher derivatives
amounts to have infinite number of conjugate momenta and thus infinite number
of degrees of freedom. In the next section we are going to construct an infinite
derivative theory such that the number of the degrees of freedom are physical
and finite.  
\section{Infinite derivative scalar field theory}
Before considering any gravitational action, it is helpful to consider a
 Lagrangian that is
constructed by infinite number of d'Alembertian operators, we build this
action in Minkowski space-time, 
\begin{equation}\label{inftscalar}
I_{}=\int d^{4}x \, \phi\mathcal{F}(\bbox)\phi, \qquad\text{with:}\qquad\mathcal{F}(\bbox)=\sum^{\
\infty}_{n=0}c_{n}\bbox^n\,,
\end{equation}
where $c_n$ are constants.
Such action is complicated and thus begs for a more technical approach, we
approach the problem by first
writing an equivalent action of the form, 
\begin{eqnarray}\label{eqvaction1}
I_{eqv}=\int d^{4}x A\mathcal{F}(\bbox)A\,,
\end{eqnarray}
Where the auxiliary field, $A$, is introduced as an equivalent scalar field
to  $\phi$, this means that the equations of the 
motion for both actions ($I$ and $I_{eqv}$) are equivalent. In the next step,
let us expand the term $\mathcal{F}(\bbox)A$, 
\begin{equation}
\mathcal{F}(\bbox)A=\sum^{\
\infty}_{n=0}c_{n}\bbox^{n}A=c_{0}A+c_{1}\bbox A+c_{2}\bbox^{2}A+c_{3}\bbox^{3}A+\cdots
\end{equation} 
Now, in order to eliminate the contribution of $\bbox A,~\bbox^{2}A$ and
so
on, we are going to introduce two auxiliary fields $\chi_n$ and $\eta_n$,
 where the $\chi_{n}$'s are dimensionless and the $\eta_{n}$'s have mass
dimension $2$ (this can be seen by parameterising $\bbox A$, $\bbox^{2} A$,
$\cdots$). We show few steps here by taking some simple examples
\begin{itemize}
\item
Let our action to be constructed  by a single box only, then, 
\begin{equation}\label{blabla1}
I_{eqv}=\int d^{4}x A\bbox A \,.
\end{equation}
Now,  to eliminate $\bbox A$ in the term $A\bbox A$, we wish to add a following
term in the
above action, 
\begin{equation}\label{additionalterm}
\int d^{4}x~\chi_1A(\eta_1-\bbox A)=\int d^{4}x\bigg[\chi_1
A\eta_1+\eta^{\mu\nu}(\partial_\mu\chi_1 A \partial_\nu A+\chi_1
\partial_\mu A
\partial_\nu A)\bigg]\,.
\end{equation}  
where we derived above as follow:
\begin{eqnarray}
\chi_1 A(\eta_1-\Box A)&=&\chi_1 A\eta_1-\chi_1 A\Box
A \nonumber\\
&=&\chi_1 A\eta_1-\eta^{\mu\nu}\chi_1 A\partial_\mu \partial_\nu A\nonumber\\&=&\chi_1
A\eta_1-\eta^{\mu\nu}\partial_\mu(\chi_1
A \partial_\nu A)+\eta^{\mu\nu}\partial_\mu\chi_1 A \partial_\nu A+\eta^{\mu\nu}\chi_1
\partial_\mu A
\partial_\nu A\nonumber\\
&=&\chi_1 A\eta_1+\eta^{\mu\nu}\partial_\mu\chi_1 A \partial_\nu A+\eta^{\mu\nu}\chi_1
\partial_\mu A
\partial_\nu A \,,
\end{eqnarray}
where it should be noted that we have dropped the total derivative and also
we have absorbed the factor of $M^{-2}$ into $\chi_1$ (the mass
dimension of $\eta_1$ is modified accordingly). Therefore, here the d'Alembertian
operator is not barred. Finally, we can write down the equivalent action
in the following form, 
\begin{equation}\label{blabla2}
I_{eqv}=\int d^{4}x\Big( A\eta_{1}+\chi_1A(\eta_1-\bbox A)\Big)\,,
\end{equation}
by solving the equation of motion for $\chi_1$, we obtain 
\begin{equation}
\eta_1=\bbox A\,,
\end{equation} and hence, Eqs. (\ref{blabla1}) and (\ref{blabla2}) are equivalent.
\item
Before generalising our method,  let us consider the following, 
\begin{equation}\label{blabla11}
I_{eqv}=\int d^{4}x~\LT A\bbox A+A\bbox^{2}A \RT\,,
\end{equation}
 in order to eliminate the term $A \bbox^{2}A$, we add the term
\begin{equation}\label{additionalterm2}
\int d^{4}x~\chi_2A(\eta_2-\bbox \eta_{1})=\int d^{4}x\bigg[\chi_2
A\eta_2+\eta^{\mu\nu}(\partial_\mu\chi_2 A \partial_\nu \eta_{1}+\chi_2
\partial_\mu A
\partial_\nu \eta_{1})\bigg]\,.
\end{equation}  
We can rewrite action Eq.~(\ref{blabla11}) as:
\begin{equation}\label{blabla3}
I_{eqv}=\int d^{4}x \Big( A(\eta_{1}+\eta_{2})+\chi_1A(\eta_1-\bbox A)+\chi_2A(\eta_2-\bbox
\eta_{1})\Big)\,.
\end{equation} 
Solving the equation of motion for $\chi_{2}$ yields $\eta_{2}=\bbox \eta_{1}=\bbox^{2}A$.
\end{itemize}
Similarly, in order to eliminate the terms $A\bbox^{n}A$ and so on,
we have to repeat the same procedure up to $\bbox^n$. Note that we have established
this by
solving the equation of motion for $\chi_{n}$, we obtain, for $n \geq 2$,
\begin{equation}
\eta_{n}=\bbox \eta_{n-1}=\bbox^{n}A.
\end{equation}
Now, we can rewrite the action Eq.~(\ref{eqvaction1}) as,
\begin{eqnarray}\label{eqvv2}
I_{eqv}&=&\int d^{4}x\Bigg\{A(c_{0}A+\sum^{\infty}_{n=1}c_n\eta_{n})
+\chi_{1}A(\eta_{1}-\Box A)+\sum
^{\infty}_{l=2}\chi_lA(\eta_l-\Box \eta_{l-1})\Bigg\}
\nonumber\\&=&
\int d^{4}x\Bigg\{A(c_{0}A+\sum^{\infty}_{n=1}c_n\eta_{n})\nonumber\\
&+&\eta^{\mu\nu}(A\partial_\mu\chi_1  \partial_\nu A+\chi_1
\partial_\mu A
\partial_\nu A)+\eta^{\mu\nu}\sum^{\infty}_{l=2}(A\partial_\mu\chi_{l}  \partial_\nu\eta_{l-1}
+\chi_l
\partial_\mu A
\partial_\nu \eta_{l-1})\nonumber\\&+&\sum ^{\infty}_{l=1}A\chi_l \eta_l 
\Bigg\}\,,\nonumber\\
&=&\int d^{4}x\Bigg\{A(c_{0}A+\sum^{\infty}_{n=1}c_n\eta_{n})+\sum
^{\infty}_{l=1}A\chi_l
\eta_l\nonumber\\
&+&\eta^{00}(A\partial_0\chi_1  \partial_0 A+\chi_1
\partial_0 A
\partial_0 A)+\eta^{ij}(A\partial_i\chi_1  \partial_j A+\chi_1
\partial_i A
\partial_j A)\nonumber\\
&+&\eta^{00}\sum^{\infty}_{l=2}(A\partial_0\chi_{l}  \partial_0\eta_{l-1}
+\chi_l
\partial_0 A
\partial_0 \eta_{l-1})+\eta^{ij}\sum^{\infty}_{l=2}(A\partial_i\chi_{l} 
\partial_j\eta_{l-1}
+\chi_l
\partial_i A
\partial_j \eta_{l-1}) 
\Bigg\}\,.\nonumber\\
\end{eqnarray}
where we have absorbed the powers of $M^{-2}$ into the $c_{n}$'s \& $\chi_{n}$'s
and the mass dimension of the $\eta_{n}$'s has been modified accordingly.
Hence, the
box operator is not barred. We shall also mention that in Eq.~(\ref{eqvv2})
we have decomposed the d'Alembertian operator to its components around
the Minkowski background: $\Box=\eta^{\mu\nu}\partial_\mu\partial_\nu=\eta^{00}\partial_0\partial_0+\eta^{ij}\partial
_{i}\partial_{j}$, where the zeroth component is the time coordinate, and
$\{i,~j\}$ are the spatial coordinates running from $1$ to $3$. The conjugate
momenta for the above action are given by:
\begin{eqnarray}
&&p_A=\frac{\partial\mathcal{L}}{\partial\dot{A}}=\Big[-(A\partial_0\chi_1
 +\chi_1
\partial_0 A)-\sum^{\infty}_{l=2}(\chi_l\partial_0 \eta_{l-1})\Big],\quad
\nonumber\\
&&p_{\chi_1}=\frac{\partial\mathcal{L}}{\partial\dot{\chi}_1}=-A\partial_0
A,\quad p_{\chi_l}=\frac{\partial\mathcal{L}}{\partial\dot{\chi}_l}=-(A\partial_0\eta_{l-1}),\quad\nonumber\\
&&p_{\eta_{l-1}}=\frac{\partial\mathcal{L}}{\partial\dot{\eta}_{l-1}}=-(A\partial_0\chi_l
+\chi_l
\partial_0 A).
\end{eqnarray}
where $\dot{A}\equiv\partial_0 A$.  Therefore, the Hamiltonian density is
given by
\begin{eqnarray}\label{hamdens}
\mathcal{H}&=&p_A\dot{A}+p_{\chi_1}\dot{\chi}_1+p_{\chi_l}\dot{\chi}_l+p_{\eta_{l-1}}\dot{\eta}_{l-1}-\mathcal{L}\nonumber\\&=&
A(c_{0}A+\sum^{\infty}_{n=1}c_n\eta_{n})-\sum ^{\infty}_{l=1}A\chi_l
\eta_l\nonumber\\
&-&(\eta^{\mu\nu}A\partial_{\mu}\chi_1  \partial_\nu A+\eta^{ij}\chi_1
\partial_i A
\partial_j A)
\nonumber\\&-&\eta^{\mu\nu}\sum^{\infty}_{l=2}(A\partial_\mu\chi_{l}  \partial_\nu\eta_{l-1}
+\chi_l
\partial_\mu A
\partial_\nu \eta_{l-1})\,.
\end{eqnarray}
See Appendix \ref{hamdensapp} for the explicit derivation of (\ref{hamdens}).
Let us recall the equivalent action (\ref{eqvv2}) before integration by parts.
That reads as, 
\be
I_{eqv}=\int d^{4}x\Bigg\{A(c_{0}A+\sum^{\infty}_{n=1}c_n\eta_{n})+\chi_{1}A(\eta_{1}-\Box
A) +\sum
^{\infty}_{l=2}\chi_lA(\eta_l-\Box \eta_{l-1})\Bigg\}\,;
\ee
we see that we
have  terms like :  $$\chi_{1}A(\eta_{1}-\Box A)$$ and $$\chi_lA(\eta_l-\Box
\eta_{l-1}),~~for~~l \geq 2.$$
Additionally,
we know that solving the equations of motion for $\chi_n$ leads to $\eta_n=\Box^{n}
A$. Therefore, it shall be concluded that the $\chi_{n}$'s are the Lagrange
multipliers,
and not dynamical as a result. From the equations of motion, we get the following
{\it primary
constraints}~\footnote{ Let us note that $\Ga_p$ is a smooth submanifold
of the phase space determined by the {\it primary constraints}; in this section,
we shall exclusively use the ``$\approx$'' notation to denote equality on
$\Ga_p$.}:
\begin{align}
\sa_{1} =\eta_{1}-\Box A & \approx 0 \,, \nonumber \\
&\vdots \\ \nonumber
\sa_{l}=\eta_{l}-\Box \eta_{l-1} & \approx 0 \,.
\end{align}
In other words, since $\chi_{n}$'s are the Lagrange multipliers, $\sa_{1}$
and $\sa_{l}$'s are {\it primary constraints}.
The time evolutions of the $\sa_n$'s fix the corresponding Lagrange multipliers
$\la^{\sa_n}$ in the total Hamiltonian (when we add the terms $\la^{\sa_n}\sa_{n}$
to the Hamiltonian density $\cH$); therefore, the $\sa_n$'s do not induce
\textit{secondary constraints}.
As a result, to classify the above constraint, we will need to show that
the Poisson bracket given by (\ref{poisdef}) is weakly vanishing:
\begin{equation}
\{\sigma _{m},\sigma _{n}\}|_{\Ga_p}= 0\,,
\end{equation}
so that $\sa_{n}$'s can be classified as {\it first-class constraints}.
However, this depends on the choice of ${\cal F}(\Box)$, whose coefficients
are hiding in $\chi$'s and $\eta$'s.
It is trivial to show that, for this case, there is no {\it second-class
constraint}, \textit{i.e.}, ${\cal B}=0$, as we do not have $\{\sigma _{m},\sigma
_{n}\}\not\approx0$. That is, the $\sa_n$'s are \textit{primary, first-class
constraints}.  
In our case, the number of  phase space variables, 
\begin{eqnarray}\label{phasespacevariables2}
2\mathcal{A}&\equiv&2\times\bigg\{(A,p_A),\underbrace{(\eta_1,p_{\eta_1}),(\eta_2,p_{\eta_2}),\cdots}_{n=1,~2,~3,\cdots
\infty}\bigg\}\equiv2\times(1+\infty)=2+\infty\,. \nonumber \\
\end{eqnarray} 
For each pair,  $(\eta_n,~p_{\eta_n})$, 
we have assigned one variable, which is multiplied by a factor of $2$, since
we
are dealing with field-conjugate momentum pairs, in the phase space. In the
next section, we will fix the form of ${\cal F}(\bar\Box)$
to estimate the number of {\it first-class constraints}, \textit{i.e.}, ${\cal
C}$ and, hence, the number of degrees of freedom. Let us also mention that
the choice of $\cF(\bbox)$ will determine the number of solutions to the
equation of motion for $A$ we will have, and consequently these solutions
can be interpreted as {\it first-class constraints} which will determine
the number of physical degrees of freedom, \textit{i.e.}  finite/infinite
number
of degrees of freedom will depend on the number of solutions of the equations
of motion for $A$. See more detail on Appendix \ref{formoffunction}. 
\subsection{Gaussian kinetic term and propagator}

Let us now consider an example of infinite derivative scalar field theory,
but with a Gaussian kinetic term in Eq. (\ref{inftscalar}), \textit{i.e.}
by exponential of an entire function,  
\begin{equation}
\label{action-entire}
I_{eqv}=\int d^{4} x~A\bigg(\Box e^{-\bbox}\bigg)A\,.
\end{equation}
For the above action, the equation of motion for $A$ is then given by:
\begin{equation}
2\bigg(\Box e^{-\bbox}\bigg)A=0\,.
\end{equation}
We observe that there is a finite number of solutions; hence, there are also
finitely many degrees of freedom~\footnote{Note that, for an infinite derivative
action of the form $I_{eqv}=\int d^{4} x~A\cos (\bbox) A$, we would have
an infinite number of solutions and, hence, infinitely many degrees of freedom. Note that the choice of $\cos (\bbox)$ leads to infinite number of solutions due to the periodicity of the cosine function. In this footnote we take $\cos (\bbox)$ to illustrate what it means by \textit{bad choice} of $F(\bbox)$.   }.
In momentum space, we obtain the following solution, 
\begin{equation}
k^{2}=0\,,
\end{equation}
and the propagator will follow as~\cite{Biswas:2011ar,Biswas:2013kla} :
\begin{equation}
\Pi(\kb^2)\sim \frac{1}{k^{2}}e^{-\kb^2}\,,
\end{equation}
where we have used the fact that in momentum space $\Box\rightarrow -k^2$,
and
we have $\bar{k}\equiv k/M$. There are some interesting properties to note
about this propagator:  

\begin{itemize}

\item{The propagator is suppressed by an {\it exponential of an entire function},
which has no zeros,  poles. 
Therefore, the only dynamical pole resides at $k^{2}=0$, \textit{i.e.}, the
massless pole in the propagator, \textit{i.e.}, degrees of freedom ${\cal
\mathcal{N}} =1$. This is to say that, even though we have   infinitely many
derivatives, but there is only one relevant degrees of freedom that is the
massless scalar field.
In fact, there are 
no new dynamical degrees of freedom. Furthermore, in the UV the propagator
is suppressed.}

\item{The propagator contains no {\it ghosts} (this is because an entire function does not give rise to poles in the infinite complex plane), which usually plagues higher
derivative theories. By virtue of this, there is no analogue of Ostr\'ogradsky
instability at classical level. Given the background equation, one can indeed
understand
the stability of the solution. }

\end{itemize}

The original action Eq.~(\ref{action-entire}) can now be recast in terms
of an equivalent action as:
\begin{eqnarray}
I_{eqv}&=&\int d^{4} x\bigg[A\big(\Box e^{-\bbox}\big)A+\chi_{1}A(\eta_{1}-\Box
A) +\sum
^{\infty}_{l=2}\chi_lA(\eta_l-\Box \eta_{l-1})\bigg]\,.\nonumber\\
\end{eqnarray}
We can  now compute the number of the physical degrees of freedom. Note that
the determinant of the phase-space dependent matrix 
$A_{mn}=\{\sa_{m},\sa_{n} \}\neq 0$, so the $\sa_n$'s do not induce further
constraints, such as {\it secondary constraints}. Therefore~\footnote{In
this case and hereafter in this chapter, one shall include the $k^2=0$ solution
when counting the number of degrees of freedom. This can be written in position
space as $\Box A=0$. Since $\Box A$ is already parameterised as $\eta_1$,
the counting remains unaffected.},
\begin{eqnarray}\label{dog}
&&2\mathcal{A}\equiv2\times\bigg\{(A,p_A),\underbrace{(\eta_1,p_{\eta_1}),(\eta_2,p_{\eta_2}),\cdots}_{n}\bigg\}
=2\times(1+\infty)=2+\infty\nonumber\\
&&\mathcal{B}=0,\nonumber\\
&&2\mathcal{C}\equiv2\times(\sigma
_{n})=2(\infty)=\infty,\nonumber\\
&&\mathcal{N}=\frac{1}{2}(2\mathcal{A}-\mathcal{B}-2\mathcal{C})=\frac{1}{2}(2+\infty-0-\infty)=1\,.
\end{eqnarray}
As expected, the conclusion of this analysis yields exactly the same dynamical
degrees of freedom as that of the Lagrangian formulation.
The coefficients  $c_i$ of ${\cal F}(\bbox)$ are all fixed by the form of
$\Box e^{-\bbox}$.
\section{IDG Hamiltonian analysis}
In this section we will take a simple action of IDG,
and study the Hamiltonian density and degrees of freedom, we proceed by 
briefly recap the ADM formalism for gravity as we will require this in our
analysis.
\subsection{ADM formalism}\label{admsec}
One of the important concepts in GR is diffeomorphism invariance, \textit{i.e.} when
one transforms coordinates at given space-time points, the physics remains
unchanged. As a result of this, one concludes that diffeomorphism is a local
transformation.   
In Hamiltonian formalism, we have to specify the direction of time. A very
useful approach to do this is ADM decomposition~\cite{Arnowitt:1962hi,Gourgoulhon:2007ue},
such decomposition permits to choose one specific time direction without
violating the diffeomorphism invariance. In other words, choosing the time
direction is nothing but gauge redundancy, or making sure that diffeomorphism
is a local transformation. We assume  that the manifold $\mathcal{M}$ is
a time orientable space-time, which
can be foliated by a family of space like hypersurfaces $\Sigma_{t}$, at
which the time is fixed to be constant $t=x^{0}$. We then introduce an induced
metric on the hypersurface as 
$$h_{ij}\equiv g_{ij}|_{_{t}}\,,$$ where the
Latin
indices run from 1 to 3 for spatial coordinates. 

In $3+1$ formalism
the line element is parameterised as, 
\begin{eqnarray}\label{metric111}
ds^{2}=-N^{2}dt^{2}+h_{ij}(dx^i+N^i dt)(dx^{j}+N^j dt)\,,
\end{eqnarray}
where $N$ is the \textit{lapse function} and $N^{i}$ is the \textit{shift
vector},
given by
\begin{equation}\label{lapseshift}
N=\frac{1}{\sqrt{-g^{00}}},\qquad N^i=-\frac{g^{0i}}{g^{00}}\,.
\end{equation} 
In terms of metric variables, we then have
\begin{eqnarray}
&&g_{00}=-N^2+h_{ij}N^iN^j,\qquad g_{0i}=N_i, \qquad g_{ij}=h_{ij}\,,\nonumber\\
&&g^{00}=-N^{-2},\qquad g^{0i}=\frac{N^i}{N^2}, \qquad g^{ij}=h^{ij}-\frac{N^{i}N^{j}}{N^{2}}\,.
\end{eqnarray} 
 Furthermore, we have a time like vector $n^{\mu}$ (\textit{i.e.} the vector
normal to the hypersurface) in Eq. (\ref{metric111}), they take the following
form:
\begin{eqnarray}\label{normvar}
n_{i}=0, \quad n^{i}=-\frac{N^{i}}{N},\quad
n_{0}=-N,\quad\ n^0=N^{-1}\,.
\end{eqnarray}
From Eq.~(\ref{metric111}), we also have $\sqrt{-g}=N\sqrt{h}$.
In addition, we are going to introduce a covariant derivative associated
with the induced
metric $h_{ij}$:
$$D_i\equiv e^{\mu}_{i}\nabla_\mu\,.$$
We will  define the extrinsic curvature as:
\begin{equation}\label{extrinsiccurvature}
K_{ij}=-\frac{1}{2N}\left(D_{i}N_{j}+D_{j}N_{i}-\partial_{t}h_{ij}\right)\,.
\end{equation} 
It is well known that the Riemannian curvatures can be written in terms of
the 3+1 variables. In the case of scalar curvature we have~\cite{Gourgoulhon:2007ue}:

\begin{eqnarray}\label{scalarcurv}
R=K_{ij}K^{ij}-K^2+\mathcal{R}+\frac{2}{\sqrt{h}}\partial_\mu(\sqrt{h}n^\mu
K)-\frac{2}{N\sqrt{h}}\partial_i(\sqrt{h}h^{ij}\partial_j N)\,,
\end{eqnarray} 
where $K=h^{ij}K_{ij}$ is the trace of the extrinsic curvature, and $\mathcal{R}$
is scalar curvature calculated using the induced metric $h_{ij}$~\footnote{We
note
that the Greek indices are $4$-dimensional while Latin indices are spatial
and $3$-dimensional.}.

One can calculate each term in~(\ref{scalarcurv})
using the information about extrinsic curvature and those provided in~(\ref{normvar}).
The decomposition of the d'Alembertian operator can be expressed as:
\begin{eqnarray}\label{boxdecomposition}
\Box&=&g^{\mu\nu}\nabla_\mu\nabla_\nu\\\nonumber
&=&(h^{\mu\nu}+\varepsilon n^\mu n^\nu)\nabla_\mu\nabla_\nu=(h^{ij}e^{\mu}_{i}e^{\nu}_{j}-
n^\mu n^\nu)\nabla_\mu\nabla_\nu\\\nonumber
&=&h^{ij}D_iD_j-n^\nu\nabla_\bn\nabla_\nu=\Box_{hyp}-n^\nu\nabla_\bn\nabla_\nu\,,
\end{eqnarray} 
where we have used the completeness relation for a space-like hypersurface,
\textit{i.e.}
$\varepsilon=-1$, and we have also defined $\nabla_\bn=n^\mu\nabla_\mu$.

\subsection{ADM decomposition of IDG}
Let us now take an IDG\ action.
We will restrict
ourselves to part of an IDG action 
which contains only the Ricci scalar,
\begin{eqnarray}\label{eq111}
S=\frac{1}{2}\int d^{4}x \, \sqrt{-g}\bigg[M^{2}_{P}R+R\mathcal{F}(\bbox)R\bigg],\qquad
\mathcal{F}(\bbox)=\sum^{\
\infty}_{n=0}f_{n}\bbox^{n}\,,
\end{eqnarray}
where $M_{P}$ is the $4$-dimensional Planck scale, given by $M^{2}_{P}=(8\pi
G_{N})^{-1}$, with $G_{N}$ is Newton's gravitational constant. The first term is
Einstein Hilbert term, with $R$ being scalar curvature in four dimensions
and the second term is the infinite derivative modification to the action,
where $\bbox\equiv\Box/M^{2}$ ,  since $\Box$ has dimension mass squared
and $\mathcal{F}(\bbox)$ will be dimensionless. Note that $\Box$ is the $4$-dimensional
d'Alembertian operator given by $\Box=g^{\mu\nu}\nabla_{\mu}\nabla_{\nu}$.
Moreover, $f_{n}$ are the dimensionless coefficients of the series expansion.

Having the $3+1$ decomposition discussed in the earlier section, we  rewrite
our
original action given in Eq.(\ref{eq111})
in its equivalent form,
\begin{eqnarray}\label{eqvaction}
S_{eqv}=\frac{1}{2}\int d^{4}x \, \sqrt{-g}\bigg[M^{2}_{P}A+A\mathcal{F}(\bbox)A+B(R-A)\bigg]\,,
\end{eqnarray}
where we have introduced two scalar fields $A$ and $B$ with mass dimension
two. Solving the equations of motion for scalar field $B$ results in $A=R$.
The equations of motion for the original action, Eq.(\ref{eq111}),  are equivalent
to the equations of motion for Eq. (\ref{eqvaction}):
\begin{eqnarray}
\delta S_{eqv}=\frac{1}{2}\delta\Bigg\{\sqrt{-g}\bigg[M^{2}_{P}A+A\mathcal{F}(\bbox)A+B(R-A)\bigg]\Bigg\}=0\Rightarrow
R=A\,.
\end{eqnarray}
Following the steps of a scalar field theory, we  expand $\mathcal{F}(\bbox)A$,
\begin{equation}
\mathcal{F}(\bbox)A=\sum^{\
\infty}_{n=0}f_{n}\bbox^{n}A=f_{0}A+f_{1}\bbox A+f_{2}\bbox^{2}A+f_{3}\bbox^{3}A+\cdots
\end{equation} 
As before, in order to eliminate  $\bbox A,~\bbox^{2}A, \cdots$, we will
introduce two new auxiliary fields $\chi_n$ and $\eta_n$ with 
the $\chi_{n}$'s being dimensionless and the $\eta_{n}$'s of mass dimension
two. 
\begin{itemize}

\item As an example, in order to eliminate $\bbox A$ in $A\bbox A$, we must

add the following terms to Eq. (\ref{eqvaction}): 
\begin{eqnarray}\label{additionaltermm}
&&\frac{1}{2}\int d^{4}x\sqrt{-g}\chi_1A(\eta_1-\bbox A)\nonumber\\
&=&\frac{1}{2}\int
d^{4}x\sqrt{-g}[\chi_1 A\eta_1-\chi_1 A\Box
A] \nonumber\\
&=&\frac{1}{2}\int
d^{4}x\sqrt{-g}[\chi_1 A\eta_1-g^{\mu\nu}\chi_1 A\partial_\mu \partial_\nu
A]\nonumber\\&=&\frac{1}{2}\int
d^{4}x\sqrt{-g}[\chi_1
A\eta_1-g^{\mu\nu}\partial_\mu(\chi_1
A \partial_\nu A)\nonumber\\&+&g^{\mu\nu}\partial_\mu\chi_1 A \partial_\nu A+g^{\mu\nu}\chi_1
\partial_\mu A
\partial_\nu A]\nonumber\\
&=&\frac{1}{2}\int
d^{4}x\sqrt{-g}[\chi_1 A\eta_1+g^{\mu\nu}\partial_\mu\chi_1 A \partial_\nu
A+g^{\mu\nu}\chi_1
\partial_\mu A
\partial_\nu A] \nonumber\\
&=&\frac{1}{2}\int
d^{4}x\sqrt{-g}\bigg[\chi_1
A\eta_1+g^{\mu\nu}(\partial_\mu\chi_1 A \partial_\nu A+\chi_1
\partial_\mu A
\partial_\nu A)\bigg]\nonumber\\
\end{eqnarray}   
Solving the equation of motion for $\chi_1$; yields $\eta_1=\bbox A$.
In the about derivation we integrated by parts on the second step and have
dropped the total derivative, also
we have absorbed the factor of $M^{-2}$ into $\chi_1$(the mass
dimension of $\eta_1$ is modified accordingly), hence, the d'Alembertian
operator is not barred.

\item 
For instance, in order to eliminate the term $A \bbox^{2}A$, we add the term
\begin{equation}\label{additionalterm26}
\frac{1}{2}\int d^{4}x~\chi_2A(\eta_2-\bbox \eta_{1})=\frac{1}{2}\int d^{4}x\bigg[\chi_2
A\eta_2+g^{\mu\nu}(\partial_\mu\chi_2 A \partial_\nu \eta_{1}+\chi_2
\partial_\mu A
\partial_\nu \eta_{1})\bigg]\,.
\end{equation}  
Solving the equation of motion for $\chi_{2}$ yields; $\eta_{2}=\bbox \eta_{1}=\bbox^{2}A$.
\end{itemize}
Similarly, in order to eliminate the terms $A\bbox^{n}A$ and so on,
we have to repeat the same procedure up to $\bbox^n$. Again, we have shown
that by
solving the equations of motion for $\chi_{n}$, we obtain,  $$\eta_{n}=\bbox
\eta_{n-1}=\bbox^{n}
A,~~for~~n \geq 2.$$
Following the above steps, we can rewrite the action Eq.~(\ref{eqvaction}),
as:
\begin{align}\label{eqv2}
S_{eqv}&=\frac{1}{2}\int d^{4}x\sqrt{-g}\Bigg\{A(M^{2}_{P}+f_{0}A+\sum^{\infty}_{n=1}f_n\eta_{n})+B(R-A)+\chi_{1}A(\eta_{1}-\Box
A)\non &
+\sum
^{\infty}_{l=2}\chi_lA(\eta_l-\Box \eta_{l-1})\Bigg\}
\nonumber\\&=
\frac{1}{2}\int d^{4}x\sqrt{-g}\Bigg\{A(M^{2}_{P}+f_{0}A+\sum^{\infty}_{n=1}f_n\eta_{n})+B(^{}R-A)\nonumber\\
&+g^{\mu\nu}(A\partial_\mu\chi_1  \partial_\nu A+\chi_1
\partial_\mu A
\partial_\nu A)+g^{\mu\nu}\sum^{\infty}_{l=2}(A\partial_\mu\chi_l  \partial_\nu\eta_{l-1}
+\chi_l
\partial_\mu A
\partial_\nu \eta_{l-1}) \nonumber \\
&+\sum ^{\infty}_{l=1}A\chi_l \eta_l 
\Bigg\}\,,
\end{align}
where we have absorbed the powers of $M^{-2}$ into the $f_{n}$'s and $\chi_{n}$'s,
 and the mass dimension of $\eta_{n}$'s has been modified accordingly, hence,
the box operator is not barred.

Note that the gravitational part of the action is simplified. 
In order to perform the ADM decomposition, let us first look at the $B(R-A)$
term, with the help of Eq. (\ref{scalarcurv})
we can write:
\begin{eqnarray}\label{bra}
B(R-A)=B\Big(K_{ij}K^{ij}-K^2+\mathcal{R}-A\Big)-2\nabla_\bn BK-\frac{2}{\sqrt{h}}\partial_j(\partial_i(B)\sqrt{h}h^{ij}
),\nonumber\\
\end{eqnarray}
where from Eq. (\ref{scalarcurv}) we expanded the following terms, 
\begin{eqnarray}
&&B\frac{2}{\sqrt{h}}\partial_\mu(\sqrt{h}n^\mu
K)\\\nonumber &&=\nabla_\mu\Big[B\frac{2}{\sqrt{h}}(\sqrt{h}n^\mu
K)\Big]-(\nabla_\mu B)\frac{2}{\sqrt{h}}(\sqrt{h}n^\mu
K)\\\nonumber &&=\nabla_\mu\Big[2Bn^\mu
K\Big]-2(\nabla_\mu B)n^\mu
K=-2n^\mu(\nabla_\mu B)K=-2\nabla_\bn BK\,,
\end{eqnarray}
and,
\begin{eqnarray}
&&-B\frac{2}{N\sqrt{h}}\partial_i(\sqrt{h}h^{ij}\partial_j N)\\ \nonumber&&=-\frac{2}{N\sqrt{h}}\partial_i(B(\sqrt{h}h^{ij}\partial_j
N))+\frac{2}{N\sqrt{h}}\partial_i(B)\sqrt{h}h^{ij}\partial_j
N\\\nonumber
&&=\frac{2}{N\sqrt{h}}\partial_i(B)\sqrt{h}h^{ij}\partial_j
N=\frac{2}{N\sqrt{h}}\partial_j(\partial_i(B)\sqrt{h}h^{ij}
N)-\frac{2}{\sqrt{h}}\partial_j(\partial_i(B)\sqrt{h}h^{ij}
)\\\nonumber
&&=-\frac{2}{\sqrt{h}}\partial_j(\partial_i(B)\sqrt{h}h^{ij}
)\,.
\end{eqnarray}
Note that we have used $n^\mu\nabla_\mu\equiv\nabla_\bn$ and dropped the
total derivatives. Furthermore, we can use the decomposition of d'Alembertian operator,
given
in (\ref{boxdecomposition}), and also in 3+1, we have 
$\sqrt{-g}=N\sqrt{h}$. Hence, the decomposition of the action (\ref{eqv2})
becomes:
\begin{eqnarray}\label{finaldecomposition}
S_{eqv}&=&\frac{1}{2}\int d^{3}xN\sqrt{h}\Bigg\{A(M^{2}_{P}+f_{0}A+\sum^{\infty}_{n=1}f_n\eta_{n})+B\Big(K_{ij}K^{ij}-K^2+\mathcal{R}-A\Big)\nonumber\\
&-&2\nabla_\bn BK-\frac{2}{\sqrt{h}}\partial_j(\partial_i(B)\sqrt{h}h^{ij}
)
\nonumber\\&+&h^{ij}(A\partial_i\chi_1  \partial_j A+\chi_1
\partial_i A
\partial_j A)-(A\nabla_\bn \chi_1  \nabla_\bn A+\chi_1
\nabla_\bn A\nabla_\bn A)\nonumber\\&+&h^{ij}\sum^{\infty}_{l=2}(A\partial_i\chi_l
 \partial_j\eta_{l-1}
+\chi_l
\partial_i A
\partial_j \eta_{l-1})-\sum^{\infty}_{l=2}(A\nabla_\bn \chi_l  \nabla_\bn
\eta_{l-1}
+\chi_l
\nabla_\bn  A\nabla_\bn \eta_{l-1})\nonumber\\&+&\sum ^{\infty}_{l=1}A\chi_l
\eta_l
\Bigg\}\,,
\end{eqnarray}
where  the Latin indices are spatial, and run from 1 to 3. Note that the
$\chi$ fields were introduced to parameterise the contribution of 
$\bbox A,~\bbox^2A,~\cdots $, and so on. Therefore, $A$ and $\eta$ are auxiliary
fields, which concludes that $\chi$ fields have no intrinsic value, and they
are  redundant. In other words they are Lagrange multiplier, when we count
the number
of phase space variables. 

The same can not be concluded regarding the $B$ field, as it is introduced
to obtain equivalence between scalar curvature, $R$, and $A$. Since $B$ field
is coupled to $R$, and the scalar curvature is physical - we must count
$B$ as a phase space variable. As we will see later in our Hamiltonian analysis,
this is a crucial point while counting the number of physical degrees of
freedom correctly. To summarize, as we will see, $B$ field is not a
Lagrange multiplier, while $\chi$ fields are.

\subsection{$f(R)$ gravity}
Before proceeding further in our analysis and count the number of degrees
of freedom for IDG, it is worth providing a well known example to test the machinery
we build so far. To this end, let us consider the action for $f(R)$ gravity,
\begin{eqnarray}\label{fraction}
S=\frac{1}{2\kappa}\int d^{4}x\sqrt{-g}f(R)\,,
\end{eqnarray}
where $f(R)$ is a function of scalar curvature and $\kappa=8\pi G_{N}$. The equivalent
action for above is then given by,  
\begin{eqnarray}
S=\frac{1}{2\kappa}\int d^{4}x\sqrt{-g}\Big(f(A)+B(R-A)\Big)\,,
\end{eqnarray}
where again solving the equations of motion for $B,$ one obtains $R=A$, and
hence it is clear that above action is equivalent 
with Eq. (\ref{fraction}). Using Eq. (\ref{bra}) we can  decompose the action
as,
\begin{eqnarray}
S=\frac{1}{2\kappa}\int d^{3}xN\sqrt{h}\Big(f(A)+B\Big(K_{ij}K^{ij}-K^2+\mathcal{R}-A\Big)-2\nabla_\bn
BK\nonumber \\
-\frac{2}{\sqrt{h}}\partial_j(\partial_i(B)\sqrt{h}h^{ij}
)\Big)\,.
\end{eqnarray}
Now that the above  action is expressed in terms of $(h_{ab},~N,~N^i,~B,~A)$,
and their time and space derivatives.
We can proceed with the Hamiltonian analysis and write down the momentum
conjugate for each of these variables:
\begin{eqnarray}\label{const-1}
&&\pi^{ij}=\frac{\partial\mathcal{L}}{\partial\dot{h}_{ij}}=\sqrt{h}B(K^{ij}-h^{ij}K)-\sqrt{h}\nabla_\bn
Bh^{ij},\qquad p_B=\frac{\partial\mathcal{L}}{\partial\dot{B}}=-2\sqrt{h}K,\nonumber\\
&&p_A=\frac{\partial\mathcal{L}}{\partial\dot{A}}\approx0, \qquad\pi_N=\frac{\partial\mathcal{L}}{\partial\dot{N}}\approx0,\qquad\
\pi_i=\frac{\partial\mathcal{L}}{\partial\dot{N^i}}\approx0\,.
\end{eqnarray}
where $\dot{A}\equiv\partial_0A$ is the time derivative of the variable.
We have used the ``$\approx$'' sign in Eq.~\eqref{const-1} to show that $(p_A,~\pi_N,~\pi_i)$
are \textit{primary constraints} satisfied on the constraint surface: 
$$
\Ga_p=(p_A
\approx 0,~\pi_N\approx 0,~\pi_i \approx 0).$$ $\Ga_p$ is defined by the
aforementioned
{\it primary constraints}. For our purposes, whether the {\it primary constraints}
vanish globally (which they do), \textit{i.e.}, throughout the phase space,
is irrelevant. Note that the Lagrangian density, $\mathcal{L}$, does not
contain  $\dot A$, $\dot N$ or $\dot N^i$, therefore, 
their conjugate momenta vanish identically. 

We can define the Hamiltonian density as: 
\begin{align}\label{hamdensity1}
\mathcal{H}&=\pi^{ij}\dot{h}_{ij}+p_B\dot{B}-\mathcal{L}\\
&\equiv\ N\mathcal{H}_N+N^i\cH_i \,,
\end{align}
where  $\mathcal{H}_N=\dot\pi_N$, and $\cH_i=\dot{\pi}_{i}$. By using Eq.
(\ref{hamdensity1}), we can write 
\begin{eqnarray}\label{h1}
&&\mathcal{H}_N=\frac{1}{\sqrt{h}B}\pi^{ij}h_{ik}h_{jl}\pi^{kl}-\frac{1}{3\sqrt{h}B}\pi^2-\frac{\pi
p_B}{3\sqrt{h}}+\frac{B}{6\sqrt{h}}p_B^{2}\nonumber\\&&-\sqrt{h}BR+\sqrt{h}BA+2\partial_j[\sqrt{h}h^{ij}\partial_i]B+f(A)\,,
\end{eqnarray}
 and, 
\begin{equation}\label{h2}
\cH_i=-2h_{ik}\nabla_l\pi^{kl}+p_B\partial_iB\,.
\end{equation}
Therefore, the total Hamiltonian can be written as,
\begin{align}
H_{tot}&=\int d^3 x \, \cH \\
&=\int d^{3}x \, \Big(N\mathcal{H}_N+N^i\cH_i+\lambda^{A}p_A+\lambda^{N}\pi_N+\lambda^{i}\pi_i\Big)\,,
\end{align}
where  $\lambda^{A},\lambda^{N},\la^{i}$ are Lagrange multipliers, and we
have  $G_A=\dot p_A$. 
\subsubsection{Classification of constraints for $f(R)$ gravity}
Having vanishing conjugate momenta means we can not express $\dot{A}$, $\dot{N}$
and $\dot{N}^i$ as a function 
of their conjugate momenta and hence $p_A\approx0,~\pi_N\approx0$ and $\pi_i\approx0$
are \textit{primary constraints},
see (\ref{const-1}). To ensure the consistency of the {\it primary constraints}
so that they are preserved under time evolution 
generated by total Hamiltonian $H_{tot}$, we need to employ the Hamiltonian
field equations and \textit{enforce} that $\cH_{N}$ and $\cH_{i}$ be zero
on the constraint surface $\Ga_p$,
\begin{equation}\label{secondaryconstraints1}
\dot{\pi}_{N}=-\frac{\delta \mathcal{H}_{tot}}{\delta N}=\cH_{N}\approx0,\qquad
\dot{\pi}_{i}=-\frac{\delta \mathcal{H}_{tot}}{\delta N^i}=\cH_{i}\approx0\,,
\end{equation}
such that $\mathcal{H}_{N}\approx0$ and $H_{i}\approx0$, and therefore they
can be treated as \textit{secondary constraints}. 

Let us also note that $\Ga_1$ is a smooth submanifold of the phase space
determined by the {\it primary} and  {\it secondary constraints}; hereafter
in this section, we shall exclusively use the ``$\approx$'' notation to denote
equality on $\Ga_1$. 
It is usual to call $\mathcal{H}_{N}$ as the \textit{Hamiltonian constraint},
and $\cH_i$ as  \textit{diffeomorphism constraint}. Note that  $\mathcal{H}_{N}$
and $\cH_{i}$ are weakly vanishing \textit{only} on the constraint surface;
this is why the r.h.s of Eqs. (\ref{h1}) and (\ref{h2}) are not  {\it identically}
zero. If $\dot{\pi}_{N}=\cH_{N}$ and $\dot{\pi}_{i}=\cH_{i}$ were \textit{identically}
zero, then there would be no {\it secondary constraints}.

Furthermore, we are going to define $G_A$, and \textit{demand} that $G_A$
be weakly zero on the constraint surface 
$\Ga_1$,
\begin{equation}
G_{A}=\partial_tp_A=\{{p_A},\mathcal{H}_{tot}\}=-\frac{\delta \mathcal{H}_{tot}
}{\delta A}=-\sqrt{h}N(B+f'(A))\approx0\,,
\end{equation}
which will act as a {\it secondary constraint} corresponding to {\it primary
constraint} $p_A\approx0$.  Hence, 
$$\Ga_1=(p_A \approx 0,~\pi_N \approx 0,~\pi_i \approx 0,~G_A \approx 0,~{\cal
H}_{N}\approx 0,~{\cal H}_{i}\approx 0).$$

Following the definition of Poisson bracket in Eq.(\ref{poisdef}), we can
see that since the constraints $\cH_N$ and $\cH_i$ are preserved under time
evolution, \textit{i.e.}, $\dot \cH_N=\{\cH_N,\mathcal{H}_{tot}\}|_{\Ga_1}=
0$ and $\dot \cH_i=\{\cH_i,\mathcal{H}_{tot}\}|_{\Ga_1}= 0$, and they fix
the Lagrange multipliers $\la^N$ and $\la^i$. That is, the expressions for
$\dot \cH_N$ and $\dot \cH_i$ include the Lagrange multipliers $\la^N$ and
$\la^i$; thus, we can solve the relations $\dot \cH_N \approx 0$ and $\dot
\cH_i \approx 0$ for $\la^N$ and $\la^i$, respectively, and compute the values
of the Lagrange multipliers. Therefore, we have no further constraints, such
as {\it tertiary} ones and so on. 
We will check the same for $G_A$, that the time evolution of $G_A$ defined
in the phase space should also vanish on the constraint surface $\Ga_1$,
\begin{eqnarray} \label{rubu}
\dot G_A\equiv\{G_{A},\mathcal{H}_{tot}\}&=&\Bigg(\frac{\delta G_{A}}{\delta
N_{}}\frac{\delta \mathcal{H}_{tot}}{\delta
\pi_N}-\frac{\delta G_{A}}{\delta
\pi_N}\frac{\delta \mathcal{H}_{tot}}{\delta N} 
\Bigg)+\Bigg(\frac{\delta  G_{A}}{\delta N^{i}}\frac{\delta \mathcal{H}_{tot}}{\delta
\pi_i}-\frac{\delta  G_{A}}{\delta
\pi_i}\frac{\delta \mathcal{H}_{tot}}{\delta N^{i}}\Bigg) \nonumber\\
&+&\Bigg(\frac{\delta  G_{A}}{\delta h_{ij}}\frac{\delta \mathcal{H}_{tot}}{\delta
\pi^{ij}}-\frac{\delta G_{A}}{\delta
\pi^{ij}}\frac{\delta \mathcal{H}_{tot}}{\delta h_{ij}}
\Bigg)+\Bigg(\frac{\delta G_{A}}{\delta A}\frac{\delta \mathcal{H}_{tot}}{\delta
p_{A}}-\frac{\delta  G_{A}}{\delta
p_{A}}\frac{\delta \mathcal{H}_{tot}}{\delta A}
\Bigg)\nonumber\\
&+&\Bigg(\frac{\delta G_{A}}{\delta B}\frac{\delta \mathcal{H}_{tot}}{\delta
p_{B}}-\frac{\delta  G_{A}}{\delta
p_{B}}\frac{\delta \mathcal{H}_{tot}}{\delta B}\Bigg)\non
&=&\frac{\delta G_{A}}{\delta A}\frac{\delta\mathcal{H}_{tot}}{\delta
p_{A}}+\frac{\delta G_{A}}{\delta B}\frac{\delta \mathcal{H}_{tot}}{\delta
p_{B}}\non
&=& N\Bigg\{\frac{N}{3}\Big(2\pi-2Bp_B\Big)-2\sqrt{h}N^i\partial_iB-\sqrt{h}f''(A)\lambda^{A}\Bigg\}
\non
&\approx&0 \,.
\end{eqnarray}
The role of Eq.~\eqref{rubu} is to fix the value of the Lagrange multiplier
$\lambda^{A}$ as long as $f''(A)\neq0$. We demand that $f''(A) \neq 0$ so
as to avoid \textit{tertiary constraints}. As a result, there are no {\it
tertiary constraints} corresponding to $G_A$. 

The next step in our Hamiltonian
analysis is to classify the constraints. As shown above, we have $3$ {\it
primary constraints} for $f(R)$ theory.
They are: $$\pi_N\approx0,~~\pi_i\approx0\,,~~p_A\approx0,$$
and there are three {\it secondary constraints}, that are: 
$$\mathcal{H}_{N}\approx0,~~\cH_{i}\approx0,~~G_A\approx0. $$ 
Following the definition of Poisson bracket in Eq. (\ref{poisdef}), we have:
\begin{eqnarray}
\{\pi_N,\pi_i\}&=&\Bigg(\frac{\delta \pi_N}{\delta N_{}}\frac{\delta \pi_i}{\delta
\pi_N}-\frac{\delta \pi_N}{\delta
\pi_N}\frac{\delta \pi_i}{\delta N}\Bigg)+\Bigg(\frac{\delta
\pi_N}{\delta N^{i}_{}}\frac{\delta \pi_i}{\delta
\pi_i}-\frac{\delta \pi_N}{\delta
\pi_i}\frac{\delta \pi_i}{\delta N^{i}}\Bigg)\nonumber\\&+&\Bigg(\frac{\delta
\pi_N}{\delta h_{ij}}\frac{\delta \pi_i}{\delta
\pi^{ij}}-\frac{\delta \pi_N}{\delta
\pi^{ij}}\frac{\delta \pi_i}{\delta h_{ij}}\Bigg)+\Bigg(\frac{\delta \pi_N}{\delta
A}\frac{\delta \pi_i}{\delta
p_{A}}-\frac{\delta \pi_N}{\delta
p_{A}}\frac{\delta \pi_i}{\delta A}\Bigg)\nonumber\\
&+&\Bigg(\frac{\delta \pi_N}{\delta B}\frac{\delta \pi_i}{\delta
p_{B}}-\frac{\delta \pi_N}{\delta
p_{B}}\frac{\delta \pi_i}{\delta B}\Bigg)\approx0\,.
\end{eqnarray}
In a similar fashion, we can prove that: 
\begin{eqnarray}
&&\{\pi_N,\pi_N\}=\{\pi_N,\pi_i\}=\{\pi_N,p_A\}=\{\pi_N,\mathcal{H}_{N}\}=\{\pi_N,\cH_{i}\}=\{\pi_N,G_{A}\}\approx0\nonumber\\
&&\{\pi_i,\pi_i\}=\{\pi_i,p_A\}=\{\pi_i,\mathcal{H}_{N}\}=\{\pi_i,\cH_{i}\}=\{\pi_i,G_{A}\}\approx0\nonumber\\
&&\{p_A,p_A\}=\{p_A,\mathcal{H}_{N}\}=\{p_A,\cH_{i}\}\approx0\nonumber\\
&&\{\mathcal{H}_{N},\mathcal{H}_{N}\}=\{\mathcal{H}_{N},\cH_{i}\}=\{\mathcal{H}_{N},G_{A}\}\approx0\nonumber\\
&&\{\cH_{i},\cH_{i}\}=\{\mathcal{H}_{i},G_{A}\}\approx0\nonumber\\
&&\{G_{A},G_{A}\}\approx0\,.
\end{eqnarray}
 The only non-vanishing Poisson bracket on $\Ga_1$ is 
\begin{equation}
\{p_A,G_A\}=-\frac{\delta p_A}{\delta
p_{A}}\frac{\delta G_A}{\delta A}
=-\frac{\delta G_A}{\delta A}=-\sqrt{h}Nf''(A)\not\approx 0\,.
\end{equation}
Having $\{p_A,G_A\}\neq 0$ for $f''(A)\neq 0$ means that both $p_A$ and $G_A$
are \textit{second-class
constraints}.  The rest of the constraints ($\pi_N,\pi_i,\mathcal{H}_N,\mathcal{H}_{i}$)
are to be counted as \textit{first-class constraints}.

\subsubsection{Number of physical degrees of freedom in $f(R)$ gravity}
Having identified the {\it primary and secondary constraints} and categorising
them into {\it first and second-class constraints}~\footnote{Having {\it
first-class and second-class constraints} means there are no arbitrary functions
in the Hamiltonian. Indeed, a set of canonical variables that satisfies the
constraint equations determines the physical state.}, we can use the formula
in  (\ref{dofcount}) to count the number of the physical degrees of freedom.
For $f(R)$ gravity, 
we have, 
\begin{eqnarray}
2\mathcal{A}&=&2\times\{(h_{ij}, \pi^{ij}),(N,\pi_N),(N^i, \pi_i),(A,p_A),(B,p_B)\}\nonumber\\
~&=& 2(6+1+3+1+1)=24,\nonumber\\
\mathcal{B}&=&(p_{A},G_{A})=(1+1)=2,\nonumber\\
2\mathcal{C}&=&2\times(\pi_N,\pi_i,\mathcal{H}_N,H_{i})=2(1+3+1+3)=16,\nonumber\\
\mathcal{N}&=&\frac{1}{2}(24-2-16)=3\,.
\end{eqnarray}
Hence $f(R)$ gravity has $3$ physical degrees of freedom in four dimensions;
that includes the physical degrees of freedom for massless graviton and also
an extra scalar degree of freedom~\footnote{We may note that the Latin indices
are running from $1$ to $3$ and are spatial. Moreover, $(h_{ij}, \pi^{ij})$
pair is symmetric therefore we get $6$ from it.}.

Let us now briefly discuss few cases of interest:

\begin{itemize}

\item{Number of degrees of freedom for $f(R)=R+\alpha R^{2}$:\\
For a specific form of 
\begin{equation}
f(R)=R+\alpha R^{2}\,,
\end{equation}
where $\alpha=(6M^{2})^{-1}$ to insure correct dimensionality. In this case
we have, 
\begin{equation}
\{p_A,G_A\}=-\sqrt{h}Nf''(A)=-2\sqrt{h}N\not\approx 0\,.
\end{equation}
The other Poisson brackets remain zero on the constraint surface $\Ga_1$,
and hence we are left with $3$ physical degrees of freedom. These 3 degrees of are corresponding to the two polarised degrees of freedom for  massless graviton, and one scalar mode. This extra degree of freedom which is  a spin-0 particle is not a ghost and is non-tachyonic. }

\item{Number of degrees of freedom for $f(R)=R$:\\
For Einstein Hilbert action $f(R)$ is simply,
\begin{equation}
f(R)=R\,,
\end{equation} 
for which, 
\begin{equation}
\{p_A,G_A\}=-\sqrt{h}Nf''(A)\approx0\,.
\end{equation}
Therefore, in this case both $p_A$ and $G_A$ are {\it {first-class constraints}}.
Hence, now our degrees of freedom counting formula in Eq. (\ref{dofcount})
takes the following form:
\begin{eqnarray}
2\mathcal{A}&=&2\times\{(h_{ij}, \pi^{ij}),(N,\pi_N),(N^i, \pi_i),(A,p_A),(B,p_B)\}\nonumber\\
&=&2(6+1+3+1+1)=24, \nonumber\\
\mathcal{B}&=&0,\nonumber\\
2\mathcal{C}&=&2\times(\pi_N,\pi_i,\mathcal{H}_N,H_{i},p_{A},G_{A})=2(1+3+1+3+1+1)=20,\nonumber\\
\mathcal{N}&=& \frac{1}{2}(24-0-20)=2\,,
\end{eqnarray}
which coincides with that of the spin-$2$ graviton as expected
from the Einstein-Hilbert action.}
\end{itemize}
\subsection{Constraints for IDG}
The action and the ADM decomposition of IDG has been explained explicitly
so far.  In this section, we will
focus on the Hamiltonian analysis for the action of the form of Eq. (\ref{eq111}).
The first step is to consider Eq. (\ref{finaldecomposition}),
and obtain the conjugate momenta, 
\begin{eqnarray}
&&\pi_N=\frac{\p \cL}{\p \dot{N}}\approx0,\quad\pi_i=\frac{\p \cL}{\p \dot{N}^i}\approx0,
\quad \pi^{ij}=\frac{\partial\mathcal{L}}{\partial\dot{h}_{ij}}=\sqrt{h}B(K^{ij}-h^{ij}K)-\sqrt{h}\nabla_\bn
Bh^{ij},\nonumber\\
&&p_A=\frac{\partial\mathcal{L}}{\partial\dot{A}}=\sqrt{h}\Big[-(A\nabla_\bn
\chi_1  +\chi_1
\nabla_\bn A)-\sum^{\infty}_{l=2}(\chi_l
\nabla_\bn \eta_{l-1})\Big],\quad p_B=\frac{\partial\mathcal{L}}{\partial\dot{B}}=-2\sqrt{h}K,\nonumber\\
&&p_{\chi_1}=\frac{\partial\mathcal{L}}{\partial\dot{\chi}_1}=-\sqrt{h}A\nabla_\bn
A,\quad p_{\chi_l}=\frac{\partial\mathcal{L}}{\partial\dot{\chi}_l}=-\sqrt{h}(A\nabla_\bn
\eta_{l-1}),\quad\nonumber\\
&&p_{\eta_{l-1}}=\frac{\partial\mathcal{L}}{\partial\dot{\eta}_{l-1}}=-\sqrt{h}(A\nabla_\bn
\chi_l
+\chi_l
\nabla_\bn A).
\end{eqnarray}
as we can see in this case, the time derivatives of the lapse, \textit{i.e.} $\dot
N$, and the shift function, $\dot N^i$, are absent. Therefore, we have two
{\it primary constraints}, 
\begin{equation}
\pi_N\approx0, \quad \pi_i\approx0\,.
\end{equation}
The total Hamiltonian is given by:
\begin{align}
H_{tot}&=\int d^{3}x \, \cH \\ \label{wowo}
&=\int d^{3}x \, \Big(N\mathcal{H}_N+N^i\cH_i+\la^{N}\pi_{N}+\la^{i}\pi_{i}\Big)\,,
\end{align}
where $\la^N$and $\la^i$ are Lagrange multipliers and the Hamiltonian density
is given by: 
\begin{align}\label{Hamilton-IDG}
\mathcal{H}&=\pi^{ij}\dot{h}_{ij}+p_A\dot{A}+p_B\dot{B}+p_{\chi_1}\dot{\chi}_1+p_{\chi_l}\dot{\chi}_l+p_{\eta_{l-1}}\dot{\eta}_{l-1}-\mathcal{L}\\
&=N\mathcal{H}_N+N^i\cH_i\,,
\end{align}
using the above equation and after some algebra we have: 
\begin{eqnarray}\label{Hamilton-IDG1}
&&\mathcal{H}_{N}=\frac{1}{\sqrt{h}B}\pi^{ij}h_{ik}h_{jl}\pi^{kl}-\frac{1}{3\sqrt{h}B}\pi^2-\frac{\pi
p_B}{3\sqrt{h}}\\\nonumber&&+\frac{B}{6\sqrt{h}}p_B^{2}-\sqrt{h}BR+\sqrt{h}BA+2\partial_j[\sqrt{h}h^{ij}\partial_i]B\\\nonumber
&&-\frac{1}{A\sqrt{h}}p_{\chi_{1}}(p_A-\frac{\chi_1}{A}p_{\chi_1})-\frac{1}{A\sqrt{h}}\sum^{n}_{l=2}p_{\chi_{l}}(p_{\eta_{l-1}}-\frac{\chi_{l}}{A}p_{\chi_{1}})\\\nonumber
&&-\sqrt{h}\sum^{n}_{l=1}A\chi_l\eta_l-\sqrt{h}\frac{1}{2}A(M^{2}_{P}+f_{0}A+\sum^{\infty}_{n=1}f_n\eta_{n})\\\nonumber
&&-\sqrt{h}h^{ij}(A\partial_i\chi_1  \partial_j A+\chi_1
\partial_i A
\partial_j A)-\sqrt{h}h^{ij}\sum^{n}_{l=2}(A\partial_i\chi_l  \partial_j\eta_{l-1}
+\chi_l
\partial_i A
\partial_j \eta_{l-1})\,,
\end{eqnarray}
and, 
\begin{eqnarray}\label{Hamilton-IDG2}
\cH_i=-2h_{ik}
\nabla_l\pi^{kl}+p_A\partial_iA+p_{\chi_1}\partial_i\chi_1+p_B\partial_iB+\sum^{n}_{l=2}(p_{\chi_l}\partial_i\chi_l+p_{\eta_{l-1}}\partial_i\eta_{l-1})\,.
\nonumber\\
\end{eqnarray}
As described before in Eq. (\ref{secondaryconstraints1}), we can determine
the
{\it secondary constraints},  by: 
\begin{equation}
\mathcal{H}_{N}\approx0,\quad \quad \cH_{i}\approx0\,.
\end{equation}
We can also show that, on the constraint surface $\Ga_1$, the time evolutions
$\dot \cH_{N}=\{\cH_{N},\mathcal{H}_{tot}\}\approx0$ and $\dot \cH_i=\{\cH_{i},\mathcal{H}_{tot}\}\approx0$
fix the Lagrange multipliers $\la^N$ and $\la^i$,  and there will be no 
{\it tertiary constraints}.

\subsubsection{Classifications of constraints for IDG } 
 
 As we have explained earlier, {\it primary and secondary constrains} can
be classified into {\it first or second-class constraints}. This is  derived
by calculating the Poisson brackets constructed out of the constraints between
themselves and each other. Vanishing Poisson brackets indicate {\it first-class
constraint} and non vanishing Poisson bracket means we have {\it second-class
constraint}.

 For IDG action,  we have two {\it primary constraints}: $\pi_N\approx0$
and $\pi_i\approx0$,  and two {\it secondary constraints}: $\mathcal{H}_{N}\approx0$,
$H_{i}\approx0$, therefore we can determine the classification of the constraints
as:
\begin{eqnarray}
\{\pi_N,\pi_i\}&=&\Bigg(\frac{\delta \pi_N}{\delta N_{}}\frac{\delta \pi_i}{\delta
\pi_N}-\frac{\delta \pi_N}{\delta
\pi_N}\frac{\delta \pi_i}{\delta N}\Bigg)+\Bigg(\frac{\delta \pi_N}{\delta
N^{i}_{}}\frac{\delta \pi_i}{\delta
\pi_i}-\frac{\delta \pi_N}{\delta
\pi_i}\frac{\delta \pi_i}{\delta N^{i}}\Bigg) 
+\Bigg(\frac{\delta \pi_N}{\delta h_{ij}}\frac{\delta \pi_i}{\delta
\pi^{ij}}-\frac{\delta \pi_N}{\delta
\pi^{ij}}\frac{\delta \pi_i}{\delta h_{ij}}\Bigg)\nonumber\\
&+&\Bigg(\frac{\delta \pi_N}{\delta A}\frac{\delta \pi_i}{\delta
p_{A}}-\frac{\delta \pi_N}{\delta
p_{A}}\frac{\delta \pi_i}{\delta A}\Bigg)+\Bigg(\frac{\delta \pi_N}{\delta
B}\frac{\delta \pi_i}{\delta
p_{B}}-\frac{\delta \pi_N}{\delta
p_{B}}\frac{\delta \pi_i}{\delta B}\Bigg)+\Bigg(\frac{\delta \pi_N}{\delta
\chi_1}\frac{\delta \pi_i}{\delta
p_{\chi_1}}-\frac{\delta \pi_N}{\delta
p_{\chi_1}}\frac{\delta \pi_i}{\delta \chi_1}\Bigg)\nonumber\\
&+&\Bigg(\frac{\delta \pi_N}{\delta \chi_l}\frac{\delta \pi_i}{\delta
p_{\chi_l}}-\frac{\delta \pi_N}{\delta
p_{\chi_l}}\frac{\delta \pi_i}{\delta \chi_l}\Bigg)+\Bigg(\frac{\delta \pi_N}{\delta
\eta_{l-1}}\frac{\delta \pi_i}{\delta
p_{\eta_{l-1}}}-\frac{\delta \pi_N}{\delta
p_{\eta_{l-1}}}\frac{\delta \pi_i}{\delta \eta_{l-1}}\Bigg)\approx 0\,.
\end{eqnarray}
In a similar manner, we can show that: 
\begin{eqnarray}
&&\{\pi_N,\pi_N\}=\{\pi_N,\pi_i\}=\{\pi_N,\mathcal{H}_{N}\}=\{\pi_N,\cH_{i}\}\approx0\nonumber\\
&&\{\pi_i,\pi_i\}=\{\pi_i,\mathcal{H}_{N}\}=\{\pi_i,\cH_{i}\}\approx0\nonumber\\
&&\{\mathcal{H}_{N},\mathcal{H}_{N}\}=\{\mathcal{H}_{N},\cH_{i}\}\approx0\nonumber\\
&&\{\cH_{i},\cH_{i}\}\approx 0\,.
\end{eqnarray}
Therefore, all of them $(\pi_N,\pi_i,\mathcal{H}_{N},\cH_{i})$ are {\it first-class
constraints}. 
We can established that by solving the 
equations of motion for $\chi_n$ yields 
$$\eta_{1}=\Box A,~~\cdots,~~\eta_l=\Box \eta_{l-1}=\Box^{l}A, $$ for $l
\geq 2$. Therefore, we can conclude that the $\chi_n$'s are 
Lagrange multipliers, and we get the following {\it primary constraints}
from equations of motion, 
\begin{align}
\Xi_{1} = \eta_{1}-\Box A & = 0 \,, \non
\Xi_{l} = \eta_{l}-\Box \eta_{l-1}& = 0 \,,
\end{align} 
where $l \geq 2$. In fact, it is sufficient to say that $ \eta_{1}-\Box A
\approx 0$ and $\eta_{l}-\Box \eta_{l-1} \approx 0$ on a constraint surface
spanned by {\it primary and secondary constraints}, \textit{i.e.}, $(\pi_N\approx
0,~\pi_{i}\approx 0,~{\cal H}_{N}\approx 0,~{\cal H}_{i}\approx 0,~\Xi_{n}\approx
0)$.
As a result, we can now show, \begin{equation}
\{\Xi _{n},\pi_N\}=\{\Xi _{n},\pi_i\}=\{\Xi _{n},\mathcal{H}_{N}\}=\{\Xi
_{n},\mathcal{H}_{i}\}=\{\Xi _{m},\Xi _{n}\}\approx 0\,;
\end{equation}
where we have used the notation $\approx$, which is a sufficient condition
to be satisfied on the 
constraint surface defined by $\Ga_1=(\pi_N\approx 0,~\pi_{i}\approx 0,~{\cal
H}_{N}\approx 0,~{\cal H}_{i}\approx 0,~\Xi_{n}\approx 0)$, which signifies
that $\Xi _{n}$'s are now part of  {\it first-class constraints}. We should
point out that we have checked that the Poisson brackets of all possible
pairs among the constraints vanish on the constraint surface $\Ga_1$; as
a result, there are no \textit{second-class constraints}. 

\subsubsection{Physical degrees of freedom for IDG }

We can again use (\ref{dofcount}) to compute the degrees of freedom for
IDG action (\ref{eq111}). First, let us establish the
number of the configuration space variables, $\mathcal{A}$. Since the auxiliary
field
$\chi_n$ are Lagrange multipliers, they are 
not dynamical and hence redundant, as we have mentioned earlier. In contrast
we have to count the $(B,~p_b)$ pair in 
the phase space as $B$ contains intrinsic value. For the IDG action Eq.~(\ref{eq111}),
we have:
\begin{eqnarray}\label{phasespacevariables}
2\mathcal{A}&\equiv&2\times\bigg\{(h_{ij}, \pi^{ij}),(N,\pi_N),(N^i,
\pi_i),(B,p_b),(A,p_A),\underbrace{(\eta_1,p_{\eta_1}),(\eta_2,p_{\eta_2}),\cdots}_{n=1,~2,~3,\cdots
\infty}\bigg\}\nonumber\\
&\equiv&2\times(6+1+3+1+1+\infty)=24+\infty\,,
\end{eqnarray}   
 we have $(\eta_n,~p_{\eta_n})$ and for each pair we have assigned
one variable, which is multiplied by a factor of $2$ since we are dealing
with field-conjugate momentum pairs in the phase space. Moreover, as we have
found from the Poisson brackets of all possible pairs among the constraints,
the number of the {\it second-class constraints}, $\mathcal{B}$, is equal
to zero. In the next sub-sections, we will show that the \textit{correct}
number of the {\it first-class constraints} depends on the choice of $\mathcal{F}(\bar
\Box)$.

\subsection{Choice of $\mathcal{F}(\bbox)$}
\label{mimi}

In this sub-section, we will focus on an appropriate choice of ${\cal F}(\Box)$
for the action Eq.(\ref{eq111}), such that the theory admits finite relevant degrees of freedom (this is what we want to satisfy when choosing the form of ${\cal F}(\Box)$). From the Lagrangian point of view, we could
analyse the propagator of the action Eq.(\ref{eq111}). It was found in Refs.~\cite{Biswas:2005qr,Biswas:2011ar}
that $\mathcal{F}(\bbox)$ can take the following form,
\begin{equation}\label{generalfbox}
\mathcal{F}(\bbox)=M^{2}_{P}\frac{c(\bbox)-1}{\Box}\,.
\end{equation}
The  choice of $c(\bbox)$ determines how many roots we have and how many
poles are present in the graviton propagator, see Refs.~\cite{Biswas:2005qr,Biswas:2011ar,Biswas:2013kla}.
Here, we will consider two choices of $c(\bbox)$, one which has infinitely
many roots, and therefore infinite poles in the propagator. 
 For instance, we can choose 
\begin{equation}
c(\bbox)=\cos(\bbox)\,,
\end{equation}
then the equivalent action would be written as: 
\begin{equation}\label{badaction}
S_{eqv}=\frac{1}{2}\int d^{4} x \, \sqrt{-g}\bigg[M^{2}_{P}\bigg(A+A\big(\frac{\cos(\bbox)-1}{\Box}\big)A\bigg
)+B(R-A)\bigg]\,.
\end{equation}
By solving the equations of motion for $A$, and subsequently solving for
$\cos(\bbox)$
we get,
\begin{eqnarray}
\cos(\bar{k}^{2})=1-\frac{k^{2}(BM^{-2}_{P}-1)}{2A}\,,
\end{eqnarray}
where in the momentum space, we have ($\Box\rightarrow -k^2$, around Minkowski
space), and also
 note $\bar{k}\equiv k/M$; where
$B$ has mass dimension $2$. From~\eqref{lklk} in appendix~\ref{appendixc},
we have that
\be
B=M^{2}_{P}\left(1+\frac{4A}{3k^{2}}\right) \,.
\ee
Therefore, solving $\cos (\kb^2)=\frac{1}{3}$, we obtain infinitely many
solutions. We observe that there is an infinite number of solutions; hence,
there are also infinitely many degrees of freedom. These, infinitely
many solutions can be written schematically as:\  
\begin{align}\label{qwqw}
&\Psi_{1}=\Box A+a_1 A=0\,,\nonumber\\
&\Psi_{2}=\Box A+a_2 A=0\,,\nonumber\\
&\Psi_{3}=\Box A +a_{3}A=0\,,\nonumber\\
&\vdots
\end{align}
or, in the momentum space,
\begin{align}
-A k^{2}+Aa_1 =0 & \Ra k^{2}=a_{1} \,,\nonumber\\
-Ak^{2}+Aa_2 =0 & \Ra k^{2}=a_{2} \,,\nonumber\\
-Ak^{2} +Aa_3=0 & \Ra k^{2}=a_{3} \,,\nonumber\\
&\vdots
\end{align}
Now, acting the $\Box$ operators on Eq.~\eqref{qwqw}, we can write
\begin{align}
\Box \Psi_{2}&=\Box^{2}A+a_{2}\Box A \,, \non
\Box^{2} \Psi_{3}&=\Box^{3}A+a_{3}\Box^{2} A \,, \non
\vdots \non
\Box^{n-1} \Psi_{n}&=\Box^{n}A+a_{n}\Box^{n-1} A \,, \non
\vdots
\end{align}
As we saw earlier it is possible to parameterize
the terms of the form $\Box A$, $\Box^{2}A$, etc, by employing the auxiliary
fields $\chi_l$,  $\eta_{l}$, for $l \geq 1$. 
Therefore, we can write the solutions $\Psi_n$ as follows:
\begin{align}
&\Psi_{1}^{'}=\eta_{1}+a_1 A=0\,,\nonumber\\
&\Psi_{2}^{'}=\eta_{2}+a_2 \eta_{1}=0  \,,\nonumber\\
&\Psi_{3}^{'}=\eta_{3} +a_{3}\eta_{2}=0  \,.\nonumber\\
&\vdots
\end{align}
We should point out that we have acted the operator $\Box$ on $\Psi_{2}$,
the operator $\Box^2$ on $\Psi_3$, etc. in order to obtain $\Psi_{2}^{'}$,
$\Psi_{3}^{'}$, etc. As a result, we can rewrite the term $A+M_{P}^{-2}A
\cF(\bbox) A$, as
\be
A+M_{P}^{-2}A \cF(\bbox)A=a_{0}\Psi_{1}^{'}\prod_{n=2}^{\infty}\Box^{-n+1}\Psi_{n}^{'}
\,.
\ee
We would also require $\phi_n$ auxiliary fields acting like Lagrange multipliers.
Now, absorbing the powers of $M^{-2}$ into
the coefficients where appropriate, 
\begin{align}
S_{eqv}&=\frac{1}{2}\int d^{4} x \, \sqrt{-g}\bigg[M^{2}_{P}a_{0}\prod_{n=1}^{\infty}\psi_{n}+B(R-A)+\chi_{1}A(\eta_{1}-\Box
A)\non
&+\sum
^{\infty}_{l=2}\chi_lA(\eta_l-\Box \eta_{l-1})+\phi_{1}(\psi_{1}-\Psi_{1}^{'})+\sum^{\infty}_{n=2}\phi_n\big(\psi_n-\Box^{-n+1}\Psi_{n}^{'}\big)\bigg]\,,
\end{align}
where  $a_0$ is a constant and, let us define $\Phi_{1}=\psi_{1}-\Psi_{1}^{'}$
and, for $n \geq 2$, $\Phi_n=\psi_n-\Box^{-n+1}\Psi_n^{'}$. 
Then the equations of motion for $\phi_n$ will yield:
\begin{equation}\label{molo}
\Phi_n=\psi_n-\Box^{-n+1}\Psi_n^{'}=0\,.
\end{equation}
Again, it is sufficient to replace $\psi_n-\Box^{-n+1}\Psi_n^{'}=0$ with
$\psi_n-\Box^{-n+1}\Psi_n^{'}\approx 0$ satisfied at the constraint surface.
As a result there are $n$ {\it primary constraints} in $\Phi_n$.  Moreover,
by taking the equations of motion for $\chi_{n}$'s and $\phi_{n}$'s simultaneously,
we will
obtain the {\it original} action, see Eq.~\eqref{badaction}. The time evolutions
of the $\Xi_n$'s \& $\Phi_n$'s fix the corresponding Lagrange multipliers
$\la^{\Xi_n}$ \& $\la^{\Phi_n}$ in the total Hamiltonian (when we add the
terms $\la^{\Xi_n}\Xi_{n}$ \& $\la^{\Phi_n}\Phi_{n}$ to the integrand in~\eqref{wowo});
hence, the $\Xi_n$'s \& $\Phi_n$'s do not induce \textit{secondary constraints}.

Now, to classify these constraints, we can show that the following Poisson
brackets involving $\Phi_n$ on the constraint surface $(\pi_N\approx0,\pi_i\approx0,\mathcal{H}_N\approx0,\mathcal{H}_{i}\approx0,\Xi_{n}\approx0,\Phi_n\approx0)$
are satisfied\footnote{ Let us note again that $\Ga_1$ is a smooth submanifold of the
phase space determined by the {\it primary} and  {\it secondary constraints};
hereafter in this section, we shall exclusively use the ``$\approx$'' notation
to denote equality on $\Ga_1$.}:
\begin{equation}
\{\Phi_n,\pi_N\}=\{\Phi_n,\pi_i\}=\{\Phi_n,\mathcal{H}_{N}\}=\{\Phi_n,\mathcal{H}_{i}\}=\{\Phi_m,\Xi
_{n}\}=\{\Phi_m,\Phi_n\}\approx0\,,
\end{equation}
which means that the $\Phi_n$'s can be treated as {\it first-class constraints}.
We should point out that we have checked that the Poisson brackets of all
possible pairs among the constraints vanish on the constraint surface $\Ga_1$;
as a result, there are no \textit{second-class constraints}. Now, from Eq.~\eqref{phasespacevariables},
we obtain:
\begin{eqnarray}
&&2\mathcal{A}\equiv2\times\bigg\{(h_{ij}, \pi^{ij}),(N,\pi_N),(N^i,
\pi_i),(B,p_b),(A,p_A),\underbrace{(\eta_1,p_{\eta_1}),(\eta_2,p_{\eta_2}),\cdots}_{n}\bigg\}
\nonumber\\&&=2\times(6+1+3+1+1+\infty)=24+\infty\nonumber\\
&&\mathcal{B}=0,\nonumber\\
&&2\mathcal{C}\equiv2\times(\pi_N,\pi_i,\mathcal{H}_N,\mathcal{H}_{i},\Xi
_{n},\Phi_n)=2(1+3+1+3+\infty+\infty)=16+\infty+\infty,\nonumber\\
&&\mathcal{N}=\frac{1}{2}(2\mathcal{A}-\mathcal{B}-2\mathcal{C})=\infty\,.
\end{eqnarray}
As we can see a injudicious choice for $\mathcal{F}(\bbox)$ can lead to infinite
number of degrees of freedom., and there are many such examples. However,
our aim is to
come up with a concrete example where IDG will be determined solely by massless
graviton and at best one massive scalar in the context of Eq.~(\ref{eq111}).

\subsection{ ${\cal F}(e^{\bbox}) $ and finite degrees of freedom}\label{dorostfunc}

In the definition of $\mathcal{F}(\bbox)$ as given in Eq. (\ref{generalfbox}),
if 
\begin{equation}
c(\bbox)= e^{-\gamma(\bbox)},
\end{equation}
where $\gamma (\bbox)$ is an entire function, we can decompose the propagator
into partial fractions and have just one extra pole apart from the spin-2
graviton. Consequently, in order to have just one extra degree of freedom,
we have to impose conditions on the coefficient in $\mathcal{F}(\bbox)$ series
expansion (The reader may also consult Appendix \ref{formoffunction}). Moreover, to avoid $\Box^{-1}$ terms appearing in the $\mathcal{F}(\bbox)$,
we must have that, 
\begin{equation}
c(\bbox)=\sum^{\infty}_{n=0}c_n\bbox^n\,,
\end{equation}
with the first coefficient $c_0=1$, therefore: 
\begin{equation}
\mathcal{F}(\bbox)=\Big(\frac{M_P}{M}\Big)^{2}\sum^{\infty}_{n=0}c_{{n+1}}\bbox^{n}\,,
\end{equation}
Suppose we have $c(\bbox)=e^{-\bbox}$, then using Eq. (\ref{generalfbox})
we have, 
 \begin{equation}
\mathcal{F}(\bbox)=\sum^{\infty}_{n=0}f_n\bbox^n\,,
\end{equation}
where the coefficient $f_n$ has the form of, 
\begin{equation}
f_n=\Big(\frac{M_P}{M}\Big)^{2}\frac{(-1)^{n+1}}{(n+1)!}\,,
\end{equation}
Indeed this particular choice of $c(\bbox)$ is very well motivated from string
field theory~\cite{Biswas:2005qr}. In fact the above choice of $\gamma(\bbox
)=-\bbox$ contains at most one extra zero in the propagator corresponding
to one extra scalar mode in the spin-0 component of the graviton propagator~\cite{Biswas:2011ar,Biswas:2013kla}.
We rewrite the action as: 
\begin{equation}\label{goodaction}
S_{eqv}=\frac{1}{2}\int d^{4} x \, \sqrt{-g}\bigg[M^{2}_{P}\bigg(A+A\big(\frac{e^{-\bbox}-1}{\Box}\big)A\bigg
)+B(R-A)\bigg]\,.
\end{equation}
The equation of motion for $A$ is then:
\begin{equation}
M^{2}_{P}\bigg(1+2\big(\frac{e^{-\bbox}-1}{\Box}\big)A\bigg
)-B=0\,.
\end{equation}
In momentum space, we can solve the equation above:
\begin{eqnarray}\label{sol-for-k}
e^{\bar{k}^{2}}=1-\frac{k^{2}(BM^{-2}_{P}-1)}{2A}\,,
\end{eqnarray}
where  in the momentum space $\Box\rightarrow -k^2$ (on Minkowski space-time)
and also $\bar{k}\equiv
k/M$. From Eq.~\eqref{toko} in the appendix~\ref{appendixc}, we have, $e^{\kb^2}=\frac{1}{3}$,
therefore solving Eq.~(\ref{sol-for-k}), we obtain
\be
B=M^{2}_{P}\left(1+\frac{4A}{3k^{2}}\right) \,.
\ee
Note that we obtain only one extra solution (apart from the one for the massless
spin-$2$ graviton). We observe that there is a finite number of real solutions;
hence, there are also finitely many degrees of freedom. The form of the solution
can be written schematically, as: 
\begin{equation}\label{Sol-IDG}
\Omega=\Box A+b_1A=0\,,
\end{equation}
or, in the momentum space,
\be
-Ak^{2}+Ab_{1}=0 \Ra k^{2}=b_{1} \,,
\ee
Now,
we can parameterize the terms like $\Box A$, $\Box^{2}A$, etc. with the help
of auxiliary fields $\chi_l$ and $\eta_{l}$, for $l \geq 1$. Therefore, equivalently,
\be
\Omega^{'}=\eta_{1}+b_{1}A=0 \,.
\ee
Consequently, we can also rewrite the term $A \cF(\bbox) A$ with the help
of auxiliary fields $\rho$ and $\omega$. Upon taking the equations of motion
for the field $\rho$, one can recast $A+M_{P}^{-2}A \cF(\bbox)A=b_{0}\omega~{\cal
G}(A,\eta_{1},\eta_{2},\dots)$. Hence, we  can recast the action, Eq. (\ref{goodaction}),
as, 
\begin{align}
S_{eqv}&=\frac{1}{2}\int d^{4} x \, \sqrt{-g}\bigg[M^{2}_{P}b_{0}\omega~{\cal
G}(A,\eta_{1},\eta_{2},\dots)+B(R-A)+\chi_{1}A(\eta_{1}-\Box A)\non
&+\sum
^{\infty}_{l=2}\chi_lA(\eta_l-\Box
\eta_{l-1})+\rho\big(\omega-\Omega^{'}\big)\bigg]\,,
\end{align} 
where $b_0$ is a constant,  and we can now take $\rho$ as a Lagrange multiplier.
The equation of motion for $\rho$ will yield:
\begin{equation}\label{pppp}
\Theta=\omega-\Omega^{'}=0\,.
\end{equation}
Note that $\Theta =\omega-\Omega^{'}\approx 0$ will suffice on the constraint
surface determined by {\it primary and secondary constraints}
$(\pi_N\approx0,\pi_i\approx0,\mathcal{H}_N\approx0,\mathcal{H}_{i}\approx0,\Xi_{n}\approx0,\Theta\approx0)$.
As a result, $\Theta$ is a {\it primary constraint}. 
The time evolutions of the $\Xi_n$'s \& $\Theta$ fix the corresponding Lagrange
multipliers $\la^{\Xi_n}$ \& $\la^{\Theta}$ in the total Hamiltonian (when
we add the terms $\la^{\Xi_n}\Xi_{n}$ \& $\la^{\Theta}\Theta$ to the integrand
in~\eqref{wowo}); hence, the $\Xi_n$'s \& $\Theta$ do not induce \textit{secondary
constraints}.

Furthermore, the function ${\cal G}(A,\eta_{1},\eta_{2},\dots)$ contains
the root corresponding to the massless spin-$2$ graviton. Furthermore, taking
the equations of motion for $\chi_{n}$'s and $\rho$ simultaneously yields
the same equation of motion as that of in Eq.~\eqref{goodaction}. The
Poisson bracket of $\Theta$ with other constraints will give rise to
\begin{equation}
\{\Theta,\pi_N\}=\{\Theta,\pi_i\}=\{\Theta,\mathcal{H}_{N}\}=\{\Theta,\mathcal{H}_{i}\}=\{\Theta,\Xi
_{n}\}=\{\Theta,\Theta\}\approx0\,,
\end{equation}
where $\approx$ would have been sufficient. This leads to $\Theta$ as a {\it
first-class constraint}. Hence, we can calculate the number of the physical
degrees of freedom  as: 
\begin{eqnarray}
&&2\mathcal{A}\equiv2\times\bigg\{(h_{ij}, \pi^{ij}),(N,\pi_N),(N^i,
\pi_i),(B,p_b),(A,p_A),\underbrace{(\eta_1,p_{\eta_1}),(\eta_2,p_{\eta_2}),\cdots}_{n}\bigg\}
\nonumber\\&&=2\times(6+1+3+1+1+\infty)=24+\infty\nonumber\\
&&\mathcal{B}=0,\nonumber\\
&&2\mathcal{C}\equiv2\times(\pi_N,\pi_i,\mathcal{H}_N,\mathcal{H}_{i},\Xi
_{n},\Theta)=2(1+3+1+3+\infty+1)=18+\infty,\nonumber\\
&&\mathcal{N}=\frac{1}{2}(2\mathcal{A}-\mathcal{B}-2\mathcal{C})=\frac{1}{2}(24+\infty-0-18-\infty)=3\,.
\end{eqnarray}
This gives $2$ degrees of freedom from the massless spin-$2$ graviton in
addition to an extra degree of freedom as expected from the propagator analysis. This extra degree of freedom which is not a ghost or tachyon comes from the choice of $\mathcal{F}(\bbox)$, in fact we have that $\mathcal{F}(\bbox)=M^2_P\frac{e^{-\bbox}-1}{\Box}$ and this function admits one pole and hence one degree of freedom. This is shown explicitly on Appendix \ref{appendixc}.
Moreover, note that the choice of $\mathcal{F}(\bbox)$ is such that we avoid $\Box^{-1}$ terms. 
\section{Summary}
In this chapter we used Hamiltonian analysis to study the number of the degrees
of freedom for an infinite derivative theory of gravity (IDG). In this gravitational modification,
IDG contains infinite number of covariant derivatives acting on the Ricci
scalar. 

In Lagrangian framework, the number of the degrees of freedom is determined
from the propagator analysis. Particularly, it hinges on the number of the
poles arising in the propagator. The results of this chapter support the
original idea that both Lagrangian and
Hamiltonian analysis will yield similar conclusions for infinite derivative
theories with Gaussian kinetic term~\cite{Biswas:2005qr}.   In case of IDG,
one can study the 
scalar and the tensor components of the propagating degrees of freedom~\cite{Biswas:2011ar,Biswas:2013kla},
and for Gaussian kinetic term which determines
 ${\cal F}(\bbox)$, there are only $2$ dynamical degrees of freedom. In order
 to make sure that there are no poles other than the original
poles (corresponding to the original degrees of freedom ) in the propagator,
one shall 
demand that the propagator be suppressed by {\it exponential of an entire
function}. An entire function does not have any poles in the finite complex
plane This choice of propagator determines the kinetic
term in Lagrangian for infinite derivative theories. For a scalar toy model
the kinetic term becomes Gaussian, \textit{i.e.}, ${\cal F}=\Box e^{-\bbox}$,
while in gravity it becomes $ {\cal F}=M_{P}^{2}\Box^{-1}(e^{-\bbox}-1)$.
 
 From the Hamiltonian perspective, the essence of finding the dynamical degrees
of freedom relies primarily on finding the total configuration space variables,
and
{\it first} and {\it second-class constraints}. As expected, infinite derivative
theories  will have infinitely many configuration space variables, and so
will be
{\it first} and {\it second-class constraints}. However,  for a Gaussian
kinetic term, ${\cal F}=M_{P}^{2}\Box^{-1}(e^{-\bbox}-1)$, the degrees of freedom are finite and the gravitational action we considered admits 3 relevant degrees of freedom, two for the massless graviton and an extra scalar mode that is not ghost or tachyonic.
\chapter{Boundary terms for higher derivative theories of gravity}\label{chahar}

In this chapter we wish to find the corresponding Gibbons-Hawking-York
term for the most general quadratic in curvature gravity
by using Coframe slicing within the ADM decomposition
of space-time in four dimensions. 

Irrespective of classical or quantum computations, one of the key features
of a covariant action is to have a well-posed 
boundary condition. In particular, in the Euclidean path integral approach
- requiring such an action to be stationary, one also requires all the boundary
terms to 
disappear on any permitted variation. Another importance of boundary terms manifests itself in  calculating the black
hole entropy using the Euclidean semiclassical approach, where the entire
contribution comes from the boundary term~\cite{Brown:1995su,hawking1,Hawking:1995fd}.

It is well known that the variation of the EH action leads
to a boundary term that depends
not just on the metric, but also on the derivatives of the metric. This is
due to
the fact that the action itself depends on the metric, along with terms that
depend linearly on the second derivatives.
Normally, in Lagrangian field theory, such linear second
derivative terms can be introduced or eliminated, by adding an  appropriate
boundary term to the action. 
In gravity, the fact that the second derivatives arise linearly
and also the existence of total derivative indicates that the second derivatives
are redundant in the sense that they can be eliminated by  integrating
by parts, or by adding an appropriate boundary  term. Indeed,
writing a boundary term for a gravitational action schematically confines
the non-covariant terms to the boundary.  For the EH action
this geometrically
transparent, boundary term is given by the Gibbons-Hawking-York (GHY) boundary
term
\cite{Gibbons:1976ue}. Adding this boundary to the bulk action results in
an elimination of the total derivative, as seen for $f(R)$ gravity~\cite{Guarnizo:2010xr,Madsen:1989rz}.

In the Hamiltonian formalism, obtaining the boundary terms for a gravitational
action is vital. This is due to the fact that the boundary term ensures that
the path integral for 
quantum gravity admits correct answers. As a result, ADM  showed \cite{Arnowitt:1962hi}
that upon decomposing space-time such that for the four dimensional Einstein
equation
we have three-dimensional surfaces (later to be defined as hypersurfaces)
and one fixed time coordinate for each slices. We can therefore formulate
and recast
the Einstein equations in terms of the Hamiltonian and hence achieve a better
insight into GR.

In the ADM decomposition, one foliates the arbitrary region $\mathcal{M}$
of the space-time manifold with a family of spacelike hypersurfaces $\Sigma_{t}$,
one for each instant in time. It has been shown by the authors of~\cite{Deruelle:2009zk}
that one can decompose a gravitational
action, using the ADM formalism and without necessarily moving into the Hamiltonian
regime, such that we obtain the total derivative of the gravitational action.
Using this powerful technique, 
one can eliminate this total derivative 
term by modifying the GHY term appropriately.

The aim of this chapter is to find the corresponding GHY boundary term for
a covariant IDG.
We start by providing a 
warm up example on how to obtain a boundary term for an infinite derivative,
massless scalar field theory. We then   briefly review
the boundary term for EH term and introduce infinite
derivative gravity. We shall  set our preliminaries by discussing the time
slicing  and reviewing how one may obtain the boundary
terms by using the $3+1$ formalism. We finally turn our attention to our gravitational action
and find the appropriate boundary terms for such theory.


\section{Warming up: Infinite derivative massless scalar field
theory}\label{sec2}

Let us consider the following action of a generic scalar field $\phi$ of
mass dimension 2:
\begin{eqnarray}\label{eq1}
S_\phi= \int d^{4} x  \, \phi \Box^{n} \phi,
\end{eqnarray}
where $\Box=\eta^{\mu\nu}\nabla_\mu\nabla_\nu$, where $\eta_{\mu\nu}$ is
the Minkowski metric~\footnote{The $\Box$ term comes with a scale $\Box/M^2$.
In our notation, we  suppress the scale $M$  in order not
to clutter our formulae for the rest of this chapter. } and $n\in \mathbb{N}_{>0}$.
Generalising, we have that $\Box^{n}=\prod_{i=1}^{n} \eta^{\mu_{i}\nu_{i}}\nabla_{\mu_{i}}\nabla_{\nu_{i}}$.
The aim is to find the total derivative term for the above action.
We may vary the scalar field $\phi$ as: $\phi \rightarrow\ \phi + \delta
\phi$.
Then
the variation of the action is given by
\begin{eqnarray}\label{eq2}
\delta S_\phi&=&\int d^{4} x \, \big[\delta\phi \Box^{n} \phi+\phi \delta(\Box^{n}
\phi)\big]\,,\nonumber\\
&=&\int d^{4} x \, \big[\delta\phi \Box^{n} \phi+\phi\Box^{n}
 \delta\phi\big]\,,\nonumber\\
 &=&\int d^{4} x \, \big[(2\Box^{n}\phi)\delta\phi+X\big]\,,
\end{eqnarray}
where now $X$ are the $2n$ total derivatives:
\begin{eqnarray}\label{eq3}
X&=&\int d^{4} x \, \big[\nabla_\mu(\phi\nabla^\mu\Box^{n-1}\delta\phi)-\nabla^\mu(\nabla_\mu\phi\Box^{n-1}\delta\phi)\nonumber\\
&+&\nabla_\lambda(\Box\phi\nabla^\lambda\Box^{n-2}\delta\phi)-\nabla^\lambda(\nabla_\lambda\phi\Box^{n-2}\delta\phi)+\cdots
\nonumber\\&+&\nabla_\sigma(\Box^{n-1}\phi\nabla^\sigma\delta\phi)-\nabla^\sigma(\nabla_\sigma\Box^{n-1}\phi\delta\phi)\big].
\end{eqnarray}
where ``$\cdots$'' in the above equation indicates the intermediate terms.

Let us now consider a more general case
\begin{eqnarray}\label{eq4}S_{\phi}=\int d^4x \, \phi\mathcal{F}(\Box)\phi,\end{eqnarray}
where $\mathcal{F}(\Box)=\sum^{\infty}_{n=0}c_n\Box^n$, where the `$c_n$'s
are dimensionless coefficients. In this case
the total derivatives are given by
\begin{eqnarray}\label{eq5}
X=\sum^{\infty}_{n=1}c_n\int d^4x\sum^{2n}_{j=1}(-1)^{j-1}\nabla_\mu(\nabla^{(j-1)}\phi\nabla^{(2n-j)}\delta\phi),
\end{eqnarray}
where the superscript $\nabla^{(j)}$ indicate the number of covariant derivatives
acting to the right. Therefore, one can always determine the total derivative
for any given action, and one can then preserve or eliminate these terms
depending on the purpose of the study. 
In the following sections we wish to address how one can obtain the total
derivative for a given gravitational action.


\section{Introducing Infinite Derivative Gravity}\label{sec3}

The gravitational action is built up of two main components, the bulk part
and the boundary part. In the simplest and the most well known
case~\cite{Gibbons:1976ue}, for the EH action, the boundary
term are the ones known as Gibbons-Hawking-York (GHY) term. We can write
the total EH action in terms of the bulk part and the boundary part simply
as, (See Appendix \ref{ehboundarytermderivation} for derivation), 
\begin{eqnarray}\label{eq6}
S_G&=&S_{EH}+S_{B}\nonumber \\
&=&\frac{1}{16\pi G_{N}}\int_{\mathcal{M}} d^{4}x \, \sqrt{-g}  \mathcal{R}+
\frac{1}{8\pi G_{N}}\oint_{\partial\mathcal{M}}
d^{3}y \, \varepsilon\left\vert h \right\vert^{1/2}K\,,  
\end{eqnarray}
where $\mathcal{R}$ is the Ricci-scalar, and  $K$ is the trace of the extrinsic
curvature with $K_{ij}\equiv -\nabla_{i}n_{j}$, $\mathcal{M}$ indicates the
4-dimensional region and $\partial\mathcal{M}$ 
denotes the $3$-dimensional boundary region. $h$ is the determinant of the
induced metric on the hypersurface $\partial\mathcal{M}$ and $\varepsilon=n^\mu
n_\mu=\pm1$, 
where $\varepsilon$ is equal to $-1$ for a spacelike hypersurface,  and is
equal to $+1$ for a timelike hypersurface when we take the metric signature
is ``mostly plus''; 
\textit{i.e.} $(-,+,+,+)$. A unit normal $n_{\mu}$ can be introduced only if the hypersurface
is not null, and $n^\mu$ is the normal vector to the hypersurface.

Indeed, one 
can derive the boundary term simply by using the variational principle. In
this case the action is varied with respect to the 
metric, and it produces a total-divergent term, which can be eliminated by
the
variation of $S_B$, \cite{Gibbons:1976ue}. Finding the boundary terms for
any action is an indication that the variation principle for the given theory
is well posed.

As mentioned earlier on, despite the many successes that the EH action brought
in understanding the universe in IR regime, the UV sector of gravity requires
corrections 
to be well behaved. We shall recall the most general {\it covariant} action
of gravity, which
is quadratic in curvature, 
\begin{eqnarray}\label{eq7}
S&=&S_{EH}+S_{UV}\nonumber\\
&=&\frac{1}{16\pi G_{N}}\int d^{4}x \, \sqrt{-g}\Big[\mathcal{R}+\alpha\big(\mathcal{R}\mathcal{F}_{1}(\Box)\mathcal{R}
+\mathcal{R}_{\mu\nu}\mathcal{F}_{2}(\Box)\mathcal{R}^{\mu\nu}\nonumber\\
&&+\mathcal{R}_{\mu\nu\rho\sigma}\mathcal{F}_{3}(\Box)\mathcal{R}^{\mu\nu\rho\sigma}\big)\Big],\quad
\text{with} \quad \mathcal{F}_{i}(\Box)=\sum^{\infty}_{n=0}f_{i_{n}}\Box^{n}
\,,
\end{eqnarray}
where $\alpha$ is a constant with mass dimension $-2$ and the `$f_{i_{n}}$'s
are dimensionless coefficients. 
For the full equations of motion of such an action, see~\cite{Biswas:2013cha}.
The aim of this chapter is to seek
the boundary terms corresponding to $S_{UV}$, while retaining the Riemann
term. 

\section{Time Slicing}\label{sec4}

Any geometric space-time can be recast in terms of time like spatial slices,
known as hypersurfaces.
How these slices are embedded in space-time, determines the extrinsic curvature
of the slices. One of the motivations of time slicing is to evolve the 
equations of motion from a well-defined set of initial conditions set at
a well-defined spacelike 
hypersurface, see~\cite{Smarr:1977uf,Baumgarte:2002jm}.

\subsection{ADM Decomposition}\label{sec4.1}

In order to define the decomposition, we first look at the foliation.
Suppose that the time orientable space-time $\mathcal{M}$ is foliated by a
family of spacelike
hypersurfaces $\Sigma_{t}$, on which time is a fixed constant $t=x^{0}$.
We then define the induced metric on the hypersurface as $h_{ij}\equiv \left.g_{ij}\right
|_{_t}$,
where the Latin indices run from $1$ to $3$~\footnote{ It should also be
noted that Greek indices run from $0$ to $3$ and Latin indices run from $1$
to $3$, that is, only spatial coordinates are considered.}. Let us remind
the  line element
as given in section \ref{admsec},~\cite{Relativiststoolkit}:
\begin{eqnarray} \label{eq8}
ds^{2}=-(N^{2}-\beta_{i}\beta^{i})dt^{2}+2\beta_{i}dx^{i}dt+h_{ij}dx^{i}dx^{j}
\end{eqnarray}
where $$N=\frac{1}{\sqrt{-g^{00}}}$$  is the ``lapse'' function, and $$\beta^{i}=-\frac{g^{0i}}{g^{00}}$$
is  the ``shift'' vector.

In the above line element Eq.~(\ref{eq8}), we also have $\sqrt{-g}=N\sqrt{h}$.
 The induced metric of
the hypersurface can be related to the $4$ dimensional full metric via the
completeness relation, where, for a spacelike hypersurface,
\begin{eqnarray}\label{eq11}
g^{\mu\nu}&=&h^{ij}e^{\mu}_{i}e^{\nu
}_{j}+\varepsilon n^{\mu}n^{\nu}\nonumber\\
&=& h^{\mu \nu}-n^{\mu}n^{\nu} \,,
\end{eqnarray}
where $\varepsilon=-1$ for a spacelike hypersurface, and $+1$ for a timelike
hypersurface, and
\begin{equation}\label{eee}
e^\mu_i=\frac{\partial x^{\mu}}{\partial y^{i}}\,,
\end{equation}
are basis vectors on the hypersurface which allow us to define tangential
tensors on the
  hypersurface\footnote{We can use $h^{\mu\nu}$ to project a tensor $A_{\mu\nu}$
onto the hypersurface: $A_{\mu\nu} e^\mu_i e^\nu_j = A_{ij}$ where $A_{ij}$
is the three-tensor associated with $A_{\mu\nu}$.}. We note $`x$'s are coordinates
on region $\mathcal{M}$, while $`y$'s are coordinates associated with the
hypersurface and we may also keep in mind that,
\begin{equation}\label{eq12}
h^{\mu\nu}=h^{ij}e^{\mu}_{i}e^{\nu
}_{j}\,,
\end{equation}
where $h^{ij}$ is the inverse of the induced metric $h_{ij}$ on the hypersurface.

The change of direction
of the normal $n$ as one moves on the hypersurface corresponds to the bending
of the hypersurface $\Sigma_{t}$  which is described by the extrinsic curvature.
The extrinsic curvature of spatial slices where time is constant is given
by:
\begin{eqnarray}\label{eq13}
K_{ij}\equiv -\nabla_{i}n_{j}=\frac{1}{2N}\left(D_{i}\beta_{j}+D_{j}\beta_{i}-\partial_{t}h_{ij}\right)\,,
\end{eqnarray}
where $D_{i}=e^{\mu}_{i}\nabla_\mu$ is the intrinsic covariant derivative
associated with the induced metric defined on the hypersurface, and $e^{\mu}_{i}$
is the appropriate basis vector which is used to transform bulk indices to
boundary ones.

Armed with this information, one can write down the Gauss, Codazzi and Ricci
equations, see \cite{Deruelle:2009zk}:
\begin{eqnarray}\label{eq14}
\mathcal{R}_{ijkl}&\equiv&K_{ik}K_{jl}-K_{il}K_{jk}+R_{ijkl}\,, \\ \label{xx1}
\mathcal{R}_{ijk\mathbf{n}}&\equiv&n^{\mu}\mathcal{R}_{ijk\mu}=-D_{i}K_{jk}+D_{j}K_{ik}\,,
\\ \label{xxx2}
\mathcal{R}_{i\mathbf{n}j\mathbf{n}}&\equiv&n^{\mu}n^{\nu}\mathcal{R}_{i\mu
j\nu}\nonumber\\&=&N^{-1}\big(\partial_{t}K_{ij}-\mathsterling_\beta
K_{ij}\big)+K_{ik}K^{\ k}_{j}+N^{-1}D_i D_jN\,,
\end{eqnarray}
where   in the left hand side of Eq.~(\ref{eq14}) we have the bulk Riemann
tensor, but where all indices are now {\it spatial} rather than both spatial
and temporal,
and $R_{ijkl}$ is the Riemann tensor constructed purely out of $h_{ij}$,
\textit{i.e.} the metric associated with the hypersurface; and $\pounds_\beta$ is
the Lie derivative with respect to shift\footnote{We
have $\pounds_\beta K_{ij}\equiv \beta^{k}D_k K_{ij}+K_{ik}D_j\beta^k+K_{jk}D_i\beta^k$.}.


\subsubsection{Coframe Slicing}\label{sec4.2}

A key feature of the 3+1 decomposition is the free choice of lapse function
and
shift vector which define the choice of foliation at the end. In this study
we stick to the {\it coframe slicing}.
The main advantage for this choice of slicing is the fact that the line element
and therefore components of the infinite derivative function in our gravitational
action will be simplified greatly. In addition, \cite{Anderson:1998cm} has
shown that such a slicing has a more transparent form of the canonical action
principle and Hamiltonian 
dynamics for gravity. This also leads to a well-posed initial-condition for
the evolution of the gravitational constraints in a vacuum by satisfying
the Bianchi identities. 
In order to map the ADM line element into the coframe slicing, we use the
convention of~\cite{Anderson:1998cm}. We define
\begin{eqnarray} \label{eq15}
\theta^{0}&=&dt\,,\nonumber\\
\theta^{i}&=&dx^{i}+\beta^{i}dt\,,
\end{eqnarray}
where $x^{i}$ and $i=1,2,3$ is the spatial and $t$ is the time coordinates.\footnote{We
note that in Eq.~(\ref{eq15}), the $``i"$ for $\theta^{i}$ is just a superscript
not a spatial index.}    The metric in the coframe takes the following form
\begin{eqnarray}
\label{eq16}
ds_{\mathrm{coframe}}^{2}=g_{\alpha\beta}\theta^{\alpha}\theta^{\beta}=-N^{2}(\theta^{0})^{2}+g_{ij}\theta^{i}\theta^{j}\,,
\end{eqnarray}
where upon substituting Eq.~(\ref{eq15}) into Eq.~(\ref{eq16})
we recover the original ADM metric given by Eq.~\eqref{eq8}. In this convention,
if we
take $\mathbf{g}$ as
the full space-time metric,
we have the following simplifications:
\begin{eqnarray}\label{eq19}
g_{ij}=h_{ij},\quad g^{ij}=h^{ij},\quad g_{0i}=g^{0i}=0.
\end{eqnarray}
 The convective derivatives $\partial_{\alpha}$ with respect to $\theta^{\alpha}$
are
\begin{align}\label{eq20}
\partial_{0}&\equiv \frac{\partial}{\partial t}-\beta^{i}\partial_{i} \,,
\nonumber\\
\partial_{i}&\equiv \frac{\partial}{\partial x^i}\,.
\end{align}
For time-dependent space tensors $T$, we can define the following derivative:
\begin{equation}\label{eq21}
\bar{\partial}_{0}\equiv \frac{\partial}{\partial t}- \pounds_{\beta}\,,
\end{equation}
where $\pounds_{\beta}$ is the Lie derivative with respect to the shift vector
$\beta^i$. This is because the off-diagonal components of the coframe metric
are zero, {\it i.e.}, $g_{0i}=g^{0i}=0$. 

We shall see later on how this time slicing helps us to simplify the calculations
when the gravitational action contains infinite derivatives.

\subsubsection{Extrinsic Curvature}\label{sec4.2.1}

A change in the choice of time slicing results
in a change of the evolution of the system. The choice of foliation also
has a direct impact on the form of the extrinsic curvature. In this section
we wish to give the form of extrinsic curvature $K_{ij}$ in the coframe slicing.
This is due to the fact that the definition of the extrinsic curvature is
an initial parameter
that describes the evolution of the system, therefore is it logical for us
to derive the extrinsic curvature in the coframe slicing as we use it throughout
the chapter.
We use \cite{Anderson:1998cm} to find the general definition for $K_{ij}$
in the coframe metric. 
In the coframe, 
\begin{align}\label{eq22}
  {\gamma^\alpha}_{\beta\gamma} &= {\Gamma^\alpha}_{\beta\gamma} + g^{\alpha\delta}
{C^\epsilon}_{\delta(\beta} g_{\gamma)\epsilon} - \frac{1}{2} {C^\alpha}_{\beta\gamma}\,,
\\ \label{eq23}
        d \theta^\alpha &= - \frac{1}{2} {C^\alpha}_{\beta\gamma} \theta^\beta
\wedge \theta^\gamma \,,
\end{align}
where $\Gamma$ is the ordinary Christoffel symbol and ``$\wedge$'' denotes
 the exterior or wedge product of vectors $\theta$. By finding the coefficients
$C$s 
and subsequently calculating the  connection coefficients $\gamma^{\alpha}_{\beta\gamma}$,
one can extract the extrinsic curvature $K_{ij}$ in the coframe setup. We
note that the expression
for $d \theta^\alpha$ is the Maurer-Cartan
structure equation~\cite{Hassani}. It is derived from the canonical 1-form
$\theta$
on a Lie group $G$ which is the left-invariant $\mathfrak{g}$-valued 1-form
uniquely determined by $\theta(\xi)=\xi$ for all $\xi\in \mathfrak{g}$. 

We
can use differential
forms (See Appendix \ref{A}) to calculate the $C$s,
the coefficients of $d \theta$ where now we can write, 
\begin{eqnarray}\label{eq24}d\theta^{k}=- \left( \partial_i \beta^k \right)
\theta^0
\wedge \theta^i + \frac{1}{2} {C^k}_{ij} \theta^j \wedge \theta^i \,,\end{eqnarray}
where $k=1,2,3$. Now when we insert the $C$s from Appendix \ref{A}, 
\begin{eqnarray}\label{eq25} d \theta^1 = d \left( d x^1 + \beta^1 dt \right)
= d \beta^1
\wedge dt \end{eqnarray}
and 
\begin{eqnarray}\label{eq26}d \theta^0 &=& d(dt) = d^{2}(t) = 0 \nonumber\\
        d \theta^i &=&  d\beta^{i} \wedge dt. \end{eqnarray}
From the definition of $d \theta^\alpha$ in Eq.~(\ref{eq23}) and using the
antisymmetric properties
of the $\wedge$ product,
\begin{eqnarray}\label{eq27}
        d \theta^\alpha = - \frac{1}{2} {C^\alpha}_{\beta\gamma} \theta^\beta
\wedge \theta^\gamma 
        = \frac{1}{2} {C^\alpha}_{\beta\gamma} \theta^\gamma
\wedge \theta^\beta = \frac{1}{2} {C^\alpha}_{\gamma\beta} \theta^\beta \wedge
\theta^\gamma,
\end{eqnarray}
we get
\begin{eqnarray}\label{eq28}
        {C^\alpha}_{\beta\gamma} =- {C^\alpha}_{\gamma\beta}\,. 
\end{eqnarray}        
Using these properties, we find that ${C^m}_{0i}
= \frac{\partial \beta^m}{\partial x^i}$, ${C^m}_{ij} = 0$ and $C^{0}{}_{ij}=0$.
Using Eq.~\eqref{eq22}, we obtain that
\begin{equation}
\gamma_{ij}^{0}=-\frac{1}{2N^2}\Big(h_{il}\partial_{j}(\beta^{l})+h_{jl}\partial_{i}(\beta^{l})-\bar
\partial_{0}h_{ij}\Big)\,.
\end{equation} 
Since from Eq.~(\ref{eq13})
\begin{eqnarray}
K_{ij}\equiv - \nabla_{i}n_{j}=\gamma_{ij}^{\mu}n_{\mu}=-N\gamma_{ij}^{0}\,,
\end{eqnarray}
the expression for the extrinsic
curvature in coframe slicing is given by:
\begin{eqnarray}\label{eq29}
K_{ij}=\frac{1}{2N}\Big(h_{il}\partial_{j}(\beta^{l})+h_{jl}\partial_{i}(\beta^{l})-\bar
\partial_{0}h_{ij}\Big)\,,
\end{eqnarray}
where $\partial_{0}$ is the time derivative and $\beta^l$ is the ``shift''
in the coframe metric Eq.~(\ref{eq16}).

\subsubsection{Riemann Tensor in the Coframe}

The fact that we move from the ADM metric into the coframe slicing has the
following implication on the form of the components of the Riemann tensor.
Essentially, since in the coframe slicing in Eq.~(\ref{eq16}) we have $g^{0i}=g_{0i}=0$,
therefore we also have, $n^i=n_i=0$. Hence the non-vanishing components
of the Riemann tensor in the coframe, namely Gauss, Codazzi and Ricci tensor,
become: \begin{eqnarray}\label{eq30}
\mathcal{R}_{ijkl}&=&K_{ik}K_{jl}-K_{il}K_{jk}+R_{ijkl}\,,\nonumber \\
\mathcal{R}_{0ijk}&=&N(-D_{k}K_{ji}+D_{j}K_{ki})\,, \nonumber
\\
\mathcal{R}_{0i0j}&=&N(\bar\partial_{0}K_{ij}+NK_{ik}K^{\ k}_{j}+D_i D_jN\,),
\end{eqnarray}
with $\bar\partial_0$ defined in Eq.~(\ref{eq21}) and $D_{j}=e^{\mu}_{j}\nabla_{\mu}$.
It can be seen that the Ricci equation, given in Eq.~(\ref{eq14}) is simplified
in above due to the definition of Eq.~(\ref{eq21}).
We note that Eq.~(\ref{eq30}) is in the coframe slicing, while Eqs. (\ref{eq14}-\ref{xxx2})
are in the ADM frame only.

\subsubsection{D'Alembertian Operator in Coframe}\label{sec4.2.2}

Since we shall be dealing with a higher-derivative theory of gravity,
it is therefore helpful to first obtain an expression for the $\Box$ operator
in this subsection.
  To do so,
we start off by writing the definition of a single box operator in the coframe,
\cite{ChoquetBruhat:1996ak}, 

\begin{eqnarray}\label{boxcoframe}
\Box&=&g^{\mu \nu}\nabla_{\mu}\nabla_{\nu}\nonumber\\
&=& (h^{\mu \nu}+\varepsilon n^{\mu}n^{\nu})\nabla_{\mu}\nabla_{\nu} \nonumber
\\
&=&-n^{\mu}n^{\nu}\nabla_{\mu}\nabla_{\nu}+h^{\mu \nu}\nabla_{\mu}\nabla_{\nu}\nonumber\\
&=&-n^{0}n^{0}\nabla_{0}\nabla_{0}+h^{ij}e^{\mu}_{i}e^{\nu
}_{j}\nabla_{\mu}\nabla_{\nu}\nonumber\\
&=& -\frac{1}{N^2}\nabla_{0}\nabla_{0}+h^{ij}D_{i}D_{j} \nonumber \\
&=&-(N^{-1}\bar{\partial_{0}})^{2}+\Box_{hyp}\,,
\end{eqnarray}
where we note that the Greek indices run from $1$ to $4$ and the Latin indices
run from
1 to $3$ ($\varepsilon=-1$ for a spacelike hypersurface). We call the spatial
box operator $\Box_{hyp}=h^{ij}D_{i}D_{j}$, which stands for
``hypersurface'' as the spatial coordinates are defined on the hypersurface
meaning $\Box_{hyp}$ is the projection of the covariant d'Alembertian operator
down to the hypersurface, \textit{i.e.} only the tangential components of the covariant
d'Alembertian operator are encapsulated by $\Box_{hyp}$. Also note that in
the coframe slicing $g^{ij}=h^{ij}.$ Generalising this result to the $n$th
power,  for our purpose, we get 
\begin{eqnarray}\label{4.18}
\mathcal{F}_{i}(\Box)=\sum_{n=0}^{\infty}f_{i_{n}}\left[-(N^{-1}\bar{\partial_{0}})^{2}+\Box_{hyp}\right]^{n}
\end{eqnarray} 
where the $f_{i_{n}}$s are the coefficients of the series.

\section{Generalised  Boundary Term}\label{sec5}

In this section, first we are going to briefly summarise the method of~\cite{Deruelle:2009zk}
for finding the boundary term. It has been shown that, given a general gravitational
action 
\begin{eqnarray}\label{5.1}S=\frac{1}{16\pi G_{N}}\int_{\mathcal{M}}d^{4}x \,
\sqrt{-g}f(\mathcal{R}_{\mu\nu\rho\sigma})\,,
\end{eqnarray}
one can introduce two auxiliary fields $\varrho_{\mu\nu\rho\sigma}$ and $\varphi^{\mu\nu\rho\sigma}$,
which are independent of each other and of the metric $g_{\mu\nu}$, while
they have all the symmetry properties of the Riemann tensor $\mathcal{R}_{\mu\nu\rho\sigma}$.
We can then write down the following equivalent action:
\textcolor{red}{}\begin{eqnarray}\label{5.2}
S=\frac{1}{16\pi G_{N}}\int_{\mathcal{M}}d^{4}x \,\sqrt{-g}\left[f(\varrho_{\mu\nu\rho\sigma})
+\varphi^{\mu\nu\rho\sigma}\left(\mathcal{R}_{\mu\nu\rho\sigma}-\varrho_{\mu\nu\rho\sigma}\right)\right]\,.
\end{eqnarray} 
 The reason we introduce these auxiliary fields is that the second derivatives
of the metric appear only linearly in~Eq.~\eqref{5.2}. Note that in Eq.~\eqref{5.2},
the terms involving the second derivatives of the metric are not  multiplied
by terms of the same type, \textit{i.e.} involving the second derivative of the metric,
so when we integrate by parts once, we are left just with the first derivatives
of the metric; we cannot eliminate the first derivatives of the metric as
well - since in this study we are keeping the boundary terms. Note that 
 the first derivatives of the metric are actually contained in these boundary
terms if we integrate by parts twice, see our toy model scalar field theory
example~in Eqs.~(\ref{eq2},\ref{eq3})~\footnote{This is because $\varrho_{\mu\nu\rho\sigma}$
and $\varphi^{\mu\nu\rho\sigma}$ 
are independent of the metric, and so although $f(\varrho_{\mu\nu\rho\sigma})$
can contain derivatives of $\varrho_{\mu\nu\rho\sigma}$, these are not derivatives
of the metric. 
$\mathcal{R}_{\mu\nu\rho\sigma}$ contains a second derivative of the metric
but this is the only place where a second derivative of the metric appears
in Eq.~\eqref{5.2}}.
Therefore,  terms which are linear in the metric can be eliminated if we
integrate by parts; moreover, the use of the auxiliary fields can prove useful
in a future Hamiltonian analysis of the action. 

From~\cite{Deruelle:2009zk}, we then decompose the above expression as 
\begin{equation}
\varphi^{\mu\nu\rho\sigma}\left(\mathcal{R}_{\mu\nu\rho\sigma}-\varrho_{\mu\nu\rho\sigma}\right)
=\phi^{ijkl}(\mathcal{R}_{ijkl}-\rho_{ijkl})-4\phi^{ijk}(\mathcal{R}_{ijk\mathbf{n}}-\rho_{ijk})-2\Psi^{ij}(\mathcal{R}_{i\mathbf{n}j\mathbf{n}}-\Omega_{ij})\,,
\end{equation}
where
\begin{eqnarray}\label{tensors1}
\mathcal{R}_{ijkl}\equiv\rho_{ijkl}\equiv\varrho_{ijkl},\qquad \mathcal{R}_{ijk\mathbf{n}}\equiv\rho_{ijk}\equiv
n^{\mu}\varrho_{ijk\mu},\qquad\mathcal{R}_{i\mathbf{n}j\mathbf{n}}\equiv\Omega_{ij}\equiv
n^{\mu}n^{\nu}\varrho_{i\mu j\nu}\nonumber\\
\end{eqnarray}
are equivalent to the components of the Gauss, Codazzi and Ricci equations
given in~Eq.~\eqref{eq14}, also,
\begin{eqnarray}
\phi^{ijkl}\equiv \varphi^{ijkl},\qquad \phi^{ijk}\equiv n_{\mu}\varphi^{ijk\mu},\qquad
\Psi^{ij}\equiv -2n_\mu n_\nu\varphi^{i\mu j\nu},
\end{eqnarray}
where $\phi^{ijkl}$, $\phi^{ijk}$ and $\Psi^{ij}$ are spatial tensors evaluated
on the hypersurface. 
The equations of motion for the auxiliary fields $\varphi^{\mu \nu \rho \sigma}$
and $\varrho_{\mu \nu \rho \sigma}$ are, respectively given by~\cite{Deruelle:2009zk},
\begin{equation}\label{5.3}
\frac{\delta S}{\delta \varphi^{\mu \nu \rho \sigma}}=0 \Rightarrow \varrho_{\mu\nu\rho\sigma}=\mathcal{R}_{\mu\nu\rho\sigma}\quad
\text{and}\quad
\frac{\delta S}{\delta \varrho_{\mu \nu \rho \sigma}}=0 \Rightarrow \varphi^{\mu\nu\rho\sigma}=\frac{\partial
f}{\partial\varrho_{\mu\nu\rho\sigma}}\,,
\end{equation}
where $\mathcal{R}_{\mu\nu\rho\sigma}$ is the four-dimensional Riemann tensor.

One can start from the action given by Eq.~\eqref{5.2}, insert the equation
of motion for $\varphi^{\mu \nu \rho \sigma}$ and recover the action given
by Eq.~\eqref{5.1}. It has been shown by~\cite{Deruelle:2009zk} that one
can find the total derivative term of 
the auxiliary action as
\begin{eqnarray}\label{eq5.4}
S=\frac{1}{16 \pi G_{N}}\int_{\mathcal{M}}d^{4}x\left( \,
\sqrt{-g}\mathcal{L}-2\partial_{\mu}[\sqrt{-g} \hspace{1mm}
n^{\mu}K \cdot\Psi]\right)\,,
\end{eqnarray}
 where $K=h^{ij}K_{ij}$, with $K_{ij}$ given by Eq.~(\ref{eq29}), and $\Psi=h^{ij}\Psi_{ij}$
, where $\Psi_{ij}$ is given in Eq.~(\ref{5.6}), are spatial tensors evaluated
on the hypersurface $\Sigma_t$ and $\mathcal{L}$ is the Lagrangian density.

In Eq.~(\ref{eq5.4}), the second term is the total derivative. It has been
shown that one may add the following action to the above action to eliminate
the total derivative appropriately. 
Indeed $\Psi$ can be seen as a modification to the GHY term, which depends
on the form of the Lagrangian density~\cite{Deruelle:2009zk}.  
\begin{equation}\label{5.5}
S_{GHY}=\frac{1}{8 \pi G_{N}} \oint_{\partial\mathcal{M}}d\Sigma_{\mu}n^{\mu}\Psi
\cdot K\,,
\end{equation} 
where
$n^{\mu}$
is the normal vector to the hypersurface and the infinitesimal vector field
\begin{eqnarray}
d\Sigma_{\mu}=\varepsilon_{\mu\alpha\beta\gamma}e^{\alpha}_{1}e^{\beta}_{2}e^{\gamma}_{3}d^{3}y\,,
\end{eqnarray}
is normal to the boundary $\partial\mathcal{M}$ and
is proportional to the volume element of $\partial\mathcal{M}$; in above
 $\varepsilon_{\mu\alpha\beta\gamma}=\sqrt{-g}[\mu\,\alpha\,\beta\,\gamma]$
is the Levi-Civita tensor and $y$ are coordinates intrinsic to the boundary~\footnote{We
shall also mention that Eq.(\ref{5.5}) is derived from Eq.(\ref{eq5.4}) by
performing Stokes theorem, that is $
\int_{\mathcal{M}}A^{\mu}_{\ ;\mu}\sqrt{-g}\hspace{1mm}d^{d}x=\oint_{\partial\mathcal{M}}A^{\mu}\hspace{1mm}d\Sigma_\mu,
$ with $A^{\mu}=n^{\mu}K \cdot\Psi$. }, and we used Eq.~(\ref{eee}). Moreover
in Eq.~(\ref{5.5}), we have: 
\begin{equation}\label{5.6}
\Psi^{ij}=-\frac{1}{2}\frac{\delta f}{\delta\Omega_{ij}}\,,
\end{equation}
where $f$ indicates the terms in the Lagrangian density and is built up of
tensors $\varrho_{\mu\nu\rho \sigma}$, $\varrho_{\mu\nu}$ and $\varrho  $
as in Eq.~(\ref{5.2}); $G_{N}$ is the universal gravitational constant and $\Omega_{ij}$
is given in Eq.~(\ref{tensors1}). Indeed, the above constraint
is extracted from the equation of motion for $\Omega_{ij}$ in the Hamiltonian
regime \cite{Deruelle:2009zk}. In the next section we are going to use the
same approach to find the boundary
terms for the most
general, covariant quadratic order action of gravity.
\section{Boundary Terms for Finite Derivative Theory of Gravity}\label{sec6}

In this section we are going to use the 3+1 decomposition and calculate the
boundary term of the EH term $\mathcal{R}$, and
$$\mathcal{R}\Box \mathcal{R},~~~\mathcal{R}_{\mu\nu}\Box \mathcal{R}^{\mu\nu},~~~\mathcal{R}_{\mu\nu\rho\sigma}\Box
\mathcal{R}^{\mu\nu\rho\sigma},$$ as prescribed in previous section, as a
warm-up exercise.

We then move on to our generalised action given in Eq.~(\ref{eq7}).  To decompose
any given term, we shall write them in terms of their auxiliary field, therefore
we have 
$\mathcal{R}=\varrho,~~~ \mathcal{R}_{\mu\nu}\equiv \varrho_{\mu\nu}$, and
$\mathcal{R}_{\mu\nu\rho\sigma}\equiv \varrho_{\mu\nu\rho\sigma}$, 
where the auxiliary fields $\varrho,\varrho_{\mu\nu}$ and  $\varrho_{\mu\nu\rho\sigma}$
have all the symmetry properties of the Riemann tensor. We shall also note
that the decomposition of the $\Box$ operator in 3+1 formalism in the coframe
setup is given by Eq.~(\ref{boxcoframe}). 
\subsection{$\mathcal{R}$}\label{sec6.1} 
For the Einstein-Hilbert term $\mathcal{R}$, in terms of the auxiliary field
$\varrho$ we
find in Appendix \ref{sec:appendixEHterm}
\begin{eqnarray}\label{6.1}
f=\varrho &=& g^{\mu\rho} g^{\nu\sigma} \varrho_{\mu\nu\rho\sigma}\,\nonumber\\
&=& \left( h^{\mu\rho} - n^\mu n^\rho \right) \left( h^{\nu\sigma}
- n^\nu n^\sigma \right) \varrho_{\mu\nu\rho\sigma}\, \nonumber\\
&=& \left( h^{\mu\rho} h^{\nu\sigma} - n^\mu n^\rho h^{\nu\sigma}
- h^{\mu\rho} n^\nu n^\sigma \right) \varrho_{\mu\nu\rho\sigma}\,  \nonumber\\
&=& \left( \rho- 2 \Omega \right),
\end{eqnarray} 
where $\Omega= h^{ij}\Omega_{ij}$ and we used $h^{ij}h^{kl}\rho_{ijkl}=\rho$,
and $h^{ij}\rho_{i\nu j\sigma}n^\nu n^\sigma= h^{ij}\Omega_{ij}$ and $\varrho\equiv
\mathcal{R}$ in the EH action and the right hand
side of Eq.~(\ref{6.1}) is the $3+1$ decomposed form of the
Lagrangian and hence $\rho$ and $\Omega$ are spatial. We may note that the
last term of the expansion on the second line of Eq.~(\ref{6.1}) vanishes
due to the symmetry properties of the Riemann tensor.
Using Eq.~(\ref{5.6}), and calculating the functional derivative, we find

\begin{eqnarray}\label{6.2}\Psi^{ij}=-\frac{1}{2}\frac{\delta f}{\delta\Omega_{ij}}=h^{ij}.\end{eqnarray}
This  verifies the result found in~\cite{Deruelle:2009zk}, and it is clear
that upon substituting this result  into Eq.~(\ref{5.5}), we recover the
well known boundary for the
EH action, as $K=h^{ij}K_{ij}$ 
and $\Psi\cdot K\equiv\Psi^{ij}K_{ij}$ where $K_{ij}$ is given by Eq.~(\ref{eq29}).
 Hence, 
\begin{equation}\label{GHY-00}
S_{GHY}\equiv S_{0}=\frac{1}{8\pi G_{N}}\oint_{\partial\mathcal{M}}d\Sigma_{\mu}n^{\mu}
K\,,
\end{equation}
where $d\Sigma_{\mu}$ is the normal to the boundary $\partial\mathcal{M}$
and
is proportional to the volume element of $\partial\mathcal{M}$ while $n^{\mu}$
is the normal vector to the hypersurface.

\subsection{$\mathcal{R}_{\mu\nu\rho\sigma}\Box \mathcal{R}^{\mu\nu\rho\sigma}$}\label{sec6.2}

Next, we start off by writing $\mathcal{R}_{\mu\nu\rho\sigma}\Box \mathcal{R}^{\mu\nu\rho\sigma}$
as its auxiliary equivalent $\varrho_{\mu\nu\rho\sigma}\Box \varrho^{\mu\nu\rho\sigma}$
 to obtain 
\begin{eqnarray}\label{eq:decompositionofriemannsquaredintermsofhsandns}
\varrho_{\mu\nu\rho\sigma}\Box \varrho^{\mu\nu\rho\sigma}
&=&\delta_{\mu}^{\alpha}\delta_{\nu}^{\beta}\delta_{\rho}^{\gamma}\delta_{\sigma}^{\lambda}\varrho_{\alpha\beta\gamma\lambda}\Box\varrho^{\mu\nu\rho\sigma}\nonumber\\
&=&\Big[h_{\mu}^{\alpha}h_{\nu}^{\beta}h_{\rho}^{\gamma}h_{\sigma}^{\lambda}-\Big(h_{\mu}^{\alpha}h_{\nu}^{\beta}h_{\rho}^{\gamma}n^{\lambda}n_{\sigma}
+h_{\mu}^{\alpha}h_{\nu}^{\beta}n^{\gamma}n_{\rho}h_{\sigma}^{\lambda}+h_{\mu}^{\alpha}n^{\beta}n_{\nu}h_{\rho}^{\gamma}h_{\sigma}^{\lambda}\nonumber\\
&&+n^{\alpha}n_{\mu}h_{\nu}^{\beta}h_{\rho}^{\gamma}h^\lambda_{\sigma}\Big)+h_{\mu}^{\alpha}n^{\beta}n_{\nu}h_{\rho}^{\gamma}n^{\lambda}n_{\sigma}
+h_{\mu}^{\alpha}n^{\beta}n_{\nu}n^{\gamma}n_{\rho}h_{\sigma}^{\lambda}+n^{\alpha}n_{\mu}h_{\nu}^{\beta}h_{\rho}^{\gamma}n^{\lambda}n_{\sigma}\nonumber\\
&+&n^{\alpha}n_{\mu}h_{\nu}^{\beta}n^{\gamma}n_{\rho}h_{\sigma}^{\lambda}\Big]
\varrho_{\alpha\beta\gamma\lambda}\left(-(N^{-1}\bar{\partial_{0}})^{2}+\Box_{hyp}\right)\varrho^{\mu\nu\rho\sigma}\,,
\end{eqnarray}
where  $\varrho_{\mu\nu\rho\sigma}=\delta_{\mu}^{\alpha}\delta_{\nu}^{\beta}\delta_{\rho}^{\gamma}\delta_{\sigma}^{\lambda}\varrho_{\alpha\beta\gamma\lambda}$
(where $\delta^\alpha_\mu$ is the Kronecker delta). This allowed us to use
the completeness relation as given in Eq.~(\ref{eq11}). In Eq.~(\ref{eq:decompositionofriemannsquaredintermsofhsandns}),
we  used the antisymmetry properties
of the Riemann tensor to eliminate irrelevant terms in the expansion. From
Eq.~(\ref{eq:decompositionofriemannsquaredintermsofhsandns}), we
have three types of terms: $$hhhh,~~~~hhhnn,~~~~hhnnnn.$$
The aim is to contract the tensors appearing in Eq.~(\ref{eq:decompositionofriemannsquaredintermsofhsandns})
and extract those terms which are $\Omega_{ij}$ dependent. This is because
we only need $\Omega_{ij}$  dependent terms to obtain $\Psi^{ij}$ as in Eq.~(\ref{5.6})
and then the boundary as prescribed in Eq.~(\ref{5.5}). 

A closer look at the expansion given in Eq.~(\ref{eq:decompositionofriemannsquaredintermsofhsandns})
leads us to know which term would admit $\Omega_{ij}$ type terms. Essentially,
as defined in Eq.~(\ref{tensors1}), $\Omega_{ij}=n^{\mu}n^{\nu}\varrho_{i\mu
j\nu}$, therefore by having two auxiliary field tensors as $\varrho_{\alpha\beta\gamma\lambda}$
and $\varrho^{\mu\nu\rho\sigma}$ in Eq.~(\ref{eq:decompositionofriemannsquaredintermsofhsandns})
(with symmetries of the Riemann tensor) we may construct $\Omega_{ij}$ dependent
terms. 
Henceforth, we can see that in this case the $\Omega_{ij}$ dependence comes
from the $hhnnnn$ term. 

To see this explicitly, note that in order to perform the appropriate contractions
in presence of the d'Alembertian operator, we first need to complete the
contractions on the left hand side 
of the $\Box$ operator. We then need to commute the rest of the tensors by
using the {\it Leibniz rule} to the right hand side of the components of
the operator,  \textit{i.e.} the $\bar{\partial}_0$'s and the 
$\Box_{hyp}$, and only then do we obtain the $\Omega_{ij}$ type terms. 

We first note that the terms that do not produce $\Omega_{ij}$ dependence
are not involved in the boundary calculation, however they might form $\rho_{ijkl}$,
$\rho_{ijk}$, 
or their contractions. These terms are equivalent to the Gauss and Codazzi
equations as shown in Eq.~(\ref{tensors1}), and we will address their formation
in Appendix \ref{RT1}. In addition, as we shall see, by performing the Leibniz
rule one produces some associated terms, the $X_{ij}$'s, which appear for
example in Eq.~(\ref{ili1}). Again we will keep them only if they are $\Omega_{ij}$
dependent, if not we will drop them.\\

\noindent
$\underline{{\bf hhnnnn}}$ terms: To this end we shall compute the $hhnnnn$
terms, hence we  commute the $h$'s and $n$'s onto the right hand side of
the $\Box$
in the $hhnnnn$ term of Eq.~(\ref{eq:decompositionofriemannsquaredintermsofhsandns}):
\begin{eqnarray}\label{ili1}
&&h_{\mu}^{\alpha}n^{\beta}n_{\nu}h_{\rho}^{\gamma}n^{\lambda}n_{\sigma}\varrho_{\alpha\beta\gamma\lambda}\left(-(N^{-1}\bar{\partial_{0}})^{2}+\Box_{hyp}\right)\varrho^{\mu\nu\rho\sigma}\nonumber\\
&&=\left(h_{x}^{i}e_{i}^{\alpha}e_{\mu}^{x}\right)n^{\beta}n_{\nu}\left(h_{y}^{j}e_{j}^{\gamma}e_{\rho}^{y}\right)n^{\lambda}n_{\sigma}\varrho_{\alpha\beta\gamma\lambda}\left(-(N^{-1}\bar{\partial_{0}})^{2}+\Box_{hyp}\right)\varrho^{\mu\nu\rho\sigma}\nonumber\\
&&=\left(h_{x}^{i}e_{\mu}^{x}\right)n_{\nu}\left(h_{y}^{j}e_{\rho}^{y}\right)n_{\sigma}\Omega_{ij}\left(-(N^{-1}\bar{\partial_{0}})^{2}+\Box_{hyp}\right)\varrho^{\mu\nu\rho\sigma}\nonumber\\
&&=-N^{-2}\Omega_{ij}\Big\{\bar{\partial_{0}^{2}}\left(\Omega^{ij}\right)\nonumber\\
&&-\bar{\partial_{0}}\left[\varrho^{\mu\nu\rho\sigma}\bar{\partial_{0}}\left(\left[\left(h_{x}^{i}e_{\mu}^{x}\right)n_{\nu}\left(h_{y}^{j}e_{\rho}^{y}\right)n_{\sigma}\right]\right)\right]-\bar{\partial_{0}}\left(\left[\left(h_{x}^{i}e_{\mu}^{x}\right)
n_{\nu}\left(h_{y}^{j}e_{\rho}^{y}\right)n_{\sigma}\right]\right)\bar{\partial_{0}}\left(\varrho^{\mu\nu\rho\sigma}\right)\Big\}\nonumber\\
&&+\Omega_{ij}\Big\{\Box_{hyp}\left[\Omega^{ij}\right]-D_{a}\left(D^{a}\left[e_{\mu}^{x}n_{\nu}e_{\rho}^{y}n_{\sigma}\right]h_{x}^{i}h_{y}^{j}\varrho^{\mu\nu\rho\sigma}\right)-D_{a}\left[e_{\mu}^{x}n_{\nu}e_{\rho}^{y}n_{\sigma}\right]D^{a}\left(h_{x}^{i}h_{y}^{j}\varrho^{\mu\nu\rho\sigma}\right)\Big\}\nonumber\\
&&=\Omega_{ij}\big(-(N^{-1}\bar{\partial_{0}})^{2}+\Box_{hyp}\big)\Omega^{ij}+\Omega_{ij}
X_{1}^{ij}\nonumber\\
&& =\Omega_{ij}\Box\Omega^{ij}+\Omega_{ij} X_{1}^{ij}
\end{eqnarray}
where $\Omega_{ij}\equiv h_{ik} e_{\kappa}^{k}
h_{jm} e_{\lambda}^{m}n_\gamma n_\delta\varrho^{\gamma\kappa\delta\lambda}=h_{ik}
h_{jm} n_\gamma n_\delta\varrho^{\gamma k\delta m}$; we note that $X_{1}^{ij}$
only appears 
because of the presence of the $\Box$ operator. \textcolor{red}{}
\begin{eqnarray}\label{x1}
X_{1}^{ij}&=&N^{-2}(\bar{\partial_{0}}\left[\varrho^{\mu\nu\rho\sigma}\bar{\partial_{0}}\left(\left[\left(h_{x}^{i}e_{\mu}^{x}\right)n_{\nu}\left(h_{y}^{j}e_{\rho}^{y}\right)n_{\sigma}\right]\right)\right]+\bar{\partial_{0}}\left(\left[\left(h_{x}^{i}e_{\mu}^{x}\right)
n_{\nu}\left(h_{y}^{j}e_{\rho}^{y}\right)n_{\sigma}\right]\right)\bar{\partial_{0}}\left(\varrho^{\mu\nu\rho\sigma}\right))
\nonumber\\
& -&D_{a}\left(D^{a}\left[e_{\mu}^{x}n_{\nu}e_{\rho}^{y}n_{\sigma}\right]h_{x}^{i}h_{y}^{j}\varrho^{\mu\nu\rho\sigma}\right)-D_{a}\left[e_{\mu}^{x}n_{\nu}e_{\rho}^{y}n_{\sigma}\right]D^{a}\left(h_{x}^{i}h_{y}^{j}\varrho^{\mu\nu\rho\sigma}\right)
 \,.
\end{eqnarray}
The term $\Omega_{rs} X_{1}^{rs}$ will yield $X_{1}^{ij}$ when functionally
differentiated with respect to $\Omega_{ij}$ as in Eq.~(\ref{5.6}). Also
note $X_{1}^{ij}$ 
does not have any $\Omega^{ij}$ dependence. Similarly for the other $X$ terms
which appear later in the chapter.
We shall note that when we take $\Box=1$ in Eq.~(\ref{eq:decompositionofriemannsquaredintermsofhsandns}),
we obtain,\begin{eqnarray}
&&h_{\mu}^{\alpha}n^{\beta}n_{\nu}h_{\rho}^{\gamma}n^{\lambda}n_{\sigma}\varrho_{\alpha\beta\gamma\lambda}\varrho^{\mu\nu\rho\sigma}\nonumber\\
&=&\left(h_{x}^{i}e_{i}^{\alpha}e_{\mu}^{x}\right)n^{\beta}n_{\nu}\left(h_{y}^{j}e_{j}^{\gamma}e_{\rho}^{y}\right)n^{\lambda}n_{\sigma}\varrho_{\alpha\beta\gamma\lambda}\varrho^{\mu\nu\rho\sigma}\nonumber\\
&=&\left(h_{x}^{i}e_{\mu}^{x}\right)n_{\nu}\left(h_{y}^{j}e_{\rho}^{y}\right)n_{\sigma}\Omega_{ij}\varrho^{\mu\nu\rho\sigma}\nonumber\\
&=&\Omega_{ij}\left(h_{x}^{i}e_{\mu}^{x}\right)n_{\nu}\left(h_{y}^{j}e_{\rho}^{y}\right)n_{\sigma}\varrho^{\mu\nu\rho\sigma}\nonumber\\
&=&\Omega_{ij}\Omega^{ij}\,,
\end{eqnarray}
where we just contract the indices and we do  \textbf{not} need to use the
Leibniz rule as we can commute any of the tensors, therefore we do not produce
any $X^{ij}$ 
terms at all~\footnote{This is the same for $\Box^2$ and $\Box^n$.}.   
Finally, one can decompose Eq.~(\ref{eq:decompositionofriemannsquaredintermsofhsandns})
as 
\begin{eqnarray}\label{dec1}\varrho_{\mu\nu\rho\sigma}\Box \varrho^{\mu\nu\rho\sigma}=4\Omega_{ij}\Box\Omega^{ij}+4\Omega_{ij}
X_{1}^{ij} + \cdots \,,
\end{eqnarray}
where ``$\cdots$'' are terms such as $\rho_{ijkl}\Box\rho^{ijkl}$, $\rho_{ijk}\Box\rho^{ijk}$
and terms that are not $\Omega^{ij}$ dependent and are the results of performing
the Leibniz rule (see Appendix \ref{RT1}). 
When we take $M^2 \to \infty$,
 \textit{i.e.}, when we set $\Box \rightarrow 0$ (recall that $\Box$ has
an associated mass scale $\Box/M^2$), which is also  equivalent to considering
$\alpha \rightarrow 0$ in Eq.~(\ref{eq7}), we recover the EH result.

When $\Box \rightarrow 1$, we recover the result for $\mathcal{R}_{\mu\nu\rho\sigma}\mathcal{R}^{\mu\nu\rho\sigma}$
found in~\cite{Deruelle:2009zk}. At both limits, $\Box \rightarrow 0$
and $\Box \rightarrow 1$, the $X_{1}^{ij}$ term is not present. To find the
boundary term, we use Eq.~(\ref{5.6}) and then Eq.~(\ref{5.5}). 
We are going to use the Euler-Lagrange equation and drop the total derivatives
as a result. We have, 
\begin{eqnarray}
\Psi_{\mathrm{Riem}}^{ij}&=&-\frac{1}{2}\frac{\delta f}{\delta\Omega_{ij}}=-\frac{4}{2}\frac{\delta
(\Omega_{ij}\Box\Omega^{ij}+\Omega_{ij} X_{1}^{ij})}{\delta \Omega_{ij}}\nonumber\\&=&-2\Bigg\{\frac{\partial
(\Omega_{ij}\Box\Omega^{ij})}{\partial\Omega_{ij}}+\Box\left(\frac{\partial(\Omega_{ij}\Box\Omega^{ij})}{\partial
(\Box\Omega_{ij})}\right)+\frac{\partial (\Omega_{ij} X_{1}^{ij})}{\partial
\Omega_{ij}}\Bigg\}\nonumber\\
&=&-2(\Box\Omega^{ij}+\Box \Omega^{ij}+X_{1}^{ij})=-4\Box\Omega^{ij}-2X_{1}^{ij}\,.
\end{eqnarray}
Hence the boundary term for $\mathcal{R}_{\mu\nu\rho\sigma}\Box \mathcal{R}^{\mu\nu\rho\sigma}$
is, 
\begin{equation}\label{eq:finalresultforriemannboxriemann}
S_{1}=-\frac{1}{4\pi G_{N}}\oint_{\partial\mathcal{M}}d\Sigma_{\mu}n^{\mu} K_{ij}(2\Box\Omega^{ij}+X_{1}^{ij}).
\end{equation}
where $K_{ij}$\ is given by Eq.~(\ref{eq29}).


\subsection{$\mathcal{R}_{\mu\nu}\Box \mathcal{R}^{\mu\nu}$}\label{sec6.3}

We start by first performing the 3+1 decomposition of $\mathcal{R}_{\mu\nu}\Box\mathcal{R}^{\mu\nu}$
in its auxiliary form $\varrho_{\mu\nu} \Box \varrho^{\mu\nu}$, 
\begin{eqnarray}\label{eq:varrhodecomposition}
&&\varrho_{\mu\nu} \Box \varrho^{\mu\nu}
= g^{\rho\sigma} \varrho_{\rho\mu\sigma\nu} \Box g^{\mu\kappa}
g^{\nu\lambda} g^{\gamma\delta} \varrho_{\gamma\kappa\delta\lambda}\nonumber\\
&&=\left(h^{\rho\sigma} - n^\rho n^\sigma \right) \left( h^{\mu\kappa}
- n^\mu n^\kappa \right) \left( h^{\nu\lambda} - n^\nu n^\lambda \right)
\left( h^{\gamma\delta} - n^\gamma n^\delta \right)  \varrho_{\rho\mu\sigma\nu}
\Box\varrho_{\gamma\kappa\delta\lambda}\nonumber\\
&&=\Big[h^{\rho\sigma} h^{\mu\kappa} h^{\nu\lambda} h^{\gamma\delta}
- \big( n^\rho n^\sigma h^{\mu\kappa} h^{\nu\lambda} h^{\gamma\delta} +
h^{\rho\sigma} n^\mu n^\kappa h^{\nu\lambda} h^{\gamma\delta} 
        + h^{\rho\sigma} h^{\mu\kappa} n^\nu n^\lambda h^{\gamma\delta} 
 \nonumber\\
        && +h^{\rho\sigma} h^{\mu\kappa} h^{\nu\lambda} n^\gamma n^\delta \big)+ n^\rho n^\sigma h^{\mu\kappa} h^{\nu\lambda} n^\gamma n^\delta
+ h^{\rho\sigma} n^\mu n^\kappa n^\nu n^\lambda h^{\gamma\delta} \Big] 
\varrho_{\rho\mu\sigma\nu} \Box \varrho_{\gamma\kappa\delta\lambda}\,,\nonumber\\
\end{eqnarray}
where we have used appropriate contractions to write the Ricci tensor in
terms of the Riemann tensor. As before, we then used the completeness relation
Eq.~(\ref{eq11}) and used the 
antisymmetric properties of the Riemann tensor to drop the vanishing terms.
We are now set to calculate each term, which we do in more detail in Appendix
\ref{RT2}.
Again our aim is to find the $\Omega_{ij}$ dependent terms, by looking at
the expansion given in Eq.~(\ref{eq:varrhodecomposition}) and the distribution
of the indices, 
the reader can see that the terms which are $\Omega_{ij}$ dependent are those
terms which have \textit{at least two} $nn$s contracted with one of the $\varrho$s
such that we form $n^{\mu}n^{\nu}\varrho_{i\mu j\nu}$.   
\begin{itemize}
\item
{\underline{\bf $hhhnn $}} terms: We start with the $hhhnn $ terms in Eq.~(\ref{eq:varrhodecomposition}).
We calculate the first of these in terms of $\Omega_{ik}$ 
and $\rho^{ik}$ also by moving the `$h$'s and `$n$'s onto the right hand
side of the $\Box$, 
\begin{eqnarray}
&&n^\rho n^\sigma h^{\mu\kappa} h^{\nu\lambda} h^{\gamma\delta}\varrho_{\rho\mu\sigma\nu}
\left( - \left( N^{-1} \bar{\partial}_0 \right)^2
+ \Box_{hyp} \right) \varrho_{\gamma\kappa\delta\lambda}\nonumber\\
&&=n^\rho n^\sigma (h^{ij} e^{\mu}_{i}e^{\kappa}_{j})(h^{kl} e^{\nu}_{k}e^{\lambda}_{l})(h^{mn}e^{\gamma}_{m}e^{\delta}_{n})\varrho_{\rho\mu\sigma\nu}\left(
- \left( N^{-1} \bar{\partial}_0 \right)^2
+ \Box_{hyp} \right) \varrho_{\gamma\kappa\delta\lambda}\nonumber\\
&&=n^\rho n^\sigma (h^{ij} e^{\kappa}_{j})(h^{kl}
e^{\lambda}_{l})(h^{mn}e^{\gamma}_{m}e^{\delta}_{n})\varrho_{\rho i\sigma
k}\left(
- \left( N^{-1} \bar{\partial}_0 \right)^2
+ \Box_{hyp} \right) \varrho_{\gamma\kappa\delta\lambda}\nonumber\\
&&= \Omega_{ik}(h^{ij} e^{\kappa}_{j})(h^{kl}
e^{\lambda}_{l})(h^{mn}e^{\gamma}_{m}e^{\delta}_{n})\left(
- \left( N^{-1} \bar{\partial}_0 \right)^2
+ \Box_{hyp} \right) \varrho_{\gamma\kappa\delta\lambda}\nonumber\\
&&=-N^{-2}\Omega_{ik}\Big\{\bar{\partial}^{2}_{0}(\rho^{ik})-\bar{\partial}_{0}\big(\varrho_{\gamma\kappa\delta\lambda}\bar{\partial}_{0}[h^{ij}
e^{\kappa}_{j}h^{kl}
e^{\lambda}_{l}h^{mn}e^{\gamma}_{m}e^{\delta}_{n}]\big)\nonumber\\&&-\bar{\partial}_{0}[h^{ij}
e^{\kappa}_{j}h^{kl}
e^{\lambda}_{l}h^{mn}e^{\gamma}_{m}e^{\delta}_{n}]\bar{\partial}_{0}\varrho_{\gamma\kappa\delta\lambda}\Big\}\nonumber\\
&&+\Omega_{ik}\Big\{\Box_{hyp}(\rho^{ik})-D_a\big(\varrho_{\gamma\kappa\delta\lambda}D^a[h^{ij}
e^{\kappa}_{j}h^{kl}
e^{\lambda}_{l}h^{mn}e^{\gamma}_{m}e^{\delta}_{n}]\big)\nonumber\\&&-D_{a}[h^{ij} e^{\kappa}_{j}h^{kl}
e^{\lambda}_{l}h^{mn}e^{\gamma}_{m}e^{\delta}_{n}]D^a\varrho_{\gamma\kappa\delta\lambda}\Big\}\nonumber\\
&&=\Omega_{ik}\Box\rho^{ik}+\Omega_{ik}X_{2(a)}^{ik}\,,
\end{eqnarray}
where the contraction is $h^{ij} e^{\kappa}_{j}h^{kl}
e^{\lambda}_{l}h^{mn}e^{\gamma}_{m}e^{\delta}_{n}\varrho_{\gamma\kappa\delta\lambda}=h^{ij}
h^{kl}
\rho_{jl}=\rho^{ik}$, and
\begin{align}\label{x2a}
X_{2(a)}^{ik}&=N^{-2}\Big\{\bar{\partial}_{0}\big(\varrho_{\gamma\kappa\delta\lambda}\bar{\partial}_{0}[h^{ij}
e^{\kappa}_{j}h^{kl}
e^{\lambda}_{l}h^{mn}e^{\gamma}_{m}e^{\delta}_{n}]\big)+\bar{\partial}_{0}[h^{ij}
e^{\kappa}_{j}h^{kl}
e^{\lambda}_{l}h^{mn}e^{\gamma}_{m}e^{\delta}_{n}]\bar{\partial}_{0}\varrho_{\gamma\kappa\delta\lambda}
\Big\}  \nonumber \\
& -D_a\big(\varrho_{\gamma\kappa\delta\lambda}D^a[h^{ij}
e^{\kappa}_{j}h^{kl}
e^{\lambda}_{l}h^{mn}e^{\gamma}_{m}e^{\delta}_{n}]\big)-D_{a}[h^{ij} e^{\kappa}_{j}h^{kl}
e^{\lambda}_{l}h^{mn}e^{\gamma}_{m}e^{\delta}_{n}]D^a\varrho_{\gamma\kappa\delta\lambda}
\,.
\end{align}
\item
{\underline{\bf $hhhnn$}} trems: The next $hhhnn$ term in Eq.~(\ref{eq:varrhodecomposition})
is
\begin{eqnarray}
&&h^{\rho\sigma} h^{\mu\kappa} h^{\nu\lambda} n^\gamma n^\delta\varrho_{\rho\mu\sigma\nu}
\left( - \left( N^{-1} \bar{\partial}_0 \right)^2
+ \Box_{hyp} \right) \varrho_{\gamma\kappa\delta\lambda}\nonumber\\
&&=(h^{ij} e^{\rho}_{i}e^{\sigma}_{j})(h^{kl} e^{\mu}_{k}e^{\kappa}_{l})(
h^{mn} e^{\nu}_{m}e^{\lambda}_{n})n^\gamma n^\delta\varrho_{\rho\mu\sigma\nu}
\left( - \left( N^{-1} \bar{\partial}_0 \right)^2
+ \Box_{hyp} \right) \varrho_{\gamma\kappa\delta\lambda}\nonumber\\
&&=\rho_{ k m}(h^{kl} e^{\kappa}_{l})(
h^{mn} e^{\lambda}_{n})n^\gamma n^\delta
\left( - \left( N^{-1} \bar{\partial}_0 \right)^2
+ \Box_{hyp} \right) \varrho_{\gamma\kappa\delta\lambda}\nonumber\\
&&=-N^{-2}\rho_{ k m}\Big\{\bar{\partial}^{2}_{0}(\Omega^{km})-\bar{\partial}_{0}\big(\varrho_{\gamma\kappa\delta\lambda}\bar{\partial}_{0}[h^{kl}
e^{\kappa}_{l}
h^{mn} e^{\lambda}_{n}n^\gamma n^\delta]\big)-\bar{\partial}_{0}[h^{kl} e^{\kappa}_{l}
h^{mn} e^{\lambda}_{n}n^\gamma n^\delta]\bar{\partial}_{0}\varrho_{\gamma\kappa\delta\lambda}\Big\}\nonumber\\
&&+\rho_{ k m}\Big\{\Box_{hyp}(\Omega^{km})-D_a\big(\varrho_{\gamma\kappa\delta\lambda}D^a[h^{kl}
e^{\kappa}_{l}
h^{mn} e^{\lambda}_{n}n^\gamma n^\delta]\big)-D_{a}[h^{kl} e^{\kappa}_{l}
h^{mn} e^{\lambda}_{n}n^\gamma n^\delta]D^a\varrho_{\gamma\kappa\delta\lambda}\Big\}
\nonumber\\
&& =\rho_{ k m}\Box\Omega^{ k m}+\cdots\,,
\end{eqnarray}
where we used $h^{kl} e^{\kappa}_{l}
h^{mn} e^{\lambda}_{n}n^\gamma n^\delta\varrho_{\gamma\kappa\delta\lambda}=h^{kl}
h^{mn} n^\gamma n^\delta\varrho_{\gamma l\delta n}=\Omega^{km}$ and we note
that ``$\cdots$'' are extra terms which do not depend on $\Omega^{km}$. 

\item
{\underline{\bf $hhnnnn$}} terms: The the next term in Eq.~(\ref{eq:varrhodecomposition})
is of the form $hhnnnn$:
\begin{eqnarray}
&& n^\rho n^\sigma h^{\mu\kappa} h^{\nu\lambda} n^\gamma n^\delta\varrho_{\rho\mu\sigma\nu}
\left( - \left( N^{-1} \bar{\partial}_0 \right)^2
+ \Box_{hyp} \right) \varrho_{\gamma\kappa\delta\lambda}\nonumber\\
&& =n^\rho n^\sigma (h^{ij} e^{\mu}_{i}e^{\kappa}_{j})(
h^{kl} e^{\nu}_{k}e^{\lambda}_{l} )n^\gamma n^\delta\varrho_{\rho\mu\sigma\nu}
\left( - \left( N^{-1} \bar{\partial}_0 \right)^2
+ \Box_{hyp} \right) \varrho_{\gamma\kappa\delta\lambda}\nonumber\\
&& =  \Omega^{jl}e^{\kappa}_{j}e^{\lambda}_{l} n^\gamma n^\delta
\left( - \left( N^{-1} \bar{\partial}_0 \right)^2
+ \Box_{hyp} \right) \varrho_{\gamma\kappa\delta\lambda}\nonumber\\
&&=-N^{-2}\Omega^{jl}\Big\{\bar{\partial}^{2}_{0}(\Omega_{jl})-\bar{\partial}_{0}\big(\varrho_{\gamma\kappa\delta\lambda}\bar{\partial}_{0}[e^{\kappa}_{j}e^{\lambda}_{l}
n^\gamma n^\delta]\big)-\bar{\partial}_{0}[e^{\kappa}_{j}e^{\lambda}_{l}
n^\gamma n^\delta]\bar{\partial}_{0}\varrho_{\gamma\kappa\delta\lambda}\Big\}\nonumber\\
&&+\Omega^{jl}\Big\{\Box_{hyp}(\Omega_{jl})-D_a\big(\varrho_{\gamma\kappa\delta\lambda}D^a[e^{\kappa}_{j}e^{\lambda}_{l}
n^\gamma n^\delta]\big)-D_{a}[e^{\kappa}_{j}e^{\lambda}_{l}
n^\gamma n^\delta]D^a\varrho_{\gamma\kappa\delta\lambda}\Big\} \nonumber\\
&& =\Omega^{jl}\Box\Omega_{jl}+\Omega^{jl}X_{2(b)jl}\,,
\end{eqnarray}
where $e^{\kappa}_{j}e^{\lambda}_{l}
n^\gamma n^\delta\varrho_{\gamma\kappa\delta\lambda}=n^\gamma n^\delta\varrho_{\gamma
j\delta l}=\Omega_{jl}$, and
\begin{align}\label{x2b}
X_{2(b)jl}&=N^{-2}\Big\{\bar{\partial}_{0}\big(\varrho_{\gamma\kappa\delta\lambda}\bar{\partial}_{0}[e^{\kappa}_{j}e^{\lambda}_{l}
n^\gamma n^\delta]\big)+\bar{\partial}_{0}[e^{\kappa}_{j}e^{\lambda}_{l}
n^\gamma n^\delta]\bar{\partial}_{0}\varrho_{\gamma\kappa\delta\lambda} \Big\}
\nonumber \\
& -D_a\big(\varrho_{\gamma\kappa\delta\lambda}D^a[e^{\kappa}_{j}e^{\lambda}_{l}
n^\gamma n^\delta]\big)-D_{a}[e^{\kappa}_{j}e^{\lambda}_{l}
n^\gamma n^\delta]D^a\varrho_{\gamma\kappa\delta\lambda} \,.
\end{align}
\item
{\underline{\bf $hhnnnn$}} terms: Finally, the  last  $hhnnnn$ terms in Eq.~(\ref{eq:varrhodecomposition})
is
\begin{eqnarray}
&& h^{\rho\sigma} n^\mu n^\kappa n^\nu n^\lambda h^{\gamma\delta}\varrho_{\rho\mu\sigma\nu}
\left( - \left( N^{-1} \bar{\partial}_0 \right)^2
+ \Box_{hyp} \right) \varrho_{\gamma\kappa\delta\lambda}\nonumber\\
&&=(h^{ij}e^{\rho}_{i}e^{\sigma}_{j}) n^\mu n^\kappa n^\nu n^\lambda (h^{mn}e^{\gamma}_{m}e^{\delta}_{n})\varrho_{\rho\mu\sigma\nu}
\left( - \left( N^{-1} \bar{\partial}_0 \right)^2
+ \Box_{hyp} \right) \varrho_{\gamma\kappa\delta\lambda}\nonumber\\
&&= \Omega n^\kappa n^\lambda h^{mn}e^{\gamma}_{m}e^{\delta}_{n}
\left( - \left( N^{-1} \bar{\partial}_0 \right)^2
+ \Box_{hyp} \right) \varrho_{\gamma\kappa\delta\lambda}\nonumber\\
&&=-N^{-2}\Omega\Big\{\bar{\partial}^{2}_{0}(\Omega)-\bar{\partial}_{0}\big(\varrho_{\gamma\kappa\delta\lambda}\bar{\partial}_{0}[n^\kappa
n^\lambda h^{mn}e^{\gamma}_{m}e^{\delta}_{n}]\big)-\bar{\partial}_{0}[n^\kappa
n^\lambda h^{mn}e^{\gamma}_{m}e^{\delta}_{n}]\bar{\partial}_{0}\varrho_{\gamma\kappa\delta\lambda}\Big\}\nonumber\\
&&+\Omega\Big\{\Box_{hyp}(\Omega)-D_a\big(\varrho_{\gamma\kappa\delta\lambda}D^a[n^\kappa
n^\lambda h^{mn}e^{\gamma}_{m}e^{\delta}_{n}]\big)-D_{a}[n^\kappa n^\lambda
h^{mn}e^{\gamma}_{m}e^{\delta}_{n}]D^a\varrho_{\gamma\kappa\delta\lambda}\Big\}
\nonumber\\
&& =\Omega\Box\Omega+\Omega X_{2(c)}\,,
\end{eqnarray}
\end{itemize}
where we used 
$ n^\kappa n^\lambda h^{mn}e^{\gamma}_{m}e^{\delta}_{n}\varrho_{\gamma\kappa\delta\lambda}=
h^{mn}\Omega_{m n}=\Omega$, and
\begin{align}\label{x2c}
X_{2(c)}&= N^{-2} \Big\{\bar{\partial}_{0}\big(\varrho_{\gamma\kappa\delta\lambda}\bar{\partial}_{0}[n^\kappa
n^\lambda h^{mn}e^{\gamma}_{m}e^{\delta}_{n}]\big)+\bar{\partial}_{0}[n^\kappa
n^\lambda h^{mn}e^{\gamma}_{m}e^{\delta}_{n}]\bar{\partial}_{0}\varrho_{\gamma\kappa\delta\lambda}
\Big\} \nonumber \\
& -D_a\big(\varrho_{\gamma\kappa\delta\lambda}D^a[n^\kappa
n^\lambda h^{mn}e^{\gamma}_{m}e^{\delta}_{n}]\big)-D_{a}[n^\kappa n^\lambda
h^{mn}e^{\gamma}_{m}e^{\delta}_{n}]D^a\varrho_{\gamma\kappa\delta\lambda}
\,.
\end{align}
Summarising this result, we can write Eq.~(\ref{eq:varrhodecomposition}),
as
\begin{eqnarray}\varrho_{\mu\nu} \Box \varrho^{\mu\nu}&=& \Omega(\Box
\Omega+X_{2(c)})+ \Omega_{ij}( \Box \Omega^{ij}+X_{2(b)}^{ij}) - \rho_{ij}
\Box
\Omega^{ij}\nonumber\\&-& \Omega_{ij}( \Box \rho^{ij}+X_{2(a)}^{ij})+\cdots
\,,
\end{eqnarray}
where ``$\cdots$'' are the contractions of $\rho_{ijkl}$ and $\rho_{ijk}$
(see Appendix \ref{RT2}) and  the terms that are the results of performing
Leibniz rule,
which have no $\Omega_{ij}$ dependence. When $\Box \rightarrow 1$, we recover
the result for $\mathcal{R}_{\mu\nu}\mathcal{R}^{\mu\nu}$ found in~\cite{Deruelle:2009zk}.

At both limits, $\Box \rightarrow 0$ and $\Box \rightarrow 1$, the $X_{2}$
terms are not present. 
Obtaining the boundary term requires us to extract $\Psi^{ij}$ as it is given
 in Eq.~(\ref{5.6}). 
Hence the boundary for $\mathcal{R}_{\mu\nu}\Box \mathcal{R}^{\mu\nu}$ is
given by,  
 \begin{align}\label{eq:finalresultforriccitensorboxriccitensor}
S_{2}&=- \frac{1}{8\pi
G_{N}}\oint_{\partial\mathcal{M}}d\Sigma_{\mu}n^{\mu}\Big[
K\Box\Omega+ K_{ij}\Box\Omega^{ij}-K_{ij}\Box\rho^{ij}\Big]\nonumber \\
&-\frac{1}{16\pi G_{N}}\oint_{\partial\mathcal{M}}d\Sigma_{\mu}n^{\mu}\Big[K
X_{2(c)}+K_{ij}(X_{2(b)}^{ij}-X_{2(a)}^{ij}) \Big]\,,
\end{align}
where $K\equiv h^{ij} K_{ij}$ and $K_{ij}$ is given by Eq.~(\ref{eq29}).


\subsection{$\mathcal{R}\Box \mathcal{R}$}\label{sec6.4}

We do not need to commute any $h$'s, or $n$'s across the $\Box$ here, we
can simply apply Eq.~(\ref{6.1}) to $\varrho \Box \varrho$, the auxiliary
equivalent of the $\mathcal{R}\Box \mathcal{R}$ 
term:
\begin{eqnarray}\label{36}
\varrho\Box \varrho = \left( \rho - 2 \Omega \right) \Box
\left( \rho - 2 \Omega \right)\,,
\end{eqnarray}
whereupon extracting $\Psi^{ij}$ using Eq.~(\ref{5.6}), and using Eq.~(\ref{5.5})
as in the previous cases, we obtain the boundary term for $\mathcal{R}\Box
\mathcal{R}$ to be
\begin{eqnarray}\label{eq:finalresultforrboxr}
S_{3}=- \frac{1}{4\pi
G_{N}}\oint_{\partial\mathcal{M}}d\Sigma_{\mu} \, n^{\mu}\Big[
2K\Box \Omega -K\Box \rho \Big],
\end{eqnarray}
where $K\equiv h^{ij} K_{ij}$ and $K_{ij}$ is given by  Eq.~(\ref{eq29}).
Again   when $\Box \rightarrow 1$, we recover the result for $\mathcal{R}^2$
found in~\cite{Deruelle:2009zk}.

\subsection{Full result} 

Summarising the results of Eq.~(\ref{eq:finalresultforriemannboxriemann}),
Eq.~(\ref{eq:finalresultforriccitensorboxriccitensor}) and Eq.~(\ref{eq:finalresultforrboxr}),
altogether we have
\begin{align}\label{eq:finalresultforallboundarytermswithonebox}
S
&=\frac{1}{16\pi G_{N}}\int _{\mathcal{M}}d^{4}x \, \sqrt{-g}\Big[\varrho+\alpha\big(\varrho\Box
\varrho+\varrho_{\mu\nu}\Box \varrho^{\mu\nu}+\varrho_{\mu\nu\rho\sigma}\Box
\varrho^{\mu\nu\rho\sigma}\big)+\varphi^{\mu\nu\rho\sigma}\left(\mathcal{R}_{\mu\nu\rho\sigma}-\varrho_{\mu\nu\rho\sigma}\right)\Big]\nonumber
\\
&-\frac{1}{8\pi G_{N}}\oint_{\partial\mathcal{M}}d\Sigma_{\mu} \, n^{\mu}\Big[-K+\alpha\big(-2K\Box
\varrho+4K\Box\Omega+K\Box \Omega+4K_{ij}\Box\Omega^{ij}-K_{ij}\Box\rho^{ij}+K_{ij}\Box\Omega^{ij}\big)\Big]\nonumber
\\
&-\frac{1}{16\pi G_{N}}\oint_{\partial\mathcal{M}}d\Sigma_{\mu}n^{\mu}\alpha\Big[K
X_{2(c)}+K_{ij}(4X_{1}^{ij}+X_{2(b)}^{ij}-X_{2(a)}^{ij}) \Big] \nonumber
\\
&=\frac{1}{16\pi G_{N}}\int _{\mathcal{M}}d^{4}x \, \sqrt{-g}\Big[\varrho+\alpha\big(\varrho\Box
\varrho+\varrho_{\mu\nu}\Box \varrho^{\mu\nu}+\varrho_{\mu\nu\rho\sigma}\Box
\varrho^{\mu\nu\rho\sigma}\big)+\varphi^{\mu\nu\rho\sigma}\left(\mathcal{R}_{\mu\nu\rho\sigma}-\varrho_{\mu\nu\rho\sigma}\right)\Big]\nonumber
\\
&-\frac{1}{8\pi G_{N}}\oint_{\partial\mathcal{M}}d\Sigma_{\mu} \, n^{\mu}\Big[-K+\alpha\big(-2K\Box\rho+5K\Box\Omega+5K_{ij}\Box\Omega^{ij}-K_{ij}\Box\rho^{ij}\Big]
\nonumber \\
& -\frac{1}{16\pi G_{N}}\oint_{\partial\mathcal{M}}d\Sigma_{\mu}n^{\mu}\alpha\Big[K
X_{2(c)}+K_{ij}(4X_{1}^{ij}+X_{2(b)}^{ij}-X_{2(a)}^{ij}) \Big]\,. 
\end{align}
 This result matches with the EH action~\cite{Deruelle:2009zk}, when we take
the limit $\Box \rightarrow 0$; that is, we are left with the same expression
for boundary 
 as in Eq.~(\ref{GHY-00}):
\begin{align}
S_{EH}&=\frac{1}{16\pi G_{N}}\int _{\mathcal{M}}d^{4}x \, \sqrt{-g}\Big[\varrho+\varphi^{\mu\nu\rho\sigma}\left(\mathcal{R}_{\mu\nu\rho\sigma}-\varrho_{\mu\nu\rho\sigma}\right)\Big]\nonumber
\\
&+\frac{1}{8\pi G_{N}}\oint_{\partial\mathcal{M}}d\Sigma_{\mu} \, n^{\mu}K \,,
\end{align}
since the $X$-type terms are not present when $\Box \rightarrow 0$.
When $\Box \rightarrow 1$, we recover the result for $\mathcal{R}+\alpha(\mathcal{R}^{2}+\mathcal{R}_{\mu
\nu}\mathcal{R}^{\mu \nu}+\mathcal{R}_{\mu\nu\rho\sigma}\mathcal{R}^{\mu\nu\rho\sigma})$
found in~\cite{Deruelle:2009zk}; that is, we are left with
\begin{align}
S &= \frac{1}{16\pi G_{N}}\int _{\mathcal{M}}d^{4}x \, \sqrt{-g}\Big[\varrho+\alpha\big(\varrho^{2}+\varrho_{\mu\nu}
\varrho^{\mu\nu}+\varrho_{\mu\nu\rho\sigma} \varrho^{\mu\nu\rho\sigma}\big)+\varphi^{\mu\nu\rho\sigma}\left(\mathcal{R}_{\mu\nu\rho\sigma}-\varrho_{\mu\nu\rho\sigma}\right)\Big]\nonumber
\\
&-\frac{1}{8\pi G_{N}}\oint_{\partial\mathcal{M}}d\Sigma_{\mu} \, n^{\mu}\Big[-K+\alpha\big(-2K\rho+5K\Omega+5K_{ij}\Omega^{ij}-K_{ij}\rho^{ij}\Big]\,;
\end{align}
again the $X$-type terms are not present when $\Box \rightarrow 1$.
We should note that the $X_1$ and $X_2$ terms are the results of having the
covariant  d'Alembertian operator so, in the absence of the d'Alembertian
operator, one does not produce them at all and hence the result found
in~\cite{Deruelle:2009zk} is guaranteed.
\\\\
We may now turn our attention to the $\mathcal{R}\Box^{2}
\mathcal{R},\mathcal{R}_{\mu\nu}\Box^{2} \mathcal{R}^{\mu\nu}$ and $\mathcal{R}_{\mu\nu\rho\sigma}\Box^{2}
\mathcal{R}^{\mu\nu\rho\sigma}$. 
Here the methodology will remain the same. One first decomposes each term
into its $3+1$ equivalent.   
Then one extracts $\Psi^{ij}$ using Eq.~(\ref{5.6}), and then the boundary
terms can be obtained using Eq.~(\ref{5.5}). In this case we will have two
operators, namely
\begin{eqnarray}\Box^2=\left( - \left( N^{-1}
\bar{\partial}_0 \right)^2
+ \Box_{hyp} \right)\left( - \left( N^{-1} \bar{\partial}_0 \right)^2
+ \Box_{hyp} \right)
\end{eqnarray} 
This means that upon expanding to $3+1$, one performs the Leibniz rule twice
and hence obtains eight total derivatives that do not produce any 
$\Omega_{ij}$s or its contractions that are relevant to the boundary calculations
and hence must be dropped.


\subsection{Generalisation to IDG Theory }\label{sec6.5}

We may now turn our attention to the infinite
derivative terms; namely, $\mathcal{R}\mathcal{F}_{1}(\Box)\mathcal{R}$, $\mathcal{R}_{\mu\nu}\mathcal{F}_{2}(\Box)\mathcal{R}^{\mu\nu}$
and $\mathcal{R}_{\mu\nu\rho\sigma}\mathcal{F}_{3}(\Box)\mathcal{R}^{\mu\nu\rho\sigma}$.
For such cases, we can write down the following relation
(see Appendix~\ref{B4}):
\begin{eqnarray}\label{33}
&&X D^{2n} Y= D^{2n}(XY)-D^{2n-1}(D(X)Y)-D^{2n-2}(D(X)D(Y))\nonumber \\
&&-D^{2n-3}(D(X)D^{2}(Y))-\dots-D(D(X)D^{2n-2}(Y))\nonumber\\&&-D(X)D^{2n-1}(Y)\,,
\end{eqnarray}
where $X$ and $Y$ are tensorial structures such as $\varrho_{\mu\nu\rho\sigma}$,
$\varrho_{\mu\nu}$, $\varrho$ and their contractions, while $D$ denotes any
operators. 
These operators do not have to be differential operators and indeed this
result can be generalised to cover the case where there are different types
of operator and a similar (albeit more complicated) structure is recovered.

From (\ref{33}), one produces $2n$ total derivatives, analogous to the scalar
toy model case, see Eqs.~(\ref{eq2},\ref{eq3}). We can then write the $3+1$
decompositions for each curvature by generalising  Eq.~(\ref{eq:finalresultforriemannboxriemann}),
~Eq.~(\ref{eq:finalresultforriccitensorboxriccitensor})
and Eq.~(\ref{eq:finalresultforrboxr}) and writing $\mathcal{R}_{\mu\nu\rho\sigma}
F_3(\Box) \mathcal{R}^{\mu\nu\rho\sigma}$, $\mathcal{R}_{\mu\nu} F_2(\Box)
\mathcal{R}^{\mu\nu}$ and $\mathcal{R} F_1(\Box) \mathcal{R}$ in terms of
their auxiliary equivalents  $\varrho_{\mu\nu\rho\sigma} F_3(\Box) \varrho^{\mu\nu\rho\sigma}$,
$\varrho_{\mu\nu} F_2(\Box) \varrho^{\mu\nu}$ and $\varrho F_1(\Box)
\varrho$.
Then 
\begin{eqnarray}
\varrho_{\mu\nu\rho\sigma}\mathcal{F}_{3}(\Box) \varrho^{\mu\nu\rho\sigma}=4\Omega_{ij}\mathcal{F}_{3}(\Box)\Omega^{ij}+4\Omega_{ij}X_{1}^{ij}+\cdots
\,,
\end{eqnarray}
\begin{eqnarray}\label{x2}
        \varrho_{\mu\nu} \mathcal{F}_{2}(\Box) \varrho^{\mu\nu}&=& \Omega\mathcal{F}_{2}(\Box)
\Omega+ \Omega_{ij} \mathcal{F}_{2}(\Box) \Omega^{ij} - \rho_{ij} \mathcal{F}_{2}(\Box)
\Omega^{ij}\nonumber\\&-& \Omega_{ij} \mathcal{F}_{2}(\Box) \rho^{ij}+\Omega_{ij}X_{2}^{ij}+\cdots
\,,
\end{eqnarray}
\begin{eqnarray}\label{x3}\varrho\mathcal{F}_{1}(\Box) \varrho = \left( \rho
- 2 \Omega \right) \mathcal{F}_{1}(\Box)
\left( \rho - 2 \Omega \right)\,,
\end{eqnarray}
where we have dropped the irrelevant terms, as we did before, while $X_{1}^{ij}$
and $X_{2}^{ij}$ are the analogues of Eqs.~\eqref{x1},~\eqref{x2a},~\eqref{x2b},~\eqref{x2c}.
We now need to use the generalised form of the Euler-Lagrange equations to
obtain the $\Psi^{ij}$ in each case:
\begin{align}\label{37}
\frac{\delta f}{\delta \Omega_{ij}}&=\frac{\partial
f}{\partial
\Omega_{ij}}-\nabla_{\mu}\left(\frac{\partial
f}{\partial
(\nabla_{\mu}\Omega_{ij})} \right)+\nabla_{\mu}\nabla_{\nu}\left(\frac{\partial
f}{\partial
(\nabla_{\mu}\nabla_{\nu}\Omega_{ij})} \right)+\cdots \nonumber\\
&=\frac{\partial
f}{\partial
\Omega_{ij}}+\sum^{\infty}_{n=1}\Box^{n}\left(\frac{\partial
f}{\partial (\Box^{n}\Omega_{ij})}\right)\,,
\end{align}
where we have imposed that $\delta \Omega_{ij}=0$ on the boundary $\partial
\mathcal{M}$.

Hence, by using $\Omega =h^{ij} \Omega_{ij}$ and $\rho=h^{ij} \rho_{ij}$,
 we find in Appendix~\ref{C} that:
\begin{eqnarray}\label{functionaldifferentation}
\frac{\delta \big(\Omega\mathcal{F}(\Box)\Omega\big)}{\delta \Omega_{ij}}=2h^{ij}\mathcal{F}(\Box)\Omega,&\quad&
\frac{\delta \big(\Omega_{ij}\mathcal{F}(\Box)\Omega^{ij}\big)}{\delta \Omega_{ij}}=2\mathcal{F}(\Box)\Omega^{ij}\nonumber\\
\frac{\delta \big(\rho\mathcal{F}(\Box)\Omega\big)}{\delta
\Omega_{ij}}=h^{ij}\mathcal{F}(\Box)\rho,&\quad& \frac{\delta \big(\rho_{ij}\mathcal{F}(\Box)\Omega^{ij}\big)}{\delta
\Omega_{ij}}=\mathcal{F}(\Box)\rho^{ij}\nonumber\\
\frac{\delta \big(\Omega\mathcal{F}(\Box)\rho\big)}{\delta
\Omega_{ij}}=h^{ij}\mathcal{F}(\Box)\rho,&\quad&
\frac{\delta \big(\Omega_{ij}\mathcal{F}(\Box)\rho^{ij}\big)}{\delta
\Omega_{ij}}=\mathcal{F}(\Box)\rho^{ij}\,,
\end{eqnarray}
and so using Eq.~(\ref{5.6}), the $\Psi^{ij}$s are:
\begin{eqnarray}\label{39}
\Psi^{ij}_{\mathrm{Riem}}&=&-4\mathcal{F}_{3}(\Box)\Omega^{ij}-2X_{1}^{ij}\nonumber\\
\Psi^{ij}_{\mathrm{Ric}}&=&\mathcal{F}_{2}(\Box)\rho^{ij}-h^{ij}\mathcal{F}_{2}(\Box)\Omega-\mathcal{F}_{2}(\Box)\Omega^{ij}-\frac{1}{2}X_{2}^{ij}\nonumber\\
\Psi^{ij}_{\mathrm{Scal}}&=&2h^{ij}\mathcal{F}_{1}(\Box)\big(-2\Omega+\rho\big)\equiv
2 h^{ij}\mathcal{F}_{1}(\Box)\varrho\,,
\end{eqnarray}
where we have used Eq.~(\ref{6.1}) in the last line. Finally, we can use
Eq.~\eqref{5.5} and write the boundary terms corresponding to our infinite-derivative
action as, 
\begin{align}\label{40}
S_{tot}&=S_{gravity}+S_{boundary}\nonumber \\&=\frac{1}{16\pi G_{N}}\int_{\mathcal{M}}
d^{4}x \, \sqrt{-g}\Big[\varrho+\alpha\big(\varrho\mathcal{F}_{1}(\Box)\varrho+\varrho_{\mu\nu}\mathcal{F}_{2}(\Box)\varrho^{\mu\nu}\nonumber
\\
&+\varrho_{\mu\nu\rho\sigma}\mathcal{F}_{3}(\Box)\varrho^{\mu\nu\rho\sigma}\big)+\varphi^{\mu\nu\rho\sigma}\left(\mathcal{R}_{\mu\nu\rho\sigma}-\varrho_{\mu\nu\rho\sigma}\right)\Big]\nonumber
\\
& +\frac{1}{8\pi
G_{N}}\oint_{\partial\mathcal{M}}d\Sigma_{\mu} \, n^{\mu}\Big[K+\alpha\big(2K\mathcal{F}_{1}(\Box)\rho-4K\mathcal{F}_{1}(\Box)\Omega
\nonumber \\
&-K\mathcal{F}_{2}(\Box)\Omega- K_{ij}\mathcal{F}_{2}(\Box)\Omega^{ij}+K_{ij}
\mathcal{F}_{2}(\Box)\rho^{ij}-4K_{ij}\mathcal{F}_{3}(\Box)\Omega^{ij}-2X_{1}^{ij}-\frac{1}{2}X_{2}^{ij}\big)\Big]\,.
\end{align}
where $\Omega_{ij}= n^\gamma n^\delta\varrho_{\gamma i\delta j}$, $\Omega=h^{ij}
\Omega_{ij}$, $\rho_{ij}=h^{km} \rho_{ijkm}$, $\rho=h^{ij} \rho_{ij}$, $K=h^{ij}
K_{ij}$ 
and $K_{ij}$ is the extrinsic curvature given by Eq.~(\ref{eq29}). 
We note that when we decompose the $\Box$, after we perform the Leibniz rule
enough times, we can reconstruct the $\Box$ in its original form, \textit{i.e.} it
is not affected by the use of the coframe. In this way, we can always reconstruct
${\cal F}_{i}(\Box)$. However, the form of the $X$-type terms will depend
on the decomposition and therefore the use of the coframe. 
In this regard, the $X$-type terms depend on the coframe but the ${\cal F}_{i}
(\Box)$ terms do not.
\section{Summary}

This chapter generalised earlier contributions for finding the boundary term
for a higher derivative theory of gravity. Our work  focused 
on seeking the boundary term or GHY contribution for a covariant infinite
derivative theory of gravity, which is quadratic in curvature.

Indeed, in this case some novel features distinctively filter through our
analysis. Since the bulk action contains non-local form factors,
${\cal F}_{i} (\Box)$, the boundary action also contains the non-locality,
as can be seen from our final expression Eq.~(\ref{40}).
Eq.~(\ref{40}) {\it also} has a smooth limit when $M\rightarrow \infty$,
or $\Box \rightarrow 0$, which is the local limit,
and our 
results then reproduce the GHY term corresponding to the EH action, and when ${\cal F}_{i}(\Box) \rightarrow
1$, our results coincide with that of~\cite{Deruelle:2009zk}.

\chapter{Thermodynamics of infinite derivative gravity}\label{panj}
In this chapter we will look at the thermodynamical aspects of the infinite
derivative gravity (IDG). In particular, we are going to study the first
law of thermodynamics \cite{Bardeen:1973gs} for number of cases. In other words, we are going to
obtain the entropy of IDG and some other theories of modified gravity for
static and spherically symmetric, $(A)dS$ and rotating background.

To proceed, we will briefly review how Wald \cite{Wald:1993nt,Iyer:1994ys} derived the entropy from an integral
over the Noether charge. We will use Wald's approach to find the entropy for
static and spherically symmetric and $(A)dS$ backgrounds. For the rotating
case, we are going to use the variation principle and obtain the generalised
Komar integrals \cite{Komar:1958wp} for gravitational actions constructed by Ricci scalar, Ricci
tensor and their derivatives. By using the Komar integrals we will obtain
the energy and the angular momentum and finally the entropy using the first law. We finally
shall use the Wald's  approach for a non-local action containing inverse
d'Alambertian operators and calculate the entropy in such case. 
\section{Wald's entropy, a brief review}
The Bekenstein-Hawking \cite{Hawking:1974sw,Bekenstein:1973ur} law states that the entropy of a black hole, $S_{BH}$,
is proportional to its horizon's area $A$ in units of Newton's constant.
A black hole in Einstein's theory of gravity has entropy of, 
\begin{equation}
S_{BH}=\frac{A}{4G_{N}}.
\end{equation}
The above relation indicates that the entropy as it stands is geometrical
and defined strictly by the black hole horizon. This relation shall satisfy
the first law of black hole mechanics, 
\begin{equation}
T_H dS = dM ,
\end{equation} 
where $M$ is the conserved or the ADM\ mass, and $\;T_H=\kappa/2\pi\;$ is
the Hawking temperature in terms
of the surface gravity, $\kappa$. For the sake of simplicity, we assumed that no charge or rotation is involved with the black hole. We shall
also note that the conserved mass and the surface gravity are well defined
for a stationary black hole and thus their definitions are free of modification
when considering various types of gravitational theories. The Wald entropy \cite{Wald:1993nt},
$S_W$, is also a geometric entropy and interpreted by Noether charge for
space-time diffeomorphisms. This entropy can be represented as a closed integral
over a cross-section of the horizon, $\mathcal{H}$,
\be\label{waldentschematic}
S_W=\oint_{\cal H} s_W dA\;,
\ee    
where $s_W$ is the entropy per unit of  horizon cross-sectional area.
For a  $D$-dimensional space-time with metric $\;ds^2 = g_{tt}dt^2
+g_{rr}dr^2+ \sum_{i,j=1}^{D-2}\sigma_{ij}dx_idx_j \;$,
 $\;dA=\sqrt{\sigma}dx_1\dots dx_{D-2}\;$.
 
 In order to derive (\ref{waldentschematic}), we start by varying a Lagrangian
density $L$ with respect to all fields $\{\psi\}$, which includes the metric.
In compact presentation, (with all tensor indices suppressed), 
\begin{equation}\label{varfield}
\delta L\;=\; {\cal E}\cdot \delta \psi \;+\; d[\theta\left(\delta \psi
\right)],
\end{equation}
where  $\;{\cal E}=0\;$ are the equations of motion and the dot denotes
a summation over all fields and  contractions of tensor indices. Also, $d$
denotes a total derivative, so that $\theta$ is a boundary term.

We shall now introduce $\pounds_{\xi}$ to be a Lie derivative operating along
some vector
field $\xi$.
Due to the diffeomorphism invariance of the theory we have, 
$$\;\delta_{\xi}\psi=\pounds_{\xi}\psi,\;$$
and $$\;\delta_{\xi}L=\pounds_{\xi} L=d\left(\xi\cdot L\right)\;.$$with the
help of the above identity and (\ref{varfield}) we can identify the
associated Noether  current, $J_{\xi}$, as:
\be\label{infivar}
J_{\xi}\;=\; \theta \left(\pounds_{\xi}\psi\right) -\xi\cdot{L}\;.
\ee
In order to satisfy the equations of motion, \textit{i.e.} ${\cal E}=0$, we should
have $dJ_{\xi}=0$. This indicates that, there should be an associated potential,
$Q_{\xi}$, such that
$J_{\xi}=dQ_{\xi}$. Now, if $D$ is the dimension of the space-time and
${\cal S}$ is a $D-1$  hypersurface with a
$D-2$ spacelike boundary $\partial{\cal S}$, then
\be
\int_{{\cal S}} J_{\xi} \;=\;\int_{\partial{\cal S}} Q_{\xi}\;,
\ee
is the associated Noether charge. 

Wald \cite{Wald:1993nt} proved in detail that the black holes' first law can be satisfied by
defining the entropy in terms of a particular form of Noether charge. This
is to choose surface $S$ as the horizon, $\cal{H}$, and the vector field,
$\xi$, as the horizon Killing vector, $\chi$ (with appropriate normalisation
to the surface gravity). Wald represented such entropy as~\footnote{Note
that since $\chi=0$ on the horizon, the most right hand side term in (\ref{infivar})
is not contributing to the Wald entropy. See \cite{Wald:1993nt}.} , 
\begin{equation}
S_W\; \equiv\; 2\pi \oint_{\cal H} Q_{\chi}\;.
\end{equation}
To understand the charge let us begin with an example. Suppose we have a
Lagrangian that is $\mathcal{L}=\mathcal{L}(g_{ab},\mathcal{R}_{abcd})$ (This
can be extended to the derivatives of the curvatures too),  
\begin{equation}
L=\sqrt{-g}{\cal L},
\end{equation}
the variation of the above Lagrangian density is, 
\begin{equation}
\delta L\;=\;-2\nabla_a\left({\cal X}^{abcd}\nabla_c \delta g_{bd}\sqrt{-g}
\right)\;+\;\cdots\;,
\end{equation}
where dots indicates that we have dropped the irrelevant terms to the entropy.
Moreover, \begin{equation}
{\cal X}^{abcd}\;\equiv\frac{\partial\cal L}{\partial {\cal R}_{abcd}}\;.
\end{equation}
The boundary can then be expressed as, 
\be
\theta \;=\; -2n_a{\cal X}^{abcd}\nabla_c \delta g_{bd}\sqrt{\gamma}\;+\;\cdots\;,
\ee
where $n^a$ is the unit normal vector and  $\gamma_{ab}$ is the induced metric
for the chosen surface
${\cal S}$.
For an arbitrary diffeomorphism
$\;\delta_{\xi} g_{ab} \;=\; \nabla_a \xi_b + \nabla_b\xi_a\;$, the associated
Noether
current is given by\be
J\;=\;-2\nabla_a\left({\cal X}^{abcd}\nabla_c\left(\nabla_b\xi_a +
\nabla_a\xi_b\right)n_a \sqrt{h}
\right)\;+\;\cdots\;.
\ee
We note that $h_{ab}$ is the induced metric corresponding to the $ \partial{\cal
S}$. Let us now assign the following: the horizon $\;{\cal S}\to{\cal H}\;$
and (normalised)  Killing vector $\;\xi^a\to\chi^a\;$; so that
$\;\ n_a\sqrt{h}\to\epsilon_a\sqrt{\sigma}\;$ with $\;\epsilon_a\equiv
\epsilon_{ab}\chi^b\;$, also we have $\epsilon_{ab} \equiv \nabla_a\chi_b$
as the binormal vector
for the horizon. By noting that $\epsilon_{ab}\;=\;-\epsilon_{ba}$ and also
using the symmetries of ${\cal X}^{abcd}$ which is due to the presence of
Riemann tensor, we get
\begin{equation}
J\;=\; -2\nabla_b\left({\cal X}^{abcd}
\nabla_c\chi_d \epsilon_a\sqrt{\sigma} \right)\;+\;\cdots\;.
\end{equation}
Finally the potential becomes: 
\begin{equation}
Q\;=\;  -{\cal X}^{abcd}
\epsilon_{ab}\epsilon_{cd}\sqrt{\sigma} \;+\;\cdots
\end{equation}
and so
\be\label{waldent1}
S_W\;=\;  -2\pi\oint_{\cal H} {\cal X}^{abcd}
\epsilon_{ab}\epsilon_{cd} dA.
\ee
\section{Spherically symmetric backgrounds}
In this section we are going to use the Wald's approach \cite{Wald:1993nt} to obtain the
entropy for number of cases where the space-time is defined by a spherically
symmetric solution. In particular we will focus on a generic and homogenous
spherically symmetric background. We then extend our calculations to the linearised
limit and then to the $(A)dS$ backgrounds. 
\subsection{Generic static and spherically symmetric background}
Let us recall the IDG action given by
(\ref{mainaction}). In $D$-dimensions we can rewrite the action as: 
\begin{eqnarray}\label{action1}
I^{tot} &=&\frac{1}{16\pi G^{(D)}_{N}}\int d^Dx \sqrt{-g}\big[ R \nonumber\\&+&
\alpha \big(R {\cal F}_1(\Box)R\big.\big.
\big. \big.+R_{\mu\nu}{\cal F}_2(\Box)R^{\mu\nu} + R_{\mu\nu\lambda\sigma}{\cal
 F}_{3}(\Box)R^{\mu\nu\lambda\sigma}\big)
\big],
\end{eqnarray}
where $G^{(D)}_{N}$ is the $D$-dimensional Newton's constant~\footnote{In $D$-dimensions
$G^{(D)}_{N}$ has dimension of $[G^{(D)}_{N}]=[G_{N}^{(4)}]L^{D-4}$ where $L$ is unit length.};
$\alpha $ is a constant~\footnote{Note that for an arbitrary choice of $\mathcal{F}(\Box)
$ at action level, $\alpha$ can be positive or negative as one can absorbs
the sign into the coefficients $f_{i_{n}}$ contained within $\mathcal{F}(\Box)$
to keep the overall action unchanged, however $\alpha$ has to be strictly
positive once we impose ghost-free condition (to be seen later).~\cite{Biswas:2005qr}
  } with
dimension of inverse mass squared; and $\mu,~\nu,~\lambda,~\sigma$
run from $0,~1,~2,\cdots  D-1$.
The form factors given by ${\cal F}_{i}(\Box)$ contain an infinite number
of covariant derivatives, of the form:
\begin{equation}
{\cal F}_{i}(\Box)\equiv\sum_{n=0}^{\infty}f_{i_{n}}\biggl(\frac{\Box}{M^2}\biggr)^{n}\,,
\end{equation}
with constants $f_{i_{n}}$, and $\Box\equiv  g^{\mu\nu}\nabla_{\mu}\nabla_{\nu}$
being the D'Alembertian operator. The reader should note that, in our presentation,
the function ${\cal F}_i(\Box)$ comes with an associated $D$-dimensional
mass scale, 
$M \leq M_P = (1/\sqrt{(8\pi G^{(D)}_{N})})$, which  determines the scale of non-locality
in a quantum sense, see~\cite{Talaganis:2014ida}.

In the framework of Lagrangian field theory, Wald \cite{Wald:1993nt} showed that one can find
the gravitational entropy by varying the Lagrangian and subsequently finding
the Noether current as
a function of an assigned vector field. By writing the corresponding Noether
charge, it has been shown that, for a static black hole, the first law of
thermodynamics can be satisfied and the entropy may be expressed by integrating
the Noether
charge over a bifurcation surface of the horizon. In so doing, one must 
choose the assigned vector field to be a horizon Killing vector, which has
been normalised to unit surface gravity.

In order to compute the gravitational entropy of the IDG theory outlined
above, we take a $D$-dimensional, static, 
homogenous  and spherically symmetric metric of the form \cite{Carroll:1997ar},
\begin{equation}\label{metric1}
ds^2= -f(r) dt^2 + f(r)^{-1}dr^2+r^2d\Omega^{2}_{D-2}\,.
\end{equation}
For a spherically symmetric metric the Wald entropy given in (\ref{waldent1})
can be written as, 
\begin{eqnarray}\label{ent2}
S_W=-2\pi \oint \frac{\delta{\cal L}}{\delta R_{abcd}}\epsilon_{ab}\epsilon_{cd}r^{D-2}d\Omega^{2}_{D-2}
\end{eqnarray}
where we shall note that $\delta$  denotes the functional differentiation
for a Lagrangian that not only does it include the metric and the curvature but
also the derivatives of the curvature, \textit{i.e.} 
\begin{equation}
{\cal L}={\cal L}(g_{\mu\nu}, R_{\mu\nu\lambda\rho},\nabla_{a_1}R_{\mu\nu\lambda\rho},\dots,\n_{(\al_1}\dots
\n_{\al_m)}R_{\mu\nu\lambda\rho   })
\end{equation}
Note that the
parentheses denote symmetrisation. Moreover, the integral in (\ref{ent2})
is over $D-2$ dimensional space-like bifurcation surface. The $\epsilon_{ab}$
is the binormal vector to the bifurcation surface. This normal vector is
antisymmetric under the exchange of $a\leftrightarrow b$ and normalised as
$\epsilon_{ab}\epsilon^{ab}=-2$. For metric (\ref{metric1}), the bifurcation
surface is at $r=r_H$ and $t=$constant. We note that $d\Omega^{2}_{D-2}$ is
the spherical element~\footnote{In four dimensions we have $d\Omega^{2}_{2}=d\theta+\sin^{2}\theta
d\phi$.}. In the case of (\ref{metric1}), the relevant Killing vector is
$\partial_t$ and $\epsilon_{tr}=1$. The $\epsilon$'s vanish for $a,b\neq
t,r$. We finally write the Wald entropy as, 
\begin{equation}\label{entropyrtrt}
S_{W}=-8\pi\oint\frac{\delta\mathcal{L}}{\delta R_{rtrt}}r^{D-2}d\Omega^{2}_{D-2}\,.
\end{equation}
Subsequently, we shall define the area of the horizon \cite{Relativiststoolkit}, that is, 
\begin{equation}
A_H=\oint r^{D-2}d\Omega^{2}_{D-2}=\frac{2\pi^{n/2}r^{n-1}}{\Gamma(\frac{n}{2})},
\end{equation}
where $n=D-1$. As an example for a 4-dimensional metric of the form given
by (\ref{metric1}), we have, 
\begin{equation}
A_H=\oint r^{2}d\Omega^{2}_{2}=\int^{2\pi}_{0} d\phi\int^{2\pi}_{0}r^{2}\sin(\theta)
d\theta=4\pi r^{2}\equiv\frac{2\pi^{3/2}r^{2}}{\Gamma(\frac{3}{2})}.
\end{equation}
It is now possible to use the generalised Euler-Lagrange equation and calculate
the functional differentiation given in (\ref{ent2}). That is, 
\begin{eqnarray}\label{genrulag}
\frac{\delta{\cal L}}{\delta R_{abcd}}&=&\frac{\partial\cal {\cal L}}{\partial
 R_{abcd}}-\nabla_{\mu_1}\Big(\frac{\partial{\cal L}}{\partial ( \nabla_{\mu_1}R_{abcd})}\Big)+\nabla_{\mu_1}\nabla_{\mu_2}\Big(\frac{\partial{\cal
L}}{\partial ( \nabla_{\mu_1}\nabla_{\mu_2}R_{abcd})}\Big)-\cdots\nonumber\\&+&(-1)^{m}\nabla_{(\mu_1\cdots}\nabla_{\mu_m)}\frac{{\cal
\partial L}}{\partial(\nabla_{(\mu_1\cdots}\nabla_{\mu_m)}R_{abcd})},
\end{eqnarray}
where we shall note that parentheses denote symmetrisation.
Now we are set to calculate the entropy for the IDG\ action, (\ref{action1})
via (\ref{entropyrtrt}). To do so, we are required to calculate the quantity
$\frac{\delta\mathcal{L}}{\delta R_{rtrt}}$. We have, 
\begin{eqnarray}
\frac{\delta{R}}{\delta R_{abcd}}&=&\frac{1}{2}(g^{ac}g^{bd}-g^{ad}g^{bc}),\\
\frac{\delta{(R {\cal F}_1(\Box)R)}}{\delta R_{abcd}}&=& {\cal F}_1(\Box)(g^{ac}g^{bd}-g^{ad}g^{bc})R,\\
\frac{\delta{(R_{\mu\nu}{\cal F}_2(\Box)R^{\mu\nu})}}{\delta R_{abcd}}&=&
\frac{1}{2}{\cal F}_2(\Box)(g^{ac}R^{bd}-g^{ad}R^{bc}\nonumber\\&-&g^{bc}R^{ad}+g^{bd}R^{ac}),
\\\frac{\delta{(R_{\mu\nu\lambda\sigma}{\cal
 F}_{3}(\Box)R^{\mu\nu\lambda\sigma})}}{\delta R_{abcd}}&=&2{\cal F}_1(\Box)R^{abcd},
\end{eqnarray}
by assigning $(a,b,c,d)\rightarrow(r,t,r,t)$ we obtain, 
\begin{eqnarray}
\frac{\delta{R}}{\delta R_{rtrt}}&=&\frac{1}{2}(g^{rr}g^{tt}-g^{rt}g^{tr})=-\frac{1}{2},\\
\frac{\delta{(R {\cal F}_1(\Box)R)}}{\delta R_{rtrt}}&=& {\cal -F}_1(\Box)R,\\
\frac{\delta{(R_{\mu\nu}{\cal F}_2(\Box)R^{\mu\nu})}}{\delta R_{rtrt}}&=&
\frac{1}{2}{\cal F}_2(\Box)(g^{rr}R^{tt}+g^{tt}R^{rr}),\\
\frac{\delta{(R_{\mu\nu\lambda\sigma}{\cal
 F}_{3}(\Box)R^{\mu\nu\lambda\sigma})}}{\delta R_{rtrt}}&=&2{\cal F}_1(\Box)R^{rtrt},
\end{eqnarray}
where we note that $g^{tt}g^{rr}=-1$, and $g^{tr}=g^{rt}=0$. See Appendix
\ref{functionalentropy} for detailed derivation of the above functional differentiation.
By using the Wald's formula given in (\ref{entropyrtrt}) we have \cite{Conroy:2015nva}, 
\begin{equation}\label{entropyeq1}
S_{W}=\frac{A_{H}}{4G^{(D)}_{N}}\bigl[1+\alpha(2 {\cal F}_1(\Box)R 
 - {\cal F}_2(\Box)\times(g^{rr}R^{tt}+g^{rr}R^{rr})-4 {\cal F}_3(\Box)R^{rtrt})\bigr]
\end{equation}
It is convenient, for illustrative purposes, to decompose the entropy equation
into its $(r,t)$ and spherical components. For the metric given in (\ref{metric1})
we denote the \(r\) and \(t\) directions by the
indices \(\{a,b\}\); and the spherical components by \(\{m,n\}\).
As such, we express the curvature scalar as follows
 \begin{equation}
R=g^{\mu\nu}R_{\mu\nu}=g^{ab}R_{ab}+g^{mn}R_{mn},
\end{equation}
where $g_{ab}$ is a $2$-dimensional metric tensor accounting for the ${r,t}$
directions and $g_{mn}$ is a $(D-2)$-dimensional metric tensor,
corresponding to the angular components, such that \begin{equation}
g^{\mu\nu}g_{\mu\nu}\equiv
g^{ab}g_{ab}+g^{mn}g_{mn}=2+(D-2)= D.
\end{equation}
Expanding the scalar curvature into Ricci and Riemann tensors, along with
the properties of the static, spherically symmetric metric (\ref{metric1}),
allows us to express the relevant components of the entropy equation as follows:
\begin{equation}
g^{rr}R^{tt}+g^{rr}R^{rr}=-g_{tt}R^{tt}-g_{rr}R^{rr}=-g^{ab}R_{ab}.
\end{equation}
Moreover, 
\begin{eqnarray}
g^{ab}R_{ab}=g^{ab}R^{\lambda}_{\ a\lambda b}=g^{ab}g^{\lambda\tau}R_{\tau
a\lambda b}=g^{ab}g^{cd}R_{ d ac b}+g^{ab}g^{mn}R_{ n am b},
\end{eqnarray}
we can write above as, 
\begin{equation}
-g^{ab}g^{cd}R_{ d ac b}=-g^{ab}R_{ab}+g^{ab}g^{mn}R_{ n am b},
\end{equation}
by assigning the coordinates we have, 
\begin{eqnarray}
&&-g^{rr}g^{rr}R_{ rrrr}-g^{tt}g^{tt}R_{ tttt}-g^{rr}g^{tt}R_{ trtr}-g^{tt}g^{rr}R_{
rtrt}\nonumber\\&&=-g^{ab}R_{ab}+g^{ab}g^{mn}R_{ n am b},
\end{eqnarray}
given that for metric (\ref{metric1}), $R_{ rrrr}=R_{ tttt}=0$, $R_{ trtr}=R_{
rtrt}$ and $g^{tt}g^{rr}=-1$, we have, 
\begin{equation}
2R_{rtrt}=-g^{ab}R_{ab}+g^{ab}g^{mn}R_{ n am b},
\end{equation}
or,
\begin{equation}\label{riemrel}
-4R_{rtrt}=2g^{ab}R_{ab}-2g^{ab}g^{
m n}R_{ ma nb}.
\end{equation}
Substitution into Eq.~(\ref{entropyeq1}), results in a decomposed $D$-dimensional
entropy equation for the action~(\ref{action1}) in a static, spherically
symmetric background :
\begin{eqnarray}\label{entropymastereq}
S_{W}&=&\frac{A_{H}}{4G^{(D)}_{N}}[1+\alpha(2 {\cal F}_1(\Box)+ {\cal F}_2(\Box)+2
{\cal F}_3(\Box))
g^{ab}R_{ab}\nonumber\\&+&2\alpha( {\cal F}_1(\Box)g^{
m n}R_{
m n}-
{\cal F}_3(\Box)g^{ab}g^{
m n}R_{ ma nb})].
\end{eqnarray}

\subsection{Linearised regime}\label{liniliaz}
In this section we shall study an interesting feature of the entropy given
in (\ref{entropymastereq}).
To begin, let us consider the perturbations around $D$-dimensional Minkowski
spacetime~\footnote{Later we shall see that in linearised regime we can take $f(r)=(1+2\Phi(r))$ and $f(r)^{-1}=(1-2\Psi(r)),$  thus (\ref{metric1}) will be of the form of (\ref{lin-met}).}  with metric tensor $\eta_{\mu\nu}$, such that $\eta_{\mu\nu}\eta^{\mu
\nu}=D$, and where the perturbations are denoted by $h_{\mu\nu}$ so that
$g_{\mu\nu}=\eta_{\mu\nu}+h_{\mu\nu}$. One should
also note that we are using {\it mostly plus} metric signature convention.

The $\mathcal{O}(h^{2})$ expressions for the Riemann tensor, Ricci tensor
and
curvature scalar in $D$-dimensions are given by \cite{Carroll:1997ar,Biswas:2013cha}:
\begin{eqnarray}\label{ident-0}
R_{\mu\nu\lambda\sigma}=\frac{1}{2}(\partial_{[\lambda}\partial_{\nu}h_{\mu\sigma]}-\partial_{[\lambda}\partial_{\mu}h_{\nu\sigma]})\nonumber
\\
R_{\mu\nu}=\frac{1}{2}(\partial_{\sigma}\partial_{(\nu}\partial^{\sigma}_{\mu)}-\partial_{\mu}\partial_{\nu}h-\Box
h_{\mu\nu})\nonumber \\
R=\partial_{\mu}\partial_{\nu}h^{\mu\nu}-\Box h .
\end{eqnarray}
Thus, the IDG\ action given in (\ref{action1}) can be written as \cite{Biswas:2014tua},  
\begin{eqnarray}\label{lin-act-0}
S_{(2)}&=& \frac{1}{32\pi G^{(D)}_{N}} \int d^D x \bigg[\frac{1}{2}h_{\mu \nu} \Box
a(\bbox) h^{\mu
\nu}+h_{\mu}^{\sigma} b(\bbox) \partial_{\sigma} \partial_{\nu} h^{\mu \nu}
\big.
\big.\nonumber \\&+&h c(\bbox)\partial_{\mu} \partial_{\nu}h^{\mu \nu} + \frac{1}{2}h
\Box
d(\bbox)h \big.
\big.+ h^{\lambda \sigma} \frac{f(\bbox)}{2\Box}\partial_{\sigma}\partial_{\lambda}\partial_{\mu}\partial_{\nu}h^{\mu
\nu}\bigg], 
\end{eqnarray}
where we have $\bbox\equiv\Box/M^2$. In above action we have \cite{Biswas:2014tua}, 
\begin{eqnarray}
R {\cal F}_1(\bbox)R= {\cal F}_1(\bbox)[h\Box^{2}h+h^{\lambda\sigma}\partial_{\sigma}\partial_{\lambda}\partial_{\mu}\partial_{\nu}h^{\mu\nu}-2h\Box\partial_{\mu}\partial_{\nu}h^{\mu\nu}],
\end{eqnarray}
\begin{eqnarray}
R_{\mu\nu} {\cal F}_2(\bbox)R^{\mu\nu} &=&{\cal F}_2(\bbox)[\frac{1}{4}h\Box^{2}h+\frac{1}{4}h^{\mu\nu}\Box^{2}h^{\mu\nu}-\frac{1}{2}h^{\sigma}_{\mu}\Box\partial_\sigma\partial_\nu
h^{\mu\nu}-\frac{1}{2}h\Box\partial_{\mu}\partial_{\nu}h^{\mu\nu}\nonumber\\&+&\frac{1}{2}h^{\lambda\sigma}\partial_{\sigma}\partial_{\lambda}\partial_{\mu}\partial_{\nu}h^{\mu\nu}],
\end{eqnarray}
\begin{eqnarray}
R_{\mu\nu\lambda\sigma} {\cal F}_3(\bbox)R^{\mu\nu\lambda\sigma}={\cal F}_3(\bbox)[h^{\mu\nu}\Box^{2}h^{\mu\nu}-2h^{\sigma}_{\mu}\Box\partial_\sigma\partial_\nu
h^{\mu\nu}+h^{\lambda\sigma}\partial_{\sigma}\partial_{\lambda}\partial_{\mu}\partial_{\nu}h^{\mu\nu}].\nonumber\\
\end{eqnarray}
As a result, $
a(\bbox),
b(\bbox),
c(\bbox),
d(\bbox)$ and $
f(\bbox)$ are given by \cite{Biswas:2014tua}, 
\begin{equation}
a(\bbox)=1+M^{-2}_{P}({\cal F}_2(\bbox)\Box+4{\cal F}_3(\bbox)\Box),
\end{equation}
\begin{equation}
b(\bbox)=-1-M^{-2}_{P}({\cal F}_2(\bbox)\Box+4{\cal F}_3(\bbox)\Box),
\end{equation}
\begin{equation}
c(\bbox)=1-M^{-2}_{P}(4{\cal F}_1(\bbox)\Box+{\cal F}_2(\bbox)\Box),
\end{equation}
\begin{equation}
d(\bbox)=-1+M^{-2}_{P}(4{\cal F}_1(\bbox)\Box+{\cal F}_2(\bbox)\Box),
\end{equation}
\begin{equation}
f(\bbox)=2M^{-2}_{P}(2{\cal F}_1(\bbox)\Box+{\cal F}_2(\bbox)\Box+2{\cal
F}_3(\bbox)\Box).
\end{equation}
It can be noted that, 
\begin{equation}
a(\bbox)+b(\bbox)=0,
\end{equation}
\begin{equation}
c(\bbox)+d(\bbox)=0,
\end{equation}
\begin{equation}
b(\bbox)+c(\bbox)+f(\bbox)=0,
\end{equation}
\begin{equation}
a(\bbox)-c(\bbox)=f(\bbox).
\end{equation}
By varying (\ref{lin-act-0}), one obtains the field equations, which can
be represented in terms of the inverse propagator. By writing down the spin
projector operators in $D$-dimensional Minkowski space and representing them
in terms of the momentum space one can obtain the graviton $D$-dimensional
propagator (around Minkowski space) as~\footnote{Obtaining the graviton propagator
for the IDG\ action is not in the scope of this thesis. Such analysis have
been done extensively and in detail by \cite{Biswas:2014tua,Conroy:2015nva}.},  
\begin{equation}
\Pi^{(D)}(-k^{2})=\frac{\mathcal P^{2}}{k^{2}a(-k^{2})}+\frac{\mathcal P^{0}_{s}}{k^{2}[a(-k^{2})-(D-1)c(-k^{2})]}.
\end{equation}
We note that, $\mathcal P^{2}$ and $\mathcal P^{0}_{s}$ are tensor and scalar
spin projector operators respectively. Since we do not wish to introduce
any extra propagating degrees of freedom apart from the massless graviton,
we are going to take $f(\bbox)=0$. Thus, 
\begin{equation}
\Pi^{(D)}(-k^{2})=\frac{1}{k^2a(-k^2)} \left( \mathcal{P}^{2}-\frac{1}{D-2}\mathcal{P}_{s}^{0}
\right).
\end{equation}
To this end, the form of $a(-k^2)$ should be such that it does not introduce
any new propagating
degree of freedom, and 
it was argued in Ref.~\cite{Biswas:2011ar,Biswas:2005qr} that the form of
$a(\Box)$ should be an {\it entire function}, so as not to introduce
any pole in the complex plane, which would result in additional degrees of
freedom in the momentum space. 

Furthermore, the form of $a(-k^2)$ should be such that in the IR, for $k\rightarrow
0,~ a(-k^2) \rightarrow 1$, therefore recovering the propagator 
of GR in the $D$-dimensions. For $D=4$, the propagator has the familiar $1/2$
factor in front of the scalar part of the propagator. One such 
example of an {\it entire function} is~\cite{Biswas:2011ar,Biswas:2005qr}:
\begin{equation}\label{choice}
a(\bbox) = e^{-\bbox}\,,
\end{equation}
which has been found to ameliorate the UV aspects of gravity while recovering
the Newtonian limit in the IR. We conclude that choosing  $f(\bbox)=0$, yields
$a(\bbox)=c(\bbox)$ and therefore we get the following constraint: 
\begin{equation}\label{constraint1}
2{\cal F}_1(\bbox)+{\cal F}_2(\bbox)+2{\cal
F}_3(\bbox)=0.
\end{equation}
At this point, the entropy found in (\ref{entropymastereq}) is very generic
prediction for the IDG\ action. Indeed, the form of entropy is irrespective
of the form of $a(\bbox) $. 
Let us assume that the $(t,~r)$ component of the original
spherically symmetric 
metric given by (\ref{metric1}) takes the form:
\begin{equation}\label{lin-met}
ds^{2}=-(1+2\Phi(r) )dt^{2}+(1-2\Psi (r)) dr^{2}+r^2d\Omega^{2}_{2}
\end{equation}
 In fact, $\Phi$ and $\Psi$ are the two
Newtonian potentials.
Note that we now took $D=4$ in the metric above for the sake of clarity.
Considering the perturbation $g_{\mu\nu}=\eta_{\mu\nu}+h_{\mu\nu}$, we have
\begin{equation}
h_{tt}=h^{tt}=-2\Phi, \quad h_{rr}=h^{rr}=-2\Psi,
\end{equation}
\begin{equation}
h_{\theta\theta}=h^{\theta\theta}=0, \quad h_{\phi\phi}=h^{\phi\phi}=0.
\end{equation}
As we are in the spherical coordinate we shall take the spherical form of
the d'Alembertian operator, 
\begin{equation}
\Box u =\frac{1}{r^2}\partial_r(r^2
\partial_r u)+\frac{1}{r^2 \sin \theta}\partial_\theta(\sin\theta \partial_\theta
u)+\frac{1}{r^2 \sin^2 \theta}\partial_\varphi^2 u-\partial_t^2
u,
\end{equation}
where $u$ is some variable at which we are operating the d'Alembertian operator
at. However, since $\Phi$ and $\Psi$ are $r$-dependent, we are only left
with the first term, \textit{i.e.}
\begin{equation}
\Box u =\frac{1}{r^2}\partial_r(r^2
\partial_r u).
\end{equation}
Now let us take the Wald entropy found in (\ref{entropymastereq}) and calculate
the relevant components in the linearised limit,
\begin{equation}
R_{ab}=R_{tt}
+R_{rr},
\end{equation}
expanding the (\ref{ident-0}) gives, 
\begin{equation}
R_{\mu\nu}=\frac{1}{2}(\partial_{\sigma}\partial_{\mu}h^{\sigma}_{\nu}+\partial_{\nu}\partial_{\sigma}h^{\sigma}_{\mu}-\partial_\nu\partial_{\mu}h-\Box
h_{\mu\nu}),
\end{equation}
hence, 
\begin{eqnarray}R_{tt}&=&\frac{1}{2}(\partial_{\sigma}\partial_{t}h^{\sigma}_{t}+\partial_{t}\partial_{\sigma}h^{\sigma}_{t}-\partial_t\partial_{t}h-\Box
h_{tt})\nonumber\\&=&-\frac{1}{2}\Box
h_{tt}=\Box\Phi=\Phi''+\frac{2\Phi'}{r},
\end{eqnarray}
where `prime' is differentiation with respect to $r$. Next we have, 
\begin{eqnarray}R_{rr}&=&\frac{1}{2}(\partial_{\sigma}\partial_{r}h^{\sigma}_{r}+\partial_{r}\partial_{\sigma}h^{\sigma}_{r}-\partial_r\partial_{r}h-\Box
h_{rr})\nonumber\\
&=&\frac{1}{2}(\partial_{r}\partial_{r}h^{r}_{r}+\partial_{r}\partial_{r}h^{r}_{r}-\partial_r\partial_{r}h-\Box
h_{rr})\nonumber\\
&=&\frac{1}{2}(2\partial^{2}_{r}h^{r}_{r}-\partial_r\partial_{r}h-\Box
h_{rr})\nonumber\\
&=&\frac{1}{2}(2\partial^{2}_{r}(\eta^{rr}h_{rr})-\partial^{2}_{r}(\eta^{tt}h_{tt}+\eta^{rr}h_{rr})-\Box
h_{rr})\nonumber\\
&=&\frac{1}{2}(-4\Psi''-2\Phi''+2\Psi''+2\Psi''+\frac{4\Psi'}{r})\nonumber\\
&=&-\Phi''+\frac{2\Psi'}{r}.
\end{eqnarray}
We note that, 
\begin{equation}
R_{mn}=R_{\theta\theta}+R_{\phi\phi}=0.
\end{equation}
Moving to the Riemann tensor, we have, 
\begin{equation}
R_{\rho\mu\sigma\nu}=\frac{1}{2}(\partial_{\sigma}\partial_{\mu}h_{\rho\nu}+\partial_{\nu}\partial_{\rho}h_{\mu\sigma}-\partial_{\nu}\partial_{\mu}h_{\rho\sigma}-\partial_{\sigma}\partial_{\rho}h_{\mu\nu}),
\end{equation}
and thus, 
\begin{equation}
R_{rtrt}=\frac{1}{2}(\partial_{r}\partial_{t}h_{rt}+\partial_{t}\partial_{r}h_{tr}-\partial_{t}\partial_{t}h_{rr}-\partial_{r}\partial_{r}h_{tt})=\Phi''.
\end{equation}
From (\ref{riemrel}), it follows that, 
\begin{eqnarray}
g^{ab}g^{
m n}R_{ ma nb}&=&2R_{rtrt}+g^{ab}R_{ab}\nonumber\\
&=&2R_{rtrt}+\eta^{tt}R_{tt}+\eta^{rr}R_{rr}\nonumber\\
&=&2\Phi''-\Phi''-\frac{2\Phi'}{r}-\Phi''+\frac{2\Psi'}{r}\nonumber\\
&=&-\frac{2\Phi'}{r}+\frac{2\Psi'}{r}.
\end{eqnarray}
Now let us look back at the entropy equation given in (\ref{entropymastereq})
, and plug in the values, 
\begin{eqnarray}
S_{W}&=&\frac{A_{H}}{4G_{N}}\Big[1-2\Phi+2\Psi-4\alpha
{\cal F}_3(\bbox)\big(\frac{\Psi'-\Phi'}{r}\big)\Big],\end{eqnarray}
where we used the constraint (\ref{constraint1}). Now if we take the Newtonian
potentials to be equal, \textit{i.e.} $\Phi(r)=\Psi(r)$, 
\begin{equation}
S_{W}=\frac{A_{H}}{4G_{N}}.
\end{equation}
To sum up we have shown that, the entropy for a spherically symmetric background
in the linearised regime and within the IDG\ framework is given only by the
area law. This is upon requiring that the massless graviton be the only propagating
mode in the Minkowski background. In other words, we required in the linearised
regime that $2{\cal F}_1(\bbox)+{\cal F}_2(\bbox)+2{\cal
F}_3(\bbox)=0$. 

\subsection{$D$-Dimensional $(A)dS$ Entropy}
We now turn our attention to another class of solutions which contain an horizon,
such as the $(A)dS$ metrics \cite{Conroy:2015nva}, 
where the $D$-dimensional non-local action Eq.~\eqref{action1} must now be
appended with a cosmological constant $\Lambda$ to ensure that $(A)dS$ is a
vacuum solution,  
\begin{eqnarray}
\label{action-main2}
I^{tot} =\frac{1}{16\pi G^{(D)}_{N}}\int d^Dx \sqrt{-g} \bigl[R-2\Lambda+\alpha
\bigl(R\mathcal{F}_1(\Box)R+R_{\mu\nu}\mathcal{F}_2(\Box)R^{\mu\nu}
 \nonumber\\+ R_{\mu\nu\lambda\sigma}
\mathcal{F}_{3}(\Box)R^{\mu\nu\lambda\sigma}\bigr)
\bigr]\,.
\end{eqnarray}
The cosmological constant is then given by
\begin{equation}
\Lambda=\pm\frac{(D-1)(D-2)}{2l^{2}}\,,
\end{equation} 
where the positive sign corresponds to $dS$, negative to $AdS$, and hereafter,
the topmost sign will refer to $dS$ and the bottom to $AdS$. $l$ denotes the
cosmological horizon.
The $(A)dS$ metric can be obtained by taking 
 \begin{equation}
f(r)=\left(1\mp\frac{r^2}{l^2}\right)\,,
\end{equation} 
in Eq.~(\ref{metric1}).
   Recalling the $D$-dimensional entropy Eq.~(\ref{entropymastereq}),   we
write, 
\begin{eqnarray}\label{bh3}
S^{(A)dS}_{W}&=&\frac{A^{(A)dS}_{H}}{4G^{(D)}_{N}}[1+\alpha(2 {\cal F}_1(\Box)+
{\cal
F}_2(\Box)+2
{\cal F}_3(\Box))g^{ab}R_{ab}\nonumber\\&+&2\alpha( {\cal F}_1(\Box)g^{\bar
m\bar n}R_{\bar
m\bar n}-
{\cal F}_3(\Box)g^{ab}g^{\bar
m\bar n}R_{\bar ma\bar nb})]\,,
\end{eqnarray} 
where now \(A^{(A)dS}_{H}\equiv l^{D-2}A_{D-2}\), with \(A_{D-2}=(2\pi^{\frac{D-1}{2}})/\Gamma[\frac{D-1}{2}]\).
Given the $D$-dimensional definitions of curvature in $(A)dS$ background, 
\begin{eqnarray}\label{curvatures}
R_{\mu\nu\lambda\sigma}=\pm\frac{1}{l^{2}}g_{[\mu\lambda}g_{\nu]\sigma},\quad
R_{\mu\nu}=\pm\frac{D-1}{l^{2}}g_{\mu\nu}, \quad R=\pm\frac{D(D-1)}{l^{2}},
\end{eqnarray} 
simple substitution reveals the gravitational entropy in $(A)dS$ can be 
expressed as:
\begin{equation}\label{bh4}
S^{(A)dS}_{W}=\frac{A^{(A)dS}_{H}
}{4G^{(D)}_{N}}(1\pm\frac{2\alpha  }{l^{2}}\{f_{1_0}D(D-1)+f_{2_0}(D-1)^{}+2f_{3_0}\}).
\end{equation} 

Note that $f_{i_{0}}$'s are now simply the leading constants of the functions
${\cal F}_i(\Box)$, due to the nature of curvature in $(A)dS$. 
In particular, in $4$-dimensions, the combination $12f_{1_0}+3f_{2_0}+2f_{3_0}$
is very different from that of the Minkowski space constraint, 
see Eq.~(\ref{constraint1}), required for the massless nature of a graviton
around Minkowski. Deriving the precise form of the
ghost-free constraint in $(A)dS$, is still an open problem for the action given by~(\ref{action-main2}).

\subsection{  Gauss-Bonnet entropy in $(A)dS$ background}

As an example, we will briefly check the entropy of Gauss-Bonnet gravity
in $D$-dimensional $(A)dS$. Recalling that the Lagrangian for the Gauss Bonnet
(GB) modification of gravity in four dimensions is given by,
\begin{equation}
\mathcal{L}_{GB}=\frac{\alpha}{16\pi
G^{(D)}_{N}}\biggl(R^{2}-4R_{\mu\nu}R^{\mu\nu}+R_{\mu\nu\lambda\sigma}R^{\mu\nu\lambda\sigma}\biggr).
\end{equation}
Hence, simply taking \( f_{1_0}=f_{3_0}=1\) and \(f_{2_0}=-4\)  in Eq.~(\ref{bh4}),
recovers the $(A)dS$ entropy of the GB modification of gravity, 
\begin{equation}
S^{(A)dS}_{W}=\frac{A^{(A)dS}_{H}
}{4G^{(D)}_{N}}(1\pm\frac{\alpha2(D-2)(D-3)  }{l^{2}}).
\end{equation}
This result is also found by \cite{Shu:2008yd}, showing the validity of our calculations.
We note that the first term corresponds to the Einstein-Hilbert term. It shall be mentioned that $ l$ denotes the cosmological radius. In the limit $l\rightarrow \infty $, the GB modification to entropy vanishes and this corresponds to the fact that the horizon is flat, so the GB term has no effect on the expression for the entropy, which is simply the area of the event horizon.

\section{Rotating black holes and entropy of modified theories of gravity}
In this section, we show how to obtain the Kerr entropy when the modification
to the general relativity contains higher order curvatures up to Ricci tensor
and also covariant derivatives by modifying the Komar integrals accordingly.
We then obtain the entropy of the Kerr black hole for a number
of modified theories of gravity. We show the corrections to the area law
which occurs due to the modification of the general relativity and present
an argument on  how these
corrections can be vanishing. 

It is well established that the black holes behave as  thermodynamical systems
\cite{Bardeen:1973gs}. The first realisation of this fact was made by Hawking,
\cite{Hawking:1974sw}.  It is discovered that quantum processes make black
holes to emit a thermal flux of particles. As a result, it is possible for
a black hole to be in thermal equilibrium with other systems.  We shall recall
the thermodynamical laws that govern black holes: The zeroth law states that
the horizon of stationary black holes have a constant surface gravity. The
first law states that when stationary black holes are being perturbed the
change in energy is related to the change of area, angular momentum and the
electric charge associated to the black hole. The second law states that,
upon satisfying the null energy condition the surface area of the black hole
can never decrease. This is the law which was realised by Hawking as the area
theorem and showed that black holes radiate. Finally, the third law states
that the black hole can not have vanishing surface gravity.

The second law of the black holes' thermodynamics requires an entropy for
black holes. It was Hawking and Bekenstein, \cite{Bekenstein:1973ur}, who
conjectured that black holes' entropy is proportional to the area of its
event horizon divided by Planck length. Perhaps, this can be seen as one
of the most striking conjectures in modern physics. Indeed, through Bekenstein
bound, \cite{Bekenstein:1972tm},  one can see that the black hole entropy,
as described by the area law, is the maximal entropy that can be achieved
and this was the main hint that led to the holographic principle, \cite{Susskind:1994vu}.

The black hole entropy can be obtained through number of ways. For instance,
Wald \cite{Wald:1993nt} has shown that the entropy for a spherically symmetric
and stationary black hole can be obtained by calculating the  Noether charge,
see the previous sections of this chapter for this approach.
Equivalently, one can obtain the change in mass and angular momentum by using
the Komar formula and subsequently use the definition of the first law of
the black holes' thermodynamics to obtain the entropy. Normally, obtaining
the entropy for non-rotating black holes is very straightforward. In this
case, one uses the  Schwarzschild metric (for a charge-less case) and follows the Wald's approach to
calculate the entropy. Also for rotating black holes that are described by
Kerr metric one can simply use the Komar integrals to find the mass and angular
momentum and finally obtain the entropy. However, when we deviate from Einstein's
theory of general relativity obtaining the conserved charge and hence the
entropy can be challenging. See  \cite{Peng:2014gha,Barnich:2004uw,Barnich:2003xg,Barnich:2001jy,Adami:2017phg}
for advancement in finding the conserved charges.

In this section, we are going to briefly review the notion of Noether and
Komar currents in variational relativity. We show how the two are identical
and then we move to calculate the entropy of Kerr black holes for a number
of examples, namely $f(R)$ gravity, $f(R,R_{\mu\nu})$ theories where the
action can contain higher order curvatures up to Ricci tensor and finally
higher derivative gravity. The entropy in each case is obtained by calculating
the modified Komar integrals. 

\subsection{Variational principle, Noether and Komar currents} 
Variational principle is a powerful tool in physics. Most of the laws in
physics are derived by using this rather simple and straightforward method.
 Given a gravitational  Lagrangian, 
 \begin{equation}\label{genl}
L=L(g_{\mu\nu}, R_{\mu\nu},\nabla_{a_1}R_{\mu\nu},\dots,\n_{(\al_1}\dots
\n_{\al_m)}R_{\mu \nu  }),
\end{equation}
where the Lagrangian is a constructed by the metric,  Ricci tensors and its
derivatives (This can be generalised to the case where Riemann tensors are
involved, however that is beyond the scope of this section). Note that the
parentheses denote symmetrisation. We can obtain the equations of motion
by simply varying the action with respect to the inverse metric, $g^{\mu\nu}$
and $R_{\mu\nu}$. In short form, this can be done by defining two \textit{covariant
momenta}~\cite{Iyer:1994ys}:
\begin{equation}
\pi_{\mu\nu}=\frac{\delta \mathcal{L}}{\delta g^{\mu\nu}},
\end{equation}
and, 
\begin{eqnarray}
P^{\mu\nu}&=&\frac{\delta
\mathcal{L}}{\delta R_{\mu\nu}}\nonumber\\&=&\frac{\p \cL}{\p
R_{\mu \nu  }}-\n_{\al_1}\frac{\p \cL}{\p \n_{\al_1}R_{\mu \nu 
}}+\dots +(-1)^{m}\n_{(\al_1}\dots \n_{\al_m)}\frac{\p \cL}{\p \n_{(\al_1}\dots
\n_{\al_m)}R_{\mu \nu  }}.\nonumber\\
\end{eqnarray}
Thus the variation of the Lagrangian would be given by \cite{Fatibene:1998rq}:

\begin{equation}
\delta \mathcal{L}=\pi_{\mu\nu}\delta g^{\mu\nu}+P^{\mu\nu}\delta R_{\mu\nu}.
\end{equation}
It is simple to see that in the example of Einstein Hilbert (EH) action,
 the first term admits the equations of motion (\textit{i.e.} $\pi_{\mu\nu}=0$)
and the second term will be the boundary term. Since we are considering gravitational
theories, the general covariance must be preserved at all time. In other
words, the Lagrangian, $L$, is covariant with respect to the action under
diffeomorphisms of space-time. Infinitesimally, the variation can be expressed
as: \begin{equation}
\delta_{\xi}L=d(i_{\xi}L)=\pi_{\mu\nu}\lie_{\xi} g^{\mu\nu}+P^{\mu\nu}\lie_{\xi}
R_{\mu\nu},
\end{equation}
where $\delta_{\xi}$ denotes an infinitesimal variation of the gravitational
action, $d$ is the exterior derivative, $i_\xi$ is interior derivative of
forms along vector field $\xi$ and $\lie_{\xi}$ is the Lie derivative with
respect to the vector field. By expanding the Lie derivative of the Ricci
tensor, and noting that:
\begin{equation}\label{metricvar}
\delta_{\xi}g_{\alpha\beta}=\pounds_{\xi}g_{\alpha\beta}=\nabla_\alpha\xi_\beta+\nabla_\beta\xi_\alpha,
\end{equation}
 the  \textit{Nother conserved current} can be obtained.  The way this
can be done for the EH\ action is  demonstrated in
Appendix \ref{ehcons} as an example. Furthermore, the conserved \textit{Noether
current associated to the general covariance of the Einstein- Hilbert action
is identical to the generalised Komar current}. This can be seen explicitly
in Appendix
\ref{generalisedkomar}. In general, we define the Komar current\footnote{As
a check it can be seen that for EH\ action we have, 
$$\label{ehpotential}
P^{\alpha\beta}_{EH}=\sqrt{-g}g^{\alpha\beta},\qquad\mathcal{U}_{EH}=\sqrt{-g}\nabla_{\alpha}\xi^{[\mu}g^{\nu]\alpha}ds_{\mu\nu}
$$
which is exactly the same as what we obtain in Eq. (\ref{noethercurrent}).}as
\cite{Fatibene:1998rq}: 
\begin{equation}\label{generalisedkomarr}
\mathcal{U}=\nabla_{\alpha}\xi^{[\mu}P^{\nu]\alpha}ds_{\mu\nu},
\end{equation}
where $ds_{\mu\nu}$ denotes the surface elements for a given background and
is the standard basis for $n-2$-forms over the manifold $M$ ($n=\mathrm{dim}(M)$).

\subsection{Thermodynamics of Kerr black hole} 
 A solution to the Einstein field equations describing  rotating black holes
was discovered by Roy Kerr. This is a solution that only describes a rotating
black hole without charge. Indeed, there is a solution for charged black
holes (\textit{i.e.} satisfies Einstein-Maxwell equations) known as Kerr-Newman.
Kerr metric can be written in number of ways and in this section we are going
to use the Boyer-Lindquist coordinate. The metric is given by \cite{Relativiststoolkit}

 \begin{align}\label{metric}
ds^{2}&=-(1-\frac{2Mr}{\rho^{2}})dt^{2}-\frac{4Mar\sin^{2}\theta}{\rho^{2}}dtd\phi+\frac{\Sigma}{\rho^{2}}\sin^{2}\theta
d\phi^{2}+\frac{\rho^{2}}{\Delta}dr^{2}+\rho^{2}d\theta^{2},
\end{align}
where, 
\begin{eqnarray}
\rho^2=r^{2}+a^2 \cos^2 \theta, \quad \Delta=r^{2}-2Mr+a^{2},\quad \Sigma=(r^{2}+a^2
)^{2}-a^{2}\Delta\sin^{2}\theta.\nonumber\\
\end{eqnarray}
The metric is singular at $\rho^2=0$. This singularity is real\footnote{This
is different
than the singularity at $\Delta=0$ which is a coordinate singularity.} and
can be checked via Kretschmann scalar\footnote{The Kretschmann scalar for
Kerr metric is given by: $R^{\alpha\beta\gamma\delta}R_{\alpha\beta\gamma\delta}=\frac{48M^{2}(r^{2}-a^{2}\cos^{2}\theta)(\rho^{4}-16a^{2}r^{2}\cos^{2}\theta)}{\rho^{12}}$.}
\footnote{We shall note that scalar curvature,
$R$, and Ricci tensor, $R_{\mu\nu}$ are vanishing for the Kerr metric and
only some components of the Riemann curvature are non-vanishing.}. The above
metric has two horizons $r_{\pm}=m\pm\sqrt{m^{2}-a^{2}}$. Furthermore, $a^2
\leq m^2$  is a length scale. Let us define the vector:
 \begin{equation}\label{comb}
\xi^{\alpha}=t^{\alpha}+\Omega\phi^{\alpha}.
\end{equation}
This vector is null at the event horizon. It is tangent to the horizon's
null generators, which wrap around the horizon with angular velocity $\Omega$.
Vector $\xi^{\alpha}$ is a Killing vector since it is equal to sum of two
Killing vectors. After all, the event horizon of the Kerr metric is a Killing
horizon.  Using Eqs. (\ref{generalisedkomarr}) and (\ref{comb}) we can define
the \textit{Komar integrals} for the general Lagrangian (\ref{genl}) describing
the energy and the angular momentum of the Kerr black hole as, 
\begin{equation}
\mathcal{E}=-\frac{1}{8\pi}\lim_{S_{t}\rightarrow \infty}\oint_{S_{t}}\nabla_{\lambda}P^{\alpha{\lambda}}\xi^{\beta}_{(t)}ds_{\alpha\beta},
\end{equation}
\begin{equation}
\mathcal{J}=\frac{1}{16\pi}\lim_{S_{t}\rightarrow \infty}\oint_{S_{t}}\nabla_{\lambda}P^{\alpha{\lambda}}\xi^{\beta}_{(\phi)}ds_{\alpha\beta},
\end{equation}
where the integral is over $S_{t}$, which is a closed two-surfaces\footnote{Note
that we can write $\lim_{S_{t}\rightarrow \infty}\oint_{S_{t}}$ as simply
$\oint_{\mathcal{H}}$ where $\mathcal{H}$ is a two dimensional cross section
of the event horizon.}. 
We shall note that $S_{t}$ is an $n-2$ surface. In above definitions $\xi^{\beta}_{(t)}$
is the space-time's time-like Killing vector and $\xi^{\beta}_{(\phi)}$ is
the rotational Killing vector and they both satisfy the Killing's equation,
$\xi_{\alpha;\beta}+\xi_{\beta;\alpha}=0$. Moreover, the sign difference
in two definition has its root in the signature of the metric. In this
thesis we are using mostly plus signature. The surface element is also given
by, 
 \begin{equation}
ds_{\alpha\beta}=-2n_{[\alpha}r_{\beta]}\sqrt{\sigma}d\theta
d\phi,
\end{equation}
where $n_{\alpha}$ and $r_{\alpha}$ are the time-like (\textit{i.e.} $n_{\alpha}n^{\alpha}=-1$)
and space-like (\textit{i.e.} $r_{\alpha}r^{\alpha}=1$) normals to $S_{t}$.
For Kerr metric in Eq. (\ref{metric})
the normal vectors are defined as: 
\begin{equation}
n_\alpha=(-\frac{1}{\sqrt{-g^{tt}}},0,0,0)=(-\sqrt{\frac{\rho^{2}\Delta}{\Sigma}},0,0,0),
\end{equation}
\begin{equation}
r_\beta=(0,\frac{1}{\sqrt{g^{rr}}},0,0)=(0,\sqrt{\frac{\rho^{2}}{\Delta}},0,0).
\end{equation}
Furthermore, the  two dimensional cross section
of the event horizon described by $t=$constant and also $r=r_+$ (\textit{i.e.}
constant), hence, from metric in Eq. (\ref{metric}) we can extract the induced
metric as: 
\begin{equation}
\sigma_{AB}d\theta^{A}d\theta^{B}=\rho^{2}d\theta^{2}+\frac{\Sigma}{\rho^{2}}\sin^{2}\theta
d\phi^{2}.
\end{equation}
Thus we can write, 
\begin{equation}\label{detinduced}
\sqrt{\sigma}=\sqrt{\Sigma}\sin\theta d\theta d\phi.
\end{equation}

\textit{First law} of black hole thermodynamics states that when a stationary
black hole at manifold $\mathcal{M}$ is perturbed slightly to $\mathcal{M}+\delta\mathcal{M}$,
 the difference in the energy, $\mathcal{E}$, angular momentum, $\mathcal{J}_{a}$,
and area, $\mathcal{A}$, of the black hole are related by:
\begin{eqnarray}\label{mainentropy}
\delta \mathcal{E}=\Omega^{a}\delta \mathcal{J}_{a}+\frac{\kappa}{8\pi}\delta
\mathcal{A}=\Omega^{a}\delta \mathcal{J}_{a}+\frac{\kappa}{2\pi}\delta
\mathcal{\mathcal{S}},
\end{eqnarray}
where  $\Omega^{a}$ are the angular velocities at the horizon. We shall note
that $\mathcal{S}$ is the associated entropy.
$\kappa$ denotes the \textit{surface gravity} of the Killing horizon 
and for the metric given in Eq. (\ref{metric}) the surface gravity is given
by
\be\label{surfacegravitykerr}
\ka=\frac{\sqrt{m^2-a^2}}{2mr_{+}} \,.
\ee
 The surface area \cite{Relativiststoolkit} of the black hole is given by\footnote{We shall
note that $\mathcal{S}=\mathcal{A}/4$
(with $G=1)$ denotes the Bekenstein-Hawking entropy.
}: 
\begin{equation}
\mathcal{A}=\oint_{\cH}\sqrt{\sa}d^2 \theta,
\end{equation}
where $d^2 \theta=d\theta d\phi$. Now by using Eq. (\ref{detinduced}), the
surface
area can be obtained as, \begin{equation}
\mathcal{A}=\oint_{\cH}\sqrt{\sa}d^2 \theta=\int_{0}^{\pi} \sin (\theta)d\theta\int_{0}^{2\pi}d\phi(r^{2}_{+}+a^{2})=4\pi(r^{2}_{+}+a^{2}).
\, 
\end{equation}
Modified theories of gravity were proposed as an attempt to describe some
of the phenomena that Einstein's theory of general relativity can not address.
Examples of these phenomena can vary from explaining the singularity to the
dark energy. In the next subsections, we obtain the entropy of the Kerr black
hole
for number of these theories.

\subsection{Einstein-Hilbert action}
As a warm up exercise let us start the calculation  for  the most well knows
case, where the action is given by: \textbf{}
\begin{equation}
S_{EH}=\frac{1}{2}\int d^{4}x\sqrt{-g}M^{2}_{P} R,
\end{equation} 
where $M^{2}_{P}$ is the Planck mass squared. For this case, as shown in footnote
\ref{ehpotential}, the Komar integrals can be found explicitly as \cite{Fatibene:1998rq},
(see Appendix \ref{boyerlinqderivation} for derivation)
\begin{eqnarray}
\mathcal{E}&=&-\frac{1}{8\pi}\oint_{\mathcal{H}}\nabla^{\alpha}t^{\beta}ds_{\alpha\beta}\nonumber\\
&=&-\frac{1}{8\pi}\int^{2\pi}_{0} d\phi\int^{\pi}_{0} d\theta\nonumber\\&&\Bigg(\frac{1}{2}
\sin (\theta ) \left(a^2 \cos (2 \theta )+a^2+2 r^2\right)\frac{8 m \left(a^2+r^2\right)
\left(a^2 \cos (2 \theta )+a^2-2 r^2\right)}{\left(a^2
\cos (2 \theta )+a^2+2 r^2\right)^3}\Bigg)=m.\nonumber\\ 
\end{eqnarray}
We took $\xi^{\alpha}=t^{\alpha}$, where $t^{\alpha}=\frac{\partial
x^{\alpha}}{\partial t}$; $x^{\alpha}$ are the space-time coordinates.
So, for instance, $g_{\mu \nu}\xi^{\mu}\xi^{\nu}=g_{\mu \nu}t^{\mu}t^{\nu}=g_{tt}$,
that is after the contraction of the metric with two Killing vectors, one
is left with the $tt$ component of the metric.
In similar manner, we can calculate the angular momentum as, 
\begin{eqnarray}
\mathcal{J}&=&\frac{1}{16\pi}\oint_{\mathcal{H}}\nabla^{\alpha}\phi^{\beta}ds_{\alpha\beta}\nonumber\\
&=&\frac{1}{16\pi}\int^{2\pi}_{0} d\phi\int^{\pi}_{0} d\theta\nonumber\\&&\Bigg(\frac{1}{2}
\sin (\theta ) \left(a^2 \cos (2 \theta )+a^2+2 r^2\right)\nonumber\\&\times&\frac{-8
a m \sin ^2(\theta ) \left(a^4-3 a^2 r^2+a^2 (a-r) (a+r) \cos (2 \theta )-6
r^4\right)}{\left(a^2 \cos (2 \theta )+a^2+2 r^2\right)^3}\Bigg)=ma.\nonumber\\
\end{eqnarray}
Now given Eq. (\ref{mainentropy}), we have, 
\begin{equation}
\frac{\kappa}{2\pi}\delta
\mathcal{\mathcal{S}}=\delta \mathcal{E}-\Omega^{a}\delta \mathcal{J}_{a}=(1-\Omega
a)^{}\delta m-\Omega^{}m^{}\delta a.
\end{equation}
By recalling the surface gravity from Eq. (\ref{surfacegravitykerr})
we have, 
\be
\mathcal{S}=2\pi m r_{+}. 
\ee
which is a well known result. 

\subsection{$f(R)$ theories of gravity} 
There are numerous ways to modify the Einstein theory of general relativity,
one of which is going to higher order curvatures. A class of theories which
attracted attention in recent years is the $f(R)$ theory of gravity \cite{Sotiriou:2008rp}.
This type of theories can be seen as  the series expansion of the scalar
curvature, $R$, and one of the very important features of them is that they
can avoid Ostrogradski instability.
The action of this gravitational theory is generally given by:
\begin{equation}\label{frgravity}
S_{f(R)}=\frac{1}{2}\int d^{4}x\sqrt{-g} f(R),
\end{equation}
where $f(R)$ is the function of scalar curvature and it can be of any order.
In this case the Komar potential can be obtained by,
 
\begin{equation}
P^{\alpha\beta}_{f(R)}=\frac{\delta f(R)}{\delta R}\frac{\delta R}{\delta
R_{\alpha\beta}}=\frac{1}{2}f'(R)\sqrt{-g}g^{\alpha\beta},
\end{equation}
and thus: 
\begin{equation}
\mathcal{U}_{f(R)}=\frac{1}{2}f'(R)\sqrt{-g}\nabla_{\alpha}\xi^{[\mu}g^{\nu]\alpha}ds_{\mu\nu}.
\end{equation}
This results in modification of the energy and angular momentum as (see Appendix
\ref{frvariationconserved} for validation), 
\begin{equation}\label{conserved1}
\mathcal{E}_{f(R)}=-\frac{1}{8\pi}\oint_{\mathcal{H}}f'(R)\nabla^{\alpha}t^{\beta}ds_{\alpha\beta}=f'(R)m,
\end{equation}
and
\begin{equation}\label{conserved2}
\mathcal{J}_{f(R)}=\frac{1}{16\pi}\oint_{\mathcal{H}}f'(R)\nabla^{\alpha}\phi^{\beta}ds_{\alpha\beta}=f'(R)ma.
\end{equation}
We know that the $f(R)$ theory of gravity is essentially the power expansion
in the scalar curvature, \begin{equation}
f(R)=M_{P}^{2} R+\alpha _{1}R^{2}+\alpha _{2}R^{3}+\cdots+\alpha _{n-1}R^{n},
\end{equation}
where $\alpha_i$ maintains the correct dimensionality, and thus, 
\begin{equation}\label{diff}
f'(R)=M_{P}^{2}+2\alpha _{1}R^{}+3\alpha _{2}R^{2}+\cdots+n\alpha _{n-1}R^{n-1}.
\end{equation}
As a result, the entropy of $f(R)$ theory of gravity is given only by the
Einstein Hilbert contribution, 
\be
\mathcal{S}_{f(R)}=\mathcal{S}_{EH}=2\pi m r_{+}. 
\ee
This is due to fact that the scalar curvature, $R$, is vanishing for the
Kerr metric given in Eq. (\ref{metric}) and so only the leading term in Eq.
(\ref{diff}) will be accountable.

\subsection{$f(R,R_{\mu\nu})$}
After considering the $f(R)$ theories of gravity, it is natural to think
about the more general form of gravitational modification. In this case:
$f(R,R_{\mu\nu}),$  the action would contain terms like $R_{\mu\nu}R^{\mu\nu}$,
$R^{\mu\alpha}R^{ \ \nu}_{\alpha}R_{\nu\mu}$ and so on. Let us take a specific
example of, 
\begin{equation}
S_{R_{\mu\nu}}=\frac{1}{2}\int d^{4}x\sqrt{-g} (M_{P}^{2} R+\lambda_{1}R_{\mu\nu}R^{\mu\nu}+\lambda_{2}R^{\mu\lambda}R^{
\ \nu}_{\lambda}R_{\nu\mu}),
\end{equation}
where $\lambda_{1}$ and $\lambda_{2}$ are coefficients of appropriate dimension
(\textit{i.e.} mass dimension $L^2$ and $L^4$ respectively where $L$ denotes
length).  The momenta would then be obtained as, 
\begin{align}
P^{\al \bt}_{R_{\mu\nu}}=\frac{\sqrt{-g}}{2}(M_{P}^{2} g^{\al \bt}+2\lambda_{1}R^{\al
\bt}+3\lambda_{2}R^{\bt \lambda}R^{\al}_{\ \lambda}).
\end{align}
As before, the only contribution comes from the EH term since the Ricci tensor
is vanishing for the Kerr metric given in  Eq.
(\ref{metric}). So, without proceeding further, we can conclude that in this
case the entropy is given by the area law only and  with no correction.

\subsection{Higher derivative gravity}

Another class of modified theories of gravity are the higher derivative theories.
We shall denote the action by $S(g,R,\nabla R,\nabla R_{\mu\nu},\cdots)$.
In this class, there are covariant derivatives acting on the curvatures.
Moreover, there are theories that contain inverse derivatives acting on the
curvatures \cite{Conroy:2014eja}. These are known as non-local theories of
gravity. 

 A well established class of higher derivative theory of gravity is given
by \cite{Biswas:2005qr} where the action contains infinite derivatives acting
on the curvatures. It has been shown that having infinite derivatives can
cure the singularity problem \cite{Biswas:2011ar}. This is achieved by replacing
the singularity with a bounce. Moreover, this class of theory preserves the
ghost freedom. This is of a very special importance, since in other classes
of modified gravity, deviating from the EH term and going to higher order
curvature terms means one will have to face the ghost states.  Having infinite
number of derivatives makes it extremely difficult to find a metric solution
which satisfies the equations of motion. Moreover,  infinite derivative theory
is associated with singularity freedom and Kerr metric is a singular one.
As a result, in this section we wish to consider a finite derivative example
as a matter of illustration,  let us define the Lagrangian of the form: 
\be\label{HD}
\cL_{HD}=\sqrt{-g} \LT M_{P}^{2}R+R \mathcal{F}_{1}(\bar{\Box})R+R_{\mu
\nu} \mathcal{F}_{2}(\bar{\Box}) R^{\mu \nu} \RT \,,
\ee
where $\mathcal{F}_{1}(\bar{\Box})=\sum_{n=1}^{m_1}f_{1_n}\bar{\Box}^n$,
$\mathcal{F}_{2}(\bar{\Box})=\sum_{n=1}^{m_2}f_{2_n}\bar{\Box}^n$ while $\Box=g^{\mu\nu}\nabla_{\mu}\nabla_{\nu}$
is d'Alembertian operator and $\bar{\Box}=\Box/M^{2}$ to ensure the correct
dimensionality. Note that $f_{i_{n}}$ are the coefficient of the expansion.
 We also note that $m_1$ and $m_2$ are some finite number. In this case,
we have the Lagrange momenta as, 
\begin{align}
P^{\al \bt}_{HD} 
=\sqrt{-g} \LT  M_{P}^{2} g^{\al \bt}+2f_{1_n}g^{\al \bt}\sum_{n=1}^{m_1}
\bar{\Box}^{n}R+2f_{2_n}\sum_{n=1}^{m_2}
\bar{\Box}^{n}R^{\al \bt} \RT  
 .
\end{align}
As mentioned previously for the Kerr metric:  $R=R^{\al \bt}=0$, this is
to conclude that the only non-vanishing term which will contribute to the
entropy will be the first term, in the above equation, which corresponds
to the EH term in the action given in Eq. (\ref{HD}).

\subsection{Kerr metric as and solution of modified gravities}
After providing some examples of the modified theories of gravity, the reader
might ask wether the Kerr metric is the solution of these theories. In this
section we are going to address this issue by considering the higher derivative
action. This is due to the fact that the higher derivative action given in
(\ref{HD}) contains  Ricci scalar and Ricci tensor and additionally their
derivatives and hence the arguments can be applied to other theories provided
in this section. 

Let us consider (\ref{HD}), the equations of motion is given by \cite{Biswas:2013cha},
\begin{eqnarray}\label{hdeom}
&&G^{\alpha\beta}+4G^{\alpha\beta}{\cal
F}_{1}(\bar \Box)R+g^{\alpha\beta}R{\cal
F}_1(\bar \Box)R-4\left(\triangledown^{\alpha}\nabla^{\beta}-g^{\alpha\beta}
\square\right){\cal F}_{1}(\bar \Box)R
\nonumber\\&&
-2\Omega_{1}^{\alpha\beta}+g^{\alpha\beta}(\Omega_{1\sigma}^{\;\sigma}+\bar{
\Omega}_{1}) +4R_{\mu}^{\alpha}{\cal F}_2(\bar \Box)R^{\mu\beta}
\nonumber\\&&
-g^{\alpha\beta}R_{\nu}^{\mu}{\cal
F}_{2}(\bar \Box)R_{\mu}^{\nu}-4\triangledown_{\mu}\triangledown^{\beta}({\cal
F}_{2}(\bar \Box)R^{\mu\alpha})
+2\square({\cal
F}_{2}(\bar \Box)R^{\alpha\beta})
\nonumber\\&&
+2g^{\alpha\beta}\triangledown_{\mu}\triangledown_{
\nu}({\cal F}_{2}(\bar \Box)R^{\mu\nu})
-2\Omega_{2}^{\alpha\beta}+g^{\alpha\beta}(\Omega_{2\sigma}^{\;\sigma}+\bar{
\Omega}_{2}) -4\Delta_{2}^{\alpha\beta}
=0\,,
\end{eqnarray}
 we have defined the following symmetric tensors \cite{Biswas:2013cha}:
\ba \label{drolo}
&&\Omega_{1}^{\alpha\beta}=\sum_{n=1}^{\infty}f_{1_{n}}\sum_{l=0}^{n-1}\nabla^{
\alpha}R^{(l)}\nabla^{\beta}R^{(n-l-1)},\quad\bar{\Omega}_{1}=\sum_{n=1}^{\infty
}f_{1_{n}}\sum_{l=0}^{n-1}R^{(l)}R^{(n-l)},
\nonumber\\
&&\Omega_{2}^{\alpha\beta}=\sum_{n=1}^{\infty}f_{2_{n}}\sum_{l=0}^{n-1}R_{\nu}^{
\mu;\alpha(l)}R_{\mu}^{\nu;\beta(n-l-1)},\quad\bar{\Omega}_{2}=\sum_{n=1}^{
\infty}f_{2_{n}}\sum_{l=0}^{n-1}R_{\nu}^{\mu(l)}R_{\mu}^{\nu(n-l)}\,,
\nonumber\\
&&\Delta_{2}^{\alpha\beta}=\frac{1}{2}\sum_{n=1}^{\infty}f_{2_{n}}\sum_{l=0}^{n-1}
[R_{
\;\sigma}^{\nu(l)}R^{(\beta|\sigma|;\alpha)(n-l-1)}-R_{\;\sigma}^{\nu;(\alpha(l)
}R^{
\beta)\sigma(n-l-1)}]_{;\nu}\,,
 \ea
Also note that $R^{(m)}\equiv\Box^{m}R$ and that we absorbed the mass dimension
in $f_{i_n}$'s where it was necessary. For the Kerr metric, the Ricci tensor
and, consequently, the Ricci scalar vanish identically. Therefore, any quantity
with covariant derivatives acting on the Ricci tensor and the Ricci scalar
would also be equal to zero; as a result, each of the five quantities defined
in~\eqref{drolo} would also vanish.
Hence, one may observe that each of the terms in the left-hand side of~\eqref{hdeom}
becomes equal to zero. Thus, the Kerr metric satisfies the equation of motion~\eqref{hdeom},
implying
that the Kerr metric can be regarded as a solution of the action~\eqref{HD}. This has been
shown explicitly by \cite{Li:2015bqa}. However, the same can not be said
in the presence of the Riemann tensors.

\section{Non-local gravity}
For higher derivative theories of gravity, it is possible to write the action
in terms of auxiliary fields. Doing so results in converting a non-local
action \cite{Conroy:2014eja} to a local one. We use this approach to find the entropy for a non-local
gravitational action. 

Indeed, Einstein's theory of general relativity can be modified in number
of ways
to address different aspects of cosmology. Non-local gravity
is constructed by inversed d'Alembertian operators that are accountable in
the IR regime. In particular, they could filter out the contribution of the
cosmological constant to the gravitating energy density, possibly providing
the key to solving one of the most notorious problems in physics \cite{ArkaniHamed:2002fu}, see
also \cite{Mitsou:2015yfa}.

Moreover, such modification to the theory of general relativity arises
naturally as quantum loop effect \cite{Conroy:2014eja} and used initially by \cite{Deser:2007jk}
to explain the cosmic acceleration. Non-local gravity further used to explain
dark energy \cite{Maggiore:2014sia}. Since such gravity is associated with
large distances, it is also possible to use it as an alternative to understand
the cosmological constant \cite{ArkaniHamed:2002fu}. Additionally, non-local
corrections arise, in the leading order, in the context of bosonic string
\cite{Aharony:2011gb}. 

It is argued in \cite{Steve} that, non-locality may have a positive rule
in understanding the black hole information problem. Recently, the non-local
effect was studied in the context of Schwarzschild black hole \cite{Mitsou:2015yfa,Calmet:2017qqa}.
In similar manner, the entropy of some non-local models were studied in \cite{Solodukhin:2012ar}.

\subsection{Higher derivative gravity reparametrisation}
Let us take the following higher derivative action, 
\begin{eqnarray}\label{local1}
&&I_{0}+I_{1}=\frac{1}{16\pi G^{(D)}_{N}}\int d^{D}x\sqrt{-g}\big[R+ R F(\bar\Box)R\big],
\nonumber\\&&\text{with:} \quad F(\bar\Box)=\sum^{m}_{n=0}f_{n}\bar\Box^{n},
\end{eqnarray}
where $G^{(D)}_{N}$ is the $D$ dimensional Newton's gravitational constant, $R$
is scalar curvature,
$\Box=\nabla_{\mu}\nabla^{\mu}$ is the d'Alembertian operator and $\bar\Box\equiv\Box/M^{2}$,
this is due to the fact that $\Box$ has dimension mass squared and we wish
to have dimensionless $F(\bar\Box)$, we shall note that $f_n$'s are dimensionless
coefficients of the series expansion. In the above action we denoted the
EH term as $I_{0}$. Finally, $m$  is some finite positive
integer. The
above action can be written as \cite{Mazumdar:2017kxr},
\begin{eqnarray}\label{param1}
I_0+\tilde I_{1}=\frac{1}{16\pi G^{(D)}_{N}}\int d^{D}x\sqrt{-g}\Big[R+
\sum^{m}_{n=0}\Big(Rf_n\eta_n+R\chi_n(\eta_n-\bar\Box^{n}R)\Big)\Big], \nonumber\\
\end{eqnarray} 
where we introduced two auxiliary fields $\chi_n$ and $\eta_n$. This is the
method which we used in the Hamiltonian chapter. By solving
the equations of motion for $\chi_n$, we obtain: 
$\eta_n=\bar\Box^{n}R$, and hence the original action given in Eq. (\ref{local1}
) can be recovered. This equivalence is also noted in \cite{Mazumdar:2017kxr}.

We are now going to use the Wald's prescription given in (\ref{entropyrtrt})
over the spherically symmetric metric  (\ref{metric1}). We know from (\ref{entropyeq1})
that the entropy for action (\ref{local1}) is given by,  
\begin{equation}
\mathcal{S}_0+\mathcal{S}_1= \frac{A_H}{4G^{(D)}_{N}}\Big(1+2 F(\bar\Box)R\Big),
\end{equation}
where we denoted the entropy by $\cal S$.
Now let us obtain the entropy for $\tilde I_{1}$ , following the entropy
Eq. (\ref{entropyrtrt}), we have, 
\begin{eqnarray}
\tilde{\mathcal{S}}_1&=&-\frac{ A_H}{2G^{(D)}_{N}}\times\sum^{m}_{n=0}(-\frac{1}{2}f_n\eta_n-\frac{1}{2}\chi_n\eta_n+\chi_n\bar\Box^{n}R)\nonumber\\
&=&-\frac{ A_H}{2G^{(D)}_{N}}\times\sum^{m}_{n=0}(-\frac{1}{2}f_n\eta_n-\frac{1}{2}\chi_n\eta_n+\chi_n\eta_n)\nonumber\\
&=&\frac{ A_H}{4G^{(D)}_{N}}\times\sum^{m}_{n=0}(2f_n\eta_n)=\frac{ A_H}{4 G}(2F(\bar\Box)R).
\end{eqnarray}
Where we fixed the lagrange multiplier as $\chi_n=-f_n$. It is clear that
both $I_{1}$ and $\tilde I_{1}$ are giving the same result for the entropy
as they should. This is to verify that it
is always possible to use the equivalent action and find the correct entropy.
This method is very advantageous in the case of non-local gravity, where
we have inversed operators.

Before proceeding to the non-local case let us consider $\tilde I_{1}$, 
\begin{equation}\label{explaination}
\tilde I_{1}=\frac{1}{16\pi G^{(D)}_{N}}\int d^{D}x\sqrt{-g}\sum^{m}_{n=0}\Big(Rf_n\eta_n+R\chi_n(\eta_n-\bar\Box^{n}R)\Big),
\end{equation}
It is mentioned that solving the equations of motion for $\chi_n$ results
in:\begin{equation}
\eta_n\equiv\bar\Box^{n}R. 
\end{equation}
We shall mention that in order to form
\begin{equation}
F(\bar\Box)=\sum^{m}_{n=0}f_{n}\bar\Box^{n},
\end{equation}
in the second term of (\ref{explaination})
we absorbed the $f_n$ into the Lagrange multiplier, $\chi_n$. Let us consider
the fixing $\chi_n=-f_n$. We do so by substituting the value of the Lagrange
multiplier,
\begin{eqnarray}
\tilde I_{1}&=&\frac{1}{16\pi G^{(D)}_{N}}\int d^{D}x\sqrt{-g}\sum^{m}_{n=0}\Big(Rf_n\eta_n-Rf_n(\eta_n-\bar\Box^{n}R)\Big)\nonumber\\
&=&\frac{1}{16\pi G^{(D)}_{N}}\int d^{D}x\sqrt{-g}\sum^{m}_{n=0}Rf_n\bar\Box^{n}R\equiv\frac{1}{16\pi
G^{(D)}_{N}}\int d^{D}x\sqrt{-g}RF(\bar\Box)R.\nonumber\\
\end{eqnarray}
As expected $\tilde I_{1}$ and $I_{1}$ are again equivalent. Thus the fixation
of the Lagrange multiplier is valid. 

\subsection{Non-local gravity's entropy}
The non-local action can be written as,

\begin{eqnarray}\label{nonloc1}
&&I_0+I_{2}=\frac{1}{16\pi G^{(D)}_{N}}\int d^{D}x\sqrt{-g}\big[R+ RG(\bar\Box)R\big],
\nonumber\\&&\text{with:} \quad G(\bar\Box)=\sum^{m}_{n=0}c_{n}\bar\Box^{-n}.
\end{eqnarray}
In this case the inversed d'Alembertian operators are acting on  the scalar
curvature. In order to localise the above action we are going to introduce
 two
auxiliary fields $\xi_n$ and $\psi_n$ and rewrite the action in its local
form as, 
\begin{eqnarray}\label{nonloc2}
I_0+\tilde{I}_{2}&=&\frac{1}{16\pi G^{(D)}_{N}}\int d^{D}x\sqrt{-g}\Big[R+
\sum^{m}_{n=0}\Big(Rc_n\psi_n+R\xi_n(\bar\Box^{n}\psi_{n}-R)\Big)\Big].\nonumber\\ 
\end{eqnarray} 
 Solving the
equations of motion for $\xi_n$, results in having: 
\begin{equation}\label{property}
\bar\Box^{n}\psi_{n}=R \quad\text{or}\quad \psi_{n}=\bar\Box^{-n}R. 
\end{equation}
Thus, the original action given in Eq. (\ref{nonloc1}) can be recovered.
This equivalence is also noted by \cite{DeFelice:2014kma,Solodukhin:2012ar}.

Finding  the Wald's entropy for the non-local action as stands in (\ref{nonloc1})
can be a challenging task, this
is due to the fact that for an action of the form Eq. (\ref{nonloc1}), the
functional differentiation contains inversed operators acting on the scalar
curvature and Wald's prescription for such case can not be applied. However,
by introducing the equivalent action and localising the gravity as given
in Eq.
(\ref{nonloc2}), one can obtain the entropy as it had been done in the previous
case. We know that the contribution of the EH\ term to the entropy is $\mathcal{S}_{0}=A_H/4G
$. Thus we shall consider the entropy of $\tilde{I}_{2}$: 
\begin{eqnarray}
\tilde{\mathcal{S}}_2&=-&\frac{ A_H}{2G^{(D)}_{N}}\times\sum^{m}_{n=0}(-\frac{1}{2}c_n\psi_n-\frac{1}{2}\xi_n\bar\Box^{n}\psi_{n}+\xi_nR)\nonumber\\
&=&-\frac{ A_H}{2G^{(D)}_{N}}\times\sum^{m}_{n=0}(-\frac{1}{2}c_n\psi_n-\frac{1}{2}\xi_n\bar\Box^{n}\psi_n+\xi_n\bar\Box^{n}\psi_n)\nonumber\\
&=&\frac{ A_H}{4G^{(D)}_{N}}\times\sum^{m}_{n=0}(c_n\psi_n+c_{n}\bar\Box^{n}\psi_n)\nonumber\\
&=&\frac{ A_H}{4G^{(D)}_{N}}\times\sum^{m}_{n=0}(c_n(\bar\Box^{-n}R)+c_{n}\bar\Box^{n}(\bar\Box^{-n}R))\nonumber\\
&=&\frac{ A_H}{4G^{(D)}_{N}}\times\sum^{m}_{n=0}(c_n\bar\Box^{-n}R+c_{n} R),
\end{eqnarray} 
where we took $\xi_n=-c_n$.
Furthermore, we used the fact that $\bar\Box^{n}(\bar\Box^{-n}R)=R$,
\cite{Conroy:2014eja}. 

As before let us check the validity of the $\xi_n=-c_n$ by considering $\tilde{I}_{2}$,
\begin{eqnarray}
\tilde{I}_{2}&=&\frac{1}{16\pi G^{(D)}_{N}}\int d^{D}x\sqrt{-g}
\sum^{m}_{n=0}\Big(Rc_n\psi_n+R\xi_n(\bar\Box^{n}\psi_{n}-R)\Big)\nonumber\\
&=&\frac{1}{16\pi G^{(D)}_{N}}\int d^{D}x\sqrt{-g}
\sum^{m}_{n=0}\Big(Rc_n\psi_n-Rc_n(\bar\Box^{n}\psi_{n}-R)\Big)
\nonumber\\
&=&\frac{1}{16\pi G^{(D)}_{N}}\int d^{D}x\sqrt{-g}
\sum^{m}_{n=0}\Big(Rc_n\psi_n-Rc_n(\psi_{n}-\bar\Box^{-n}R)\Big)\nonumber\\
&\equiv&\frac{1}{16\pi G^{(D)}_{N}}\int d^{D}x\sqrt{-g}
RG(\bar\Box)R,
\end{eqnarray}
where used the property of (\ref{property}) and recovered the non-local action
in (\ref{nonloc1}). Thus the fixation of $\xi_n=-c_n$ is valid. 
\section{Summary}

In this chapter we have shown how  Wald's approach can be used to find the entropy
for a static, spherically symmetric metric. It is shown that deviation from
 GR results in having correction to the entropy. However, in
the framework of IDG, we have shown that one can recover the area law by
going to the linearised regime for a spherically symmetric background. Linearisation
means perturbation around Minkowski background and obtaining a constraint,
required to have the massless graviton as the only propagating degree of
freedom. 

We then used Wald's formulation of entropy to find the corrections around
the $(A)dS$ background. We verified our result by providing the entropy of
the Gauss-Bonnet gravity as an example. It has been shown that the constraint
found  in  the linearised regime (around Minkowski background) is not applicable
to the $(A)dS$ case. This is due to the fact that the form of the propagator
for $(A)dS$ background is an open problem and thus there is no known ghost
free constraint. 

We continued our study to a rotating background described by the Kerr metric.
For this case,  we modified Komar integrals appropriately to calculate the
entropy for the Kerr background in various  examples. It has been shown that
deviating from the EH\ gravity up to Ricci tensor will have no effect in
the amount of entropy, and the entropy is given solely by the area law. This
is because the scalar curvature and Ricci tensor are vanishing for the Kerr
metric given in Eq.
(\ref{metric}) (see \cite{Li:2015bqa} on rigorous derivation of this). 

In the presence of the Riemann tensor and its derivatives, the same conclusion
can not be made. This is due to number of reasons: i)  The Riemann tensor
for the Kerr metric is non-vanishing, ii) In the presence of the Riemann
tensor and its derivatives, the Ricci-flat ansatz illustrated by \cite{Li:2015bqa}
may not hold and thus Kerr background may not be an exact solution to the
modified theories of gravity. iii) Variation of the action and obtaining
the appropriate form of Komar integral is technically
 demanding and requires further studies. 

Finally, we have provided a method to obtain the entropy of a non-local action.
Gravitational non-local action is constructed by inversed d'Alembertian operators
acting on the scalar curvature. The Wald approach to find the entropy can
not be applied to a non-local action. Thus, we introduced an equivalent action
via auxiliary fields and localised the non-local action. We then obtained
the entropy using the standard method provided by Wald. In the case of higher
derivative gravity we have checked that both, the original higher derivative
action and its equivalent action, are producing the same results for the entropy.
However, such check can not be done in the non-local case. As a future work,
it would be interesting to obtain the Noether charge such that one can calculate
the entropy for a non-local action without the need of localisation and
check wether the entropies agree after localisation.

\chapter{Conclusion}\label{shish}
In this thesis some of the classical aspects of infinite derivative gravitational
(IDG) theories were considered. The aim was mainly to build an appropriate
machinery which can be used later to build upon and further understanding of infinite
derivative theories of gravity. The main focus of this thesis was a ghost-
and singularity-free infinite derivative theory of gravity. This theory is made up of covariant derivatives acting on Riemannian curvatures, we represented this infinite series of derivatives as function $F(\bar\Box)$. We  found that upon choosing specific form of $F(\bar\Box) $, that is an exponential of an entire function,  the theory is ghost free and singularity free.  
\\\\
\textit{Outline of results}
\\\\

In Chapter \ref{dou} we performed the Hamiltonian analysis for wide range of higher
derivatives and infinite derivatives theories.
We started our analysis by considering some toy models: homogeneous case
and infinite derivative scalar model and then we moved on and applied the
Hamiltonian analysis to infinite derivative gravitational theory  (IDG).
The aim of our analysis were to find the physical degrees of freedom for
higher derivative theories from the Hamiltonian formalism. As for the IDG\
action, we truncated the theory such that the only modification would be
$RF_1(\bar\Box)R$ term. Such action is simpler than the general IDG\ containing
higher order terms such as $R_{\mu\nu}F_2(\bar\Box)R^{\mu\nu}$ and $R_{\mu\nu\lambda\sigma}F_3(\bar\Box)R^{\mu\nu\lambda\sigma}$.
Adding higher curvature terms would lead to further complexities when it
comes to the ADM decomposition and we shall leave this for future studies.

From Lagrangian formalism, the number of degrees of freedom is determined
via propagator analysis. In other words, calculating number of degrees of
freedom is associated with the number of poles arising in the propagator
for a given theory. As for the case of IDG\ is it known that for a Gaussian
kinetic term in the Lagrangian, the theory admits two dynamical degrees of
freedom. This can be readily obtained by considering the spin-0 and spin-2
components of the propagator. In order to maintain the original dynamical
degrees of freedom and avoiding extra poles (and thus extra propagating degrees
of freedom), one shall demand that the propagator be suppressed by the exponential
of an entire function. This is due to the fact that an entire function does
not produce poles in the infinite complex plain. Thus, it is reasonable to
modify the kinetic term in the Lagrangian for infinite derivative theories.
Such modification in the case of scalar toy model would take the form of
$F(\bar\Box)=\Box e ^{-\bar\Box}$ and in the context of gravity the modification
would be $F(\bar\Box)=M^{2}_{P}\Box^{-1}( e ^{-\bar\Box}-1)$. It is clear
that one must expect the same physical results from Lagrangian and Hamiltonian
analysis for a given theory. 

To obtain the number of degrees of freedom in Hamiltonian regime, one
starts with first identifying the configuration space variables and computing
the first class and second class constraints. In the case of IDG, there exist
infinite number of configuration space variables and thus first class and
second class constraints. However, for a Gaussian kinetic term, $F(\bar\Box)$,
the number of degrees of freedom are finite. This holds for both scalar toy
models and gravitational Hamiltonian densities. 
\\\\
In Chapter \ref{chahar}, the generalised Gibbons-Hawking-York (GHY) boundary term,
for the IDG\ theory, was obtained. It has been shown that in order to find
the boundary term for the IDG\ theory one shall use the ADM\ formalism and
in particular coframe slicing to obtain the appropriate form of the extrinsic
curvature. Moreover, in coframe slicing d'Alembertian operators are fairly
easy to be handled when in comes to commutation between derivatives and tensorial components. It should be noted that the conventional way of finding the surface
contribution is using the variation principle. However, for an infinite derivative
term that would not be a suitable approach. This is due to the fact that
for a theory with $n$ number of covariant derivatives we will have $2n$ total
derivatives, and clearly extracting a GHY\ type surface term to cancel these
total derivatives is not a trivial task. Indeed, it is not clear, in the
case of IDG, how from the variation principle one would be extracting a neat
extrinsic curvature to cancel out the surface contribution. To this end,
we took another approach, namely to recast the IDG\ action to an equivalent
form where now we have auxiliary fields. We then decomposed the equivalent
action and used it to calculate the generalised GHY term for the IDG\ theory.
To validate our method it can be seen that, for the case of $\Box\rightarrow
0$ our result would recover the GR's boundary term\ as given by the GHY action
and for $\Box\rightarrow 1$ one shall recover the well known results of Gauss-Bonnet
gravity upon substituting the right coefficients.  
\\\\
In Chapter \ref{panj} we considered the thermodynamical aspects of the IDG theory.
In GR\ it is well known that the entropy of a stationary black hole is given
by the area law. Given different solutions to the Einstein-Hilbert action
the area law would be modified yet the proportionality of the entropy to the
area remains valid. The deviation from GR results in correction to the entropy.
In the context of IDG we performed entropy calculation and obtained the corresponding
corrections. We began our analysis by considering a static and spherically
symmetric background. We shall note that in this metric we have
not defined any value for $f(r)$. This is to keep the metric general. We
then used Wald's description and obtained the corrections. We extended our
discussion to the linearised regime and found that in the weak field limit
the entropy of the IDG action is solely given by the area law and thus the
higher order corrections are not affecting the entropy. It is important to
remind that in order to achieve this result we imposed the constraint $2F_1(\bar\Box)+F_2(\bar\Box)+2F_3(\bar\Box)=0$,
which ensures that the only propagating degree of freedom is the massless
graviton. Moreover, we imposed $\Phi(r)=\Psi(r)$, in other words we demanded
that the Newtonian potentials to be the same.

We then moved to the $(A)dS$ backgrounds and compute the entropy for the
IDG theory using the Wald's approach and obtained the corrections. $(A)dS$
backgrounds admit constant curvatures. This leads to have constant
corrections to the entropy. In this regime, we have shown that upon choosing
the appropriate coefficients, the IDG entropy reduces to the corresponding
Gauss-Bonnet gravity.  

After, we turned our attention to a rotating background and used variational
principle to find the generalised Komar integrals for theories that are constructed
by the metric tensor, Ricci scalar, Ricci tensor and their derivatives. We
then used the first law of thermodynamics and computed the entropy for number
of cases: $f(R)$, $f(R,R_{\mu\nu})$ and finally higher derivative theories
of gravity. We used the Ricci flatness ansatz found in \cite{Li:2015bqa} and concluded that
for a rotating background described by the Kerr metric we have $R=R_{\mu\nu}=0$
regardless of the modified theory of gravity one is considering; thus
the only contribution to the entropy comes from the Einstein-Hilbert term.
This holds true as long as we do not involve the Riemann contribution to
the gravitational action. Furthermore, the generalised form of Komar integrals
are unknown for the case where the gravitational action contains Riemann
tensor and its derivatives. 

Finally, we wrapped up the chapter by considering an infinite derivative
action where we have inverse d'Alembertian operators ($\Box^{-1}$), the entropy
of such action using the Wald approach can not be found. This is due to the
fact that the functional differentiation is not known for inverse derivatives.
For instance, terms like $\delta (R(\bar\Box)^{-1}R)/\delta R_{\mu\nu\lambda\sigma}$
can not be differentiated using the normal Euler-Lagrange functional differentiation.
The non-locality of such action can be localised by introducing auxiliary
fields and rewriting the non-local action in its localised equivalent form.
This allows to compute the entropy using the Wald's prescription. We verified
our method for an already known theory where the Lagrangian density is given
by $\mathcal{L}\sim R+RF_{1}(\bar\Box)R$.  
\\\\
\textit{Future work}
\\\\
\begin{itemize}
\item IDG is now know to address the black hole singularity in the weak field
regime. It would be interesting to see wether the singularities can be avoided
in the case of astrophysical black holes. 
\item The form of Wald entropy for non-local theories of gravity, where there
is $\Box^{-1}$ in the action, is not known. It would be interesting to formulate
the charge for such theories and to check wether they reproduce the same
results as if one was to localise the theory by introducing auxiliary fields.
\item  Addressing the cosmological singularity issue in the presence of matter
source is an open question. The exact cosmological solutions were
only obtained in the presence of a cosmological constant. A realistic cosmological
scenario must include an appropriate exit from the inflationary phase. So
far
such transition is unknown and any progress in that direction is useful.
\item So far IDG is studied around Minkowski background. It would be interesting
to discover the classical and quantum aspects of IDG\ over other backgrounds.
\item Establishing unitarity within the framework of IDG is another open
problem where a novel prescription shall be found. This would be a major
step towards construction of a fully satisfactory theory of quantum gravity.

\item Deriving  IDG or any other modified theories of gravity from string
theory is another open problem, despite the fact that they are stringy inspired.
It is of great interest to know how any of
the modified theories of gravity can be derived from string theory and not
just by writing down an effective gravitational action.

\item There are many other aspects of IDG\ which can be studied using the
current knowledge built by others and us. An example of that would be obtaining
the holographic entanglement entropy for an IDG\ in the context of AdS/CFT
(see for instance \cite{Dong:2013qoa,Miao:2014nxa}). Another example would
be studying the IDG\ action in other dualities such as Kerr/CFT, where one
can gain great knowledge about the CFT\ in the context of higher derivative
theories (see the Gauss-Bonnet example \cite{Krishnan:2009tj}). 
\item In the cosmological sense, non-local theories can be promising in studying
some phenomenological aspects
such as dark energy (see \cite{Maggiore:2014sia} as an example). It would
be interesting to look at other phenomenological aspects of IDG. 
\item As for Hamiltonian formalism, it would be interesting to obtain the
physical degrees of freedom for the full IDG\ action, containing the higher
order curvatures. This would be a good check to make sure one recovers the
same results from the Lagrangian analysis.
\end{itemize}


\begin{appendices}

\chapter{}
\label{appA:appa}
\section{Useful formulas, notations and conventions}
The metric signature used in this thesis is, 
\begin{equation}
g_{\mu\nu}=(-,+,+,+).
\end{equation}
In natural units $(\hbar=c=1)$. We also have, 
\begin{equation}
M_P=\kappa^{-1/2}=\sqrt{\frac{\hbar c}{8\pi G^{(D)}_{N}}},
\end{equation}
where $M_P$ is the Planck mass and $G^{(D)}_{N}$ is Newton's gravitational
constant in D-dimensional space-time. 

The relevant mass dimensions are: 
\begin{equation}
[dx]=[x]=[t]=M^{-1},
\end{equation}
\begin{equation}
[\partial_\mu]=[p_\mu]=[k_\mu]=M^{1},
\end{equation}
\begin{equation}
[velocity]=\frac{[x]}{[t]}=M^{0}.
\end{equation}
As a result, 
\begin{equation}
[d^{4}x]=M^{-4}.
\end{equation}
The action is a dimensionless quantity: 
\begin{equation}
[S]=[\int d^{4}x\mathcal{L}]=M^{0}. 
\end{equation}
Therefore, 
\begin{equation}
[\mathcal{L}]=M^{4}.
\end{equation}
\section{Curvature}
Christoffel symbol is, 
\begin{equation}
\Gamma^{\lambda}_{\mu\nu}=\frac{1}{2}g^{\lambda\tau}(\partial_\mu g_{\nu\tau}+\partial_\nu
g_{\mu\tau}-\partial_\tau g_{\mu\nu}).
\end{equation}
The Riemann tensor is, 
\begin{equation}
R^{\lambda}_{\ \mu\sigma\nu}=\partial_\sigma \Gamma^{\lambda}_{\mu\nu}-\partial_\nu
\Gamma^{\lambda}_{\mu\sigma}+\Gamma^{\lambda}_{\sigma\rho}\Gamma^{\rho}_{\nu\mu}-\Gamma^{\lambda}_{\nu\rho}\Gamma^{\rho}_{\sigma\mu},
\end{equation}
\begin{equation}
R_{\rho\sigma\mu\nu}=g_{\rho\lambda}R^{\lambda}_{\ \sigma\mu\nu}=g_{\rho\lambda}(\partial_\mu\Gamma^{\lambda}_{\nu\sigma}-\partial_\nu\Gamma^{\lambda}_{\mu\sigma}),
\end{equation}
\begin{equation}
R_{\mu\nu\lambda\sigma}=-R_{\nu\mu\lambda\sigma}=-R_{\mu\nu\sigma\lambda}=R_{\lambda\sigma\mu\nu},
\end{equation}
\begin{equation}
R_{\mu\nu\lambda\sigma}+R_{\mu\lambda\sigma\nu}+R_{\mu\sigma\nu\lambda}=0.
\end{equation}
The Ricci tensor is given by, 
\begin{equation}
R_{\mu\nu}=R^{\lambda}_{\ \mu\lambda\nu}=\partial_{\lambda}\Gamma_{\mu\nu}^{\lambda}-\partial_{\nu}\Gamma_{\mu\lambda}^{\lambda}+\Gamma_{\lambda\rho}^{\lambda}\Gamma_{\nu\mu}^{\rho}-\Gamma_{\nu\rho}^{\lambda}\Gamma_{\lambda\mu}^{\rho},
\end{equation}
The Ricci tensor associated with the Christoffel connection is symmetric,
\begin{equation}
R_{\mu\nu}=R_{\nu\mu}.
\end{equation}
The Ricci scalar is given by, 
\begin{equation}
R=R^{\mu}_{\ \mu}=g^{\mu\nu} R_{\mu\nu}=g^{\mu\nu}\partial_{\lambda}\Gamma_{\mu\nu}^{\lambda}-\partial^{\mu}\Gamma_{\mu\lambda}^{\lambda}+g^{\mu\nu}\Gamma_{\lambda\rho}^{\lambda}\Gamma_{\nu\mu}^{\rho}-g^{\mu\nu}\Gamma_{\nu\rho}^{\lambda}\Gamma_{\lambda\mu}^{\rho}.
\end{equation}
The Weyl tensor is given by,
\begin{equation}
C^{\mu}{ }_{\alpha\nu\beta}\equiv R^{\mu}_{\alpha\nu\beta}-\frac{1}{2}(\delta_{\nu}^{
\mu}R_{\alpha\beta}-\delta_{\beta}^{\mu}R_{\alpha\nu}+R_{\nu}^{\mu}g_{
\alpha\beta}-R_{\beta}^{\mu}g_{\alpha\nu})+\frac{R}{6}(\delta_{\nu}^{\mu}g_{
\alpha\beta}-\delta_{\beta}^{\mu}g_{\alpha\nu}),
\end{equation}
\begin{equation}
C^\lambda_{\ \mu\lambda\nu}=0.
\end{equation}
The Einstein tensor is given by, 
\begin{equation}
G_{\mu\nu}=R_{\mu\nu}-\frac{1}{2}g_{\mu\nu}R.
\end{equation}
Varying the Einstein-Hilbert action, 
\begin{equation}
S_{EH}=\frac{1}{2}\int d^{4}x
\sqrt{-g}(M^{2}_{P}R-2\Lambda),
\end{equation}
where $\Lambda$ is the cosmological constant of mass dimension 4, leads to
the Einstein equation,
\begin{equation}
M^{2}_{P}G_{\mu\nu}+g_{\mu\nu}\Lambda=T_{\mu\nu},
\end{equation}
where $T_{\mu\nu}$ is the energy-momentum tensor. When considering the perturbations
around the Minkowski space-time, the cosmological constant is set to be zero.

The Bianchi identity is given by, 
\begin{equation}\label{bianchi1}
\nabla_{\kappa}R_{\mu\nu\lambda\sigma}+\nabla_{\sigma}R_{\mu\nu\kappa\lambda}+\nabla_{\lambda}R_{\mu\nu\sigma\kappa}=0.
\end{equation}
This results from the sum of cyclic permutations of the first three indices.
We note that the antisymmetry
properties of Riemann tensor allows this to be written as, 
\begin{equation}
\nabla_{[\kappa}R_{\mu\nu]\lambda\sigma}=0.
\end{equation}
Contracting (\ref{bianchi1}) with $g^{\mu\lambda}$ results in the contracted
Bianchi identity, 
\begin{equation}\label{bianchi2}
\nabla_{\kappa}R_{\nu\sigma}-\nabla_{\sigma}R_{\nu\kappa}+\nabla^{\lambda}R_{\lambda\nu\sigma\kappa}=0.
\end{equation}
Contracting (\ref{bianchi2}) with $g^{\nu\kappa}$, we obtain, 
\begin{equation}
\nabla_{\kappa}R_{\sigma}^{\kappa}=\frac{1}{2}\nabla_{\sigma}R,
\end{equation}
which similarly implies, 
\begin{equation}
\nabla^{\sigma}\nabla_{\kappa}R_{\sigma}^{\kappa}=\frac{1}{2}\Box R,
\end{equation}
and, 
\begin{equation}
\nabla_\mu G^\mu_\nu=0.
\end{equation}
\section{Useful formulas}
The commutation of covariant derivatives acting on a tensor of arbitrary
rank is given by: 
\begin{eqnarray}
[\nabla_\rho,\nabla_\sigma]X^{\mu_1\dots\mu_k}_{\ \ \ \nu_1\dots\nu_l}&=&-T^{\
\ \lambda}_{\rho\sigma}\nabla_{\lambda}X^{\mu_1\dots\mu_k}_{\ \ \ \nu_1\dots\nu_l}\nonumber\\
&+&R^{\mu_1}_{\ \lambda\rho\sigma}X^{\lambda\mu_2\dots\mu_k}_{\ \ \ \nu_1\dots\nu_k}+R^{\mu_2}_{\
\lambda\rho\sigma}X^{\mu_1\lambda\dots\mu_k}_{\ \ \ \nu_1\dots\nu_k}+\cdots\nonumber\\
&-&R^{\lambda}_{\ \nu_1\rho\sigma} X^{\mu_1\dots\mu_k}_{\ \ \ \lambda\nu_2\dots\nu_l}-R^{\lambda}_{\
\nu_2\rho\sigma}X^{\mu_1\dots\mu_k}_{\ \ \ \nu_1\lambda\dots\nu_l}-\cdots
,\end{eqnarray}
where the Torsion tensor is given by, 
\begin{equation}
T^{\ \lambda}_{\mu\nu}=\Gamma^{\lambda}_{\mu\nu}-\Gamma^{\lambda}_{\nu\mu}=2\Gamma^{\lambda}_{[\mu\nu]}.
\end{equation}
Covariant derivative action on a tensor of arbitrary rank is given by, 
\begin{eqnarray}
\nabla_{\sigma}X^{\mu_1\mu_2\dots\mu_k}_{\ \ \ \nu_1\nu_2\dots\nu_l}&=&\partial_{\sigma}X^{\mu_1\mu_2\dots\mu_k}_{\
\ \ \nu_1\nu_2\dots\nu_l}\nonumber\\
&+&\Gamma^{\mu_1}_{\sigma\lambda}X^{\lambda\mu_2\dots\mu_k}_{\ \ \nu_1\nu_2\dots\nu_l}+\Gamma^{\mu_2}_{\sigma\lambda}X^{\mu_1\lambda\dots\mu_k}_{\
\ \nu_1\nu_2\dots\nu_l}+\cdots\nonumber\\
&-&\Gamma^{\lambda}_{\sigma\nu_1}X^{\mu_1\mu_2\dots\mu_k}_{\
\ \ \lambda\nu_2\dots\nu_l}-\Gamma^{\lambda}_{\sigma\nu_2}X^{\mu_1\mu_2\dots\mu_k}_{\
\ \ \nu_1\lambda\dots\nu_l}-\cdots.
\end{eqnarray}
The Lie derivative along $V$ on some arbitrary ranked tensor is given by,
\begin{eqnarray}
\pounds_V X^{\mu_1\mu_2\dots\mu_k}_{\ \ \ \nu_1\nu_2\dots\nu_l}&=&V^{\sigma}\partial_\sigma
X^{\mu_1\mu_2\dots\mu_k}_{\ \ \ \nu_1\nu_2\dots\nu_l}\nonumber\\
&-&(\partial_\lambda V^{\mu_1})X^{\lambda\mu_2\dots\mu_k}_{\ \ \ \nu_1\nu_2\dots\nu_l
}-(\partial_\lambda V^{\mu_2})X^{\mu_1\lambda\dots\mu_k}_{\ \ \ \nu_1\nu_2\dots\nu_l
}-\cdots\nonumber\\
&+&(\partial_{\nu_1}V^{\lambda})X^{\mu_1\mu_2\dots\mu_k}_{\ \ \ \lambda\nu_2\dots\nu_l}+(\partial_{\nu_2}V^{\lambda})X^{\mu_1\mu_2\dots\mu_k}_{\
\ \ \nu_1\lambda\dots\nu_l}+\cdots.
\end{eqnarray}
In similar manner the Lie derivative of the metric would be, 
\begin{eqnarray}
\pounds_V g_{\mu\nu}=V^{\sigma}\nabla_\sigma g_{\mu\nu}+(\nabla_\mu V^{\lambda})g_{\lambda\nu}+(\nabla_\nu
V^{\lambda})g_{\mu\lambda}=2\nabla_{(\mu}V_{\nu)}.
\end{eqnarray}
Symmetric and anti-symmetric properties, respectively, are, 
\begin{equation}
X_{(ij)k\dots}=\frac{1}{2}(X_{ijk\dots}+X_{jik\dots}),
\end{equation}
\begin{equation}
X_{[ij]k\dots}=\frac{1}{2}(X_{ijk\dots}-X_{jik\dots}).
\end{equation}

\chapter{Newtonian potential}\label{NewtonianPotential}
\begin{figure}[h!]
\centering
\includegraphics[scale=0.5]{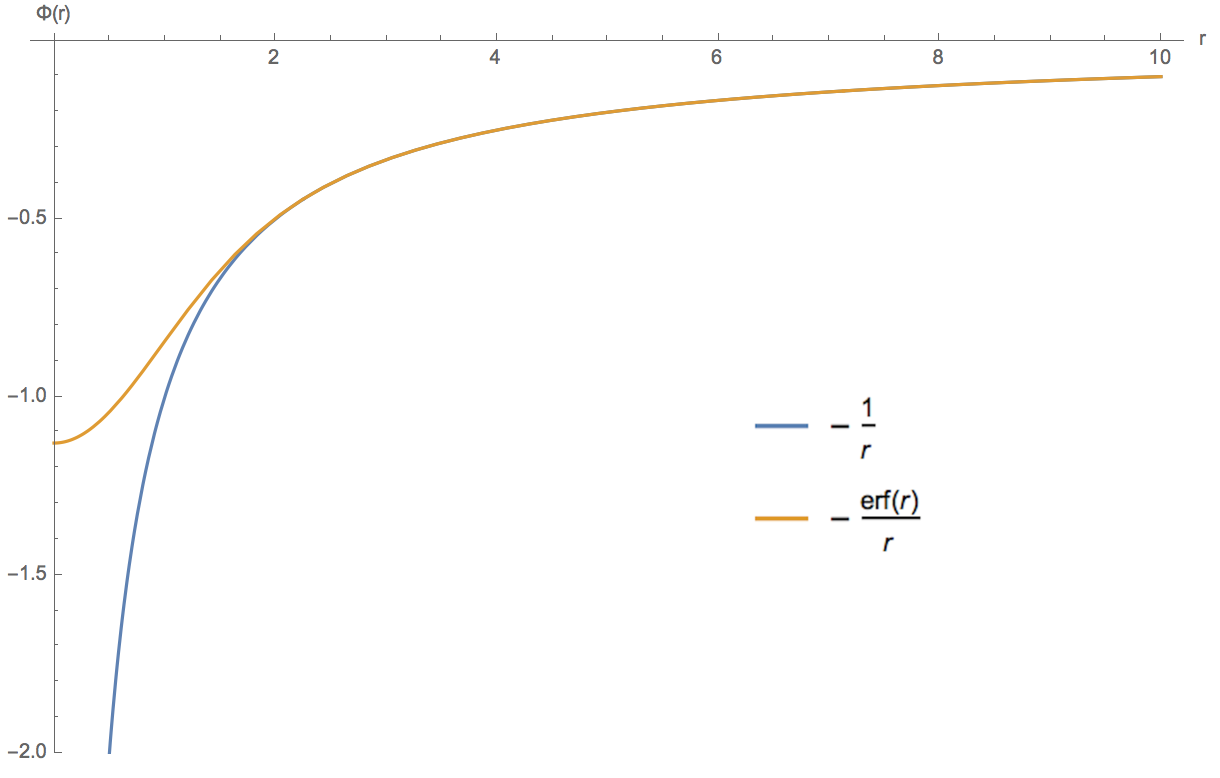}
\caption{Newtonian potentials. The orange line denotes the non-singular 
potential while the blue line indicates the original GR potential.}
\label{fig:newt}
\end{figure}

Let us consider the Newtonian potential in the weak-field regime. The Newtonian
approximation of a perturbed metric for a static point source is given by
the following line element, 
\begin{eqnarray}
ds^2&=&(\eta_{\mu\nu}+h_{\mu\nu})dx^{\mu}dx^{\nu}\nonumber\\
&=&-[1+2\Phi(r)]dt^{2}+[1-2\Psi(r)](dx^{2}+dy^{2}+dz^{2})
\end{eqnarray}
where the perturbation is given by, 
\begin{equation}
h_{\mu\nu}=\begin{pmatrix}-2\Phi(r) & 0 & 0 & 0 \\
0 & -2\Psi(r) & 0 & 0 \\
0 & 0 & -2\Psi(r) & 0 \\
0 & 0 & 0 & -2\Psi(r) \\
\end{pmatrix}.
\end{equation}
The field equations for the infinite derivative theory of gravity is given
by, 
\begin{eqnarray}
-\kappa T_{\mu\nu}&=&\frac{1}{2}[a(\bar \Box)\Box h_{\mu\nu}+b(\bar \Box)\partial_\sigma(\partial_\mu
h^{\sigma}_{\nu}+\partial_\nu
h^{\sigma}_{\mu})+c(\bar \Box)(\partial_\nu\partial_\mu
h+\eta_{\mu\nu}\partial_\sigma\partial_\tau h^{\sigma\tau})\nonumber\\
&+&d(\bar \Box)\eta_{\mu\nu}\Box h+\frac{f(\bar \Box)}{\Box}\partial_\mu
\partial_\nu\partial_\sigma\partial_\tau h^{\sigma\tau}],
\end{eqnarray}
where $\kappa = 8\pi G^{(D)}_{N}=M^{-2}_{P}$ and $T_{\mu\nu}$ is the stress-energy
tensor. By using the definitions of the linearised curvature, we can rewrite the
field equations as, 
\begin{eqnarray}
\kappa T_{\mu\nu}=a(\bar \Box)R_{\mu\nu}-\frac{1}{2}\eta_{\mu\nu}c(\bar \Box)R-\frac{f(\bar
\Box)}{2}\partial_\mu
\partial_\nu R.
\end{eqnarray} 
It is apparent how the modification of Einstein-Hilbert action changed the
field equations. The trace and 00-component of the field equations are given
by,  
\begin{eqnarray}
-\kappa T_{00}= \frac{1}{2}[a(\bar \Box)-3c(\bar \Box)]R,\nonumber\\
\kappa T_{00}=a(\bar \Box)R_{00}+\frac{1}{2}c(\bar \Box) R,
\end{eqnarray}
where $ T_{00}$ gives the energy density. In the static, linearised limit,
$\Box = \nabla^2 = \partial_i \partial^i$. In other words, the flat space
d'Alembertian operator becomes the Laplace operator. This leads to, 
\begin{eqnarray}
-\kappa T_{00}=(a(\bar \Box)-3c(\bar \Box))(2\nabla^2 \Psi -\nabla^2 \Phi
),\nonumber\\
\kappa T_{00} = (a(\bar \Box)-c(\bar \Box))\nabla^2 \Phi+2c(\bar \Box)\nabla^2
\Psi .
\end{eqnarray}
We note that, 
\begin{equation}
R= 2(2\nabla^2 \Psi -\nabla^2 \Phi), \quad R_{00}=\nabla^2 \Phi.
\end{equation}
The Newtonian potentials can be related as, 
\begin{equation}
\nabla^2 \Phi=-\frac{a(\bar \Box)-2c(\bar \Box)}{c(\bar \Box)}\nabla^2 \Psi,
\end{equation}
and therefore, 
\begin{equation}\label{eq82}
\kappa T_{00}=\frac{a(\bar \Box)(a(\bar \Box)-3c(\bar \Box))}{a(\bar \Box)-2c(\bar
\Box)}\nabla^2 \Phi=\kappa m_g \delta^{3}(\vec r).
\end{equation}
We note that $ T_{00}$ is the point source and $ T_{00}= m_g \delta^{3}(\vec
r)$, $ m_g $ is the mass of the object generating the gravitational potential
and $\delta^{3}$ is the three-dimensional Dirac delta-function; this is given
by, 
\begin{equation}
\delta^{3}(\vec r)=\int \frac{d^{3}k}{(2\pi)^{3}}e^{ik\vec r}.
\end{equation}
Thus, with noting that $\Box\rightarrow -k^2$, we can take the Fourier components
of (\ref{eq82}) and obtain, 
\begin{eqnarray}
\Phi(r)&=&-\frac{\kappa m_g}{(2\pi)^{3}}\int^{\infty}_{-\infty} d^{3}k\frac{a(-\bar
k^{2})-2c(-\bar
k^{2})}{a(-\bar k^{2})(a(-\bar k^{2})-3c(-\bar k^{2}))}\frac{e^{ik\vec r}}{k^{2}}\nonumber\\
&=&-\frac{\kappa m_g}{2\pi^{2}r^{}}\int^{\infty}_{0} d^{}k\frac{a(-\bar
k^{2})-2c(-\bar
k^{2})}{a(-\bar k^{2})(a(-\bar k^{2})-3c(-\bar k^{2}))}\frac{\sin(kr)}{k},
\end{eqnarray}
and, 
\begin{equation}
\Psi(r)=-\frac{\kappa m_g}{2\pi^{2}r^{}}\int^{\infty}_{0} d^{}k\frac{c(-\bar
k^{2})}{a(-\bar k^{2})(a(-\bar k^{2})-3c(-\bar k^{2}))}\frac{\sin(kr)}{k}.
\end{equation}
In order to avoid extra degrees of freedom in the scalar sector of the propagator
and to maintain massless graviton as the only propagating degree of freedom,
we shall set $a(\bar
k^{2})=c(\bar
k^{2})$, this leads to, 
\begin{equation}
\Phi(r)=\Psi(r)=-\frac{\kappa m_g}{(2\pi)^{2}r^{}}\int^{\infty}_{0} d^{}k\frac{\sin(kr)}{a(-\bar
k^{2})k}.
\end{equation}
Taking $a(\bar \Box)=e^{-\bar \Box}$, we obtain, 
\begin{equation}
\Phi(r)=\Psi(r)=-\frac{\kappa m_g}{(2\pi)^{2}r^{}}\int^{\infty}_{0} d^{}k\frac{\sin(kr)}{e^{\frac{k^{2}}{M^{2}}}k}=-\frac{\kappa
m_g\text{Erf}[\frac{M r}{2}]}{8\pi r}.
\end{equation}
As $r\rightarrow \infty$, then $\text{Erf}[\frac{M r}{2}]\rightarrow1$ and
we recover the $-r^{-1}$ divergence of GR. When $r\rightarrow 0$, we have,
\begin{equation}
\lim_{r\rightarrow0}\Phi(r)=\lim_{r\rightarrow0}\Psi(r)=-\frac{\kappa
m_gM}{8\pi ^{3/2}}
\end{equation}
which is constant. Thus the Newtonian potential is non-singular. See Fig.
\ref{fig:newt}.
\chapter{Gibbons-York-Hawking\\ boundary term}\label{ehboundarytermderivation}

Let us take the Einstein Hilbert action, 
\begin{equation}
S = \frac{1}{2\kappa}\bigl(S_{EH} + S_{GYH}\bigr), 
\end{equation}
where, 
\begin{equation}\label{eh1}
S_{EH} = \int_{\mathcal{V}} d^4x\, \sqrt{-g}R,
\end{equation}
\begin{equation}
S_{GYH} = 2\oint_{\partial \mathcal{V}} d^3y \, \varepsilon\sqrt{|h|}K,
\end{equation}
Gibbons-York-Hawking boundary term is denoted by $S_{GYH}$. Also, $\kappa=8\pi
G_{N}$. We are considering the space-time as a pair $(\mathcal{M},g)$ with $\mathcal{M}$
a four-dimensional manifold and $g$ a metric on $\mathcal{M}$.
Thus, $\mathcal{V}$ is a hyper-volume on manifold $\mathcal{M}$, and $\partial
\mathcal{V}$
 is its boundary. $h$ the determinant of the induced metric, $K$ is the trace
of the extrinsic curvature of the boundary $\partial \mathcal{V}$, and $\varepsilon$
is equal to $+1$ if $\partial \mathcal{V}$ is time-like and $-1$ if $\partial
\mathcal{V}$ is space-like. We shall derive the $S_{GYH}$ in this section.
To start we fix the following condition, 
\begin{equation}\label{frontera}
\delta g_{\alpha\beta}\biggl|_{\partial \mathcal{V}} =0,
\end{equation}
We also have the following useful formulas, 
\begin{equation}\label{varmet1}
\delta g_{\alpha\beta} = -g_{\alpha\mu}g_{\beta\nu}\delta g^{\mu\nu}, \qquad
\delta g^{\alpha\beta} = -g^{\alpha\mu}g^{\beta\nu}\delta g_{\mu\nu},
\end{equation}
\begin{equation}
\delta \sqrt{-g} =  -\frac{1}{2}\sqrt{-g} g_{\alpha\beta}\delta g^{\alpha\beta},
\end{equation}
\begin{equation}
\delta R_{\beta\gamma\delta}^{\alpha} = \nabla_{\gamma}(\delta\Gamma_{\delta\beta}^{\alpha})
- \nabla_{\delta}(\delta\Gamma_{\gamma\beta}^{\alpha}),
\end{equation}
\begin{equation}\label{palatini}
\delta R_{\alpha\beta} = \nabla_{\gamma}(\delta\Gamma_{\beta\alpha}^{\gamma})
- \nabla_{\beta}(\delta\Gamma_{\gamma\alpha}^{\gamma}).
\end{equation}
Let us now vary the Einstein-Hilbert action,
\begin{equation}\label{varaccion1}
\delta S_{EH} = \int_{\mathcal{V}} d^4x \, \bigl(R\delta\sqrt{-g} + \sqrt{-g}\,
\delta R\bigr).
\end{equation}
The variation of the Ricci scalar is given by, 
\begin{equation}
\delta R = \delta g^{\alpha\beta}R_{\alpha\beta} + g^{\alpha\beta}\delta
R_{\alpha\beta}.
\end{equation}
using the Palatini's identity (\ref{palatini}),
\begin{align}
\delta R &= \delta g^{\alpha\beta}R_{\alpha\beta} + g^{\alpha\beta}\bigl(\nabla_{\gamma}(\delta\Gamma_{\beta\alpha}^{\gamma})
- \nabla_{\beta}(\delta\Gamma_{\alpha\gamma}^{\gamma})\bigr),\nonumber \\
&= \delta g^{\alpha\beta}R_{\alpha\beta} + \nabla_{\sigma
}\bigl(g^{\alpha\beta}(\delta\Gamma_{\beta\alpha}^{\sigma}) - g^{\alpha\sigma}
(\delta\Gamma_{\alpha\gamma}^{\gamma})\bigr),
\end{align}
where  the metric compatibility indicates $\nabla_{\gamma}g_{\alpha\beta}\equiv
0$ and dummy indices were relabeled. Plugging back this result into the action
variation, we obtain, 
\begin{align}
\delta S_{EH} &= \int_{\mathcal{V}} d^4x \, \bigl(R\delta\sqrt{-g} + \sqrt{-g}\,
\delta R\bigr),\nonumber \\
&= \int_{\mathcal{V}} d^4x \, \biggl(-\frac{1}{2}Rg_{\alpha\beta}\sqrt{-g}\,
\delta g^{\alpha\beta} + R_{\alpha\beta}\sqrt{-g}\delta g^{\alpha\beta} +
\sqrt{-g}\nabla_{\sigma}\bigl(g^{\alpha\beta}(\delta\Gamma_{\beta\alpha}^{\sigma})
- g^{\alpha\sigma}
(\delta\Gamma_{\alpha\gamma}^{\gamma})\bigr)\biggr),\nonumber \\
&= \int_{\mathcal{V}}  d^4x \, \sqrt{-g} \biggl(R_{\alpha\beta}-\frac{1}{2}Rg_{\alpha\beta}\biggr)\delta
g^{\alpha\beta} + \int_{\mathcal{V}}  d^4x\sqrt{-g} \nabla_{\sigma}\bigl(g^{\alpha\beta}(\delta\Gamma_{\beta\alpha}^{\sigma})
- g^{\alpha\sigma}
(\delta\Gamma_{\alpha\gamma}^{\gamma})\bigr).
\end{align}
We are going to name the divergence term as $\delta S_{B}$,
\textit{i.e.}\begin{equation}
\delta S_B = \int_{\mathcal{V}} d^4x \, \sqrt{-g}\,  \nabla_{\sigma}\bigl(g^{\alpha\beta}(\delta\Gamma_{\beta\alpha}^{\sigma})
- g^{\alpha\sigma}
(\delta\Gamma_{\alpha\gamma}^{\gamma})\bigr),
\end{equation}
and define,
\begin{equation}\label{V}
V^{\sigma} = g^{\alpha\beta}(\delta\Gamma_{\beta\alpha}^{\sigma}) - g^{\alpha\sigma}
(\delta\Gamma_{\alpha\gamma}^{\gamma}),
\end{equation}
yielding,
\begin{equation}\label{boundary}
\delta S_B = \int_{\mathcal{V}} d^4x \, \sqrt{-g}\,  \nabla_{\sigma}V^{\sigma}.
\end{equation}The Gauss-Stokes theorem is given by,
\begin{equation}\label{gauss}
\int_{\mathcal{V}} d^{n}x\, \sqrt{|g|}\nabla_{\mu}A^{\mu} = \oint_{\partial
\mathcal{V}}d^{n-1}y\, \varepsilon\sqrt{|h|}n_{\mu}A^{\mu},
\end{equation}
where $n_{\mu}$ is the unit normal to $\partial \mathcal{V}$. W shall use
this and rewrite the  boundary term as,
\begin{equation}
\delta S_B = \oint_{\partial \mathcal{V}} d^{3}y\, \varepsilon \sqrt{|h|}n_{\sigma}V^{\sigma},
\end{equation}
with $V^{\sigma}$ given in (\ref{V}). The variation of the Christoffel symbol
is given by, 
\begin{align}\label{varsim1}
\delta \Gamma_{\beta\alpha}^{\sigma} &= \delta \biggl(\frac{1}{2}g^{\sigma\gamma}\bigl[\partial_{\beta}g_{\gamma\alpha}
+ \partial_{\alpha}g_{\gamma\beta} - \partial_{\gamma}g_{\beta\alpha}\bigr]\biggr),\nonumber
\\
&= \frac{1}{2}\delta g^{\sigma\gamma}\bigl[\partial_{\beta}g_{\gamma\alpha}
+ \partial_{\alpha}g_{\gamma\beta} - \partial_{\gamma}g_{\beta\alpha}\bigr]
+ \frac{1}{2}g^{\sigma\gamma}\bigl[\partial_{\beta}(\delta g_{\gamma\alpha})
+ \partial_{\alpha}(\delta g_{\gamma\beta}) -  \partial_{\gamma}(\delta g_{\beta\alpha})\bigr].
\end{align}
From the boundary conditions $\delta g_{\alpha\beta} = \delta g^{\alpha\beta}=0
$. Thus, 
\begin{equation}
\delta \Gamma_{\beta\alpha}^{\sigma}\Bigl|_{\partial \mathcal{V}} = \frac{1}{2}g^{\sigma\gamma}\bigl[\partial_{\beta}(\delta
g_{\gamma\alpha}) + \partial_{\alpha}(\delta g_{\gamma\beta}) -  \partial_{\gamma}(\delta
g_{\beta\alpha})\bigr],
\end{equation}
and so,
\begin{equation}
V^{\mu}\Bigl|_{\partial \mathcal{V}} = g^{\alpha\beta}\biggl[\frac{1}{2}g^{\mu\gamma}\bigl[\partial_{\beta}(\delta
g_{\gamma\alpha}) + \partial_{\alpha}(\delta g_{\gamma\beta}) -  \partial_{\gamma}(\delta
g_{\beta\alpha})\bigr]\biggr] - g^{\alpha\mu}\biggl[\frac{1}{2}g^{\nu\gamma}\partial_{\alpha}(\delta
g_{\nu\gamma}) \biggr],
\end{equation}
we can write
\begin{align}
V_{\sigma}\Bigl|_{\partial \mathcal{V}} = g_{\sigma\mu}V^{\mu}\Bigl|_{\partial
\mathcal{V}} &= g_{\sigma\mu}g^{\alpha\beta}\biggl[\frac{1}{2}g^{\mu\gamma}\bigl[\partial_{\beta}(\delta
g_{\gamma\alpha}) + \partial_{\alpha}(\delta g_{\gamma\beta}) -  \partial_{\gamma}(\delta
g_{\beta\alpha})\bigr]\biggr] \nonumber \\&- g_{\sigma\mu}g^{\alpha\mu}\biggl[\frac{1}{2}g^{\nu\gamma}\partial_{\alpha}(\delta
g_{\nu\gamma}) \biggr] \nonumber \\
&= \frac{1}{2}\delta_{\sigma}^{\gamma}g^{\alpha\beta}\bigl[\partial_{\beta}(\delta
g_{\gamma\alpha}) + \partial_{\alpha}(\delta g_{\gamma\beta}) -  \partial_{\gamma}(\delta
g_{\beta\alpha})\bigr] - \frac{1}{2}\delta_{\sigma}^{\alpha}g^{\nu\gamma}\bigl[\partial_{\alpha}(\delta
g_{\nu\gamma}) \bigr]\nonumber \\
&= g^{\alpha\beta}\bigl[\partial_{\beta}(\delta g_{\sigma\alpha}) -  \partial_{\sigma}(\delta
g_{\beta\alpha})\bigr].
\end{align}
Let us now calculate the term $n^{\sigma}V_{\sigma}\bigl|_{\partial \mathcal{V}}$.
We note that, 
\begin{equation}
g^{\alpha\beta} = h^{\alpha\beta} + \varepsilon n^{\alpha}n^{\beta},
\end{equation}
then
\begin{align}
n^{\sigma}V_{\sigma}\Bigl|_{\partial \mathcal{V}} &= n^{\sigma}(h^{\alpha\beta}+\varepsilon
n^{\alpha}n^{\beta})[\partial_{\beta}(\delta g_{\sigma\alpha}) -  \partial_{\sigma}(\delta
g_{\beta\alpha})], \nonumber \\
&= n^{\sigma}h^{\alpha\beta}[\partial_{\beta}(\delta g_{\sigma\alpha}) -
 \partial_{\sigma}(\delta g_{\beta\alpha})],
\end{align}
where we use the antisymmetric part of $\varepsilon n^{\alpha}n^{\beta}$
with $\varepsilon = n^{\mu}n_{\mu}=\pm 1$. Since $\delta g_{\alpha\beta}=0$
on the boundary  we have $h^{\alpha\beta}\partial_{\beta}(\delta g_{\sigma\alpha})=0$.
Finally we have, \begin{equation}
n^{\sigma}V_{\sigma}\Bigl|_{\partial \mathcal{V}} = -n^{\sigma}h^{\alpha\beta}\partial_{\sigma}(\delta
g_{\beta\alpha}).
\end{equation}
Hence, the variation of the Einstein-Hilbert action is, 
\begin{equation}\label{secondterm}
\delta S_{EH} = \int_{\mathcal{V}}  d^4x \, \sqrt{-g} \biggl(R_{\alpha\beta}-\frac{1}{2}Rg_{\alpha\beta}\biggr)\delta
g^{\alpha\beta}-\oint_{\partial \mathcal{V}} d^{3}y\, \varepsilon \sqrt{|h|}
h^{\alpha\beta}\partial_{\sigma}(\delta g_{\beta\alpha})n^{\sigma}.
\end{equation}
The variation of the Gibbons-York-Hawking boundary term is,\\
\begin{equation}
\delta S_{GYH} = 2\oint_{\partial \mathcal{V}} d^3y\, \varepsilon\sqrt{|h|}\delta
K.
\end{equation}
Using the definition of the  trace of extrinsic curvature,
\begin{align}
K &= \nabla_{\alpha}n^{\alpha}, \nonumber \\
&= g^{\alpha\beta}\nabla_{\beta}n_{\alpha}, \nonumber \\
&= (h^{\alpha\beta}+\varepsilon n^{\alpha}n^{\beta})\nabla_{\beta}n_{\alpha},
\nonumber \\
&= h^{\alpha\beta}\nabla_{\beta}n_{\alpha}, \nonumber \\
&= h^{\alpha\beta}(\partial_{\beta}n_{\alpha}-\Gamma_{\beta\alpha}^{\gamma}n_{\gamma}),
\end{align}
and subsequently its variation, 
\begin{align}\label{deltaK}
\delta K &= -h^{\alpha\beta}\delta\Gamma_{\beta\alpha}^{\gamma}n_{\gamma},
\nonumber \\
&= -\frac{1}{2}h^{\alpha\beta}g^{\sigma\gamma}\bigl[\partial_{\beta}(\delta
g_{\sigma\alpha}) + \partial_{\alpha}(\delta g_{\sigma\beta}) -  \partial_{\sigma}(\delta
g_{\beta\alpha})\bigr]n_{\gamma}, \nonumber \\
&= -\frac{1}{2}h^{\alpha\beta}\bigl[\partial_{\beta}(\delta g_{\sigma\alpha})
+ \partial_{\alpha}(\delta g_{\sigma\beta}) -  \partial_{\sigma}(\delta g_{\beta\alpha})\bigr]n^{\sigma},
\nonumber \\
&= \frac{1}{2}h^{\alpha\beta}\partial_{\sigma}(\delta g_{\beta\alpha})n^{\sigma},
\end{align}
where we used $h^{\alpha\beta}\partial_{\beta}(\delta
g_{\sigma\alpha})=0$, $h^{\alpha\beta}\partial_{\alpha}(\delta g_{\sigma\beta})=0$
on the boundary; The variation of the Gibbons-York-Hawking becomes, 
\begin{equation}\label{vargyh}
\delta S_{GYH} = \oint_{\partial \mathcal{V}} d^3y\, \varepsilon\sqrt{|h|}h^{\alpha\beta}\partial_{\sigma}(\delta
g_{\beta\alpha})n^{\sigma}.
\end{equation}
We see that the second term of (\ref{secondterm}) is matching with (\ref{vargyh}).
In other words, this term exactly cancel the boundary contribution of the
Einstein-Hilbert
term.
\chapter{Simplification example \\in IDG action}\label{simplificationexampleidg}
Let us consider the following terms from (\ref{actionexpansion}), 
\begin{equation}
RF_1(\Box)R+RF_2(\Box)\nabla_\nu\nabla_\mu
R^{\mu\nu}
+R^{\nu}_{\ \mu}F_4(\Box)\nabla_\nu\nabla_\lambda R^{\mu\lambda},
\end{equation}
we can recast above as, 
\begin{equation}
RF_1(\Box)R+\frac{1}{2}RF_2(\Box)\Box
R^{}+\frac{1}{2}R^{\nu}_{\ \mu}F_4(\Box)\nabla_\nu\nabla^\mu R,
\end{equation}
by having the identity $\nabla_\mu R^{\mu\nu}=\frac{1}{2}\nabla^\nu R$ and
also $\nabla_\nu\nabla_\mu R^{\mu\nu}=\frac{1}{2}\Box R$, which occurs due
to contraction of Bianchi identity, we can perform integration by parts on
the final term and obtain, 
\begin{eqnarray}
&&RF_1(\Box)R+\frac{1}{2}RF_2(\Box)\Box
R^{}+\frac{1}{2}\nabla^\mu\nabla_\nu R^{\nu}_{\ \mu}F_4(\Box) R\nonumber\\
=&&RF_1(\Box)R+\frac{1}{2}RF_2(\Box)\Box
R^{}+\frac{1}{4}RF_4(\Box)\Box R\nonumber\\
\equiv&&RF_1(\Box)R,
\end{eqnarray}
where we redefined the arbitrary function $F_1(\Box)$ to absorb $F_2(\Box)$
and $F_4(\Box)$. We simplified (\ref{actionexpansion}) in similar manner
to reach (\ref{simaction}). 
\chapter{Hamiltonian density}\label{hamdensapp}
Hamiltonian density corresponding to action Eq.~(\ref{hamdens}) is explicitly
given by, 
\begin{eqnarray}\label{hamapp-1}
\mathcal{H}&=&p_A\dot{A}+p_{\chi_1}\dot{\chi}_1+p_{\chi_l}\dot{\chi}_l+p_{\eta_{l-1}}\dot{\eta}_{l-1}-\mathcal{L}\nonumber\\
&=&-(A\dot{\chi_1}
 \dot{A}+\dot{A}\chi_1
\dot{A})-\sum^{\infty}_{l=2}(\dot{A}\chi_l\dot{\eta}_{l-1})\nonumber\\
&-&(A\dot{
A)}\dot{\chi}_1-(A\dot{\eta}_{l-1})\dot{\chi}_l-(A\dot{\chi}_l
\dot{\eta}_{l-1}+\chi_l
 \dot{A}\dot{\eta}_{l-1})\nonumber\\
 &-&\Bigg(A(f_{0}A+\sum^{\infty}_{n=1}f_n\eta_{n})+\sum ^{\infty}_{l=1}A\chi_l
\eta_l\nonumber\\
&-&(A\partial_0\chi_1  \partial_0 A+\chi_1
\partial_0 A
\partial_0 A)+g^{ij}(A\partial_i\chi_1  \partial_j A+\chi_1
\partial_i A
\partial_j A)\nonumber\\
&-&\sum^{\infty}_{l=2}(A\partial_0\chi_1  \partial_0\eta_{l-1}
+\chi_l
\partial_0 A
\partial_0 \eta_{l-1})+g^{ij}\sum^{\infty}_{l=2}(A\partial_i\chi_l  \partial_j\eta_{l-1}
+\chi_l
\partial_i A
\partial_j \eta_{l-1})\Bigg)\nonumber\\
&=&-\sum^{\infty}_{l=2}(\dot{A}\chi_l\dot{\eta}_{l-1})
-(A\dot{
A)}\dot{\chi}_1-(A\dot{\eta}_{l-1})\dot{\chi}_l
 +\Bigg(A(f_{0}A+\sum^{\infty}_{n=1}f_n\eta_{n})-\sum ^{\infty}_{l=1}A\chi_l
\eta_l\nonumber\\ \\
&-&g^{ij}(A\partial_i\chi_1  \partial_j A+\chi_1
\partial_i A
\partial_j A)
-g^{ij}\sum^{\infty}_{l=2}(A\partial_i\chi_l  \partial_j\eta_{l-1}
+\chi_l
\partial_i A
\partial_j \eta_{l-1})\Bigg)\nonumber\\ \\
&=&
A(f_{0}A+\sum^{\infty}_{n=1}f_n\eta_{n})-\sum ^{\infty}_{l=1}A\chi_l
\eta_l\nonumber\\
&-&(g^{\mu\nu}A\partial_{\mu}\chi_1  \partial_\nu A+g^{ij}\chi_1
\partial_i A
\partial_j A)
-g^{\mu\nu}\sum^{\infty}_{l=2}(A\partial_\mu\chi_l  \partial_\nu\eta_{l-1}
+\chi_l
\partial_\mu A
\partial_\nu \eta_{l-1})\,.\nonumber\\
\end{eqnarray}
 
\chapter{Physical degrees of freedom via propagator
analysis}\label{appendixc}

We have an action of the form~\cite{Biswas:2005qr}
\be
S=\frac{1}{2} \int d^{4}x \, \sqrt{-g} \LT M_{P}^{2}R +R\cF\left(\bbox\right)R
\RT
\ee
or, equivalently,
\be \label{luwo}
S=\frac{1}{2} \int d^{4}x \, \sqrt{-g} \LT M_{P}^{2}A +A\cF\left( \bbox
\right)A +B(R-A)\RT \,.
\ee
$A$ and $B$ have mass dimension $2$.

The propagator around Minkowski space-time is of the form~\cite{Biswas:2011ar,Biswas:2013kla}
\be
\Pi (-k^2)=\frac{\cP ^ 2}{k^{2}a(-k^2)}+\frac{\cP_{s} ^ 0}{k^{2}(a(-k^2)-3c(-k^2))}
\,,
\ee
where $a(\Box)=1$ and $c(\Box)=1+M_{P}^{-2}\cF \left( \bbox \right)\Box$.
Hence,
\be \label{ol}
\Pi (-k^2)=\frac{\cP ^ 2}{k^{2}}+\frac{\cP_{s} ^ 0}{k^{2}(-2+3M_{P}^{-2}k^{2}\cF(-k^{2}/M^{2}))}
\ee
We know that~\cite{Biswas:2011ar,Biswas:2013kla}
\be
\cF (\bar{\Box})=M_{P}^{2}\frac{c(\bar{\Box})-1}{\Box} \,.
\ee
Only if $c(\Box)$ is the exponent of an entire function can we decompose
into partial fractions and have just one extra pole.

The upshot is that, in order to have just one extra degree of freedom, we
have to impose conditions on the coefficients in $\cF \LF \bbox \RF$. In
order to avoid $\Box^{-1}$ terms appearing in $\cF{(\bar{\Box})}$, we must
have that
\be
c(\bar{\Box})=\sum_{n=0}^{\infty}c_{n}\bar{\Box}^{n} 
\ee
and $c_{0}=1$. Hence,
\be
\cF(\bar{\Box})=\LF \frac{M_{P}}{M} \RF^{2}\sum_{n=0}^{\infty}c_{n+1}\bar{\Box}^{n}
\,.
\ee
To get infinitely many poles and, hence, degrees of freedom, one could have,
for instance, that
\be
c(\bar{\Box})= \cos (\bar{\Box}) \,,
\ee
so that $c_{0}=1$. Then Eq.~\eqref{luwo} becomes
\be 
S=\frac{1}{2} \int d^{4}x \, \sqrt{-g} \LT M_{P}^{2}A +M_{P}^{2}A\left(\frac{\cos
(\bar{\Box})-1}{\Box}\right)A +B(R-A)\RT \,.
\ee
Using~\eqref{ol}, apart from the $k^2=0$ pole,  we have poles when
\be \label{lklk}
\cos \left(\frac{k^{2}}{M^2}\right)=\frac{1}{3}\,.
\ee 
Eq.~\eqref{lklk} has infinitely many solutions due to the periodicity of
the cosine function and, therefore, the propagator has infinitely many poles
and, hence, degrees of freedom.
We can
write
the solutions as $\bar{k}^{2}=2m\pi$, where $m=0,1,2,\cdots$, one can also
write: 
\begin{equation}
\cos(\bar{k}^{2})=\prod^{\infty}_{l=1}\Bigg(1-\frac{4\bar{k}^{4}}{(2l-1)^{2}\pi^{2}}\Bigg)
\end{equation}
or 
\begin{equation}
\cos(\bbox)=\prod^{\infty}_{l=1}\Bigg(1-\frac{4\bbox^{2}}{(2l-1)^{2}\pi^{2}}\Bigg)
\end{equation}
Now, to get just one extra degrees of freedom, one can make, for instance,
the choice $c(\bar{\Box})=e^{-\bar{\Box}}$, then
\be
\cF(\bar{\Box})=\sum_{n=0}^{\infty}f_{n}\bar{\Box}^{n} \,,
\ee
where
\be
f_{n}=\LF \frac{M_P}{M} \RF^{2}\frac{(-1)^{n+1}}{(n+1)!} \,.
\ee
Using~\eqref{ol}, apart from the $k^2=0$ pole, we have poles when
\be \label{toko}
e^{k^{2}/M^{2}}=\frac{1}{3}\,.
\ee
There is just one extra pole and, hence, degrees of freedom. In total, there
are $3$ degrees of freedom.
\chapter{Form of $\mathcal{F}(\bar \Box)$ and constraints}\label{formoffunction}
We have shown in section \ref{dorostfunc} that the primary constraints are built as follow:

\begin{equation}
\sigma_1=\eta_1-\Box A\approx 0,\cdots, \sigma_l=\eta_l-\Box \eta_{l-1}\approx
0.
\end{equation}
We also mentioned that they are first class constraints due to their Poission
brackets vanishing weakly. The number of the degrees of freedom are related
to the form of $\mathcal{F}(\bar\Box)$. Expanding on this, we shall consider
the case when: $\mathcal{F}(\bar\Box)=\Box e^{-\bar \Box}$. Then for an action
of the form, 
\be
S=\int d^4 x A \mathcal{F}(\bbox)A.
\ee
The equation of motion for $A$ is given by
\be
2\cF(\bbox)A=0.
\ee
For the case when $\cF(\bbox)=\Box e^{-\bbox}$, the equation of motion becomes,
\be
\Theta=\eta_{1}=0.
\ee
Moreover, $\cF(\bbox)$ cannot be written in the form \be\cF(\bbox)=(\Box+m_{2}^{2})(\Box+m_{1}^{2})\cG_{1}(\bbox),\ee
where $m_{1}^{2}$, $m_{2}^{2}$ are arbitrarily chosen parameters and $\cG_{1}$
has no roots.
This is a constraint, which can be written as follows:
\be\label{s2}
\Xi_{2}=\cF(\bbox)A-(\Box+m_{2}^{2})(\Box+m_{1}^{2})\cG_{1}(\bbox)A \not
\approx 0.
\ee
Moreover, we have the constraint
\be
\Xi_{3}=\cF(\bbox)A-(\Box+m_{3}^{2})(\Box+m_{2}^{2})(\Box+m_{1}^{2})\cG_{2}(\bbox)A
\not \approx 0,
\ee
where  $m_{1}^{2}$, $m_{2}^{2}$, $m_{3}^{2}$ are arbitrarily chosen parameters
and $\cG_{2}$ has no roots. This goes on and on.

Regarding degrees of freedom and given that  the constraints
are first-class, we have,
\begin{align}
2\mathcal{A}&\equiv2\times\left\{(A,p_A),(\eta_1,p_{\eta_1}),(\eta_2,p_{\eta_2}),\dots\right\}=2\times(2+\infty)=4+\infty
, \non
\mathcal{B}&=0 ,\non
2\mathcal{C}&\equiv2\times(\Theta,\Xi_{2},\Xi_{3},\dots)=2(1+\infty)=2+\infty
 ,\non
\mathcal{N}&=\frac{1}{2}(2\mathcal{A}-\mathcal{B}-2\mathcal{C})=\frac{1}{2}(4+\infty-2-\infty)=1,
\end{align}
as expected.

A similar prescription can be applied in infinite derivative gravity.
Now as a final clarification we shall reparametrise these constraints (\textit{i.e.}
$\Theta,\Xi_{2},\Xi_{3},\dots$), into $\sigma$'s. From (\ref{s2}) we have,
\be\Xi_{2}=\cF(\bbox)A-(\Box+m_{2}^{2})(\Box+m_{1}^{2})\cR_{1}(\bbox)A  \approx
0,
\ee
where $\cR_{1}(\bbox)$ has no roots and contains $\Box^{-1}$ terms.
Then
\be\label{s3}
\frac{\cF(\bbox)}{\cR_{1}(\bbox)}A\approx(\Box+m_{2}^{2})(\Box+m_{1}^{2})A=\eta_{2}+(m_{1}^{2}+m_{2}^{2})\eta_{1}+m_{1}^{2}m_{2}^{2}A
\ee
Redefining $\eta_{2}$ and $\eta_{1}$ appropriately, (\ref{s3}) can be written
in
the form $\eta_{2}-\Box \eta_{1}=0$. Similarly for $\Xi_3$ and so on.
Hence, the constraints are equivalent.
\chapter{$K_{ij}$ in the Coframe Metric}\label{A}

In this section we wish to use the approach of~\cite{Anderson:1998cm} and
find the general definition for $K_{ij}$ in the coframe metric. 
Given,
\begin{eqnarray} \label{eq:defitionofextrinsiccurvaturegivenbyArlen}
  {\gamma^\alpha}_{\beta\gamma} &=& {\Gamma^\alpha}_{\beta\gamma} + g^{\alpha\delta}
{C^\epsilon}_{\delta(\beta} g_{\gamma)\epsilon} - \frac{1}{2} {C^\alpha}_{\beta\gamma}\,,
\\ \label{zuzu}
        d \theta^\alpha &=& - \frac{1}{2} {C^\alpha}_{\beta\gamma} \theta^\beta
\wedge \theta^\gamma \,,
\end{eqnarray}
where $\Gamma$ is the ordinary Christoffel symbol, $\wedge$ is the ordinary
wedge product and the $C$s are coefficients
to be found. 
By comparing the values given in \cite{Anderson:1998cm} with the ordinary
Christoffel symbols, we
can see that 
\begin{eqnarray}\label{eq:differentCs}
{C^i}_{00} = {C^0}_{0i} = {C^0}_{i0} = {C^0}_{00} = 0\,, \nonumber \\
        {C^i}_{0k} = {C^i}_{k0} + 2 \partial_k \beta^i\,, \nonumber \\
        {C^i}_{jk} = {{C_k}^i}_{j} + {{C_j}^i}_{k}\,, 
\end{eqnarray}
Now in the coframe metric in Eq.~(\ref{eq16}),  
\begin{eqnarray}
 && g^{0\delta} {C^\epsilon}_{\delta(i} g_{j)\epsilon} \,, \nonumber \\
        &=& - \frac{1}{N^2} \left[ {C^\epsilon}_{0(i} g_{j)\epsilon} \right]\,,
\nonumber \\
        &=& - \frac{1}{2N^2} \left[ {C^\epsilon}_{0i} g_{j\epsilon} + {C^\epsilon}_{0j}
g_{i\epsilon} \right]\,, \nonumber \\
        &=& - \frac{1}{2N^2} \left[ {C^m}_{0i} g_{jm} + {C^m}_{0j} g_{im}
\right] \,,\nonumber \\
&=& -\frac{1}{2N^2} \left[C_{j0i}+C_{i0j} \right]
\end{eqnarray}
and $C_{j0i} = g_{\alpha j} {C^\alpha}_{0i}=g_{jk}{C^k}_{0i}$. 
In the coframe
and using  the conventions in~\cite{Anderson:1998cm}
\begin{eqnarray}K_{ij} = -\nabla_i n_j = \frac{1}{2N} \left( D_i \beta_j
+ D_j \beta_i -\dot{h}_{ij} \right)\,,\end{eqnarray}
 and 
\begin{eqnarray}  
\partial_0 h_{ij} = \partial_t h_{ij} - \beta^l \partial_l
h_{ij} \,.
\end{eqnarray}

In general, for a $p$-form $\alpha$ and a $q$-form $\beta,$
\begin{align}
\alpha \wedge \beta &= (-1)^{pq} \beta \wedge \alpha \,, \\
d(\alpha \wedge \beta) &= (d \alpha)\wedge \beta + (-1)^{p}\alpha \wedge
(d \beta) \,.
\end{align}
Hence, if $p$ is odd,
\begin{eqnarray} \label{eq:antisymmetricpropertiesofwedgeproduct}
\alpha \wedge \alpha = (-1)^{p^2} \alpha \wedge \alpha =-\alpha
\wedge \alpha= 0 \,.
\end{eqnarray}
From Eq.~(\ref{zuzu}) and Eq.~(\ref{eq:differentCs}) we can see that 
\begin{eqnarray}
d \theta^1 &=& - \frac{1}{2} {C^1}_{\beta\gamma} \theta^\beta \wedge \theta^\gamma\,,
 \nonumber\\
        &=& - \frac{1}{2} {C^1}_{0i} \theta^0 \wedge \theta^i - \frac{1}{2}
{C^1}_{i0} \theta^i \wedge \theta^0 - \frac{1}{2} {C^1}_{ij} \theta^i \wedge
\theta^j \,,\nonumber\\
        &=& - \frac{1}{2} \left[ {C^1}_{i0} + 2 \partial_i \beta^1 \right]
\theta^0 \wedge \theta^i + \frac{1}{2} {C^1}_{i0} \theta^0 \wedge \theta^i+
\frac{1}{2} {C^1}_{ij} \theta^j \wedge \theta^i \,, \nonumber\\
        &=& - \left( \partial_i \beta^1 \right) \theta^0 \wedge \theta^i
+ \frac{1}{2} {C^1}_{ij} \theta^j \wedge \theta^i\,.
\end{eqnarray}
We get a similar result for $d\theta^2$ and $d\theta^3$, so we can say that
\begin{eqnarray}\label{eq:definitionofdthetak}
d\theta^{k}=- \left( \partial_i \beta^k \right) \theta^0
\wedge \theta^i + \frac{1}{2} {C^k}_{ij} \theta^j \wedge \theta^i \,,
\end{eqnarray}
where $k=1,2,3$. Now from the definition of $\theta$ in Eq.~(\ref{eq23}),

\begin{eqnarray}\label{eq:defntheta1} 
d \theta^1 = d \left( d x^1 + \beta^1 dt \right) = d \beta^1
\wedge dt \,,
\end{eqnarray}
and 
\begin{eqnarray}\label{eq:dthetazerovanishesanddefnofdthetai}
        d \theta^0 &=& d(dt) = d^{2}(t) = 0 \,,\nonumber\\
        d \theta^i &=& d \left( dx^i + \beta^i dt \right)\,, \nonumber\\
        &=& d \left(dx^i \right) + d \left( \beta^i \wedge dt \right) \,,\nonumber\\
        &=& d \left( \beta^i \wedge dt \right)\,, \nonumber \\
        &=& d\beta^{i} \wedge dt \,.
        \end{eqnarray}
        Let us point out that $\beta^{i}dt=\beta^{i} \wedge dt$. 
\begin{eqnarray} \theta^0 \wedge \theta^i &=& dt \wedge \left( dx^i + \beta^i
\wedge dt \right) \,, \nonumber\\
&=& dt \wedge dx^i \nonumber\\
        \theta^i \wedge \theta^j &=& \left( dx^i + \beta^i dt \right) \wedge
\left( d x^j + \beta^j dt \right) \,, \nonumber\\
        &=& dx^i \wedge dx^j + dx^i \wedge \left( \beta^j dt \right) + \left(
\beta^i dt \right) \wedge dx^j + \left( \beta^i dt \right) \wedge \left(
\beta^j \wedge dt \right) \,,\nonumber\\
        &=& dx^i \wedge dx^j + dx^i \wedge \beta^j \wedge dt + \beta^i \wedge
dt  \wedge dx^j \,,\nonumber\\
        &=& dx^i \wedge dx^j + \beta^j \wedge dx^i \wedge dt - \beta^i \wedge
dx^j \wedge dt\,.
\end{eqnarray}
Now using Eq.~(\ref{eq:definitionofdthetak}) and  Eq.~(\ref{eq:dthetazerovanishesanddefnofdthetai}),

\begin{eqnarray} \label{eq:dthetakintermsofwedge}
d \theta^k &=& d \beta^k \wedge dt  \,,\nonumber\\
        &=& \left( \frac{\partial \beta^k}{\partial x^1} dx^1 + \frac{\partial
\beta^k}{\partial x^2} dx^2 + \frac{\partial \beta^k}{\partial x^3} dx^3
\right) \wedge dt \,,\nonumber\\
        &=& - \left( \partial_i \beta^k \right) dt \wedge dx^i - \frac{1}{2}
{C^k}_{ij} \left[ dx^i \wedge dx^j + \beta^j \wedge dx^i \wedge dt - \beta^i
\wedge dx^j \wedge dt \right],\nonumber\\\end{eqnarray}
where $k=1,2,3$. 
From the definition of $d \theta^\alpha$ in Eq.~(\ref{zuzu}) and using the
antisymmetric properties
of the $\wedge$ product from Eq.~(\ref{eq:antisymmetricpropertiesofwedgeproduct}),
\begin{eqnarray}
        d \theta^\alpha &=& - \frac{1}{2} {C^\alpha}_{\beta\gamma} \theta^\beta
\wedge \theta^\gamma \,,\nonumber\\
        &=& - \frac{1}{2} {C^\alpha}_{\gamma\beta} \theta^\gamma \wedge \theta^\beta\,,
\nonumber\\
        &=& \frac{1}{2} {C^\alpha}_{\gamma\beta} \theta^\beta \wedge \theta^\gamma\,,
\end{eqnarray}
and therefore
\begin{eqnarray}\label{eq:symmetryofC}
        {C^\alpha}_{\beta\gamma} = -{C^\alpha}_{\gamma\beta} \,,
\end{eqnarray}
we can then write
\begin{eqnarray}
        && {C^\alpha}_{00} = {C^\alpha}_{11} = {C^\alpha}_{22} = {C^\alpha}_{33}
= 0 \,,\nonumber\\
        && {C^\alpha}_{0i} = - {C^\alpha}_{i0} \,.
\end{eqnarray}
Combining Eq.~(\ref{zuzu}), Eq.~(\ref{eq:dthetazerovanishesanddefnofdthetai}),
Eq.~(\ref{eq:dthetakintermsofwedge})
and utilising Eq.~(\ref{eq:symmetryofC})
\begin{eqnarray}
        0 &=& d\theta^0 = - \frac{1}{2} {C^0}_{\beta\gamma} \theta^\gamma
\wedge \theta^\beta \,,\nonumber\\
        &=& - \frac{1}{2} {C^0}_{0i} \theta^0 \wedge \theta^i - \frac{1}{2}
{C^0}_{i0} \theta^i \wedge \theta^0 - \frac{1}{2} {C^0}_{ij} \theta^i \wedge
\theta^j \text{(for } i \neq j) \,,\nonumber\\
        &=& - {C^0}_{0i} \theta^0 \wedge \theta^i - {C^0}_{ij} \theta^i \wedge
\theta^j \text{(for } i < j) \,,\nonumber\\
        &=& - {C^0}_{01} \theta^0 \wedge \theta^1- - {C^0}_{02} \theta^0
\wedge \theta^2 - {C^0}_{03} \theta^0 \wedge \theta^3 
        - {C^0}_{12} \theta^1 \wedge \theta^2 - {C^0}_{13} \theta^1 \wedge
\theta^3 \nonumber\\&&- {C^0}_{23} \theta^2 \wedge \theta^3 \,,\nonumber\\
        &=& - {C^0}_{01} dt \wedge dx^1 - {C^0}_{02} dt \wedge dx^2 - {C^0}_{03}
dt \wedge dx^3 \,, \nonumber\\
        &&- {C^0}_{12} \left[ dx^1 \wedge dx^2 + \beta^2 dx^1 \wedge dt -
\beta^1 dx^2 \wedge dt \right]\,, \nonumber\\
        && - {C^0}_{13} \left[ dx^1 \wedge dx^3 + \beta^3 dx^1 \wedge dt
- \beta^1 dx^3 \wedge dt \right] \,,\nonumber\\
        && - {C^0}_{23} \left[ dx^2 \wedge dx^3 + \beta^3 dx^2 \wedge dt
- \beta^2 dx^3 \wedge dt \right]\,.
\end{eqnarray}   
In order for this to be satisfied,  each term must vanish separately as the
$dx^\alpha
\wedge dx^j$ are linearly independent and so the coefficient
of each must be zero
and thus ${C^0}_{12} =  {C^0}_{13} = {C^0}_{23}={C^0}_{01}={C^0}_{02} ={C^0}_{03}=0$
and thus ${C^0}_{\alpha\beta}=0$. Similarly using Eqs.~(\ref{zuzu}), (\ref{eq:dthetazerovanishesanddefnofdthetai}),
(\ref{eq:dthetakintermsofwedge}) and (\ref{eq:symmetryofC})
\begin{eqnarray}
        d \beta^1 \wedge dt &=& \frac{\partial \beta^1}{\partial dx^1} dx^1
+  \frac{\partial \beta^1}{\partial dx^2} dx^2 +  \frac{\partial \beta^1}{\partial
dx^3} dx^3 \,,\nonumber\\
        &=& d \theta^1 = - {C^1}_{0i} \theta^0 \wedge \theta^i - {C^1}_{ij}
\theta^i \wedge \theta^j \,,\nonumber\\
        &=& - {C^1}_{01} dt \wedge dx^1 - {C^1}_{02} dt \wedge dx^2 - {C^1}_{03}
dt \wedge dx^3 \,,\nonumber\\
        &&- {C^1}_{12} \left[ dx^1 \wedge dx^2 + \beta^2 dx^1 \wedge dt -
\beta^1 dx^2 \wedge dt \right] \,,\nonumber\\
        && - {C^1}_{13} \left[ dx^1 \wedge dx^3 + \beta^3 dx^1 \wedge dt
- \beta^1 dx^3 \wedge dt \right] \,,\nonumber\\
        && - {C^1}_{23} \left[ dx^2 \wedge dx^3 + \beta^3 dx^2 \wedge dt
- \beta^2 dx^3 \wedge dt \right]\,.
\end{eqnarray}
Again, in order for this relation to be satisfied, ${C^1}_{12} = {C^1}_{13}
= {C^1}_{23} = 0$ and ${C^1}_{01} = \frac{\partial \beta^1}{\partial x^1}$,
${C^1}_{02} = \frac{\partial \beta^1}{\partial x^2}$, ${C^1}_{03} = \frac{\partial
\beta^1}{\partial x^3}$. We deduce that ${C^m}_{0i}
= \frac{\partial \beta^m}{\partial x^i}$, ${C^m}_{ij} = 0$ and $C^{0}{}_{ij}=0$.
Using Eq.~\eqref{zuzu} and that in the coframe 
$\Gamma^0_{ij} = \frac{1}{2} \frac{1}{N^2} \bar \partial_0 h_{ij}$, we obtain
that
\begin{eqnarray}
\gamma_{ij}^{0}=-\frac{1}{2N^2}\Big(h_{il}\partial_{j}(\beta^{l})+h_{jl}\partial_{i}(\beta^{l})-\bar
\partial_{0}h_{ij}\Big)\,.
\end{eqnarray}
Since from Eq.~(\ref{eq13})
\begin{eqnarray}
K_{ij}\equiv - \nabla_{i}n_{j}=\gamma_{ij}^{\mu}n_{\mu}=-N\gamma_{ij}^{0}\,,
\end{eqnarray}
Eq. \eqref{eq:defitionofextrinsiccurvaturegivenbyArlen}
becomes 
\begin{eqnarray}K_{ij}=\frac{1}{2N}(h_{il}\partial_{j}(\beta^{l})+h_{jl}\partial_{i}(\beta^{l})-\bar
\partial_{0}h_{ij})\,.\end{eqnarray}

\chapter{3+1 Decompositions}\label{B}

\section{Einstein-Hilbert term}\label{sec:appendixEHterm}

We can write the Einstein-Hilbert term $\mathcal{R}$ as its auxiliary equivalent
$\varrho$. Then we can use the completeness relation Eq.~(\ref{eq11}) to
show that
\begin{eqnarray}\label{eq:decompositionofthericciscalarauxequiv}
\varrho &=& g^{\mu\rho} g^{\nu\sigma} \varrho_{\mu\nu\rho\sigma} \,,\nonumber\\
        &=& \left( h^{\mu\rho} - n^\mu n^\rho \right) \left( h^{\nu\sigma}
- n^\nu n^\sigma \right) \varrho_{\mu\nu\rho\sigma} \,,\nonumber\\
        &=& \left( h^{\mu\rho} h^{\nu\sigma} - n^\mu n^\rho h^{\nu\sigma}
- h^{\mu\rho} n^\nu n^\sigma + n^\mu n^\rho n^\nu n^\sigma \right) \varrho_{\mu\nu\rho\sigma}\,,
\nonumber\\
        &=& \left( h^{\mu\rho} h^{\nu\sigma} - n^\mu n^\rho h^{\nu\sigma}
- h^{\mu\rho} n^\nu n^\sigma \right) \varrho_{\mu\nu\rho\sigma} \,,\nonumber\\
        &=& \left( \rho - 2 \Omega \right)\,,
\end{eqnarray}
noting that the term with four $n^\alpha$s vanishes due to the antisymmetry
of the Riemann tensor in the first and last pair of indices (recall that
$\varrho_{\mu\nu\rho\sigma}$ has the same symmetry properties as the Riemann
tensor)
\section{Riemann Tensor}\label{RT1}
In this section we wish to show the contraction of the rest of the terms
in Eq.~(\ref{eq:decompositionofriemannsquaredintermsofhsandns}) for the sake
of completeness. We have, from $hhhh,$ 
\begin{eqnarray}
&&h_{\mu}^{\alpha}h_{\nu}^{\beta}h_{\rho}^{\gamma}h_{\sigma}^{\lambda}\varrho_{\alpha\beta\gamma\lambda}\left[-(N^{-1}\bar{\partial_{0}})^{2}+\Box_{hyp}\right]\varrho^{\mu\nu\rho\sigma}\nonumber\\
&&=\left(h_{\mu}^{i}e_{i}^{\alpha}\right)\left(h_{\nu}^{j}e_{j}^{\beta}\right)\left(h_{\rho}^{k}e_{k}^{\gamma}\right)\left(h_{\sigma}^{l}e_{l}^{\lambda}\right)\varrho_{\alpha\beta\gamma\lambda}\Big[-(N^{-1}\bar{\partial_{0}})^{2}+\Box_{hyp}\Big]\varrho^{\mu\nu\rho\sigma}\nonumber\\
&&=\left(h_{\mu}^{i}\right)\left(h_{\nu}^{j}\right)\left(h_{\rho}^{k}\right)\left(h_{\sigma}^{l}\right)\rho_{ijkl}\Big[-(N^{-1}\bar{\partial_{0}})^{2}+\Box_{hyp}\Big]\varrho^{\mu\nu\rho\sigma}\nonumber\\
&&=-N^{-2}\left[\left(h_{m}^{i}e_{\mu}^{m}\right)\left(h_{n}^{j}e_{\nu}^{n}\right)\left(h_{x}^{k}e_{\rho}^{x}\right)\left(h_{y}^{l}e_{\sigma}^{y}\right)\right]\rho_{ijkl}\Big[-(N^{-1}\bar{\partial_{0}})^{2}+\Box_{hyp}\Big]\varrho^{\mu\nu\rho\sigma}\nonumber\\
&&=-N^{-2}\rho_{ijkl}\Big\{\bar{\partial_{0}^{2}}\left(\rho^{ijkl}\right)\nonumber\\
&&-\bar{\partial_{0}}\left[\varrho^{\mu\nu\rho\sigma}\bar{\partial_{0}}\left(\left[\left(h_{m}^{i}e_{\mu}^{m}\right)\left(h_{n}^{j}e_{\nu}^{n}\right)\left(h_{x}^{k}e_{\rho}^{x}\right)\left(h_{y}^{l}e_{\sigma}^{y}\right)\right]\right)\right]\nonumber\\
&&-\bar{\partial_{0}}\left(\left[\left(h_{m}^{i}e_{\mu}^{m}\right)\left(h_{n}^{j}e_{\nu}^{n}\right)\left(h_{x}^{k}e_{\rho}^{x}\right)\left(h_{y}^{l}e_{\sigma}^{y}\right)\right]\right)\bar{\partial_{0}}\left(\varrho^{\mu\nu\rho\sigma}\right)\Big\}\nonumber\\
&&+\rho_{ijkl}\Big\{\Box_{hyp}\left[\rho^{ijkl}\right]-D_{a}\left(D^{a}\left[e_{\mu}^{m}e_{\nu}^{n}e_{\rho}^{x}e_{\sigma}^{y}\right]h_{m}^{i}h_{n}^{j}h_{x}^{k}h_{y}^{l}\varrho^{\mu\nu\rho\sigma}\right)\nonumber\\
&&-D_{a}\left[e_{\mu}^{m}e_{\nu}^{n}e_{\rho}^{x}e_{\sigma}^{y}\right]D^{a}\left(h_{m}^{i}h_{n}^{j}h_{x}^{k}h_{y}^{l}\varrho^{\mu\nu\rho\sigma}\right)\Big\}
\end{eqnarray}
which produced $\rho_{ijkl}\Box\rho^{ijkl}$ and the terms which are the results
of Leibniz rule. Next in Eq.~(\ref{eq:decompositionofriemannsquaredintermsofhsandns})
is, 
\begin{eqnarray}
&&h_{\mu}^{\alpha}h_{\nu}^{\beta}h_{\rho}^{\gamma}n^{\lambda}n_{\sigma}\varrho_{\alpha\beta\gamma\lambda}\left(-(N^{-1}\bar{\partial_{0}})^{2}+\Box_{hyp}\right)\varrho^{\mu\nu\rho\sigma}\nonumber\\
&&=\left(h_{\mu}^{i}e_{i}^{\alpha}\right)\left(h_{\nu}^{j}e_{j}^{\beta}\right)\left(h_{\rho}^{k}e_{k}^{\gamma}\right)n^{\lambda}n_{\sigma}\varrho_{\alpha\beta\gamma\lambda}\left(-(N^{-1}\bar{\partial_{0}})^{2}+\Box_{hyp}\right)\varrho^{\mu\nu\rho\sigma}\nonumber\\
&&=\left(h_{m}^{i}e_{\mu}^{m}\right)\left(h_{n}^{j}e_{\nu}^{n}\right)\left(h_{x}^{k}e_{\rho}^{x}\right)n^{\lambda}n_{\sigma}\varrho_{ijk\lambda}\left(-(N^{-1}\bar{\partial_{0}})^{2}+\Box_{hyp}\right)\varrho^{\mu\nu\rho\sigma}\nonumber\\
&&=\left(h_{m}^{i}e_{\mu}^{m}\right)\left(h_{n}^{j}e_{\nu}^{n}\right)\left(h_{x}^{k}e_{\rho}^{x}\right)n_{\sigma}\rho_{ijk}\left(-(N^{-1}\bar{\partial_{0}})^{2}+\Box_{hyp}\right)\varrho^{\mu\nu\rho\sigma}\nonumber\\
&&=-N^{-2}\rho_{ijk}\Big\{\bar{\partial_{0}^{2}}\left(\rho^{ijk}\right)\nonumber\\
&&-\bar{\partial_{0}}\left[\varrho^{\mu\nu\rho\sigma}\bar{\partial_{0}}\left(\left[h_{m}^{i}e_{\mu}^{m}h_{n}^{j}e_{\nu}^{n}h_{x}^{k}e_{\rho}^{x}n_{\sigma}\right]\right)\right]-\bar{\partial_{0}}\left(\left[h_{m}^{i}e_{\mu}^{m}
h_{n}^{j}e_{\nu}^{n}h_{x}^{k}e_{\rho}^{x}n_{\sigma}\right]\right)\bar{\partial_{0}}\left(\varrho^{\mu\nu\rho\sigma}\right)\Big\}\nonumber\\
&&+\rho_{ijk}\Big\{\Box_{hyp}\left[\rho^{ijk}\right]-D_{a}\left(D^{a}\left[e_{\mu}^{m}e_{\nu}^{n}e_{\rho}^{x}n_{\sigma}\right]h_{m}^{i}h_{n}^{j}h_{x}^{k}\varrho^{\mu\nu\rho\sigma}\right)\nonumber\\
&&-D_{a}\left[e_{\mu}^{m}e_{\nu}^{n}e_{\rho}^{x}n_{\sigma}\right]D^{a}\left(h_{m}^{i}h_{n}^{j}h_{x}^{k}\varrho^{\mu\nu\rho\sigma}\right)\Big\}\end{eqnarray}
with $\rho_{ijk}\equiv n^\mu \rho_{ijk\mu}$. Here we produced $\rho_{ijk}\Box\rho^{ijk}$
and the extra terms which are the results
of the Leibniz rule.
Similarly we can find the contractions for different terms in Eq.~(\ref{eq:decompositionofriemannsquaredintermsofhsandns}).
\section{Ricci Tensor}\label{RT2}
In similar way as we did in the Riemann case we can find all the other contractions
in the expansion of Eq.~(\ref{eq:varrhodecomposition}) which we omitted.
They are:
\begin{eqnarray}
&&h^{\rho\sigma} h^{\mu\kappa} h^{\nu\lambda} h^{\gamma\delta}
\varrho_{\rho\mu\sigma\nu} \Box \varrho_{\gamma\kappa\delta\lambda} \nonumber\\
&&=  (h^{im}e^{\rho}_{i}e^{\sigma}_{m})(h^{jn}e^{\mu}_{j}e^{\kappa}_{n})(h^{kx}e^{\nu}_{k}e^{\lambda}_{x})(h^{ly}e^{\gamma}_{l}e^{\delta}_{y})\varrho_{\rho\mu\sigma\nu}
 \left( - \left( N^{-1} \bar{\partial}_0 \right)^2
+ \Box_{hyp} \right) \varrho_{\gamma\kappa\delta\lambda}\nonumber\\
&&= (h^{jn}e^{\kappa}_{n})(h^{kx}e^{\lambda}_{x})(h^{ly}e^{\gamma}_{l}e^{\delta}_{y})\rho_{jk}
 \left( - \left( N^{-1} \bar{\partial}_0 \right)^2
+ \Box_{hyp} \right) \varrho_{\gamma\kappa\delta\lambda}\nonumber\\
&&=-N^{-2}\rho_{jk}\Big\{\bar{\partial}^{2}_{0}(\rho^{jk})-\bar{\partial}_{0}(\varrho_{\gamma\kappa\delta\lambda}\bar{\partial}_{0}[(h^{jn}e^{\kappa}_{n})(h^{kx}e^{\lambda}_{x})(h^{ly}e^{\gamma}_{l}e^{\delta}_{y})])\nonumber\\
&&-\bar{\partial}_{0}[(h^{jn}e^{\kappa}_{n})(h^{kx}e^{\lambda}_{x})(h^{ly}e^{\gamma}_{l}e^{\delta}_{y})]\bar{\partial}_{0}\varrho_{\gamma\kappa\delta\lambda}\Big\}\nonumber\\
&&+\rho_{jk}\Big\{\Box_{hyp}(\rho^{jk})-D_{a}(\varrho_{\gamma\kappa\delta\lambda}D^{a}[(h^{jn}e^{\kappa}_{n})(h^{kx}e^{\lambda}_{x})(h^{ly}e^{\gamma}_{l}e^{\delta}_{y})])\nonumber\\
&&-D_{a}[(h^{jn}e^{\kappa}_{n})(h^{kx}e^{\lambda}_{x})(h^{ly}e^{\gamma}_{l}e^{\delta}_{y})]D^{a}\varrho_{\gamma\kappa\delta\lambda}\Big\}
\end{eqnarray} 
with $(h^{jn}e^{\kappa}_{n})(h^{kx}e^{\lambda}_{x})(h^{ly}e^{\gamma}_{l}e^{\delta}_{y})\varrho_{\gamma\kappa\delta\lambda}=\rho^{jk}$.
Above we produced $\rho_{jk}\Box\rho^{jk}$ plus other terms that are results
of the Leibniz rule. And, 
\begin{eqnarray}
&&h^{\rho\sigma} n^\mu n^\kappa h^{\nu\lambda} h^{\gamma\delta}\varrho_{\rho\mu\sigma\nu}
\left( - \left( N^{-1} \bar{\partial}_0 \right)^2
+ \Box_{hyp} \right) \varrho_{\gamma\kappa\delta\lambda}\nonumber\\
&&= (h^{ij} e^{\rho}_{i}e^{\sigma}_{j})n^\mu n^\kappa (h^{kl} e^{\nu}_{k}e^{\lambda}_{l})(h^{mn}e^{\gamma}_{m}e^{\delta}_{n})\varrho_{\rho\mu\sigma\nu}
\left( - \left( N^{-1} \bar{\partial}_0 \right)^2
+ \Box_{hyp} \right) \varrho_{\gamma\kappa\delta\lambda}\nonumber\\
&&=  n^\kappa (h^{kl} e^{\lambda}_{l})(h^{mn}e^{\gamma}_{m}e^{\delta}_{n})\rho_{
k}
\left( - \left( N^{-1} \bar{\partial}_0 \right)^2
+ \Box_{hyp} \right) \varrho_{\gamma\kappa\delta\lambda}\nonumber\\
&&=-N^{-2}\rho_{
k}\Big\{\bar{\partial}^{2}_{0}(\rho^{k})-\bar{\partial}_{0}\big(\varrho_{\gamma\kappa\delta\lambda}\bar{\partial}_{0}[n^\kappa
h^{kl} e^{\lambda}_{l}h^{mn}e^{\gamma}_{m}e^{\delta}_{n}]\big)-\bar{\partial}_{0}[n^\kappa
h^{kl} e^{\lambda}_{l}h^{mn}e^{\gamma}_{m}e^{\delta}_{n}]\bar{\partial}_{0}\varrho_{\gamma\kappa\delta\lambda}\Big\}\nonumber\\
&&+\rho_{
k}\Big\{\Box_{hyp}(\rho^{k})-D_a\big(\varrho_{\gamma\kappa\delta\lambda}D^a[n^\kappa
h^{kl} e^{\lambda}_{l}h^{mn}e^{\gamma}_{m}e^{\delta}_{n}]\big)-D_{a}[n^\kappa
h^{kl} e^{\lambda}_{l}h^{mn}e^{\gamma}_{m}e^{\delta}_{n}]D^a\varrho_{\gamma\kappa\delta\lambda}\Big\}\nonumber\\
\end{eqnarray}
where we used $n^\mu\varrho_{\mu k}=\rho_k$ and $n^\kappa h^{kl} e^{\lambda}_{l}h^{mn}e^{\gamma}_{m}e^{\delta}_{n}\varrho_{\gamma\kappa\delta\lambda}=n^\kappa
h^{kl} \varrho_{\kappa l}=\rho^k$.
We produced $\rho_{k}\Box\rho^{k}$ plus other terms that are results
of the Leibniz rule.
We may also note that one can write, $\rho_{ij}\equiv h^{kl}\rho_{ikjl}$,
$\rho\equiv h^{ik}h^{il}\rho_{ijkl}$
and $\rho_{i}\equiv h^{jk}\rho_{jik}$.
\section{Generalisation from $\Box$ to $\mathcal{F}(\Box)$}\label{B4}
In Eq.~(\ref{ili1}) for $\Box^2$, we have, 
\begin{eqnarray}
&&\Omega_{ij}\left[h_{x}^{i}e_{\mu}^{x}n_{\nu}h_{y}^{j}e_{\rho}^{y}n_{\sigma}\right]\Box^{2}\varrho^{\mu\nu\rho\sigma}\nonumber\\
&&=\Omega_{ij}\left[h_{x}^{i}e_{\mu}^{x}n_{\nu}h_{y}^{j}e_{\rho}^{y}n_{\sigma}\right]\left(-(N^{-1}\bar{\partial_{0}})^{2}+\Box_{hyp}\right)\left(-(N^{-1}\bar{\partial_{0}})^{2}+\Box_{hyp}\right)\varrho^{\mu\nu\rho\sigma}\,,\nonumber\\
&&=N^{-4}\Omega_{ij}\left[h_{x}^{i}e_{\mu}^{x}n_{\nu}h_{y}^{j}e_{\rho}^{y}n_{\sigma}\right]\bar{\partial_{0}^{4}}\varrho^{\mu\nu\rho\sigma}\nonumber\\
&&-N^{-2}\Omega_{ij}\left[h_{x}^{i}e_{\mu}^{x}n_{\nu}h_{y}^{j}e_{\rho}^{y}n_{\sigma}\right]\bar{\partial_{0}^{2}}D_{a}D^{a}\varrho^{\mu\nu\rho\sigma}\nonumber\\
&&+\Omega_{ij}\left[h_{x}^{i}e_{\mu}^{x}n_{\nu}h_{y}^{j}e_{\rho}^{y}n_{\sigma}\right]D_{a}D^{a}\left[-\left(N^{-1}\bar{\partial_{0}}\right)^{2}\right]\varrho^{\mu\nu\rho\sigma}\nonumber\\
&&+\Omega_{ij}\left[h_{x}^{i}e_{\mu}^{x}n_{\nu}h_{y}^{j}e_{\rho}^{y}n_{\sigma}\right]D_{a}D^{a}D_{b}D^{a}\varrho^{\mu\nu\rho\sigma}\,.
\end{eqnarray}
As a general rule we can write, 
\begin{eqnarray}\label{eq:generalrelationforcommutingoperator1}
&&XDDDDY=D\left(XDDDY\right)-D(X)DDD(Y) \,,\nonumber\\
&&=D\left(D(XDD(Y))-D(X)DD(Y)\right)-D(X)DDD(Y)\,, \nonumber\\
&&=DD(XDD(Y))-D(D(X)DD(Y))-D(X)DDD(Y)\,, \nonumber\\
&&=DD\left(D(XD(Y))-D(X)D(Y)\right)-D(D(X)DD(Y))-D(X)DDD(Y)\,, \nonumber\\
&&=DDD(XD(Y))-DD(D(X)D(Y))-D(D(X)DD(Y))-D(X)DDD(Y)\,, \nonumber\\
&&=DDD\left(D(XY)-D(X)Y\right)-DD(D(X)D(Y))-D(D(X)DD(Y))\nonumber\\&&-D(X)DDD(Y)\,, \nonumber\\
&&=DDDD(XY)-DDD(D(X)Y)-DD(D(X)D(Y))\,, \nonumber\\
&&-D(D(X)DD(Y))-D(X)DDD(Y)\,,
\end{eqnarray} 
where $X$ and $Y$ are some tensors and $D$ is some operator.
Applying this we can write, 
\begin{eqnarray}\label{eq:boundarytermspapereqnwithomegaij}
&&N^{-4}\Omega_{ij}\left[h_{x}^{i}e_{\mu}^{x}n_{\nu}h_{y}^{j}e_{\rho}^{y}n_{\sigma}\right]\bar{\partial_{0}^{4}}\varrho^{\mu\nu\rho\sigma}
\nonumber\\
&&=N^{-4}\Omega_{ij}\{\bar{\partial_{0}^{4}}(\Omega^{ij})-\bar{\partial_{0}^{4}}\left[h_{x}^{i}e_{\mu}^{x}n_{\nu}h_{y}^{j}e_{\rho}^{y}n_{\sigma}\right]\varrho^{\mu\nu\rho\sigma}-\bar{\partial_{0}^{3}}\left[h_{x}^{i}e_{\mu}^{x}n_{\nu}h_{y}^{j}e_{\rho}^{y}n_{\sigma}\right]\bar{\partial_{0}}\varrho^{\mu\nu\rho\sigma}
\nonumber\\
&&-\bar{\partial_{0}^{2}}\left[h_{x}^{i}e_{\mu}^{x}n_{\nu}h_{y}^{j}e_{\rho}^{y}n_{\sigma}\right]\bar{\partial_{0}^{2}}\varrho^{\mu\nu\rho\sigma}-\bar{\partial_{0}}\left[h_{x}^{i}e_{\mu}^{x}n_{\nu}h_{y}^{j}e_{\rho}^{y}n_{\sigma}\right]\bar{\partial_{0}^{3}}\varrho^{\mu\nu\rho\sigma}\}+\cdots
\,,
\end{eqnarray} 
where we dropped the irrelevant terms.
We moreover can generalise the result of (\ref{eq:generalrelationforcommutingoperator1})
and write, 
\begin{eqnarray}
&&X D^{2n} Y= D^{2n}(XY)-D^{2n-1}(D(X)Y)-D^{2n-2}(D(X)D(Y))\,, \nonumber\\
&&-D^{2n-3}(D(X)D^{2}(Y))-\dots-D(D(X)D^{2n-2}(Y))-D(X)D^{2n-1}(Y)
\,.\nonumber\\
\end{eqnarray}

\chapter{Functional Differentiation}\label{C}

Given the constraint equation 
\begin{equation}
2\Psi^{ij}+\frac{\delta f}{\delta \Omega_{ij}}=0,
\end{equation}
suppose that $f=\Omega \mathcal{F}(\Box)\Omega$ and $\mathcal{F}(\Box)=\sum^{\infty}_{n=0}f_{n}\Box^{n}$,
where the coefficients $f_n$ are massless~\footnote{Recall that the $\Box$
term comes
with an associated scale $\Box/M^2$.
}. Then, using the
generalised Euler-Lagrange equations, we have in the coframe (and imposing
the condition that $\delta \Omega_{ij}=0$ on the boundary $\partial \mathcal{M}$)
\begin{eqnarray}
\frac{\delta f}{\delta \Omega_{ij}}&=&\frac{\partial
f}{\partial
\Omega_{ij}}-\nabla_{\mu}\left(\frac{\partial
f}{\partial
(\nabla_{\mu}\Omega_{ij})} \right)+\nabla_{\mu}\nabla_{\nu}\left(\frac{\partial
f}{\partial
(\nabla_{\mu}\nabla_{\nu}\Omega_{ij})} \right)+\cdots \nonumber \\
&=&\frac{\partial f}{\partial \Omega_{ij}}+\Box\left(\frac{\partial
f}{\partial (\Box\Omega_{ij})}\right)+\Box^{2}\left(\frac{\partial
f}{\partial (\Box^{2}\Omega_{ij})}\right)+\cdots \nonumber \\
&=&\frac{\partial f}{\partial \Omega_{ij}}+\sum^{\infty}_{n=1}\Box^{n}\left(\frac{\partial
f}{\partial (\Box^{n}\Omega_{ij})}\right) \nonumber \\
&=&f_{0}\frac{\partial(\Omega^{2})}{\partial\Omega_{ij}}+f_{1}\frac{\partial
(\Omega\Box\Omega)}{\partial\Omega_{ij}}+f_{1}\Box\left(\frac{\partial(
\Omega\Box^{}\Omega)}{\partial (\Box\Omega_{ij})}\right)+f_{2}\Box^{2}\left(\frac{\partial
\Omega\Box^{2}\Omega}{\partial (\Box^{2}\Omega_{ij})}\right)+\cdots \nonumber
\\
&=&2f_0h^{ij}\Omega+f_{1}h^{ij}\Box\Omega+f_{1}\Box (h^{ij}\Omega)+\cdots
\nonumber \\
&=&2f_0h^{ij}\Omega+f_1\Big[ h^{ij}\Box\Omega+\left(\Box \Omega\right) h^{ij}\Big]+\cdots
\nonumber \\
&=&2f_0h^{ij}\Omega+2f_1h^{ij}\Box\Omega+\cdots \nonumber \\
&=&2h^{ij}\left( f_0 + f_1 \Box +\cdots\right)\Omega \nonumber \\
&=&2h^{ij}\mathcal{F}(\Box)\Omega \,,
\end{eqnarray}
where we have used that $\Box g^{ij}=\Box h^{ij}=0$.
Note also that:\begin{eqnarray}
f_{1}\Box\left(\frac{\partial(
\Omega\Box^{}\Omega)}{\partial (\Box\Omega_{ij})}\right)=f_{1}\Box \Big(\frac{\partial(
\Omega\Box^{}[h^{mn}\Omega_{mn}])}{\partial (\Box\Omega_{ij})}\Big)\,.
\end{eqnarray}
So we can summarise the results and write, 
\begin{eqnarray}
\frac{\delta (\Omega\Box\Omega)}{\delta \Omega_{ij}}&=&\frac{\partial
(\Omega\Box\Omega)}{\partial\Omega_{ij}}+\Box\left(\frac{\partial(
\Omega\Box^{}\Omega)}{\partial (\Box\Omega_{ij})}\right)\nonumber\\
&=&h^{ij}\Box\Omega+\Box (h^{ij}\Omega)=\Big[ h^{ij}\Box\Omega+\Box\Omega
h^{ij}\Big]=2h^{ij}\Box\Omega\,.
\end{eqnarray}
\begin{eqnarray}
\frac{\delta 
(\Omega_{ij}\Box\Omega^{ij})}{\delta \Omega_{ij}}&=&\frac{\partial
(\Omega_{ij}\Box\Omega^{ij})}{\partial\Omega_{ij}}+\Box\left(\frac{\partial(\Omega_{ij}\Box\Omega^{ij})}{\partial
(\Box\Omega_{ij})}\right)\nonumber\\
&=&\Box\Omega^{ij}+\Box \Omega^{ij}=2\Box\Omega^{ij}\,.
\end{eqnarray}
\begin{eqnarray}
\frac{\delta (\rho\Box\Omega)}{\delta \Omega_{ij}}&=&\Box\left(\frac{\partial(\rho\Box\Omega)}{\partial
(\Box\Omega_{ij})}\right)
=\Box (\rho h^{ij})=h^{ij}\Box\rho\,.
\end{eqnarray}
\begin{eqnarray}
\frac{\delta (\rho_{ij}\Box\Omega^{ij})}{\delta \Omega_{ij}}&=&\Box\left(\frac{\partial(\rho_{ij}\Box\Omega^{ij})}{\partial
(\Box\Omega_{ij})}\right)=\Box \rho^{ij}\,.
\end{eqnarray}
\begin{eqnarray}
\frac{\delta (\Omega\Box\rho)}{\delta \Omega_{ij}}&=&\frac{\partial
(\Omega\Box\rho)}{\partial\Omega_{ij}}=h^{ij}\Box\rho\,.
\end{eqnarray}
\begin{eqnarray}
\frac{\delta (\Omega_{ij}\Box\rho^{ij})}{\delta \Omega_{ij}}&=&\frac{\partial
(\Omega_{ij}\Box\rho^{ij})}{\partial\Omega_{ij}}=\Box \rho^{ij}\,.
\end{eqnarray}
and generalise this to: 
\begin{eqnarray}\label{eq:functionaldifferentationofdifferentterms}
\frac{\delta \big(\Omega\mathcal{F}(\Box)\Omega\big)}{\delta \Omega_{ij}}=2h^{ij}\mathcal{F}(\Box)\Omega,&\quad&
\frac{\delta \big(\Omega_{ij}\mathcal{F}(\Box)\Omega^{ij}\big)}{\delta \Omega_{ij}}=2\mathcal{F}(\Box)\Omega^{ij}\,,\\
\frac{\delta \big(\rho\mathcal{F}(\Box)\Omega\big)}{\delta
\Omega_{ij}}=h^{ij}\mathcal{F}(\Box)\rho,&\quad& \frac{\delta \big(\rho_{ij}\mathcal{F}(\Box)\Omega^{ij}\big)}{\delta
\Omega_{ij}}=\mathcal{F}(\Box)\rho^{ij}\,,\\
\frac{\delta \big(\Omega\mathcal{F}(\Box)\rho\big)}{\delta
\Omega_{ij}}=h^{ij}\mathcal{F}(\Box)\rho,&\quad&
\frac{\delta \big(\Omega_{ij}\mathcal{F}(\Box)\rho^{ij}\big)}{\delta
\Omega_{ij}}=\mathcal{F}(\Box)\rho^{ij}\,.
\end{eqnarray}
\chapter{Riemann tensor components in ADM gravity}\label{riemtenscomps}
Using the method of \cite{{Golovnev:2013fj}}, we can find the Riemann tensor
components. The Christoffel symbols for the ADM metric in Eq.~(\ref{eq8})
are
\begin{eqnarray}\label{eq:ChristoffelforADM}
       \Gamma_{ij0} &=& \Gamma_{i0j} = - N K_{ij} + D_j \beta_i \nonumber\\
       \Gamma_{ijk} &=& {}^{(3)}\Gamma_{ijk}\nonumber\\
       \Gamma_{00}^0 &=& \frac{1}{N} \left(\dot{N} + \beta^i \partial_i N
- \beta^i \beta^j K_{ij} \right)\nonumber\\
        \Gamma^0_{0i} = \Gamma^0_{i0} &=& \frac{1}{N} \left(\partial_i N
- \beta^j K_{ij} \right) \nonumber\\
        \Gamma^i_{0j} = \Gamma^i_{j0} &=& - \frac{\beta^i \partial_j N}{N}
- N \left( h^{ik} - \frac{\beta^i \beta^k}{N^2} \right) K_{kj} + D_j \beta^i
\nonumber\\
        \Gamma^0_{ij} &=& - \frac{1}{N} K_{ij} \nonumber\\
        \Gamma^i_{jk} &=& {}^{(3)} \Gamma^i_{jk} + \frac{\beta^i}{N} K_{jk}
\end{eqnarray}  
where $K_{ij}$ is the extrinsic curvature given by (\ref{eq13}) and in the
ADM metric, $N$ is the lapse, $\beta_i$ is the shift and $h_{ij}$ is the
induced metric on the hypersurface. Now we can find the Riemann tensor components
\begin{eqnarray}\label{eq:RiemanntensorcompsforADM}
        \mathcal{R}_{ijkl} &=& g_{i\rho} \partial_k \Gamma^\rho_{lj} - g_{i\rho}
\partial_l \Gamma^\rho_{kj} + \Gamma_{ik\rho} \Gamma^\rho_{lj} - \Gamma_{il\rho}
\Gamma^\rho_{kj} \nonumber\\
        &=& -\beta_i \partial_k \left( \frac{1}{N} K_{jl} \right) + h_{im}
\partial_k \left( {}^{(3)} \Gamma^m_{jl} + \frac{\beta^m}{N} K_{jl} \right)
- \frac{1}{N} K_{jl} \left( - N K_{ik} + D_k \beta_i \right)\nonumber\\
        && + {}^{(3)} \Gamma_{ikm} \left( {}^{(3)} \Gamma^m_{lj} + \frac{\beta^m}{N}
K_{lj} \right) - \left(k \leftrightarrow l \right) \nonumber\\
        &=& R_{ijkl} + K_{ik} K_{jl} - K_{il} K_{jk} \nonumber\\  \end{eqnarray}
 where $R_{ijkl}$ is the Riemann tensor of the induced metric on the hypersurface.
Then
 \begin{eqnarray}
        n_\mu {\mathcal{R}^\mu}_{ijk} &=& - N \left( \partial_j \Gamma^0_{ki}
+ \Gamma^0_{j\rho}\Gamma^\rho_{ki}\right) - \left( j \leftrightarrow k \right)
\nonumber\\
        &=& \partial_j K_{ki} + {}^{(3)} \Gamma^m_{ki} K_{jm} - \left( j
\leftrightarrow k \right) \nonumber\\
        &=& D_j K_{ki} - D_k K_{ji} 
\end{eqnarray}
Relabelling the indices, we obtain that
\begin{equation}
n^{\mu}\mathcal{R}_{ijk \mu}= D_{j}K_{ki}-D_{i}K_{jk}
\end{equation}
Finally, we have that 
\begin{eqnarray}
        n_\mu {\mathcal{R}^\mu}_{i0j} &=& n_\mu \left( \partial_0 \Gamma^\mu_{ji}
- \partial_j \Gamma^\mu_{0i} + \Gamma^\mu_{0\rho} \Gamma^\rho_{ji} - \Gamma^\mu_{j\rho}
\Gamma^\rho_{0i} \right) \nonumber\\
        &=& \dot{K}_{ij} + D_i D_j N + N {K_i}^k K_{kj} - D_j \left( K_{ik}
\beta^k \right) - K_{kj} D_i \beta^k        
\end{eqnarray}

Hence
\begin{eqnarray} 
        n^\mu n^\nu \mathcal{R}_{\mu i \nu j} &=& n^0 n^\mu \mathcal{R}_{\mu
i0 j}+ n^k n^\mu \mathcal{R}_{\mu i kj } \nonumber\\
        &=& \frac{1}{N} \left( \dot{K}_{ij} + D_i D_j N + N {K_i}^k K_{kj}-
D_j \left( K_{ik} \beta^k \right) - K_{kj} D_i \beta^k \right)\nonumber\\
        && + \frac{\beta^k}{N^2} \left(D_j K_{ki} - D_k K_{ji}\right)\nonumber\\
        &=&\frac{1}{N} \left( \dot{K}_{ij} + D_i D_j N + N {K_i}^k K_{kj}
- \pounds_\beta K_{ij} \right)    
\end{eqnarray}
where $\pounds_\beta K_{ij}\equiv \beta^{k}D_k K_{ij}+K_{ik}D_j\beta^k+K_{jk}D_i\beta^k$.\
Therefore overall, we have 
\begin{eqnarray}
\mathcal{R}_{ijkl}&\equiv&K_{ik}K_{jl}-K_{il}K_{jk}+R_{ijkl}\,, \\ \label{xx5}
\mathcal{R}_{ijk\mathbf{n}}&\equiv&n^{\mu}\mathcal{R}_{ijk\mu}=D_{j}K_{ik}-D_{i}K_{jk}\,,
\\ \label{xx2}
\mathcal{R}_{i\mathbf{n}j\mathbf{n}}&\equiv&n^{\mu}n^{\nu}\mathcal{R}_{i\mu
j\nu}=N^{-1}\big(\partial_{t}K_{ij}-\mathsterling_\beta
K_{ij}\big)+K_{ik}K^{\ k}_{j}+N^{-1}D_i D_jN\,,\nonumber\\
\end{eqnarray}
\section{Coframe}\label{coframeee}

Since in the coframe slicing Eq.~(\ref{eq16}) we have $g^{0i}=g_{0i}=0$,
therefore from  $n^i=n_i=0$ . Then the Christoffel symbols become~\footnote{In
the coframe slicing when we write $\partial_\mu$ we mean that $\partial_\mu$
is $\bar\partial_0$ when $\mu=0$ and $\partial_\mu$
is $\partial_i$ when $\mu=i$. }
\begin{eqnarray}
        \Gamma^0_{00} &=& \frac{1}{2} g^{0\mu} \left(\bar \partial_0 g_{\mu
0}
+ \bar \partial_0 g_{0 \mu} - \partial_\mu g_{00} \right) \nonumber \\
        &=& \frac{1}{2} g^{00}  \partial_0 g_{0 0} 
        = \frac{1}{2} \left( - \frac{1}{N^2} \right) \partial_0 \left(
- N^2 \right) \nonumber \\
        &=& \frac{\partial_0 N}{N}, 
\end{eqnarray}
\begin{eqnarray}
        \Gamma^0_{0i} = \Gamma^0_{i0} &=& \frac{1}{2} g^{0\mu} \left(\bar
\partial_0
g_{\mu i} + \partial_i g_{\mu 0} - \partial_\mu g_{i0} \right) \nonumber
\\
        &=& \frac{1}{2} g^{00} \partial_i g_{00} 
        = \frac{1}{2} \left( \frac{-1}{N^2} \right) \partial_i \left( - N^2
 \right) \\
\nonumber
        &=& \frac{\partial_i N }{N}, 
\end{eqnarray}
\begin{eqnarray}
        \Gamma^i_{00} &=& \frac{1}{2} g^{i\mu} \left(\bar \partial_0 g_{\mu
0}
+\bar \partial_0 g_{\mu 0} - \partial_\mu g_{00} \right) \nonumber \\
        &=& - \frac{1}{2} g^{ij} \partial_j g_{00} 
        = - \frac{1}{2} h^{ij} \partial_j \left( - N^2 \right) \nonumber\\
        &=&  N h^{ij} \partial_j N,
\end{eqnarray}
\begin{eqnarray}
        \Gamma^i_{j0} &=& \frac{1}{2} g^{i \mu} \left(\bar \partial_0 g_{\mu
j} 
        + \partial_j g_{\mu 0} - \partial_\mu g_{j 0} \right) \\  \nonumber
        &=& \frac{1}{2} g^{ik} \left(\bar \partial_0 g_{kj} \right) \nonumber\\
        &=& \frac{1}{2} h^{ik}\bar \partial_0 h_{jk}, 
\end{eqnarray}
\begin{eqnarray}
        \Gamma^0_{ij} &=& \frac{1}{2} g^{0\mu} \left( \partial_j g_{\mu i}
+ \partial_i g_{\mu j} - \partial_\mu g_{ij} \right) \nonumber \\
        &=& - \frac{1}{2} g^{00}\bar \partial_0 g_{ij} 
        = - \frac{1}{2}\left( \frac{-1}{N^2} \right)\bar \partial_0 h_{ij}
\nonumber \\
        &=& \frac{1}{2} \frac{1}{N^2}\bar \partial_0 h_{ij},
\end{eqnarray}
\begin{eqnarray}
        \Gamma^i_{jk} &=& \frac{1}{2} g^{i\mu} \left( \partial_j g_{\mu k}
+ \partial_k g_{\mu j} - \partial_\mu g_{kj} \right) \nonumber \\
        &=& \frac{1}{2} h^{il} \left( \partial_j h_{lk} + \partial_k h_{lj}
- \partial_l h_{jk} \right).
\end{eqnarray}
To summarise
\begin{eqnarray}
        \Gamma^0_{00} &=& \frac{\partial_0 N}{N}, \quad \quad \quad \quad
        \Gamma^0_{0i} =  \frac{\partial_i N}{N}, \nonumber \\
        \Gamma^i_{00} &=&  N h^{ij} \partial_j N, \quad \quad \quad
        \Gamma^i_{j0} = \frac{1}{2} h^{ik}\bar \partial_0 h_{jk}, \nonumber\\
        \Gamma^0_{ij} &=& \frac{1}{2} \frac{1}{N^2}\bar \partial_0 h_{ij},
\quad \quad
        \Gamma^i_{jk} =  \frac{1}{2} h^{il} \left( \partial_j h_{lk} +
\partial_k h_{lj} - \partial_l h_{jk} \right).
\end{eqnarray}
Then using Eq.~(\ref{eq22}) and Eq.~(\ref{eq23}), we can find the $\gamma^\mu_{\nu\rho}$s,
the analogues of the Christoffel symbols in the coframe.
\begin{eqnarray}
        \gamma^{i}_{jk} = \Gamma^i_{jk}, 
        \quad \gamma^i_{0k} = - N {K^i}_k,
        \quad \gamma^{i}_{j0} = - N {K^i}_k + \partial_j \beta^i,
        \quad \gamma^{0}_{ij} = - N^{-1} K_{ij},\nonumber \\
        \gamma^{i}_{00} = N \partial^i N, 
        \quad \gamma^{0}_{0i} = \gamma^{0}_{i0} = \partial_i \log N,
        \quad \gamma^{0}_{00} = \partial_0 \log N \quad
\end{eqnarray}
Then using the same method as in Eq.~(\ref{eq:RiemanntensorcompsforADM}),
\begin{eqnarray}
        \mathcal{R}_{ijkl} &=& g_{i\rho} \partial_k \gamma^{\rho}_{lj} -
g_{i\rho} \partial_l \gamma^{\rho}_{kj} + \gamma_{ik\rho} 
        \gamma^\rho_{lj} -  \gamma_{il\rho} \gamma^{\rho}_{kj}\nonumber\\
        &=& R_{ijkl} + K_{ik} K_{jl} - K_{il} K_{jk} 
\end{eqnarray}
 Next
 \begin{eqnarray}
        \mathcal{R}_{0ijk} &=& - N^{2} \left( \partial_j \gamma^0_{ki} +
\gamma^0_{j\rho}\gamma^\rho_{ki}\right)
- \left( j \leftrightarrow k \right) \nonumber\\
       &=& N \left( D_j K_{ki} - D_k K_{ji} \right)
\end{eqnarray}
Finally we have that in the coframe,
\begin{eqnarray} 
         \mathcal{R}_{0 i 0 j} &=&-N^{2} \left( \bar \partial_0 \gamma_{ji}^{0}
-\partial_j \gamma_{0i}^{0} + \gamma_{0\rho}^{0} \gamma^\rho_{ji} - \gamma_{j\rho}^{0}
\gamma^\rho_{0i} \right) \nonumber\\
        &=& N \left( \bar \partial_{0}K_{ij}  + N {K_i}^k K_{kj}+ D_i D_j
N\right) 
\end{eqnarray}
Hence the non-vanishing components of the Riemann
tensor in the coframe, namely the Gauss, Codazzi and Ricci tensor, become:
\begin{eqnarray}
\mathcal{R}_{ijkl}&=&K_{ik}K_{jl}-K_{il}K_{jk}+R_{ijkl}\,,\nonumber \\
\mathcal{R}_{0ijk}&=&N(D_{j}K_{ki}-D_{k}K_{ji})\,, \nonumber
\\
\mathcal{R}_{0i0j}&=&N(\bar \partial_{0} K_{ij}+NK_{ik}K^{\ k}_{j}+D_i D_jN\,),
\end{eqnarray}    
where $K_{ij}$ is the extrinsic curvature of the hypersurface, given in the
coframe by Eq.~(\ref{eq29}) and $R_{ijkl}$ is the Riemann tensor of the induced
metric on the hypersurface.

\chapter{Entropy and functional differentiation}\label{functionalentropy}
For the scalar curvature which corresponds to the Einstein-Hilbert term we
have, 
\begin{eqnarray}
\frac{\delta R}{\delta R_{\mu\nu\rho\sigma}}&=&\frac{\delta(g^{\beta\xi}g^{\alpha\gamma}
R_{\alpha\beta\gamma\xi})}{\delta R_{\mu\nu\rho\sigma}}\nonumber\\
&=&g^{\beta\xi}g^{\alpha\gamma}\delta^{[\mu}_{[\alpha}\delta^{\nu]}_{\beta]}\delta^{[\rho}_{[\gamma}\delta^{\sigma]}_{\xi]}\nonumber\\
&=&g^{\beta\xi}g^{\alpha\gamma}\delta^{[\mu}_{\alpha}\delta^{\nu]}_{\beta}\delta^{[\rho}_{\gamma}\delta^{\sigma]}_{\xi}\nonumber\\
&=&g^{\rho[\mu}g^{\nu]\sigma}.
\end{eqnarray}
The next term we shall consider is $ RF(\Box)R$, to do so we shall use the
generalised Euler-Lagrange equation given in (\ref{genrulag}),
\begin{eqnarray}
\frac{\delta( RF(\Box)R)}{\delta R_{\mu\nu\rho\sigma}}&=&f_0\frac{\partial
(R^{2})}{\partial R_{\mu\nu\rho\sigma}}+f_{1}\frac{\partial
(R_{}\Box R^{})}{\partial R_{\mu\nu\rho\sigma}}\nonumber\\&+&f_1\Box\frac{\partial
(R\Box R^{})}{\partial(\Box R_{\mu\nu\rho\sigma})}+f_{2}\Box^{2}\frac{\partial
(R_{}\Box ^{2}R^{})}{\partial (\Box ^{2}R_{\mu\nu\rho\sigma})}+\cdots
,\end{eqnarray}
where $\cdots$ are the terms up to infinity. Term by term we have, 
\begin{eqnarray}
f_0\frac{\partial
(R^{2})}{\partial R_{\mu\nu\rho\sigma}}&=&f_0\frac{\partial
(g^{\beta\xi}g^{\alpha\gamma}
g^{bd}g^{ac}R_{\alpha\beta\gamma\xi }R_{abcd})}{\partial R_{\mu\nu\rho\sigma}}\nonumber\\
&=&g^{\beta\xi}g^{\alpha\gamma}g^{bd}g^{ac}\delta^{[\mu}_{[\alpha}\delta^{\nu]}_{\beta]}\delta^{[\rho}_{[\gamma}\delta^{\sigma]}_{\xi]}R_{abcd}\nonumber\\
&+&g^{\beta\xi}g^{\alpha\gamma}g^{bd}g^{ac}\delta^{[\mu}_{[a}\delta^{\nu]}_{b]}\delta^{[\rho}_{[c}\delta^{\sigma]}_{d]}R_{\alpha\beta\gamma\xi
}\nonumber\\
&=&2g^{\rho[\mu}g^{\nu]\sigma}R,
\end{eqnarray}
\begin{eqnarray}
f_{1}\frac{\partial
(R\Box R)}{\partial R_{\mu\nu\rho\sigma}}&=&f_{1}\frac{\partial
(g^{ac}g^{bd}R_{abcd}\Box R^{})}{\partial R_{\mu\nu\rho\sigma}}\nonumber\\
&=&f_1g^{ac}g^{bd}\delta^{[\mu}_{a}\delta^{\nu]}_{b}\delta^{[\rho}_{c}\delta^{\sigma]}_{d}\Box
R\nonumber\\
&=&f_1g^{\rho[\mu}g^{\nu]\sigma}\Box R,
\end{eqnarray}
\begin{eqnarray}
f_1\Box\frac{\partial
(R\Box R^{})}{\partial(\Box R_{\mu\nu\rho\sigma})}&=&f_1\Box\frac{\partial
(g^{ac}g^{bd}R\Box R_{abcd}^{})}{\partial(\Box R_{\mu\nu\rho\sigma})}\nonumber\\
&=&f_1\Box(g^{ac}g^{bd}\delta^{[\mu}_{a}\delta^{\nu]}_{b}\delta^{[\rho}_{c}\delta^{\sigma]}_{d}
R)\nonumber\\
&=&f_1g^{\rho[\mu}g^{\nu]\sigma}\Box R.
\end{eqnarray}
Thus, we can summarise as, 
\begin{eqnarray}
\frac{\delta( RF(\Box)R)}{\delta R_{\mu\nu\rho\sigma}}&=&2g^{\rho[\mu}g^{\nu]\sigma}R+f_1g^{\rho[\mu}g^{\nu]\sigma}\Box
R+f_1g^{\rho[\mu}g^{\nu]\sigma}\Box R+\cdots\nonumber\\
&=&2g^{\rho[\mu}g^{\nu]\sigma}(f_0+f_{1}\Box+f_2\Box^2+\cdots)R=2g^{\rho[\mu}g^{\nu]\sigma}F(\Box)R.\nonumber\\
\end{eqnarray}
In similar manner we shall consider the next term, 
\begin{eqnarray}
\frac{\delta (R_{\alpha\beta}F(\Box)R^{\alpha\beta})}{\delta R_{\mu\nu\rho\sigma}}&=&f_0\frac{\partial
(R_{\alpha\beta}R^{\alpha\beta})}{\partial R_{\mu\nu\rho\sigma}}+f_{1}\frac{\partial
(R_{\alpha\beta}\Box R^{\alpha\beta})}{\partial R_{\mu\nu\rho\sigma}}\nonumber\\&+&f_1\Box\frac{\partial
(R_{\alpha\beta}\Box R^{\alpha\beta})}{\partial(\Box R_{\mu\nu\rho\sigma})}+f_{2}\Box^{2}\frac{\partial
(R_{\alpha\beta}\Box ^{2}R^{\alpha\beta})}{\partial (\Box ^{2}R_{\mu\nu\rho\sigma})}+\cdots,\nonumber\\
\end{eqnarray}
again, term by term we have: 
\begin{eqnarray}
f_0\frac{\partial
(R_{\alpha\beta}R^{\alpha\beta})}{\partial R_{\mu\nu\rho\sigma}}&=&f_0\frac{\partial
(g^{\eta\zeta}g^{\gamma\lambda}g^{\alpha\kappa}g^{\beta\omega}R_{\eta\alpha\zeta\beta}R_{\gamma\kappa\lambda\omega})}{\partial
R_{\mu\nu\rho\sigma}}\nonumber\\
&=&f_0g^{\eta\zeta}g^{\gamma\lambda}g^{\alpha\kappa}g^{\beta\omega}\delta^{[\mu}_{[\eta}\delta^{\nu]}_{\alpha]}\delta^{[\rho}_{[\zeta}\delta^{\sigma]}_{\beta]}
R_{\gamma\kappa\lambda\omega}\nonumber\\
&+&f_0g^{\eta\zeta}g^{\gamma\lambda}g^{\alpha\kappa}g^{\beta\omega}\delta^{[\mu}_{[\gamma}\delta^{\nu]}_{\kappa]}\delta^{[\rho}_{[\lambda}\delta^{\sigma]}_{\omega]}
R_{\eta\alpha\zeta\beta}\nonumber\\&=&2f_0g^{\eta\zeta}g^{\alpha\kappa}g^{\beta\omega}\delta^{[\mu}_{\eta}\delta^{\nu]}_{\alpha}\delta^{[\rho}_{\zeta}\delta^{\sigma]}_{\beta}
R_{\kappa\omega}\nonumber\\
&=&2f_0g^{\kappa[\nu}g^{\mu][\rho}g^{\sigma]\omega}R_{\kappa\omega},
\end{eqnarray}
\begin{eqnarray}
f_{1}\frac{\partial
(R_{\alpha\beta}\Box R^{\alpha\beta})}{\partial R_{\mu\nu\rho\sigma}}&=&f_1\frac{\partial
(g^{\eta\zeta}g^{\gamma\lambda}g^{\alpha\kappa}g^{\beta\omega}R_{\eta\alpha\zeta\beta}\Box
R_{\gamma\kappa\lambda\omega})}{\partial
R_{\mu\nu\rho\sigma}}\nonumber\\
&=&f_1\frac{\partial
(g^{\eta\zeta}g^{\alpha\kappa}g^{\beta\omega}R_{\eta\alpha\zeta\beta}\Box
R_{\kappa\omega})}{\partial
R_{\mu\nu\rho\sigma}}\nonumber\\
&=&f_1g^{\eta\zeta}g^{\alpha\kappa}g^{\beta\omega}\delta^{[\mu}_{\eta}\delta^{\nu]}_{\alpha}\delta^{[\rho}_{\zeta}\delta^{\sigma]}_{\beta}\Box
R_{\kappa\omega}\nonumber\\
&=&f_1g^{\kappa[\nu}g^{\mu][\rho}g^{\sigma]\omega}\Box
R_{\kappa\omega},
\end{eqnarray}
\begin{eqnarray}
f_1\Box\frac{\partial
(R_{\alpha\beta}\Box R^{\alpha\beta})}{\partial(\Box R_{\mu\nu\rho\sigma})}&=&f_1\Box\frac{\partial
(g^{\eta\zeta}g^{\gamma\lambda}g^{\alpha\kappa}g^{\beta\omega}R_{\eta\alpha\zeta\beta}\Box
R_{\gamma\kappa\lambda\omega})}{\partial
(\Box R_{\mu\nu\rho\sigma})}\nonumber\\
&=&f_1\Box( g^{\gamma\lambda}g^{\alpha\kappa}g^{\beta\omega}\delta^{[\mu}_{\gamma}\delta^{\nu]}_{\kappa}\delta^{[\rho}_{\lambda}\delta^{\sigma]}_{\omega}R_{\alpha\beta})\nonumber\\
&=&f_1\Box(g^{\alpha[\nu}g^{\mu][\rho}g^{\sigma]{\alpha\beta}}R_{\alpha\beta}),
\end{eqnarray}
thus: 
\begin{eqnarray}
\frac{\delta (R_{\alpha\beta}F(\Box)R^{\alpha\beta})}{\delta R_{\mu\nu\rho\sigma}}&=&2f_0g^{\kappa[\nu}g^{\mu][\rho}g^{\sigma]\omega}R_{\kappa\omega}+2f_1g^{\kappa[\nu}g^{\mu][\rho}g^{\sigma]\omega}\Box
R_{\kappa\omega}+\cdots\nonumber\\
&=&2g^{\kappa[\nu}g^{\mu][\rho}g^{\sigma]\omega}(f_0+f_{1}\Box+f_2\Box^2+\cdots)
R_{\kappa\omega}\nonumber\\
&=&2g^{\kappa[\nu}g^{\mu][\rho}g^{\sigma]\omega}F(\Box)
R_{\kappa\omega}.
\end{eqnarray}
Finally we can consider the Riemann tensor contribution, 
\begin{eqnarray}
\frac{\delta (R_{\alpha\beta\gamma\eta}F(\Box)R^{\alpha\beta\gamma\eta})}{\delta
R_{\mu\nu\rho\sigma}}&=&f_0\frac{\partial
(R_{\alpha\beta\gamma\eta}R^{\alpha\beta\gamma\eta})}{\partial R_{\mu\nu\rho\sigma}}+f_{1}\frac{\partial
(R_{\alpha\beta\gamma\eta}\Box R^{\alpha\beta\gamma\eta})}{\partial R_{\mu\nu\rho\sigma}}\nonumber\\&+&f_1\Box\frac{\partial
(R_{\alpha\beta\gamma\eta}\Box R^{\alpha\beta\gamma\eta})}{\partial(\Box
R_{\mu\nu\rho\sigma})}+f_{2}\Box^{2}\frac{\partial
(R_{\alpha\beta\gamma\eta}\Box ^{2}R^{\alpha\beta\gamma\eta})}{\partial (\Box
^{2}R_{\mu\nu\rho\sigma})}+\cdots,\nonumber\\
\end{eqnarray}
as before, we consider the leading order terms and then generalise the results:
\begin{eqnarray}
f_0\frac{\partial
(R_{\alpha\beta\gamma\eta}R^{\alpha\beta\gamma\eta})}{\partial R_{\mu\nu\rho\sigma}}&=&f_0\frac{\partial
(g^{\alpha\xi}g^{\beta\lambda}g^{\gamma\kappa}g^{\eta\omega}R_{\alpha\beta\gamma\eta}R_{\xi\lambda\kappa\omega})}{\partial
R_{\mu\nu\rho\sigma}}\nonumber\\
&=&f_0g^{\alpha\xi}g^{\beta\lambda}g^{\gamma\kappa}g^{\eta\omega}\delta^{[\mu}_{[\alpha}\delta^{\nu]}_{\beta]}\delta^{[\rho}_{[\gamma}\delta^{\sigma]}_{\eta]}R_{\xi\lambda\kappa\omega}\nonumber\\
&+&f_0g^{\alpha\xi}g^{\beta\lambda}g^{\gamma\kappa}g^{\eta\omega}\delta^{[\mu}_{[\xi}\delta^{\nu]}_{\lambda]}\delta^{[\rho}_{[\kappa}\delta^{\sigma]}_{\omega]}R_{\alpha\beta\gamma\eta}\nonumber\\
&=&2f_0\delta^{[\mu}_{\alpha}\delta^{\nu]}_{\beta}\delta^{[\rho}_{\gamma}\delta^{\sigma]}_{\eta}R^{\alpha\beta\gamma\eta}=2f_0R^{\mu\nu\rho\sigma},
\end{eqnarray}
\begin{eqnarray}
f_{1}\frac{\partial
(R_{\alpha\beta\gamma\eta}\Box R^{\alpha\beta\gamma\eta})}{\partial R_{\mu\nu\rho\sigma}}&=&f_{1}\delta^{[\mu}_{\alpha}\delta^{\nu]}_{\beta}\delta^{[\rho}_{\gamma}\delta^{\sigma]}_{\eta}\Box
R^{\alpha\beta\gamma\eta}=f_1\Box R^{\mu\nu\rho\sigma},
\end{eqnarray}
\begin{eqnarray}
f_1\Box\frac{\partial
(R_{\alpha\beta\gamma\eta}\Box R^{\alpha\beta\gamma\eta})}{\partial(\Box
R_{\mu\nu\rho\sigma})}&=&f_1\Box\frac{\partial
(g^{\alpha\xi}g^{\beta\lambda}g^{\gamma\kappa}g^{\eta\omega}R_{\alpha\beta\gamma\eta}\Box
R_{\xi\lambda\kappa\omega})}{\partial(\Box
R_{\mu\nu\rho\sigma})}\nonumber\\
&=&f_1\Box\frac{\partial
(
R^{\xi\lambda\kappa\omega}\Box
R_{\xi\lambda\kappa\omega})}{\partial(\Box
R_{\mu\nu\rho\sigma})}\nonumber\\
&=&f_1\Box(\delta^{[\mu}_{\xi}\delta^{\nu]}_{\lambda}\delta^{[\rho}_{\kappa}\delta^{\sigma]}_{\omega}
R^{\xi\lambda\kappa\omega})=f_1\Box R^{\mu\nu\rho\sigma}.
\end{eqnarray}
We can conclude that,
\begin{eqnarray}
\frac{\delta (R_{\alpha\beta\gamma\eta}F(\Box)R^{\alpha\beta\gamma\eta})}{\delta
R_{\mu\nu\rho\sigma}}&=&2f_0R^{\mu\nu\rho\sigma}+f_1\Box R^{\mu\nu\rho\sigma}+f_1\Box
R^{\mu\nu\rho\sigma}+\cdots\nonumber\\
&=&2(f_0+f_{1}\Box+f_2\Box^2+\cdots)R^{\mu\nu\rho\sigma}=2F(\Box)R^{\mu\nu\rho\sigma}.\nonumber\\
\end{eqnarray}

\chapter{Conserved current for Einstein-Hilbert gravity}\label{ehcons}
Given the EH action to be of the form, 
\begin{equation}
S_{EH}=\frac{M_P^2}{2}\int d^{4}x\sqrt{-g} R.
\end{equation}
we can imply the variation principle infinitesimally by writing,
\begin{eqnarray}\label{ehvar}
\delta_{\xi}S_{EH}=\frac{M_P^2}{2}\int d^{4}x\delta_{\xi}(\sqrt{g} R)
&=&\frac{M_P^2}{2}\int d^{4}x\sqrt{g}\Big(G_{\mu\nu}\delta_{\xi}g^{\mu\nu}+g^{\mu\nu}\delta_{\xi}(R_{\mu\nu})\Big)
\nonumber\\&=&\frac{M_P^2}{2}\int d^{4}x\sqrt{g}\nabla_\alpha(\xi^\alpha
R)=0,
\end{eqnarray}
where $G_{\mu\nu}$ is the Einstein tensor and given by $G_{\mu\nu}=R_{\mu\nu}-\frac{1}{2}g_{\mu\nu}R$.
The term involving the Einstein tensor can be expanded further as, 
\begin{eqnarray}\label{term1}
G_{\mu\nu}\delta_{\xi}g^{\mu\nu}=G_{\mu\nu}(\nabla^\mu\xi^\nu+\nabla^\nu\xi^\mu)=2G_{\mu\nu}\nabla^\mu\xi^\nu=\nabla_\mu(-2R^{\mu}_{\nu}+\delta^{\mu}_{\nu}R)\xi^\nu,
\end{eqnarray}
where we used Eq. (\ref{metricvar}) and performed integration by parts. Then
we move on to the next term and expand it as, 
\begin{equation}\label{term2}
g^{\mu\nu}\delta_{\xi} R_{\mu\nu}=(\nabla^{\mu}\nabla^{\nu}-g^{\mu\nu}\Box)\delta_{\xi}
g_{\mu\nu}=\nabla_{\lambda}\Big((g^{\lambda\alpha}g^{\nu\beta}-g^{\lambda\nu}g^{\alpha\beta})\nabla_{\nu}(\nabla_\alpha\xi_\beta+\nabla_\beta\xi_\alpha)\Big),
\end{equation}
by substituting Eq's. (\ref{term1}) and (\ref{term2}) into (\ref{ehvar})
we obtain, 
\begin{eqnarray}
\delta_{\xi}S_{EH}
=\frac{M_P^2}{2}\int d^{4}x\sqrt{-g}\nabla_\mu\Big(-2R^{\mu}_{\nu}\xi^\nu+(g^{\mu\alpha}g^{\nu\beta}-g^{\mu\nu}g^{\alpha\beta})\nabla_{\nu}(\nabla_\alpha\xi_\beta+\nabla_\beta\xi_\alpha)\Big)=0,\nonumber\\
\end{eqnarray}
and hence for any vector field $\xi^{\mu}$ one obtains the conserved N\"oether
current, 
\begin{equation}\label{noethercurrent}
J^{\mu}(\xi)=R^{\mu}_{\nu}\xi^\nu+\frac{1}{2}(g^{\mu\alpha}g^{\nu\beta}-g^{\mu\nu}g^{\alpha\beta})\nabla_{\nu}(\nabla_\alpha\xi_\beta+\nabla_\beta\xi_\alpha)\equiv\nabla_\nu(\nabla^{[\mu}\xi^{\nu]}).
\end{equation}
\chapter{Generalised Komar current}\label{generalisedkomar}
It can be shown that the N\"oether current that was obtained in Eq. (\ref{noethercurrent})
is identical to generalised Komar current via
\begin{eqnarray}
J^{\mu}(\xi)&=&\frac{1}{2}\nabla_\nu(\nabla^{\mu}\xi^{\nu}-\nabla^{\nu}\xi^{\mu})=\nabla_\nu\nabla^{\mu}\xi^{\nu}-\frac{1}{2}\nabla_\nu(\nabla^{\nu}\xi^{\mu}+\nabla^{\mu}\xi^{\nu})\nonumber\\
&=&[\nabla^{\nu},\nabla^\mu]\xi_\nu+\nabla^\mu(\nabla_\nu\xi^\nu)-\frac{1}{2}\nabla_\nu(\nabla^{\nu}\xi^{\mu}+\nabla^{\mu}\xi^{\nu})\nonumber\\
&=&R^{\mu}_{\nu}\xi^\nu+\frac{1}{2}(g^{\mu\alpha}g^{\nu\beta}-g^{\mu\nu}g^{\alpha\beta})\nabla_{\nu}(\nabla_\alpha\xi_\beta+\nabla_\nu\xi_\beta),
\end{eqnarray}
where we\ used: $[\nabla^{\nu},\nabla^\mu]\xi_\nu=R^{\mu}_{\lambda\mu\nu}\xi^{\lambda}=R_{\lambda\nu}\xi^{\lambda}$.
\chapter{Komar integrals in Boyer-Linquist coordinate}\label{boyerlinqderivation}
In Boyer-Linquist coordinate the kerr metric is given by,   
\begin{footnotesize}\begin{eqnarray}
g_{\mu\nu}=\left(
\begin{array}{cccc}
 \frac{2 M r}{r^2+a^2 \cos ^2(\theta )}-1 & 0 & 0 & -\frac{2 a M r \sin ^2(\theta
)}{r^2+a^2 \cos ^2(\theta )} \\
 0 & \frac{r^2+a^2 \cos ^2(\theta )}{a^2+r^2-2 M r} & 0 & 0 \\
 0 & 0 & r^2+a^2 \cos ^2(\theta ) & 0 \\
 -\frac{2 a M r \sin ^2(\theta )}{r^2+a^2 \cos ^2(\theta )} & 0 & 0 & \frac{\sin
^2(\theta ) \left(\left(a^2+r^2\right)^2-a^2 \left(a^2+r^2-2 M r\right) \sin
^2(\theta )\right)}{r^2+a^2 \cos ^2(\theta )} \\
\end{array}
\right)\nonumber\\
\end{eqnarray}\end{footnotesize}
Given, 
\begin{equation}
n^1=(1,0,0,0), \qquad n^2=(0,1,0,0).
\end{equation}
the only surviving components of normal vectors would be, 
\begin{equation}
n^{1}_{[\alpha}n^{2}_{\beta]}=n^{1}_{[t}n^{2}_{r]}=\frac{1}{2}(n^{1}_{t}n^{2}_{r}-n^{1}_{r}n^{2}_{t})=\frac{1}{2},
\end{equation}
\begin{equation}
n^{1}_{[\alpha}n^{2}_{\beta]}=n^{1}_{[r}n^{2}_{t]}=\frac{1}{2}(n^{1}_{r}n^{2}_{t}-n^{1}_{t}n^{2}_{r})=-\frac{1}{2}.
\end{equation}
We now want to calculate the Komar integrals:\ 
\begin{equation}
M=-\frac{1}{8\pi}\oint_{\mathcal{H}}\nabla^{\alpha}t^{\beta}ds_{\alpha\beta},
\end{equation}
lets us take:\begin{eqnarray}\nabla^{\alpha}t^{\beta}ds_{\alpha\beta}&=&\sqrt{-g}\nabla^{\alpha}t^{\beta}n^{1}_{[\alpha}n^{2}_{\beta]}d\theta
d\phi\nonumber\\
&=&\sqrt{-g}(g^{\alpha\lambda}\nabla_{\lambda}t^{\beta})n^{1}_{[\alpha}n^{2}_{\beta]}d\theta
d\phi\nonumber\\
&=&\sqrt{-g}g^{\alpha\lambda}(\partial_\lambda t^{\beta}+\Gamma^{\beta}_{\lambda\rho}t^{\rho})n^{1}_{[\alpha}n^{2}_{\beta]}d\theta
d\phi\nonumber\\
&=&\sqrt{-g}g^{\alpha\lambda}\Gamma^{\beta}_{\lambda\rho}t^{\rho}n^{1}_{[\alpha}n^{2}_{\beta]}d\theta
d\phi\nonumber\\
&=&\sqrt{-g}\Big(g^{t\lambda}\Gamma^{r}_{\lambda t}n^{1}_{[t}n^{2}_{r]}
+g^{r\lambda}\Gamma^{t}_{\lambda t}n^{1}_{[r}n^{2}_{t]}\Big)d\theta
d\phi\nonumber\\
&=&\frac{1}{2}\sqrt{-g}\Big(g^{t\lambda}\Gamma^{r}_{\lambda t}-g^{r\lambda}\Gamma^{t}_{\lambda
t}\Big)d\theta
d\phi\nonumber\\
&=&\frac{1}{2}\sqrt{-g}\Big(g^{tt}\Gamma^{r}_{t t}+g^{t\phi}\Gamma^{r}_{\phi
t}-g^{rr}\Gamma^{t}_{r
t}\Big)d\theta
d\phi.\nonumber\\
\end{eqnarray}
We have: 
\begin{equation}
g^{tt}\Gamma^{r}_{t t}+g^{t\phi}\Gamma^{r}_{\phi
t}-g^{rr}\Gamma^{t}_{r
t}=\frac{8 m \left(a^2+r^2\right) \left(a^2 \cos (2 \theta )+a^2-2 r^2\right)}{\left(a^2
\cos (2 \theta )+a^2+2 r^2\right)^3},
\end{equation}
\begin{equation}
\sqrt{-g}=\frac{1}{2} \sin (\theta ) \left(a^2 \cos (2 \theta )+a^2+2 r^2\right).
\end{equation}
Thus, 
\begin{eqnarray}
M&=&-\frac{1}{8\pi}\oint_{\mathcal{H}}\nabla^{\alpha}t^{\beta}ds_{\alpha\beta}\nonumber\\
&=&-\frac{1}{8\pi}\int^{2\pi}_{0} d\phi\int^{\pi}_{0} d\theta\nonumber\\&&\Bigg(\frac{1}{2}
\sin (\theta ) \left(a^2 \cos (2 \theta )+a^2+2 r^2\right)\frac{8 m \left(a^2+r^2\right)
\left(a^2 \cos (2 \theta )+a^2-2 r^2\right)}{\left(a^2
\cos (2 \theta )+a^2+2 r^2\right)^3}\Bigg)=m.\nonumber\\ 
\end{eqnarray} 
Now let us look at the angular momentum:
\begin{equation}
J=\frac{1}{16\pi}\oint_{\mathcal{H}}\nabla^{\alpha}\phi^{\beta}ds_{\alpha\beta},
\end{equation}
\begin{eqnarray}\nabla^{\alpha}\phi^{\beta}ds_{\alpha\beta}&=&\sqrt{-g}\nabla^{\alpha}\phi^{\beta}n^{1}_{[\alpha}n^{2}_{\beta]}d\theta
d\phi\nonumber\\
&=&\sqrt{-g}(g^{\alpha\lambda}\nabla_{\lambda}\phi^{\beta})n^{1}_{[\alpha}n^{2}_{\beta]}d\theta
d\phi\nonumber\\
&=&\sqrt{-g}g^{\alpha\lambda}(\partial_\lambda \phi^{\beta}+\Gamma^{\beta}_{\lambda\rho}\phi^{\rho})n^{1}_{[\alpha}n^{2}_{\beta]}d\theta
d\phi\nonumber\\
&=&\sqrt{-g}g^{\alpha\lambda}\Gamma^{\beta}_{\lambda\rho}\phi^{\rho}n^{1}_{[\alpha}n^{2}_{\beta]}d\theta
d\phi\nonumber\\
&=&\sqrt{-g}\Big(g^{t\lambda}\Gamma^{r}_{\lambda \phi}n^{1}_{[t}n^{2}_{r]}
+g^{r\lambda}\Gamma^{t}_{\lambda \phi}n^{1}_{[r}n^{2}_{t]}\Big)d\theta
d\phi\nonumber\\
&=&\frac{1}{2}\sqrt{-g}\Big(g^{t\lambda}\Gamma^{r}_{\lambda \phi}-g^{r\lambda}\Gamma^{t}_{\lambda
\phi}\Big)d\theta
d\phi\nonumber\\
&=&\frac{1}{2}\sqrt{-g}\Big(g^{tt}\Gamma^{r}_{t \phi}+g^{t \phi}\Gamma^{r}_{
\phi \phi}-g^{rr}\Gamma^{t}_{r
\phi}\Big)d\theta
d\phi,
\end{eqnarray}
we have: 
\begin{eqnarray}
&&g^{tt}\Gamma^{r}_{t \phi}+g^{t \phi}\Gamma^{r}_{
\phi \phi}-g^{rr}\Gamma^{t}_{r
\phi}\nonumber\\&=&-\frac{8 a m \sin ^2(\theta ) \left(a^4-3 a^2 r^2+a^2
(a-r) (a+r) \cos (2 \theta )-6 r^4\right)}{\left(a^2 \cos (2 \theta )+a^2+2
r^2\right)^3}.\nonumber\\
\end{eqnarray}
Hence, 
\begin{eqnarray}
J&=&\frac{1}{16\pi}\oint_{\mathcal{H}}\nabla^{\alpha}\phi^{\beta}ds_{\alpha\beta}\nonumber\\
&=&\frac{1}{16\pi}\int^{2\pi}_{0} d\phi\int^{\pi}_{0} d\theta\nonumber\\&&\Bigg(\frac{1}{2}
\sin (\theta ) \left(a^2 \cos (2 \theta )+a^2+2 r^2\right)\nonumber\\&\times&\frac{-8
a m \sin ^2(\theta ) \left(a^4-3 a^2 r^2+a^2 (a-r) (a+r) \cos (2 \theta )-6
r^4\right)}{\left(a^2 \cos (2 \theta )+a^2+2 r^2\right)^3}\Bigg)=ma.\nonumber\\
\end{eqnarray}
Note: $\xi^{\alpha}=t^{\alpha}+\Omega_H\phi^{\alpha}$.
\chapter{ $f(R)$ gravity conserved current}\label{frvariationconserved}
Variation of the $f(R)$ action given in (\ref{frgravity}) would follow as,

\begin{eqnarray}
\delta_\xi I&=&\int d^{4}x\delta_\xi\Big(\sqrt{-g}f(R)\Big)\nonumber\\
&=&\int d^{4}x\delta_{\xi}\Big(\delta_{\xi}(\sqrt{-g})f(R)+\sqrt{-g}\delta_{\xi}(f(R))\Big)\nonumber\\
&=&\int d^{4}x\sqrt{-g}\Big([f'(R)R_{\mu\nu}-\frac{1}{2}g_{\mu\nu}f(R)]\delta_{\xi}g^{\mu\nu}+f'(R)g^{\mu\nu}\delta_{\xi}(R_{\mu\nu})\Big)\nonumber\\
&=&\int d^{4}x\sqrt{-g} \nabla_{\alpha}\Big(\xi^{\alpha} f(R)\Big).
\end{eqnarray}
Now let $\mathcal{G}_{\mu\nu}=f'(R)R_{\mu\nu}-\frac{1}{2}g_{\mu\nu}f(R)$,
then:
\begin{eqnarray}
\mathcal{G}_{\mu\nu}\delta_{\xi}g^{\mu\nu}&=&\mathcal{G}_{\mu\nu}\pounds_{\xi}g^{\mu\nu}=\mathcal{G}_{\mu\nu}(\nabla^\mu\xi^\nu+\nabla^\nu\xi^\mu)=2\mathcal{G}_{\mu\nu}\nabla^\mu\xi^\nu\nonumber\\
&=&-2\nabla_\mu
\mathcal{G}^{\mu\nu}\xi_\nu=-2\nabla_\mu
\mathcal{G}^{\mu}_{\nu}\xi^\nu=\nabla_\mu(-2f'(R)R^{\mu}_{\nu}+\delta^{\mu}_{\nu}f(R)R)\xi^\nu.\nonumber\\
\end{eqnarray}
And,
\begin{eqnarray}
f'(R)g^{\mu\nu}\delta_{\xi} R_{\mu\nu}&=&f'(R)(\nabla_{\lambda}(g^{\mu\nu}\delta_{\xi}\Gamma^{\lambda}_{\mu\nu})-\nabla_{\nu}(g^{\mu\nu}\delta_{\xi}\Gamma^{\lambda}_{\lambda\mu}))\nonumber\\
&=&f'(R)\nabla_{\lambda}\big(g^{\mu\nu}\delta_{\xi}\Gamma^{\lambda}_{\mu\nu}-g^{\mu\lambda}\delta_{\xi}\Gamma^{\nu}_{\nu\mu}\big)\nonumber\\
&=&f'(R)(\nabla^{\mu}\nabla^{\nu}-g^{\mu\nu}\Box)\delta_{\xi} g_{\mu\nu}\nonumber\\
&=&f'(R)(g^{\mu\alpha}g^{\nu\beta}-g^{\mu\nu}g^{\alpha\beta})\nabla_{\mu}\nabla_{\nu}\delta_{\xi}
g_{\alpha\beta}\nonumber\\
&=&f'(R)\nabla_{\lambda}\Big((g^{\lambda\alpha}g^{\nu\beta}-g^{\lambda\nu}g^{\alpha\beta})\nabla_{\nu}\delta_{\xi}
g_{\alpha\beta}\Big)\nonumber\\
&=&f'(R)\nabla_{\lambda}\Big((g^{\lambda\alpha}g^{\nu\beta}-g^{\lambda\nu}g^{\alpha\beta})\nabla_{\nu}(\nabla_\alpha\xi_\beta+\nabla_\beta\xi_\alpha)\Big)
.\nonumber\\\end{eqnarray}
Thus,
\begin{eqnarray}
\delta_{\xi}I&=&\int d^{4}x\sqrt{g}\Big(\mathcal{G}_{\mu\nu}\delta_{\xi}g^{\mu\nu}+g^{\mu\nu}\delta_{\xi}(R_{\mu\nu})\Big)\nonumber\\
&=&\int d^{4}x\sqrt{g}\Big(\nabla_\mu(-2f'(R)R^{\mu}_{\nu}\xi^\nu+\delta^{\mu}_{\nu}f(R)R\xi^\nu)\nonumber\\&+&f'(R)\nabla_{\mu}\Big((g^{\mu\alpha}g^{\nu\beta}-g^{\mu\nu}g^{\alpha\beta})\nabla_{\nu}(\nabla_\alpha\xi_\beta+\nabla_\beta\xi_\alpha)\Big)\Big)
\nonumber\\
&=&\int d^{4}x\sqrt{g}\nabla_\alpha(\xi^\alpha f(R))=0.
\end{eqnarray}
Which reduces to 
\begin{eqnarray}
\delta_{\xi}I_{EH}
=\int d^{4}x\sqrt{g}\nabla_\mu\Big(-2f'(R)R^{\mu}_{\nu}\xi^\nu+f'(R)(g^{\mu\alpha}g^{\nu\beta}-g^{\mu\nu}g^{\alpha\beta})\nabla_{\nu}(\nabla_\alpha\xi_\beta+\nabla_\beta\xi_\alpha)\Big)=0.\nonumber\\
\end{eqnarray}
Thus the conserved Noether current is:
\begin{equation}
J^{\mu}(\xi)=f'(R)R^{\mu}_{\nu}\xi^\nu+\frac{1}{2}f'(R)(g^{\mu\alpha}g^{\nu\beta}-g^{\mu\nu}g^{\alpha\beta})\nabla_{\nu}(\nabla_\alpha\xi_\beta+\nabla_\beta\xi_\alpha)=f'(R)\nabla_\nu(\nabla^{[\mu}\xi^{\nu]}).
\end{equation}
Moreover, 
\begin{eqnarray}
J^{\mu}(\xi)&=&\frac{1}{2}f'(R)\nabla_\nu(\nabla^{\mu}\xi^{\nu}-\nabla^{\nu}\xi^{\mu})=f'(R)\nabla_\nu\nabla^{\mu}\xi^{\nu}-\frac{1}{2}f'(R)\nabla_\nu(\nabla^{\nu}\xi^{\mu}+\nabla^{\mu}\xi^{\nu})\nonumber\\
&=&f'(R)[\nabla^{\nu},\nabla^\mu]\xi_\nu+f'(R)\nabla^\mu(\nabla_\nu\xi^\nu)-\frac{1}{2}f'(R)\nabla_\nu(\nabla^{\nu}\xi^{\mu}+\nabla^{\mu}\xi^{\nu})\nonumber\\
&=&f'(R)R^{\mu}_{\nu}\xi^\nu+\frac{1}{2}f'(R)(g^{\mu\alpha}g^{\nu\beta}-g^{\mu\nu}g^{\alpha\beta})\nabla_{\nu}(\nabla_\alpha\xi_\beta+\nabla_\nu\xi_\beta).
\end{eqnarray}
Now the Komar Integrals modified by,
\begin{equation}
M=-\frac{f'(R)}{8\pi}\oint_{\mathcal{H}}\nabla^{\alpha}t^{\alpha}ds_{\alpha\beta}=f'(R)m,
\end{equation}
\begin{equation}
J=\frac{f'(R)}{16\pi}\oint_{\mathcal{H}}\nabla^{\alpha}\phi^{\alpha}ds_{\alpha\beta}=f'(R)ma.
\end{equation}
As appeared in (\ref{conserved1}) and (\ref{conserved2}).

\end{appendices}

\backmatter

\nocite{*} 

\renewcommand{\bibname}{References}




\end{document}